\documentclass[12pt]{article}
\usepackage{epsfig}

\newcommand\pubnumber{SLAC--PUB--11687 \\ LBNL-59634}
\newcommand\pubdate{February, 2006}
\newcommand\hepnumber{hep-ph/0602187}

\def\LBL{Department of Physics and Lawrence Berkeley Laboratory \\
     University of California, Berkeley, California 94720 USA}
\def\lblack{\footnote{Work supported by the US Department of Energy,
                     contract DE--AC02--05CH11231.}}
\def\KAVLI{Kavli Institute for Particle Astrophysics and Cosmology\\
    Stanford University, Stanford, California 94309 USA}
\def\SLAC{Stanford Linear Accelerator Center\\
    Stanford University, Stanford, California 94309 USA}
\def\doeack{\footnote{Work supported by the US Department of Energy,
                     contract DE--AC02--76SF00515.}}

\def\Title#1{\begin{center} {\Large #1 } \end{center}}
\def\Author#1{\begin{center}{ \sc #1} \end{center}}
\def\Address#1{\begin{center}{ \it #1} \end{center}}

\def\submit#1{\begin{center}Submitted to {\sl #1} \end{center}}
\newcommand\pubblock{\rightline{\begin{tabular}{l} \pubnumber\\
         \pubdate \\ \hepnumber \end{tabular}}}
\newenvironment{Abstract}{\begin{quotation} \begin{center}
                       ABSTRACT
     \end{center}\bigskip  }{\end{quotation}}

\def\submit#1{\begin{center}Submitted to {\sl #1} \end{center}}
\def\Acknowledgements{\bigskip  \bigskip \begin{center} \begin{large}
             \bf ACKNOWLEDGEMENTS \end{large}\end{center}}

\def\beq{\begin{equation}}
\def\eeq#1{\label{#1}\end{equation}}
\def\eeqn{\end{equation}}

\newenvironment{Eqnarray}%
   {\arraycolsep 0.14em\begin{eqnarray}}{\end{eqnarray}}
\def\beqa{\begin{Eqnarray}}
\def\eeqa#1{\label{#1}\end{Eqnarray}}
\def\eeqan{\end{Eqnarray}}

\def\leqn#1{(\ref{#1})}

\let\bar=\overbar

\def\etal{{\it et al.}}

\def\VEV#1{\left\langle{ #1} \right\rangle}
\def\bra#1{\left\langle{ #1} \right|}
\def\ket#1{\left| {#1} \right\rangle}

\def\lsim{\mathrel{\raise.3ex\hbox{$<$\kern-.75em\lower1ex\hbox{$\sim$}}}}
\def\gsim{\mathrel{\raise.3ex\hbox{$>$\kern-.75em\lower1ex\hbox{$\sim$}}}}

\def\L{{\cal L}}

\def\half{\frac{1}{2}}

\def\del{\partial}
\def\Dslash{\not{\hbox{\kern-4pt $D$}}}
\def\dslash{\not{\hbox{\kern-2pt $\del$}}}

\def\Pl{{\mbox{\scriptsize Pl}}}

\def\ee{e^+e^-}

\def\mz{m_Z}

\def\msb{{\bar{\scriptsize M \kern -1pt S}}}

\def\drb{{\bar{\scriptsize D \kern -1pt R}}}

\def\s#1{\widetilde{#1}}

\makeatletter
\def\section{\@startsection{section}{0}{\z@}{5.5ex plus .5ex minus
 1.5ex}{2.3ex plus .2ex}{\large\bf}}
\def\subsection{\@startsection{subsection}{1}{\z@}{3.5ex plus .5ex minus
 1.5ex}{1.3ex plus .2ex}{\normalsize\bf}}
\def\subsubsection{\@startsection{subsubsection}{2}{\z@}{-3.5ex plus
-1ex minus  -.2ex}{2.3ex plus .2ex}{\normalsize\sl}}

\renewcommand{\@makecaption}[2]{%
   \vskip 10pt
   \setbox\@tempboxa\hbox{\small #1: #2}
   \ifdim \wd\@tempboxa >\hsize     
       \small #1: #2\par          
     \else                        
       \hbox to\hsize{\hfil\box\@tempboxa\hfil}
   \fi}

 \def\citenum#1{{\def\@cite##1##2{##1}\cite{#1}}}
 
\newcount\@tempcntc
\def\@citex[#1]#2{\if@filesw\immediate\write\@auxout{\string\citation{#2}}\fi
  \@tempcnta\z@\@tempcntb\m@ne\def\@citea{}\@cite{\@for\@citeb:=#2\do
    {\@ifundefined
       {b@\@citeb}{\@citeo\@tempcntb\m@ne\@citea\def\@citea{,}{\bf ?}\@warning
       {Citation `\@citeb' on page \thepage \space undefined}}%
    {\setbox\z@\hbox{\global\@tempcntc0\csname b@\@citeb\endcsname\relax}%
     \ifnum\@tempcntc=\z@ \@citeo\@tempcntb\m@ne
       \@citea\def\@citea{,}\hbox{\csname b@\@citeb\endcsname}%
     \else
      \advance\@tempcntb\@ne
      \ifnum\@tempcntb=\@tempcntc
      \else\advance\@tempcntb\m@ne\@citeo
      \@tempcnta\@tempcntc\@tempcntb\@tempcntc\fi\fi}}\@citeo}{#1}}
\def\@citeo{\ifnum\@tempcnta>\@tempcntb\else\@citea\def\@citea{,}%
  \ifnum\@tempcnta=\@tempcntb\the\@tempcnta\else
  {\advance\@tempcnta\@ne\ifnum\@tempcnta=\@tempcntb \else\def\@citea{--}\fi
    \advance\@tempcnta\m@ne\the\@tempcnta\@citea\the\@tempcntb}\fi\fi}
\makeatother

\begin{document}
\begin{titlepage}
\pubblock

\vfill
\Title{Determination of Dark Matter Properties\\ at High-Energy Colliders}
\vfill
\Author{Edward A. Baltz\doeack}
\Address{\KAVLI}
\Author{Marco Battaglia\lblack}
\Address{\LBL}
\Author{Michael E. Peskin and Tommer Wizansky$^1$}
\Address{\SLAC}
\vfill
\begin{Abstract}
If the cosmic dark matter consists of weakly-interacting massive particles,
these particles should be produced in reactions at the next generation of
high-energy accelerators.  Measurements at these accelerators can then be used
to determine the microscopic properties of the dark matter.  From this, we can
predict the cosmic density, the annihilation cross sections, and the cross
sections relevant to direct detection.  In this paper, we present studies in
supersymmetry models with neutralino dark matter that give quantitative
estimates of the accuracy that can be expected.  We show that these are well
matched to the requirements of anticipated astrophysical observations of dark
matter.  The capabilities of the proposed International Linear Collider (ILC)
are expected to play a particularly important role in this study.
\end{Abstract}
\newpage
\tableofcontents
\vfill
\submit{Physical Review {\bf D}}
\vfill
\end{titlepage}
\newpage
\def\thefootnote{\fnsymbol{footnote}}
\setcounter{footnote}{0}

\section{Introduction}

It is now well established that roughly 20\% of the energy density of the
universe consists of neutral weakly interacting non-baryonic matter, `dark
matter'~\cite{cosmoreview,JKG,Bertone}.  The picture of structure formation by
the growth of fluctuations in weakly interacting matter explains the elements
of structure in the universe from the fluctuations in the cosmic microwave
background down almost to the scale of galaxies. However, many mysteries
remain.  From the viewpoint of particle physics, we have no idea what dark
matter is made of.  The possibilities range in mass from axions (mass $10^{-5}$
eV) to primordial black holes (mass $10^{-5} M_\odot$).  From the viewpoint of
astrophysics, it is still controversial how dark matter is distributed in
galaxies and even whether the picture of weakly interacting dark matter
adequately explains the structure of galaxies.  Furthermore, a much larger
range of masses is allowed, bounded only by the quantum limit ($10^{-22}$ eV)
for bosons \cite{HuFuzzy} and the discreteness limit ($10^3\,M_\odot$) above
which galactic globular clusters would be disrupted by the dark matter
``particles'' \cite{MooreGlobular}.

To improve this situation, we need more experimental measurements.  
Unfortunately, precisely because dark matter is weakly interacting and 
elusive, any single new piece of data has multiple interpretations.
If we improve the upper limit on the direct detection of dark matter, does
this mean that the microscopic cross section is small or that our detector
is located in a trough of the galactic dark matter distribution?  If 
we see a signal of dark matter annihilation at the center of the galaxy, does 
this measure the annihilation cross section, or does it measure the 
clustering of dark matter associated with the galaxy's formation?  If we 
observe a massive weakly-interacting elementary particle in a high-energy
physics experiment, can we demonstrate that this particle is a constituent of
dark matter?  For any single question, there are no definite answers.  It is
only by carrying out a program of experiments that include both particle
physics and astrophysics measurements and marshalling all of the
resulting information that we could reach definite conclusions.

It is our belief that the role that particle physics measurements will 
play in this program has been underestimated in the literature.  Much of the
particle astrophysics literature on dark matter particle detection is written
as if we should not expect strong constraints from particle physics on the 
microscopic cross sections.  This leaves a confusing situation, in which 
one needs to determine the basic properties of dark matter in the face of 
large uncertainties in its galactic distribution, or vice versa.  To make
matters worse, many of these discussions use models of dark matter with an 
artificially small number of parameters, precisely to limit the possible
modes of variation in the microscopic properties of dark matter.

We see good reason to be more optimistic.  Among the many possible models
of dark matter, we believe that there are strong reasons to concentrate  
on the particular class of models in which the dark matter particle is a 
massive neutral particle with a mass of the order of 100 GeV.  We will 
refer to the particles in this special class of models as WIMPs.  We will
define the class more precisely in Section 2.  In this class of models, 
the WIMP should be discovered in high-energy physics experiments 
just a few years from now at the CERN
Large Hadron Collider (LHC).  Over the next ten years, the LHC experiments
and experiments at the planned International Linear Collider (ILC) will make
precision measurements that will constrain the properties of the 
WIMP.  This in turn will lead to very precise
determinations of the microscopic cross sections that enter the dark 
matter abundance and detection rates.

In the best situation, these experiments could apply to  the study of dark 
matter the strategy used in more  familiar  areas of astrophysics. 
 When we study stars and the 
visible components of galaxies and clusters, every measurement is 
determined by a convolution of microscopic cross sections with 
astrophysical densities.  We go into the laboratory to measure atomic and
nuclear transition rates and then apply this information to learn the 
species and conditions in the object we are observing.  We might hope that the
LHC and ILC experiments on dark matter would provide the basic data for 
this type of analysis of experiments that observe galactic dark matter.

Our main goal in this paper is to demonstrate that this objective 
can be achieved.  To show this, we need to realistically evaluate the 
power of the 
LHC and ILC experiments to determine the cross sections of direct 
astrophysical interest.  Dark matter particles are invisible to 
high-energy physics experiments, and so such determinations are necessarily 
indirect.  On the other hand, the large number of specific and precise
measurements that can be expected will allow us to determine the model
of which the WIMP is a part.  We will show that this information 
indeed constrains the elusive WIMP cross sections, and we will estimate 
the accuracy with which those cross sections can be predicted from the
LHC and ILC data.

One might describe the actual calculations in our analysis as merely 
a simple exercise in error propagation.  This description is correct,
except that the exercise is not simple.  The WIMP cross sections have a 
complicated dependence on the underlying spectrum parameters, and many of
those parameters cannot realistically be measured in high-energy physics
experiments.  We address these problems by using as our starting point
the results of 
detailed high-energy physics simulations and applying to these 
results a statistical method that is robust with respect to incomplete
information.  We believe that the results that we are presenting will 
be of interest to high-energy physicists who are planning experiments at
future accelerators and as well as to astrophysicists looking into the
future of dark matter detection experiments.

Dark matter measurements also have the potential to feed information back 
to particle physics.  Today, the very existence of
 dark matter is the strongest piece of evidence for physics beyond the 
Standard Model.  The cosmic density of dark matter is already quite
well known.  This density has been determined to 6\% accuracy by the 
cosmological data, especially by the 
WMAP measurement of the cosmic microwave background (CMB) \cite{WMAP}.
Later in this decade, the Planck satellite should improve this determination
to the 0.5\% level~\cite{BondE}.  If it should become attractive to assume
that a WIMP observed at the LHC accounts for all of the dark matter, 
these measurements can be used to give precision determinations of some
particle physics parameters.  Over time, this assumption could be tested
with higher-precision high-energy physics experiments.  In some cases, 
measurements of direct signals of astrophysical dark matter could also 
provide interesting constraints on particle physics.  We will give some 
illustrations 
of this interplay between astrophysical and microscopic constraints
in the context of our examples.

Here is an outline of our analysis:  In Section 2, we will specify 
the WIMP class of dark matter models, and we will review the set of 
WIMP properties that should be determined by microscopic experiments.
To perform specific calculations of the ability of LHC and ILC to 
determine these cross sections, we will study in detail the case of 
supersymmetry models in which the dark matter particle is the lightest
neutralino.  In Section 3, we will review the various physical mechanisms
that can be responsible for setting the dark matter relic density in 
these models, and we will choose four benchmark models for detailed
study.  We will also explain how we evaluate the model uncertainty in 
the predictions from the collider measurements, using an 
exploration of the parameter space by Markov Chain Monte Carlo techniques.
In Sections 4-7, we present the results of our Monte Carlo study for each 
of the benchmark points. In Section 8, we will present some general 
observations on the determination of dark matter annihilation cross sections.
Finally, in 
Section 9, we review the results of our study and present the general
conclusions that we draw from them.

Our calculations make heavy use of the ISAJET~\cite{ISAJET} and 
DarkSUSY computer programs~\cite{DarkSUSY}
to evaluate the neutralino dark matter properties from the underlying
supersymmetry parameters. We thank the authors for making these 
useful tools available.

The determination of the cosmic dark matter density from collider
data has also been studied recently by Allanach, Belanger, Boudjema,
and Pukhov~\cite{ABBP} and by Nojiri, Polesello, and Tovey~\cite{NPT}.
We will compare our strategies and results in Sections 3 and 4.  
A  first version of this analysis has been presented in 
\cite{BPLCWS}; this work supersedes the results presented in that 
paper.

\section{Preliminaries}

Before beginning our study of specific WIMP models of dark matter, 
we would like to review some general aspects of dark matter and its
observation.  In this section, we will define what we mean by the WIMP
scenario, give an overview of how WIMPs can be studied at high-energy
colliders, and review the set of cross sections needed to analyze
WIMP detection.  All of the material in this section is review, 
intended to introduce the questions that we will answer in the specific
model analyses of Section 4--7.

\subsection{Why the WIMP model of dark matter deserves special attention}

Among the particle physics candidates for dark matter, many share a set of
common properties.  They are heavy, neutral, weakly-interacting particles
with interaction cross sections nevertheless large enough that they were
in thermal equilibrium for some period in the early universe.  It is these
particles that we refer to collectively as WIMPs.

The assumption of thermal equilibrium allows a precise prediction of the 
cosmic density of the WIMP.  We must of course also assume that standard
cosmology can be extrapolated back to this era.  Given these assumptions,
it is straightforward to integrate the Boltzmann equation for the WIMP
density through the time at which the WIMP drops out of equilibrium.  The 
resulting density is the `relic density' of the WIMP.  To 10\% accuracy,
the ratio of this relic density to the closure density is given 
by the formula~\cite{TurnerSch}
\beq
\Omega_{\chi}h^2 = {s_0\over \rho_c/h^2} \left( {45\over \pi g_*}\right)^{1/2}
          { x_f \over m_\Pl }{1\over \VEV{\sigma v}}
\eeq{Omegachi}
where $s_0$ is the current entropy density of the universe, $\rho_c$ is the
critical density, $h$ is the (scaled) Hubble constant, $g_*$ is the number of 
relativistic degrees of freedom at the time that the dark matter particle
goes out of thermal equilibrium, $m_\Pl$ is the Planck mass, $x_f \approx 25$,
and  $\VEV{\sigma v}$ is the thermal average of the 
dark matter pair annihilation cross section times the relative velocity.
Most of these quantities are numbers with
large exponents.  However, combining them and equating the result to 
$\Omega_N \sim 0.2$~\cite{cosmoreview}, we obtain
\beq
        \VEV{\sigma v} \sim 0.9 \ \mbox{pb}
\eeq{findsigmav}
Interpreting this in terms of a mass, using $\VEV{\sigma v} = 
\pi \alpha^2/8m^2$, we find $m \sim 100$~GeV.

This argument gives only the order of magnitude of the dark matter
particle mass.  Still, it 
is remarkable that the estimate points to a mass scale
where we expect for other reasons to find new physics 
beyond the Standard Model of particle physics.  Our current understanding
of the weak interaction is that this arises from a gauge theory of 
the group $SU(2)\times U(1)$ that is spontaneously broken at the 
hundred-GeV energy scale.  An astronomer might note this as a remarkable 
coincidence.  A particle theorist would go further.  There are many 
possible, and competing, models of weak interaction symmetry breaking.
In any of these models, it is possible to add a discrete symmetry that makes
the lightest newly introduced particle stable.  Generically, this particle
is heavy and neutral and meets the definition of a WIMP that we have given 
above.  In many cases, the discrete symmetry in question is actually 
required for the consistency of the theory or arises naturally from its
geometry.  For example, in models with supersymmetry, imposing a 
 discrete symmetry called 
R-parity is the most straightforward way to eliminate dangerous baryon-number
violating interactions.  Thus, as particle theorists, we are almost justified
in saying that the problem of electroweak symmetry breaking predicts the 
existence of WIMP dark matter.  This statement also has a striking 
experimental implication, which we will discuss in the next section.

The fact that models of electroweak symmetry breaking predict WIMP dark matter
was recognized very early for the important illustrative example of
supersymmetry.  Dark matter was discussed as a consequence of the theory in
some of the earliest papers on supersymmetry
phenomenology~\cite{PagelsPrimack,Weinberg1,Weinberg2,Goldberg,Ellis}.
However, it is important to realize that the logic of this connection is not
special to supersymmetry; it is completely general.  This has been emphasized
recently by the introduction of dark matter candidates associated with
extra-dimensional and little Higgs models of electroweak symmetry
breaking~\cite{UEDDM,moreUEDDM,WarpedDM,KatzNelson,Tparity}.

If WIMP dark matter is preferred by theory, it is also preferred by 
experiment, or at least, by experimenters.  Almost every technique that 
has been discussed for the detection of dark matter requires that the 
dark matter is composed of heavy neutral particles with weak-interaction
cross sections.   There are a few counterexamples to this statement; axion
dark matter is searched for by a technique special to that 
particle~\cite{Livermore}, and 
axion and gravitino warm dark matter are searched for in low-energy
gamma rays~\cite{BoehmHooper}.  But the direct and indirect detection 
experiments that we will discuss later in this section, which are 
considered generic search methods, assume that the dark matter particle 
is a WIMP.

The arguments we have given do not rule out additional dark matter particles
 in other
mass regions.  It might well be true that WIMPs exist as a consequence of
our models of weak interaction symmetry breaking, but that they make up only
a small part of the dark matter.  The only way to find this out is
to carry out the experiments that define the properties of the  WIMPs,
predict their relic density and detection cross sections, and find
discrepancies with astrophysical observation.  Either way, we must
continue to the steps described in the following sections.

\subsection{WIMPs at high-energy colliders}

There is a further assumption that, when added to the properties of a WIMP
just listed, has dramatic implications.  Models of electroweak symmetry
breaking typically contain new heavy particles with QCD color.  These appear
as partners of the quarks to provide new physics associated with the 
generation of the large top quark mass.  In supersymmetry and in many other
models, electroweak symmetry breaking arises as a result of 
radiative corrections
due to these particles, enhanced by the large coupling of the Higgs 
boson to the top quark.   Thus, we would like to add to the structure of 
WIMP models the assumption that there exists a new particle that carries
the conserved discrete symmetry and couples to QCD.  This particle should
have a mass of the 
same order of magnitude as the WIMP, below 1000~GeV.

Any particle with these properties will be pair-produced at the CERN
Large Hadron Collider (LHC) with a cross section of tens of pb.  The
particle will decay to quark or gluon jets and a WIMP that exits a 
particle physics detector unseen.  Thus, any model satisfying these 
assumptions predicts that the LHC experiments will observe 
events with many hadronic jets and an imbalance
of measured momentum.  These `missing energy' events
are well-known to be a signature of models of supersymmetry.  In 
fact, they should be seen in any model (subject to the assumption just given)
that contains a WIMP dark matter candidate.

\begin{figure}
\begin{center}
\epsfig{file=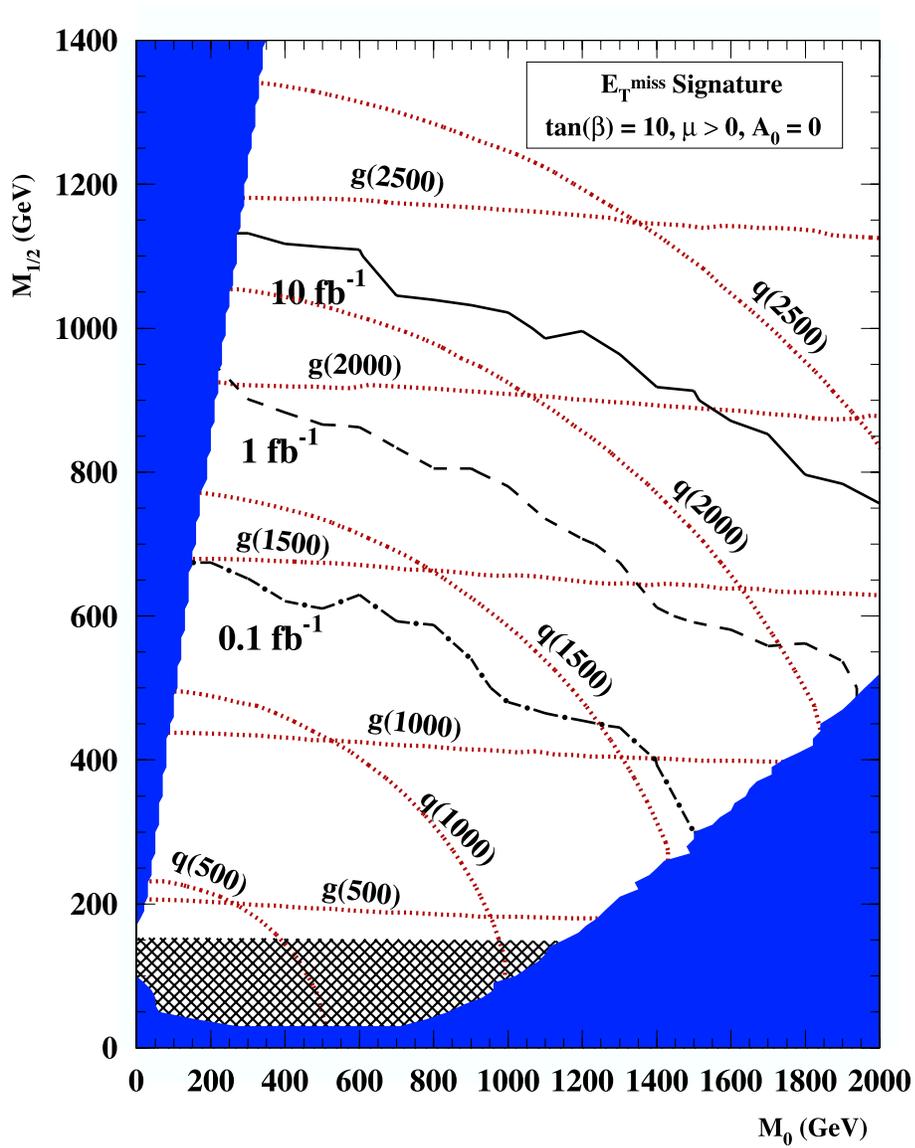,height=6.0in}
\caption{Contours in a parameter space of supersymmetry models for the 
    discovery of the missing energy plus jets signature of new physics
      by the ATLAS experiment at the LHC.
    The three sets of contours correspond to levels of integrated 
     luminosity at the LHC (in fb$^{-1}$), contours of constant squark
      mass, and contours of constant gluino mass.  From \cite{Tovey}. } 
\label{fig:ATLAS}
\end{center}
\end{figure}

The rate of such missing energy events
depends strongly on the mass of the colored particle that is produced and
only weakly on other properties of the model.  So it is reasonable to 
estimate the discovery potential of the LHC by looking at the predictions
for the special case of supersymmetry.  In Fig.~\ref{fig:ATLAS}, we show
the estimates of the ATLAS collaboration for the discovery of missing
energy events at various levels of the LHC integrated 
luminosity~\cite{Tovey}.  For 
the purpose of this discussion, it suffices to follow the contours of 
mass for the squarks and gluinos that are the primary colored particles 
produced.  According to the figure, if either the squark or the gluino
has a mass below 1000~GeV, the missing energy events can be discovered with 
an integrated luminosity of 100 pb$^{-1}$, about 1\% of the LHC first-year
design luminosity.  Thus, we will know very early in the LHC program that
the LHC is producing a WIMP candidate.  This will open the way to detailed
studies of the role of this WIMP in astrophysics.

\subsection{Qualitative determination of WIMP parameters}

For reasons that we will detail in the next section, it is very important 
after the discovery of the WIMP to identify it qualitatively, that is,
to single out what theory gives rise to this particle and what its 
basic interactions are.  This next step may turn out to be very difficult
at the LHC.  

The reason for this is just the converse of the argument 
that the characteristic signature of the WIMP is observed missing
momentum.  At a proton collider such as the LHC, reactions that produce
heavy particles are initiated by quarks and gluons inside the proton. 
We do not know {\it a priori} how much of the momentum of the proton
each initial particle carries.  Since we do not observe the final-state
WIMPs, we also cannot learn the energies and momenta of the produced 
particles from the final state. If we cannot find the rest frame of the 
massive particles, it is very difficult to determine the spins of these
particles or to specifically identify their decay modes.  

\begin{figure}
\begin{center}
\epsfig{file=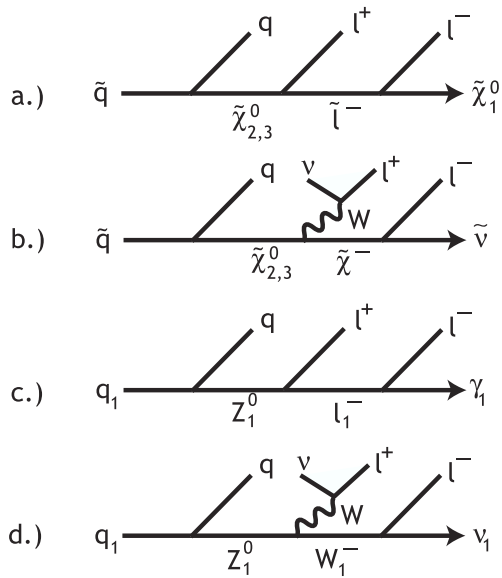,height=3.0in}
\caption{Four scenarios for decay chains observed at LHC.  Each exhibits jets,
hard leptons, and missing energy.  Distinguishing between these cases by
detailed study of energy distributions may not be possible with LHC alone.}
\label{fig:confused}
\end{center}
\end{figure}

As a concrete illustration of this argument, consider the four models of 
the decay of a colored primary particle shown in Fig.~\ref{fig:confused}.
Examples (a) and (b) are drawn from models of supersymmetry in which the 
WIMP is the supersymmetric partner of the photon or neutrino.  Examples
(c) and (d) are drawn from models of extra dimensions in which the 
WIMP is, similarly, a higher-dimensional excitation of a photon or a 
neutrino.  The observed particles in all four decays are the same; the 
subtle differences in their momentum distributions are obscured by the 
uncertainty in reconstructing the frame of the primary colored particle.
It is possible to make use of more model-dependent features.  In the
recent papers \cite{Datta,Dattatwo,Webber}
specific features of the models
have been identified that can distinguish the cases of supersymmetry and
extra dimensions.  
Still, it is likely that, from the
LHC experiments alone we will be left with several competing possibilities
for the qualitative identity of the WIMP.

\begin{figure}
\begin{center}
\epsfig{file=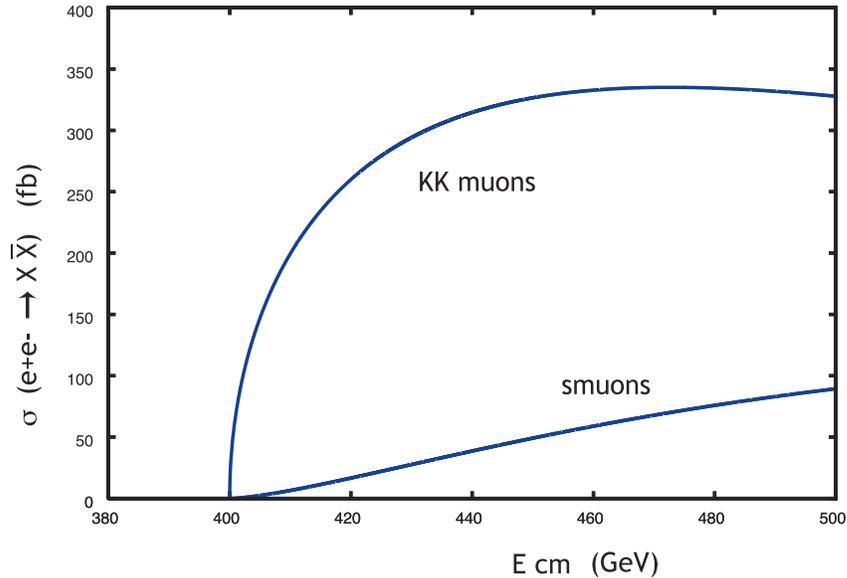,height=3.0in}
\caption{Threshold behavior of pair production cross sections for spin 1/2 (KK
muon) and spin 0 (smuon) counterpart to the Standard Model muon.  These
distributions are easily distinguished by an $e^+e^-$ collider.}
\label{fig:thresholds}
\end{center}
\end{figure}

Fortunately, another tool is likely to be available to particle physicists.
At an elec\-tron-pos\-i\-tron col\-lider, the pair production process $\ee \to
X\bar X$ is gives an ex\-qui\-site diagnostic of the quantum numbers of a 
massive particle $X$. As long as only the diagrams with annihilation 
through $\gamma$ and $Z$ are relevant, the angular distribution and
threshold shape of the reaction are characteristic for each spin, and the 
normalization of the cross section directly determines the $SU(2)\times U(1)$
quantum numbers.  These tests can be applied to any particles with electric
or weak charge whose pair-production thresholds lie in the range of the 
collider.  In Figure~\ref{fig:thresholds}, we show one example of such a test
for models (a) and (c) of Fig~\ref{fig:confused}, 
by plotting the threshold behavior of the pair-production cross sections
for the supersymmetry or extra-dimensional muon partner that appears at the 
last stage of the decay process.  This single 
measurement would already pin down 
the spin and quantum numbers of the particle and bring us a long way toward
the qualitative identification of the model.  Particle physicists are now
designing an electron-positron collider, the International Linear Collider
(ILC), which will reach 500~GeV in the center of mass in its initial
stage and will be upgradable in energy to about 1000~GeV.

The discussion of this and the previous section highlights the contrasting
strengths of the LHC and the ILC, and of the technologies of proton and 
electron colliders.  The LHC can more easily reach high energies and offers
very large cross sections for specific states of a model of new physics.
The ILC typically reaches fewer states in the new particle
spectrum,  but it gives
extremely incisive measurements of the properties of the particles that are
available to it.  Also, as we will see, these particles are typically 
the ones on which the dark matter density depends most strongly.
Both LHC and ILC can make precision measurements, but the 
measurements at the ILC typically have a more direct 
interpretation in terms of particle masses and couplings. 
 In our discussion in Section 4--7, we will see many 
examples in which the greater energy reach of the LHC contrasts with the 
greater specificity of the measurements from the ILC.

\subsection{Quantitative determination of WIMP parameters}

For the purpose of understanding dark matter, what we actually want from an
understanding of the WIMP in particle physics  
is the ability to predict the WIMP's relic density and detection
cross sections.  Given that the WIMP is not observable in the high-energy
physics experiments, it is not so obvious how to make these predictions,
or, even, that the predictions can be made.  The only strategy available to
us is to understand the underlying particle physics model well enough to 
fix the interactions of the WIMP.  To do this, we must determine its 
couplings and the masses and properties
 of the  observable particles to which it couples.

As a matter of principle, this is a very difficult undertaking.  The one 
advantage that we have is that the cross sections of the WIMP that are
the most important in astrophysics involve very low energies.  The 
relic density is determined by the WIMP annihilation cross section at 
temperatures such that $T/m_\chi \sim 1/25$, corresponding to nonrelativistic
motion~\cite{TurnerSch}.  When we observe the WIMP through its 
annihilation processes, the annihilation energy is very close to threshold.
In direct detection of WIMPs, or in the capture of WIMPs into the earth
or the sun, the WIMPs are moving with a velocity  $v/c \sim 10^{-3}$.
Though we will see some exceptions to this, it is typical that the most 
important diagrams for computing these cross sections involve the 
lightest particles in the model.  If we can characterize these particles
and measure their properties with precision, we can reach the goal of 
making microscopic predictions of the WIMP properties.

We do not know a way to give a general proof of this claim, but we can 
illustrate its validity through studies of models.  In Section 3, we will
explain in detail how we will test this claim for supersymmetry models 
with neutralino dark matter.

\subsection{Astrophysical dark matter measurements: relic density}

In the next three sections, we will review the astrophysical measurements
that will require cross sections and other particle properties that might 
be determined by particle physics measurements. The first of these is the 
most basic property of dark matter, its cosmic mass density.

The mass density of dark matter is already known today to impressive 
accuracy, and this accuracy is expected to improve significantly before
the end of the decade.   The analysis of fluctuations in the cosmic microwave 
background (CMB)---in
particular, the measurement of the acoustic peaks that reflect oscillations
in the plasma that filled the universe at temperatures just below 1~eV---allow
determinations of the density of baryonic and non-baryonic matter.  The 
current value of the dark matter density, dominated by the data from 
the WMAP experiment~\cite{WMAP}, is~\cite{cosmoreview}
\beq
     \Omega_\chi h^2 = 0.111 \pm 0.006
\eeq{WMAPOmega}
This is already a determination to 6\% accuracy.  In 2007, we expect the
launch of the Planck satellite, which will give an even more precise
measurement of the properties of the CMB.   From
 this experiment, we can expect an improvement in the accuracy of 
$\Omega_\chi h^2$ to 0.4\%~\cite{BondE}.

It will be very difficult for microscopic predictions of the WIMP density
to match this level of precision.  But we will see that, in the 
specific models that we will consider, it is possible to give a microscopic
prediction of the WIMP density to an accuracy of 20\% or better. 
 Thus, it will 
be a quite nontrivial test to compare 
the microscopic prediction to the density
determined from the CMB.

A discrepancy between the microscopic and CMB values could arise for many 
reasons.  The WIMP could provide only a portion of the dark matter, with 
other portions arising from different types of particles.  The WIMP could
decay to a lighter and even more weakly interacting particle, a 
`superWIMP'~\cite{FengSuper}.  In this case, experiments on astrophysical
WIMP detection should see nothing, but particle physics experiments might
find evidence for the WIMP instability both from the new particle spectrum
and from direct observation of the 
decay~\cite{moreFengSuper,stillmoreFengSuper,Hamaguchi,moreHamaguchi}.  
The density
of WIMPs could be diluted between the temperature of WIMP decoupling
($T\sim$ GeV) and the temperature of primordial nucleosynthesis by 
some mechanism of entropy production, due to a phase transition or late
particle decay.  On the other side, the WIMP density could mainly be 
generated out of equilibrium, during reheating to TeV temperatures or 
from the decay of heavy particles to WIMPs.  In supersymmetry models, this
scenario has been studied in models of anomaly-mediated 
supersymmetry breaking~\cite{MoroiRandall,Kitano}.  These models contain
large annihilation cross sections and so   predict large astrophysical 
signals of WIMP annihilation.

In the study of primordial nucleosynthesis, both late entropy production
and nonthermal processes have been considered, along with more exotic 
effects from new physics.  But, in fact, the predictions of primordial 
nucleosynthesis are in remarkable agreement with the predictions based on 
standard cosmology combined with detailed laboratory measurements of 
low-energy nuclear cross sections~\cite{primordialNucl}.  This gives us 
confidence that our cosmological model is correct back to times of the 
order of 1 minute after the Big Bang.  From this experience, we conclude
that it is possible also that the measured and predicted WIMP density might
turn out to be in excellent agreement, verifying standard cosmology back 
to times of $10^{-9}$ seconds.

If the CMB and microscopic determinations of the WIMP density within their
individual accuracies, it will be tempting to impose the more stringent 
astrophysical constraint on the particle physics model.  At present, 
we do not know the particle physics model, and imposing the constraint
\leqn{WMAPOmega} does not seem to exclude any qualitative
possibilities, though it does
narrow the parameter space if a given model is assumed.  After the particle
physics model is known, the stringent constraint on the dark matter density
from Planck can have more interesting consequences.  In Sections 
4.6, 5.5, 6.5, and 7.5, we will give specific examples of how a precise
value of the dark matter density can be combined with LHC data to
predict parameters of the supersymmetry spectrum that are difficult to
measure at the LHC.

\subsection{Astrophysical dark matter measurements: direct detection}

If the microscopic cross sections measured at colliders agree with the CMB
measurements of the dark matter density, a crucial uncertainty remains, alluded
to in the previous section.  Does the dark matter candidate in addition have a
lifetime longer than the age of the universe, or is the consistency merely a
coincidence?  A complete picture of dark matter requires that the candidate
particle be observed as the major constituent of our galaxy.  This can be
accomplished by direct or indirect detection, discussed in this section and the
next.

Direct detection of WIMPs in sensitive low-background experiments involves
the cross sections for WIMP scattering from nucleons near threshold.  The 
expressions for these cross sections naturally divide into spin-dependent
and spin-independent isoscalar and isovector contributions.  The 
spin-independent isoscalar term is enhanced in WIMP-nucleus cross sections
by factors of $A^2$, so we will emphasize the prediction of this 
contribution.  The predictions for direct detection cross sections in
supersymmetry models are typically displayed on scatter plots that 
cover five orders of magnitude~\cite{JKG}. 
 As we will see, data from the LHC and the
ILC can significantly narrow this range.

The technologies for direct detection of WIMPs and 
experimental limits have recently been
reviewed by Gaitskell\cite{Gaitskell}.  The strongest current 
limits, from the CDMS \cite{CDMS}, CRESST~\cite{CRESST},
 EDELWEISS~\cite{Edelweiss}, and 
ZEPLIN~\cite{ZEPLIN} experiments, give an upper bound to the 
cross section of a 100~GeV WIMP at about $\sigma(\chi p) 
< 2\times 10^{-7}$~pb.
The DAMA experiment~\cite{DAMA} reported the observation of annual 
modulation in a low-background NaI detector of a size consistent with
a WIMP with $\sigma(\chi p) \sim 5\times 10^{-6}$~pb,
but unfortunately the experiments listed previously 
contradict this interpretation~\cite{DAMAOK}.  For the purposes of 
this paper, we will choose reference models of WIMP dark matter 
with $\sigma(\chi p) <  10^{-7}$~pb.

Later in the paper, we will present
analyses in which  WIMP direct detection rates are combined
with LHC and ILC results.  In these analyses, we will use the predicted
counting rates from the proposed SuperCDMS experiment~\cite{superCDMS}.
This particular
experiment is chosen simply as a concrete illustration of our methods.
Alternative technologies, for example, large noble gas 
detectors~\cite{XENON,WARP}, promise higher counting rates and might 
well improve on these results.

Predictions for the rate in direct detection experiments rely on the 
assumption that the average density of dark mater in 
the halo of the galaxy can be used to estimate the 
flux of dark matter that impinges on a detector on the earth. This density 
is uncertain, but, in addition,  we do not expect the density of dark matter 
to be constant over the halo.  In models of cold dark matter, the 
galaxy is assembled from smaller clusters of dark matter.  The 
initial situation is inhomogeneous, and these inhomogeneities are not 
expected to be smoothed out in the time since the galaxy was formed.

 The local halo density is inferred by fitting to models of the 
galactic halo.  These models are constrained by a variety of 
observations, including the rotational speed at the solar circle 
and other locations, the total projected mass density 
(estimated by considering the motion of stars perpendicular 
to the galactic disk), peak to trough variations in 
the rotation curve (`flatness constraint'), and 
microlensing. Gates, Gyuk, and Turner~\cite{GGTurner}
 have collected these constraints and estimated the local halo density to 
lie between 4 $\times$ 10$^{-25}$ g/cm$^{-3}$ and 13 $\times$ 10$^{-25}$ 
g/cm$^{-3}$.  Limits on the density of MACHO microlensing objects
imply that at least 80\% of this is cold dark matter.   The velocity of the
WIMPs would be close to the galactic rotation velocity,  $230 \pm 20$ 
km/sec~\cite{Drukier}.   The effects of varying the model of the WIMP
halo are illustrated in a recent paper of the Torino group that 
shows this variation in terms of the region of the $m_\chi$ vs. $\sigma$
plane consistent with the DAMA result~\cite{Torino}.

These constraints rely on the assumption that the dark matter has a 
smooth distribution in the  galactic halo.  However, the general 
conclusion from high resolution N-body simulations is that
the dark matter distribution is highly irregular. Using a 
hierarchical clustering model of galactic structure,
Stiff, Widrow, and Frieman have argued that the solar neighborhood 
might be expected to be located within a clump of dark matter with only
slightly higher local density but a velocity distribution peaked at 
a relatively large value~\cite{StiffW}.
Other authors have argued for larger density variation in the galactic
distribution of dark matter.
Sikivie and Ipser~\cite{Sikivie} 
have proposed  that spherical infall
of dark matter on to developing galaxies will tend to accumulate
along singular surfaces or caustics, leading to very large local
fluctuations. An even more extreme model was recently put forward
by Diemand, Moore, and Stadel~\cite{DMS}, who argued that WIMPs are 
likely to appear in clusters of mass comparable to the mass of 
the earth and densities roughly 10$^3$ larger than the average 
density of the disk. In addition to these models that rely on the 
general features of galaxy formation,  the local geography of 
our region of the galaxy might affect the dark matter density.
 For example, Freese, Gondolo and Newberg
have proposed that the Sagittarius stream should add up to 23\% to the
local dark matter density, with a characteristic annual 
modulation~\cite{Freese}.    These models illustrate not only that there is a 
large uncertainty in the value of the local dark matter flux
at the earth  but also that 
the understanding of this value relative to the overall
average density of dark matter is an interesting astrophysical question.

If we view these questions as uncertainties, we must say that direct
detection experiments cannot by themselves
put constraints on the microscopic scattering cross sections of WIMPs.
On the other hand, if we could obtain the microscopic 
cross sections from particle
physics, the event rate in direct detection experiments
would directly measure the local flux of WIMP dark matter.
If direct detection experiments failed to detect dark matter, it would be
even more important to have the microscopic cross section in order
to establish strong upper bounds on the local WIMP density.

There is another source of uncertainty in this program that also needs 
to be addressed~\cite{thanks}.  
When the dark matter direct detection cross section 
is calculated from particle physics models, even if the parameters of 
these models are assumed to be precisely known, there is an uncertainty
in the prediction of the cross section that can be as large as a factor
of 4 coming from a poorly understood effect of low-energy QCD.  In 
many WIMP models, including the SUSY models that we will take as reference
points in this study, $\sigma(\chi p)$ is dominated by $t$-channel
Higgs exchange.  The coupling of the Higgs boson to the proton receives
its dominant
contributions from two sources, the coupling of the Higgs to gluons
through a heavy quark loop and the direct coupling of the Higgs to strange
quarks~\cite{complete}.  That means that this coupling depends on the parameter
\beq
         f_{Ts} =  {\bra{p}  m_s \bar s s \ket{p} \over \bra{p}  H_{QCD}
                           \ket{p}} \ ,
\eeq{fTsdefin}
that is, the fraction of the mass of the proton that arises from the mass
of the non-valence strange quarks in the proton wavefunction.  In 
1987, Kaplan and Nelson argued that this quantity is larger and more 
uncertain than previously thought~\cite{KaplanNelson},
\beq
         f_{Ts} =  0.36 \pm 0.14 
\eeq{KMvalue}
In the intervening twenty years, there has been essentially no progress
in improving our knowledge of this quantity.  Several recent papers have
highlighted the uncertainty in $f_{Ts}$ as a major uncertainty in WIMP
directly detection cross sections~\cite{Bottino,EllisUpdate}.  Indeed,
for a heavy SUSY Higgs boson in a model with a large value of the
parameter $\tan\beta$, the Higgs-proton coupling is given quite accurately
by
\beq
        \lambda_{Hpp} =   {m_p\over 250\ {GeV}} \left[ {2\over 27} + 
          {25\over 27} f_{Ts} \right] \tan \beta + \cdots
\eeq{Higgscoupling}
so that $\sigma(\chi p)$ is almost proportional to $f_{Ts}^2$.  This 
can produce an uncertainty in the direct detection cross section of
a factor of 4 or even larger~\cite{Bottino}.

How could we resolve this problem?  We consider it unlikely that 
an improvement of the data  on which the estimate \leqn{KMvalue} is 
based will improve the error.  However, it should eventually be possible
to compute $f_{Ts}$ in lattice gauge theory.  Because $f_{Ts}$ is a 
non-valence quantity, this is beyond the current state of the art.
For example, the current result from the UKQCD collaboration is 
$f_{Ts} = -0.20\pm 0.23$~\cite{UKQCD}.   However, the valence
pion nucleon sigma term is already under control; a recent 
analysis gives $\sigma_{\pi N} = (49\pm 3)$~MeV~\cite{Weise}.
For the non-valence case, the problem seems mainly one
of obtaining computer power to generate high statistics, 
and this should increase dramatically over the next ten years~\cite{BUthanks}.

For the analyses of this paper, we will ignore uncertainties
in the calculation of direct detection cross sections that are not 
associated with variation of the parameters of the new physics model.
This means that we will ignore 
the uncertainty in $f_{Ts}$, corresponding to the assumption that 
lattice calculations will eventually solve
this problem.  Other sources of theoretical error seem to be under control
at the 10\% level.  We will assume that $f_{Ts} = 0.14$, the default value
in the DarkSUSY code~\cite{DarkSUSY}, so in any event we will underestimate
the experimental counting rates.

Parenthetically, we would like to call attention to an aspect of direct
detection experiments that is very important but is not often emphasized.
It is very likely that direct detection experiments will see evidence
of WIMPs in the same time frame, before the end of the decade, that the
LHC observes missing energy events.  In this case, it will be essential 
to compare the mass of the WIMP observed in each setting.  In supersymmetry
models,  as we will discuss in Sections 4-7,
the kinematics of events with squark production and decay can
determine the mass of the WIMP to about 10\% accuracy.
We believe that this type of analysis will give the WIMP mass to 
similar accuracy in typical models of the general class discussed 
in Section 2.2.   

Direct detection experiments can determine the mass of the WIMP
by measuring the recoil energy
$E_R$.  This varies with the mass of the WIMP, with a 
resonance where the WIMP mass equals the target mass.  Roughly, one 
expects
\beq
        \VEV{E_R} \approx  { 2 v^2 m_T\over (1 + m_T/m_\chi)^2 }, 
\eeq{recoilenergy}
where $m_T$ is the target mass and $v$ is the WIMP velocity, 
with corrections depending on the precise target material
and the properties of
the detector~\cite{LewinSmith}.   Assuming the
standard velocity distribution in smooth halo models, 
with the 10\% uncertainty quoted above,
an experiment with a Xenon or Germanium target 
that detects 100 signal events for 
a WIMP of mass $m_\chi = 100$ GeV
can expect to measure the mass of this particle to 20--25\%.  A 
discrepancy between the value of the WIMP mass observed in direct
detection and that found at the LHC could signal a nonstandard 
velocity distribution.   At a later stage, this could be checked by 
comparing the detection rate, which is proportional to the flux of 
WIMPs, to the cross section determined from high-energy collider data.

\begin{figure}
\begin{center}
\epsfig{file=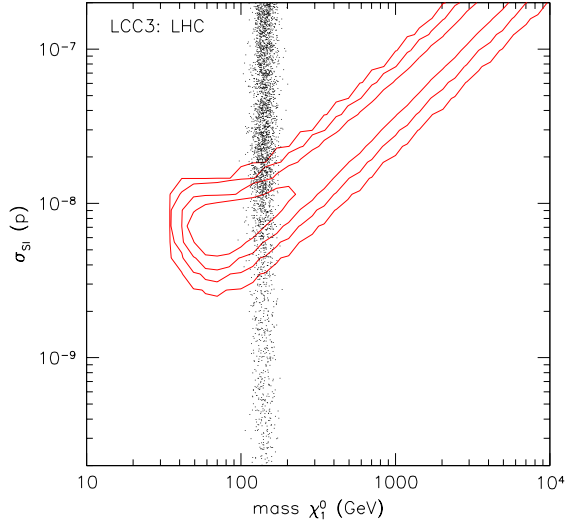,height=3.0in}
\caption{Projected significance contours ($1,2,3,4 \sigma$) in the plane of
WIMP mass versus cross section for an observation of dark matter by the
SuperCDMS experiment (25 kg target, two year dataset)~\cite{superCDMS,Schnee},
compared to the determination of these parameters from data from the LHC.  The
projections are done using the model LCC3, to be defined in Section 3.2.}
\label{fig:CDMScontours}
\end{center}
\end{figure}

In Fig. \ref{fig:CDMScontours}, we show a
comparison of the determination expected
for the WIMP mass from the LHC data and from the analysis of data from 
the SuperCDMS detector~\cite{superCDMS,moresCDMS} for one of the supersymmetry
models that we introduce in Section 3.2.  The contours are based on 
a sample of 27 events and so are statistically limited.
Still, it is clear that a nontrivial comparison
of WIMP masses between accelerator and astrophysical experiments will be
possible. 

More powerful strategies for measuring the WIMP mass from direct
detection data have been proposed by Primack, Seckel, and 
Sadoulet~\cite{Primackmass}
and by Bourjaily and Kane~\cite{BandK}.
  However, this would require a  sample of direct
detection events about 100 times larger than those in the illustrative examples
we will present here.

Finally, we comment on the possibility that WIMPs measured at colliders make up
only a fraction of the dark matter.  In this case, the annihilation cross
sections tend to be larger than the expected 1 pb, and also the direct
detection cross sections tend to be large.  In fact, these effects broadly
speaking cancel: a WIMP making up 10\% of the cosmic dark matter tends to have
cross sections 10 times as large as expected, thus the direct detection rate is
roughly independent of the inferred relic density~\cite{DudaGelminiGondolo}.

\subsection{Astrophysical dark matter measurements: WIMP annihilation}

WIMP annihilation could potentially be observed through gamma ray, positron,
antiproton, antideuteron, and neutrino signals.  Of these, the observations
through gamma rays is the simplest and most robust.  There are already 
claims that an excess of gamma rays from the galactic center gives evidence
for WIMP dark matter~\cite{deBoer}.  We will examine how this study will 
be aided by collider physics determinations of the WIMP annihilation
cross section.

The flux of gamma rays observed on earth from WIMP annihilation is given 
by the formula
\beq
       E_\gamma {d \Phi_\gamma\over d E_\gamma d \Omega} = \frac{1}{2}
  (\sigma_{\chi\chi} v) \cdot {E_\gamma\over \sigma_{\chi\chi}}
        {d \sigma_{\chi\chi}\over d E_\gamma} \cdot 
    {1\over 4\pi m_\chi^2} \cdot \int dz \ \rho^2(z) \ ,
\eeq{gammaflux}
where $\sigma_{\chi\chi}/2$ is the annihilation cross section near threshold, 
(which 
typically behaves as $1/v$), $z$ is a coordinate along the line of sight,
and $\rho$ is the WIMP mass density. The factors of $1/2$ are appropriate
assuming that the dark matter particles are self-conjugate; we will later
apply this equation to the neutralino WIMP in SUSY models.
The first three factors come from 
microscopic physics; the final factor is a question of  astrophysics.
The density integral is commonly written
\beq
    \int dz \rho^2(z)  =     r_0 \rho_0^2  J(\Omega) \ , 
\eeq{Jdef}
where $r_0 = 8.5$ kpc is the distance to the center of the galaxy and $\rho_0 =
0.3$ GeV cm$^{-3}$ $(5.34\times 10^{-25}$~g/cm$^3)$ is a reference value
of the local density of dark matter.  The quantity $J(\Omega)$ is then
dimensionless.

It is likely that the microscopic quantities that appear in \leqn{gammaflux}
could be estimated to an accuracy of 20\% even in the early stages of
the study of WIMPs in particle physics experiments.  The mass
$m_\chi$ will be determined to better than 10\%  accuracy from the 
kinematics of missing energy events at the LHC.  We will show in 
Section 8.1 that the second factor in \leqn{gammaflux}, the shape of the
gamma-ray spectrum, is almost completely independent 
of the specific annihilation process and can be obtained 
accurately from  particle physics simulations. 
For the first factor, the total annihilation
cross section, it is tempting to insert the value \leqn{findsigmav}
from the relic density.  This is sometimes, but not always, a good
approximation, depending on the qualitative properties of the 
supersymmetry spectrum.  We will discuss the systematics of this quantity in 
Section 8.1.  In many physics scenarios, all three quantities will be 
known well enough already from the LHC data
 to quantitatively interpret gamma ray observations.

\begin{figure}
\begin{center}
\epsfig{file=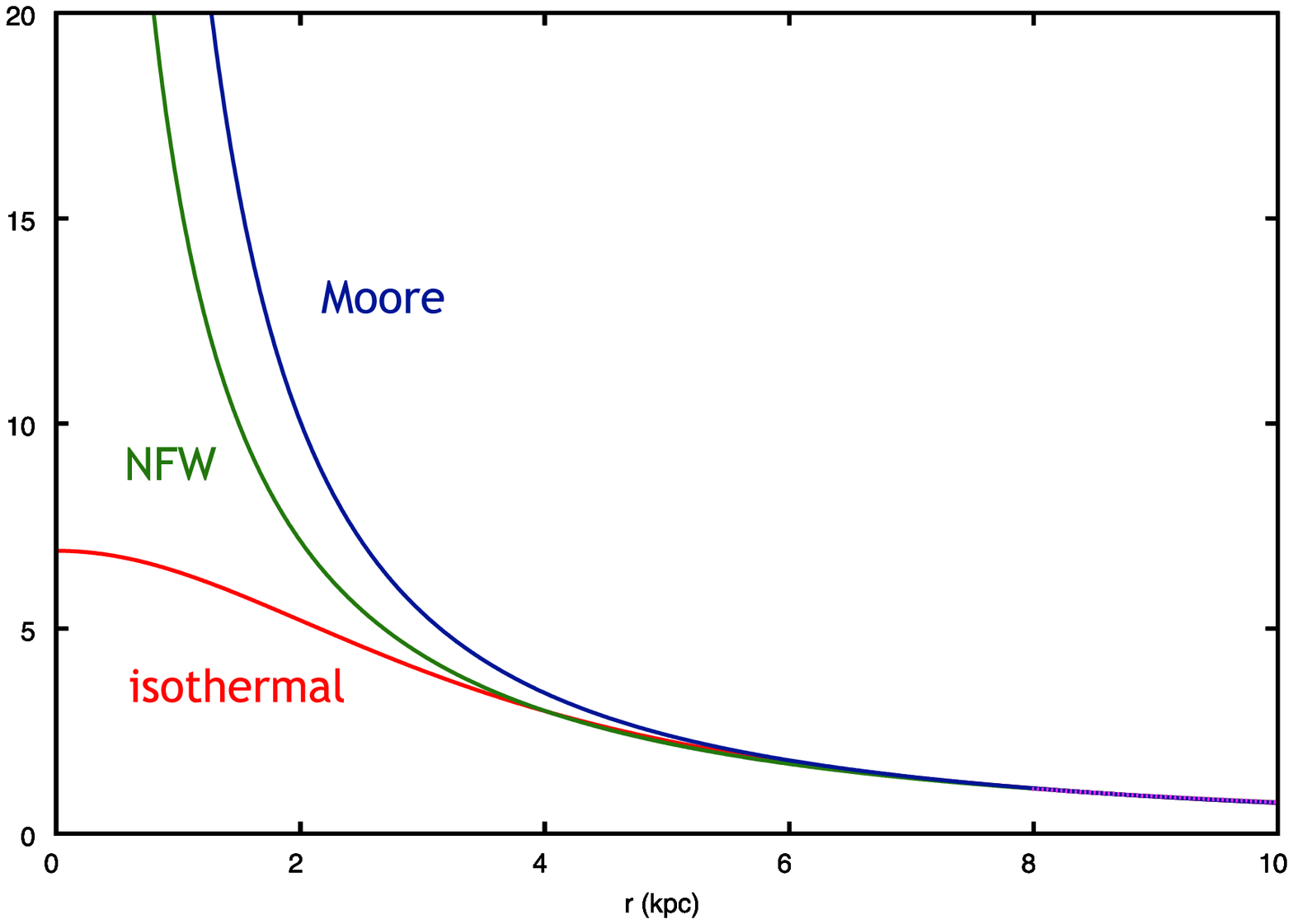,height=2.5in}\\ 
\epsfig{file=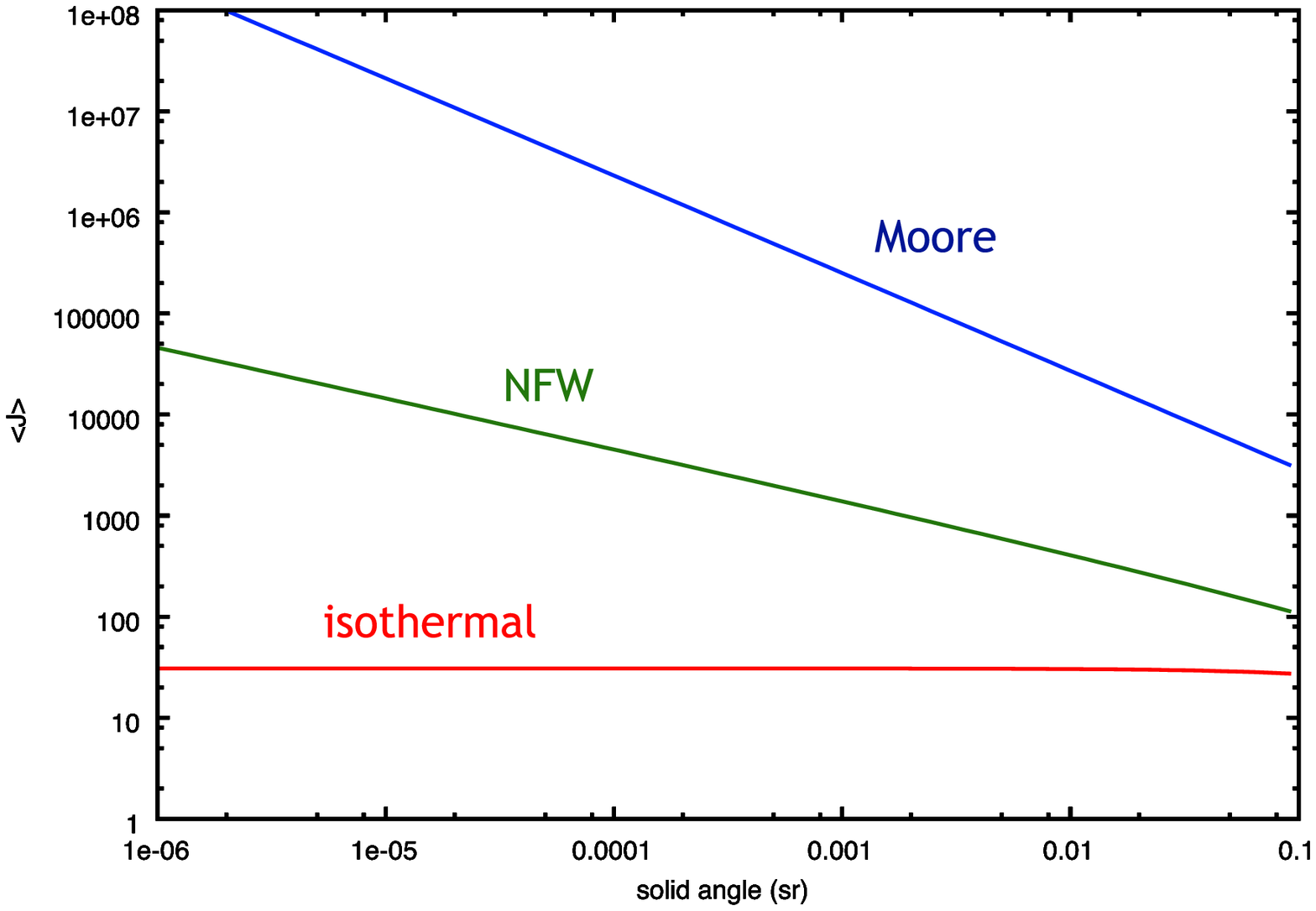,height=2.5in}
\end{center}
\caption{(a) Suggested profiles of the dark matter mass density 
 near the galactic center \cite{NFW,Moore}.
(b)  Corresponding values of $\VEV{J}$ at the galactic center, 
as a function of the angular resolution indicated as a circle size 
in sr.}
\label{fig:profiles}
\end{figure}

It is very important that the particle physics 
factors in \leqn{gammaflux} can be
determined from microscopic measurements, because the last factor raises
major questions about the structure of the dark matter halo of the galaxy.
Let us first discuss the dark matter profile at the galactic center.
A typical profile of the smoothed dark matter halo is written
\beq
     \rho(r) =     \rho_0 (r/r_0)^\gamma \left({ 1 + 
          (r_0/a)^\alpha\over  1 + 
          (r/a)^\alpha} \right)^{(\beta-\gamma)/\alpha} \ ,   
\eeq{DMprofile}
where $\alpha$, $\beta$, $\gamma$  are parameters and the scale size $a$ 
should be about 20 kpc.
Navarro, Frenk, and White (NFW) have argued that that
numerical simulations of galaxy formation by cold dark matter
suggest the profile
$(\alpha, \beta, \gamma) = (1,3,1)$~\cite{NFW}.
  In the explicit examples that 
we will present in Section 4.4, we will assume the NFW profile as
a canonical choice.  However, other groups have drawn different 
conclusions from the simulation data.  Moore {\it et al.}
have argued for a much steeper profile at the galactic center:
 $(\alpha, \beta, \gamma) = (1.5,3, 1.5)$~\cite{Moore}.  
 For this profile, the integral over $\rho^2(z)$ 
formally 
diverges at the galactic center~\cite{logdivergence}.  
Wechsler and collaborators have argued that the halos found in a 
given simulation can actually have different  shapes depending on their
history, roughly covering the range between the NFW and Moore 
profiles~\cite{Wechsler,Primack}.   We should note that the simulations that
we are discussing include only dark matter, and that the addition of
a dissipative component could also alter the profile.
  The variation in \leqn{gammaflux} near the galactic center
 for the NFW and Moore profiles, and for an isothermal profile with
$(\alpha, \beta, \gamma) = (2,2,0)$, is shown in Fig.~\ref{fig:profiles}. 
We present both the profile functions themselves and the value of 
$J(\Omega)$ averaged over a disk of varying solid angle centered on 
the galactic center.
The predictions for $\VEV{J(\Omega)}$ span six orders of magnitude 
between models.
Further, these estimates are all smooth profiles, and we have already
seen that the distribution of dark matter might be clumpy.  Such small-scale
structure would enhance indirect detection rates by the ratio
\beq
        B =  \VEV{\rho^2}/\VEV{\rho}^2 \ ,
\eeq{boostfactor}
sometimes called the `boost factor'.  The final conclusion from all
of these considerations is that
indirect dark matter
detection rates depend on distributions that are highly uncertain and touch
on major questions of astrophysics.   It would be interesting to 
extract these distributions from the experiments, or even to obtain upper
limits on them.  For this, we need to know the microscopic cross sections
from particle physics.

This conclusion applies also to other possible sources
of gamma rays from WIMP annihilation.  For gamma rays from the centers 
of local group and other nearby galaxies, the considerations are
very similar to those for the galactic center~\cite{Baltzgalactic}. 
Simulations of galaxy formation with cold dark matter predict many more
dwarf companions of the Milky Way than are actually observed.  Probably,
many of these have been tidally disrupted or absorbed.  However, it is 
likely that some of these unobserved dwarf galaxy are actually present
as pure dark matter halos from which the baryonic gas has been 
blown out~\cite{TaylorBabul}.
The GLAST telescope, with its $\pi$ angular coverage for gamma rays,
has the ability to search for these objects.  Even upper limits are 
interesting, but their interpretation will depend critically on knowledge 
of the microscopic factors in \leqn{gammaflux}.

The expectations for observation of positrons, neutrinos, antiprotons and
antideuterons from WIMP annihilation depend on more specific details of the
physics model.  We will discuss the subject of positron signals further
in Section 8.2.     For the other cases, the analysis of the interplay 
between collider physics data and indirect detection signals is more
complex and its analysis is beyond 
the scope of this paper.  We should note, however, that the four 
reference models that we will introduce in Section 3.2 are consistent
with all current constraints from indirect detection of WIMPs.

\subsection{Summary: steps toward an understanding of WIMP dark matter}

The program of experiments that we have described in this section has the
potential to give us a complete understanding of the nature of WIMP
dark matter.  The steps are the following:
\begin{enumerate}
\item  Discover missing-energy events at a collider and estimate the
   mass of the WIMP.
\item  Observe dark matter particles in direct detection experiments         
                and determine whether their mass
     is the same as that observed in collider experiments.
\item  Determine the qualitative physics model 
          that leads to missing-energy events
\item  Determine the parameters of this model that predict the relic density.
\item Determine the parameters of this model that predict the direct and 
  indirect detection cross sections
\item   Measure products of cross sections and densities from astrophysical
 observations to  reconstruct the density distribution of dark matter.
\end{enumerate}
If dark matter is composed of a single type of WIMP, this program of 
measurements should reveal what this particle
is and how it is distributed in the galaxy.  If the composition of dark 
matter is more complex, we will only learn that by carrying out this 
program and finding that it does not sum to a complete picture.  Hopefully,
further evidence from the microscopic theory will suggest other necessary
ingredients.

The main goal of the remainder of this paper is to show that the 
experiments foreseen for the LHC and ILC will be able to predict the 
microscopic dark matter cross sections with sufficient accuracy that 
we can carry out this program.  In the next section, we will describe
our strategy for addressing this question.

\section{Models of neutralino dark matter}

As we have already noted in Section 2.3, the annihilation and detection 
cross sections needed to interpret observations of WIMP dark matter
cannot be measured directly in high-energy physics experiments.  To 
predict these cross sections, we must interpret experimental data on the 
spectra and parameters of the underlying physics model.   To do this, we
must understand, at a qualitative level, what the correct model is. 
We must then convert measurements of the spectrum of new particles
into constraints on the underlying model parameters.  Some care should be
taken in the choice of the model.  If we  
work in too restrictive a model context, this procedure will artificially
restrict the solutions, and we will claim an unjustified small accuracy 
for our predictions.  Thus, to evaluate how accurately collider data will
predict the dark matter cross section, we need to work within a model that,
under overall restrictions from spin and quantum number measurements,
has a large parameter space and allows a wide variety of physical effects
to come into play.

Among models of physics beyond the Standard Model, the only one in which 
dark matter properties have been studied over such a large parameter
space is supersymmetry~\cite{UEDdarkmatter}.  The Minimal Supersymmetric
Standard Model (MSSM) introduces a very large number of 
 new parameters and allows many 
physically distinct possibilities for the mass spectrum of new 
particles.  Thus, our strategy for evaluating the implications of collider
data for dark matter cross sections will be to study a set of MSSM
parameter points which illustrate the variety of physics scenario that this
general model can contain.  In 
each case, we will systematically scan the 
parameter space of the MSSM for models that are consistent with the 
expected collider measurements.  We hope that the insights obtained from this
study will lead us to conclusions of broader applicability 
about the power to high-energy
physics measurements to restrict the properties of dark matter.

\subsection{Mechanisms of neutralino annihilation}

\begin{figure}
\begin{center}
\epsfig{file=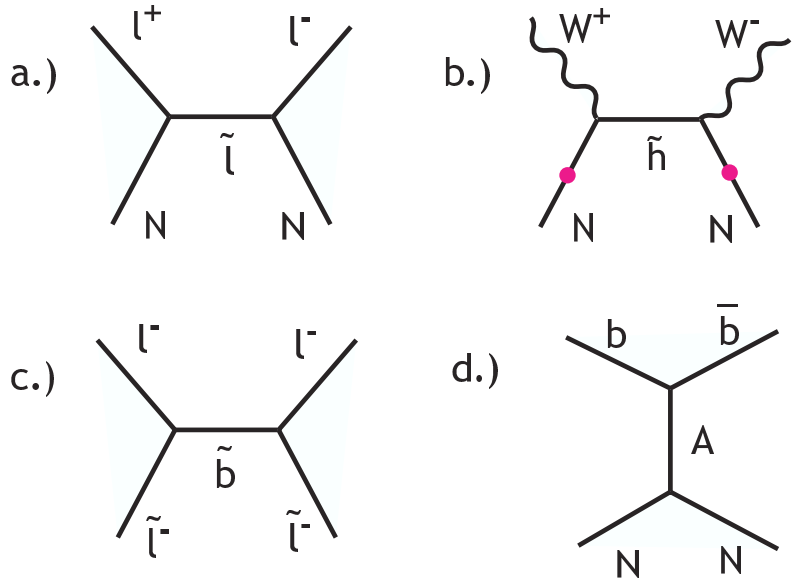,height=3.0in}
\caption{Four neutralino annihilation reactions that are important in 
       different regions of the MSSM parameter space: (a) annihilation 
      to leptons, (b) annihilation to $W^+W^-$, (c)  coannihilation  
            with $\s\tau$, (d) annihilation through
       the $A^0$ resonance.}
\label{fig:NNann}
\end{center}
\end{figure}

From here on, then, we restrict our attention to models with supersymmetry
in which the  role of the WIMP $\chi$ 
is taken by the lightest `neutralino'---a 
mixture of the superpartners of $\gamma$ and  $Z$ (`gauginos') 
and the superpartners of the neutral Higgs
 bosons (`Higgsinos').  Depending on
the spectrum and couplings of the superpartners, several different reactions
can dominate the process of neutralino pair annihilation.
Some of the most important possibilities are illustrated in 
Fig.~\ref{fig:NNann}.

The simplest possibility (Fig.~\ref{fig:NNann}(a)) 
is that neutralinos annihilate to Standard Model
fermions by exchanging their scalar superpartners.  Sleptons are typically
lighter than squarks, so the dominant reactions are $\chi\chi \to \ell^+
\ell^-$.  It turns out, however, that this reaction is less important
than one might expect over most of the supersymmetry parameter space.
Because neutralinos are Majorana particles, they annihilate in the S-wave
only in a configuration of total spin 0.  However, light fermions are 
naturally produced in a spin-1 configuration, and the spin-0 state is 
helicity-suppressed by a factor $(m_\ell/m_\chi)^2$.  The dominant 
annihilation is then in the P-wave.  Since the relic density is 
determined at a temperature for which the neutralinos are 
nonrelativistic, the annihilation cross section is suppressed and the 
prediction for the relic density is, typically, too large.  To obtain
values for the relic density that agree with the WMAP determination, we 
need light sleptons, with masses below 200 GeV.

Neutralinos can also annihilate to Standard Model vector bosons.
A pure $U(1)$ gaugino (`bino')
cannot annihilate to $W^+W^-$ or $Z^0Z^0$.  However, these
annihilation channels open up if the gaugino contains an admixture of
$SU(2)$ gaugino (`wino') or Higgsino content (Fig.~\ref{fig:NNann}(b)).  
The annihilation cross 
sections to vector bosons are  large, so only a relatively small mixing
is needed.

The annihilation to third-generation fermions can be enhanced by a resonance
close to threshold.  In particular, if mass of the CP odd Higgs boson $A^0$
is close to $2 m_\chi$, the resonance produced by this particle can 
enhance the S-wave amplitude for neutralino annihilation to $b\bar b$
and $\tau^+\tau^-$ (Fig.~\ref{fig:NNann}(d)).

If other superparticles are close in mass to the neutralino,
these particles can have significant densities when the neutralinos 
decouple, and their annihilation cross sections can also contribute to the 
determination of the relic density through a coannihilation process.
If the sleptons are only slightly heavier than the neutralino, the 
reactions $\s\ell \chi \to \gamma \ell$ and $\s \ell \s\ell \to \ell \ell$
can proceed in the S-wave and dominate the annihilation 
(Fig.~\ref{fig:NNann}(c)).  Coannihilation with $W^+$ partners (`charginos')
and with top squarks can also be important in some regions of the MSSM
parameter space.

A common feature of all four mechanisms is that the annihilation cross 
section depends strongly both on the masses of the lightest supersymmetric
particles and on the mixing angles that relate the original 
gaugino and Higgsino 
states to the neutralino mass eigenstates.  Both sets of parameters must
be fixed in order to obtain a precise prediction for the relic density.

A supersymmetry model that produces a relic density of neutralinos in the
range required by the WMAP data should implement one of these mechanisms.
To the extent that the operation of the mechanism requires special 
conditions on the supersymmetry spectrum, the neutralino relic density will
depend more sensitively on the spectrum parameters than we might at first
have estimated.  We will see the effects of this observation in our 
model studies.

\subsection{Choice of benchmark models}

We would now like to choose four specific supersymmetry models that 
illustrate the four mechanisms described in the previous section.

Let us first discuss the parameters of the MSSM.  The most general 
formulation of the MSSM has 108 parameters beyond those of the Standard
Model.  However, many of these parameters violate CP or induce flavor-changing
neutral current processes and are thus tightly constrained.  If we 
include only interactions that conserve CP and flavor, we find the 
following set of parameters:
\begin{itemize}
 \item gaugino and Higgsino masses: $m_1$, $m_2$, $m_3$, $\mu$
 \item slepton masses:  $m^2(L_i)$, $m^2(\bar e_i)$,  $i = 1, 2,3$
 \item squark masses:   $m^2(Q_i)$, $m^2(\bar u_i)$, $m^2(\bar d_i)$, 
       $ i = 1,2, 3$
 \item Higgs potential terms:   $m_A$, $\tan\beta$
  \item $A$ terms:   $A_\tau$, $A_b$, $A_t$
\end{itemize}
a total of 24 new physics parameters.

Most papers on neutralino dark matter adopt additional, {\it ad hoc}, 
assumptions to reduce this parameter list to a smaller set.  Most typically,
they assume unification of the various mass terms at the grand unification
scale.  This assumption, called `mSUGRA' or `cMSSM', reduces the parameter
set to four, plus a choice of sign: $m_0$, $m_{1/2}$, $A_0$, $\tan\beta$,
sign($\mu$).   The restricted parameter space of mSUGRA is very interesting
to illustrate the various possibilities for neutralino dark matter.  In 
particular, this subspace contains examples of all four mechanisms that 
we have presented in the previous section~\cite{ElliscMSSM,Gondolo,Baer}.

This makes it very convenient to choose parameter points from the 
mSUGRA subspace.  In Table~\ref{tab:LCCpoints}, we list the four parameter
sets that define the supersymmetry models that we 
will study in detail~\cite{ILCstudy}.
The  spectra for these points, and for the more general supersymmetry
parameter points that we will study, are computed with 
ISAJET 7.69~\cite{ISAJET}.  Results for the relic density and for 
neutralino detection cross sections are computed with 
DarkSUSY-4.1~\cite{DarkSUSY}.  We have checked that 
Micromegas~1.3~\cite{Micromegas} gives similar results for the relic
density. 

\begin{table}
\centering
\begin{tabular}{l|cccccc|c|| r r} 
   Point &  $m_0$ & $m_\half$ & $\tan\beta$ & $A_0$ & $\mbox{sign}\ mu$ & $m_t$
       &    reference    &    $\Omega_\chi h^2$ & \\ \hline
   LCC1 &   100 &  250 &   10  & $-100$ & $+$ & 175 & \cite{LHCILC} & 
                      0.192 \\
   LCC2 & 3280 & 300 & 10 & 0 &  $+$ & 175 &   \cite{Gray}  & 
                        0.109 \\ 
   LCC3 &  213 & 360 &  40 &  0 & $+$ & 175 & \cite{Dutta} & 
                         0.101 \\
   LCC4 &  380 & 420 &  53 & 0   & $+$ & 178 & \cite{BattagliaParis} &
                0.114 \\ \hline
   SPS1a$'$ &   70 &  250 &   10  & $-300$ & $+$ & 175 & \cite{SPA} & 
                      0.115
\end{tabular}
\caption{mSUGRA parameter sets for four illustrative models of neutralino
     dark matter.  Masses are given in GeV. The table also lists the 
value of $\Omega_\chi h^2$.  The references given are the primary references
for simulation studies of the accuracy of spectrum measurements at colliders.
The point SPS1a$'$ has a phenomenology similar to that of LCC1 but gives 
a more correct value of the relic density.}
\label{tab:LCCpoints}
\end{table}

The four points listed in Table~\ref{tab:LCCpoints} illustrate the 
four scenarios described in the previous section.  Point LCC1 is identical
to the point SPS1a~\cite{Allanach} whose collider phenomenology
is studied  in some detail in \cite{LHCILC}.   This point has 
light sleptons, with neutralino annihilation dominated by roughly 
equal annihilation cross sections to $e^+e^-$, $\mu^+\mu^-$, and 
$\tau^+\tau^-$.  The sleptons are not quite light enough; the spectrum
achieves a relic density $\Omega h^2 = 0.19$, almost doubly the 
WMAP value.   Point LCC2 is chosen as a point with substantial 
gaugino-Higgsino mixing at which the neutralino annihilation is dominated
by annihilation to $W^+W^-$, $Z^0Z^0$,  and $Z^0 h^0$.  Point LCC3 is 
chosen in the region where coannihilation with the $\s \tau$ plays an
important role.  Point LCC4 is chosen in a region where the $A^0$ resonance
makes an important contribution to the neutralino annihilation cross section.

The four points are intentionally chosen
so that the lightest particles of the supersymmetry spectrum can be observed
at the ILC at its initial center of mass energy of 500 GeV.  This is also
the most probable region of the parameter space, since in this region
the dynamics of supersymmetry can generate electroweak symmetry breaking
without extensive fine-tuning of parameters~\cite{FengMoroiMatchev}.

It will be important to us to have well-justified estimates for the accuracy
with which high-energy physics experiments can measure the supersymmetry
spectrum parameter at these points.  For these reason, we have chosen to 
analyze points in the MSSM parameter space at which simulations
have been carried out to estimate the ability of colliders to measure
parameters of the supersymmetry spectrum~\cite{morebenchi}.
For the point LCC1 or SPS1a, an 
extensive set of simulation studies for both LHC and ILC is described in 
\cite{LHCILC}.  These studies assume a luminosity sample of 300~fb$^{-1}$
and LHC and 500 fb$^{-1}$ for ILC.
They include realistic modeling of 
particle detection and Standard Model backgrounds. 
 For the other LCC points, similarly detailed studies have been 
performed only for the ILC~\cite{otherLHC}.   However, as we will see,
the conclusions of \cite{LHCILC} and other LHC studies such as those 
reported in \cite{ATLAS} can be used to estimate the LHC capabilities at these
points.  The point LCC2 has been studied for the ILC by Alexander, \etal, in 
\cite{Gray}.  The point LCC3 has been studied by Khotilovich, \etal, in 
\cite{Dutta,ourchange}.  
The  analysis of this point requires a precision measurement
of the  $\s\tau$ mass, and the question of how well that can be done has
also been studied at related point by Bambade, \etal\ in \cite{Bambade}.
The point LCC4 has been studied in the ILC environment in 
\cite{BattagliaParis}.  

The reader will note that the point LCC1
gives a value of $\Omega_\chi h^2$ outside
the range preferred by the current data.   We are not troubled by 
this, except to note that the appropriate figure of merit is the 
relative accuracy, rather than the absolute accuracy,
 with which $\Omega_\chi h^2$ can be determined.  Other authors, however,
have been concerned about this and have extrapolated the simulation results
obtained at LCC1 (SPS1a) to a nearby point SPS1a$'$ which, because it 
pushes into the stau coannihilation region, predicts a relic density
$\Omega_\chi h^2 = 0.115$~\cite{SPA}.
  In this paper, we have done our main analysis
at the original point in order to cleanly separate examples in which different
physics determines the relic density.  However, Nojiri, Polesello, and 
Tovey~\cite{NPT} have performed a detailed analysis of the LHC prediction of 
the relic density at SPS1a$'$, and so, for comparison, we will also present
the corresponding results for this point in our framework.

The four points are chosen within the limited subspace of mSUGRA models.
However, when we interpret measurements from the colliders, we will not 
want to assume that the true supersymmetry model lives in this subspace.
There is no
compelling physics argument for the mSUGRA assumptions.  More importantly,
the restriction 
to a four-parameter model induces correlations between parameters
of the supersymmetry spectrum, for example,
$\tan\beta$, $\mu$, and the top squark mass, that should properly be 
considered independent.  Some interesting studies have been performed
in which the capability of the LHC to predict $\Omega_\chi h^2$ has been 
assessed by scanning over the parameter space of 
mSUGRA~\cite{Polesello,BattagliaandIan}.  However, we believe that that
this restriction makes the conclusions excessively optimistic.

Instead, it is our strategy to study the implications of measurements at
the LHC and the ILC by comparing to models over the full 24-dimensional 
parameter space of the MSSM described at the beginning of this section.
The MSSM is also a restricted subspace of the space of all, completely 
general, supersymmetry models.   However, as we will see, the large number
of parameters of the MSSM allows a very wide range of scenarios
to appear, with sufficient independence of parameters that
measurements specific to the particles involved 
are needed to constraint the important physical effects.  We believe that 
the exploration of this space is a reasonable way to estimate 
the capabilities
of colliders to extract model-independent conclusions within the overall
category of supersymmetry models.   Many of the dependences of the
neutralino relic density on individual MSSM parameters are displayed in 
\cite{ABBP}, even though the conclusions are stated within the mSUGRA
framework.  Theoretical errors in the relic density calculation, which
we do not consider in this paper, are also highlighted there.   These 
errors become relevant when the relic density is determined to the 
1\% level.   The 
dependence of the relic density on individual MSSM parameters at several
of the reference points is 
also studied in \cite{BirkedalLCWS}.
Nojiri, Polesello and Tovey have 
recently redone the study \cite{Polesello}
in the context of the full MSSM~\cite{NPT}.  These results can be 
compared directly to ours, and we will do that in Section 4.3.

We conclude this section with a  more technical explanation of our
use of the MSSM parameters.  First, there is ambiguity in how one would
assign values to the 24 MSSM parameters from the output of ISAJET. We do
this as follows:
  We first run
ISAJET with the parameters in Table~\ref{tab:LCCpoints}, and we run 
DarkSUSY at
each point using {\tt roption='isasu'} to use the running Yukawa 
couplings from
ISAJET.  From this,  we extract the various running parameters 
evaluated at the
mass scale $Q = (m(\s t_1)m(\s t_2))^{1/2}$, which ISAJET uses as the mass
scale of supersymmetry.   These include the top, bottom and tau masses 
(MTQ, MBQ, MLQ), the Higgs vacuum expectation values
(VUQ$^2$+VDQ$^2$)$^{1/2}$, the strong coupling constant GSS(3), 
and the Yukawa
couplings GSS(4,5,6). We set the ISAJET parameter NSTEP0 = 10000 to 
improve the accuracy of the RG integration, especially in the focus
point region.
 The 24 MSSM parameters listed above are extracted at
the SUSY scale, with the exception of $m_A$ which we take as the 
physical mass,
$\tan\beta$ which we take as the input value, and $m_3$ which we take as the
physical gluino mass.  We now run the benchmark model again using a 
low energy
treatment in DarkSUSY.  The spectrum is calculated according to the ISAJET
function {\tt ssmass}, from which we extract the neutralino, chargino, 
sfermion,
and Higgs masses and mixings.  Then DarkSUSY computes the relic density and
other quantities.  This is the benchmark point that we use. 
This procedure defines the benchmark point, yielding the reference 
spectra and the values of $\Omega_\chi h^2$ listed in 
Table~\ref{tab:LCCpoints}.  

We note that, because we 
recompute the spectrum from the low-energy parameters as just defined, 
our spectrum
calculation differs from that of the implementation of 
mSUGRA in ISAJET.  Thus, we find 
results for the mass spectrum that differ slightly from those 
obtained by inserting the original mSUGRA parameters into ISAJET.
For 
example, using ISAJET 7.69 directly with the parameters of LCC3 in 
the Table 1 gives a mass
splitting $m(\s\tau_1)-m(\s\chi^0_1) = 12$ GeV, while we find a 
splitting of 10.8~GeV.  This shift, less than 1\% in the $\s\tau$ mass,
makes a significant difference in the neutralino relic density. 
However, we calculate the 
neutralino relic density from the 
spectrum defined by the low-energy parameters, and that calculation
is correct, for that spectrum, to the percent level.

 To describe a point in the more general parameter space of the MSSM,
we compute the spectrum from  ISAJET with {\tt ssmass} using the
new values of the 24 MSSM parameters, taking 
 MTQ, MBQ, MLQ, the Higgs VEV, and the strong coupling constant GSS(3)
to be fixed at the benchmark value.  We take the Yukawa couplings
GSS(4,5,6) scaled by $\tan\beta$ from the benchmark values. 
For example, for LCC3 we take GSS(4) to be the benchmark value times
$((1+\tan^2\beta)/(1+(40)^2))^{1/2}$.  With these choices, we describe
the MSSM with low energy parameters in such a way that a point in our 
parameter space gives a close match to 
the input mSUGRA model.

\subsection{Scanning of parameter space}

We have now reduced the prediction  of the properties of WIMPs
to the following problem:  Given a parameter space of $n$ coordinates $x$, 
and given a set of measurements $m_i$, each with standard deviation 
$\sigma_i$, what is the expectation for the prediction for additional
 observable 
quantities $O_j$ that depend on the parameters $x$?  In Bayesian statistics,
the probability distribution of $x$ is given by the likelihood function
\beq
 d^n x \ \L(x) =  
      d^n x \ \prod_i   \exp\left[ - { ( M_i(x) - m_i)^2\over 2 \sigma_i^2 } 
           \right]  \ .  
\eeq{likelihood}
We have made the assumption of a flat  {\it a priori} distribution 
of the values of $x$, and we have assumed that the distribution of 
measurement errors is Gaussian.  The 
prediction for $O_j(x)$ is then given by the expectation value of this 
function of $x$ in the measure \leqn{likelihood}.  More generally, we can 
consider the distribution of the values of $O_j(x)$ induced by this 
distribution of the parameters $x$.  In the next few sections, we will 
use this method to 
present the distributions of the WIMP relic densities and cross sections
that follow from the constraints imposed by supersymmetry spectrum 
measurements.

A simple way to generate the distributions \leqn{likelihood} is to choose
points $x$ randomly in the parameter space and assign each one the indicated
weight.  However, if some of our measurements are very precise, the Gaussian
distributions in \leqn{likelihood} will be very steep and important points will
be selected only rarely.  A much more effective method for sampling points is
the Markov Chain Monte Carlo (MCMC)
method~\cite{Gilks,Dunkley,BGMCMC,AllanachLester,dATR}.  From an initial
starting configuration $x_0$, we generate a sequence of points $x_I$ by the
following algorithm: Choose a nearby point $x'$. If $\L(x') \geq \L(x_I)$, let
$x_{I+1} = x'$.  If  $\L(x') < \L(x_I)$, set $x_{I+1}$ 
either to $x'$ or to $x_I$, choosing
$x'$ with probability $\L(x')/\L(x_I)$.
 This process satisfies detailed balance and
therefore should equilibrate to an ensemble of points in which each point
appears with the relative weight $\L(x)$.

There are many possible ways to choose the distribution of initial
conditions, step sizes, and criteria for convergence in the MCMC
algorithm.  In our use of this algorithm, we have chosen 
a reference point $X_*$ to be one of the LCC points, we have set the 
central values of measurements $m_i$ in the likelihood function to be the
values $M_i(X_*)$ predicted at this point, and we have taken the initial
point of the chain to be the reference point: $x_0 = X_*$.   In 
choosing step sizes, we have 
attempted to measure the size of the distribution of points $\{ x_I\}$
and to readjust the step size so that it is comparable to this size
in each dimension.   For each reference point, we have generated 25 
independent chains, with 160,000 points $x_I$ per chain, for a total of 4
million points.

We have tested our method by performing MCMC scans
in which the step size has been preset 
(to 1 GeV for the masses of color singlet particles and to 10 GeV
for the masses of colored particles), and 
by performing
flat scans of the parameter space in the vicinity of the reference points. 
The distributions near the reference points are  similar in the three 
methods.
The final errors estimated by 
the MCMC method are larger, but we are convinced that this is due
to the fact that the MCMC scans explore the parameter space more deeply.

The full details of our MCMC algorithm are presented in Appendix A.  It is easy
to obtain apparently excellent but incorrect results with MCMC when the chains
do not come to equilibrium.  In Appendix A, we describe direct tests and
cross-checks that have convinced us that our MCMC chains have correctly
converged to the true likelihood distribution.

\subsection{Parameters and constraints}

Now that we have described the general idea of MCMC scanning, we would
like to present the explicit manner in which we have applied this
algorithm to the scanning of the MSSM parameter space.   We need to 
specify how the variables $x$ with flat {\it a priori} distributions
are related to the MSSM parameters.  If the likelihood function
 \leqn{likelihood} does not constrain a particular parameter, the 
MCMC chain will run to infinity in that direction.  Thus, it is  also 
necessary to constrain the range of the parameters, at least within
some broad limits.  We will now explain the choices that we have made.

For most of the 24 MSSM parameters, we have taken $x$ to be the 
logarithm of the corresponding parameter.  Thus, for example, we have
taken each MSSM soft mass parameter to have the {\it a priori} 
distribution $d m/m$.  The parameters $A_\tau$, $A_b$, $A_t$, and 
$\mu$, and $m_1$ can take either sign, so a pure logarithmic
distribution is 
inappropriate for these cases.  We have taken instead the 
formula  $ A =  {\cal M} \sinh(x)$, with ${\cal M} = 50$ GeV.
This corresponds to an  {\it a priori} distribution $ dA/\sqrt{
 A^2 + {\cal M}^2}$.  When the results of the MCMC cluster in a 
small region, these results are almost independent of the prior 
distribution.  This is true for our scans when the fractional 
error is less than about 30\%.

Given a set of MSSM parameters,
 we have used ISAJET 7.69 to compute the supersymmetry spectrum and
mixing angles, in the manner that we have described at the end of Section 3.2.
The ISAJET computation includes finite one-loop radiative corrections, assuming
the MSSM parameters to be renormalized parameters at the scale $Q$ defined in
that discussion.  When we have required cross sections for supersymmetric
particle production, we have computed these from the ISAJET output parameters
using the tree level formulae.

 For slepton masses and 
color-singlet gaugino masses,  
we have imposed a lower bound
 $m > 100$ GeV; for squark and gluino masses, we have imposed 
$m > 250$ GeV, except that $m(\s t) > 150$ GeV.  These lower limits
were almost always irrelevant, since measurements at the LHC and ILC
would  provide stronger lower bounds.  However, we  also needed to provide 
upper bounds to the parameter space; to do  this, we  restricted all 
mass parameters to be less than 5 TeV in absolute value.
We also imposed the following 
restrictions:  For the parameter $\tan\beta$:  $2 < \tan\beta < 100$;
for the chargino mass:   $m(\chi_1^+) > 125$ GeV, assuming that an 
excess of trilepton events is not observed at the Tevatron. For the 
$A$ parameters, we have used 
\beq  
A_t^2 + \mu^2 < 7.5( m^2(\s t_1) + m^2(\s t_2))  \ , 
\eeq{chargecolor}
as an approximate criterion to forbid charge- and color-breaking vacuum 
states~\cite{KLSegre}.  We imposed no additional bounds on $A_b$ and $A_\tau$
since the top constraint is the most stringent.   As we explained in 
Section 3.2, we set terms in the
soft masses and the $A$ terms that are off-diagonal in flavor equal to 
zero; this corresponds to the assumption of `minimal flavor violation'.
We did not implement the constraint from $b\to s\gamma$, since the predictions
for $b\to s\gamma$ depend on flavor-mixing parameters that we are not including
in our analysis.  In our examples, however, the collider measurements of the
supersymmetry spectrum exclude the region in which a large deviation in $b\to s
\gamma$ is expected in models with minimal flavor violation.

In the results presented in the body of this paper, we further 
restrict the range of $\mu$, $m_1$, and $m_2$ to positive values.
  It is possible to choose $m_2$ to be positive
by convention, but then $m_1$ and $\mu$ can in principle have
either sign.  It is usually very difficult to determine these signs from 
the LHC data.  This is a special case of a more general problem, that 
spectrum constraints from  the LHC often allow a number of interpretations 
in terms of the underlying supersymmetry parameters, related by discrete
interchange operations.   These `discrete ambiguities' have been 
highlighted in \cite{Osland,Thaler}.  We will see many examples of such
ambiguities in the simulation results that we will present below, and we
have studied in detail that our simulations count the multiple solutions
correctly.  By studying scans in which $m_1$ and $\mu$ can take values 
of either
sign, we have concluded that allowing negative signs makes little 
difference to the results that we 
will show for the LHC, and that the ILC can typically distinguish the
solution with positive signs.  We will discuss this issue in more detail
is Appendix B.

At the LHC, cross sections depend mainly on particle masses and so are
useful only as cross-checks on 
the mass determinations.  However, the electroweak cross sections at the
ILC are often sensitive to mixing angles and provide additional 
information about the spectrum parameters.  We have therefore included in 
our sets of measurements a number of ILC cross sections.  The simulation
studies for LCC1, LCC2, and LCC3 include estimates of the error that 
would be obtained on the total cross sections for some of the 
interesting reactions at these points.  

The ILC will have polarized 
electron beams and probably also polarized positron beams.  The cross 
sections have strong polarization-dependence.  With this in mind,
we have considered separately the cross sections for two 
different initial beam configurations: left-handed electron beam polarization
with right-handed positron beam polarization and right-handed electron 
beam polarization with left-handed positron beam polarization, assuming
polarizations of 80\% for the electron beam and 60\% for the positron beam.
(Almost identical results are obtained with 95\% electron polarization and
zero positron polarization.)
For certain processes, especially,
$\ee \to \chi_1^+\chi_1^-$, there is additional information in the 
forward-backward asymmetry.  In these cases, we have divided the statistics
between forward production ($\cos\theta > 0$) and backward production 
($\cos \theta < 0$) and assigned errors to the two cross sections accordingly.

We have computed cross sections ignoring 
beamstrahlung and initial state radiation, and, for the cases with 
forward-backward asymmetries, ignoring possible problems 
in determining $\cos\theta$ from the decay final states.
Where the error  $\Delta\sigma$ was not available from simulation studies, we 
took the estimated error on the total cross section for a given
polarization setting to be 
\beq
    \Delta\sigma/\sigma =  1/( 0.02 \times 250 \ \mbox{fb}^{-1} \times \
 \sigma)^{1/2}\ ,  
\eeq{erroronsig}
that is, the purely statistical error assuming 2\% acceptance and a data
sample of 250 fb$^{-1}$ per polarization state. This estimate is typically
conservative with respect to the results of the simulation studies. 

\subsection{Importance sampling}

It is often useful to apply new constraints on a sample set after the
calculation is complete.  This amounts to adding new terms to the likelihood
function ${\cal L}$.  While the points in the Markov chain are distributed
properly according to the original likelihood function, it is possible to
re-weight them so that they are distributed according to a new likelihood
function ${\cal L}'$.  This procedure is called importance sampling.  If the
two likelihood functions are very different, the reweighted chains will have
poor statistics, so care must be taken.  The basic procedure is very simple.
Rather than counting each point in the chain once in computing statistics,
count it ${\cal L}'/{\cal L}$ times.  Points that are highly unlikely according
to ${\cal L}'$ are thus counted with a very small weight.

We use importance sampling to apply astrophysical constraints to our datasets.
We have considered constraining the relic density to agree with the 
dark matter density as measured from the CMB, and constraining the 
direction cross section to that corresponding to a direct detection signal
and the assumption of a canonical smooth halo model.

For the relic density,
we take the reweighting  factor to be
\begin{equation}
\frac{{\cal L}'}{{\cal L}} = \exp\left(-\frac{(\Omega h^2-
\Omega h^2_{\rm CMB})^2}{2\sigma^2}\right),
\end{equation}
with a maximum value of unity.  In the examples of the following sections, we
have used a 3\% constraint on relic density, corresponding to the design
sensitivity of WMAP.  This illustrates the effect of relic density constraints
without having too harsh an effect on the statistical significance of the
results.

When we
impose constraints from direct detection, we are trying to extract
information about the supersymmetry particle masses. For this,
we must convert the observed number of events $n_{\rm ev}$ into a
cross section.  To do that, we must divide by the effective 
luminosity assumed in the analysis.   In the analyses of the SuperCDMS
experiment described in Section 4.5, the effective integrated luminosities 
for the four benchmark models are
 $L_{\rm DD}=$ 3.80~zb$^{-1}$,
3.60~zb$^{-1}$, 3.12~zb$^{-1}$, 2.79~zb$^{-1}$, respectively.
  It is appropriate
to take different values for the four LCC points because the 
detection efficiency depends on the mass of the WIMP.

We use this information as follows to compute the new likelihood 
function:  The expected number of events is
$\tau = \sigma L_{\rm DD}$. We assume that the detection 
is a Poisson process and
thus
\begin{equation}
{\cal L}'(n_{\rm ev}|\tau)\propto e^{-\tau}\tau^{n_{\rm ev}}.
\end{equation}
We use Bayes' theorem to derive
\begin{equation}
{\cal L}'(\tau|n_{\rm ev})\propto e^{-\tau}\tau^{n_{\rm ev}} P(\tau),
\end{equation}
where here $P(\tau)$ is the prior on $\tau$, which we assume to be constant.
The reweighting factor can thus be taken as
\begin{equation}
\frac{{\cal L}'}{{\cal L}} = e^{n_{\rm ev}-\tau}\left(\frac{\tau}{n_{\rm ev}}
\right)^{n_{\rm ev}},
\end{equation}
which again has a maximum value of unity.

\section{Benchmark point LCC1}

In this and the next few sections, we describe the predictions for 
the properties of the neutralino WIMP that would follow from collider
measurements for the four LCC reference points.  We will present the 
qualitative features of the analysis at each point and display relevant
projections of the MCMC data.  Our final quantitative results for 
the predictions of $\Omega_\chi h^2$ and other WIMP properties are 
collected in Tables~\ref{tab:HEPresults} and  \ref{tab:Astroresults}
in Section 9.

\begin{figure}
\begin{center}
\epsfig{file=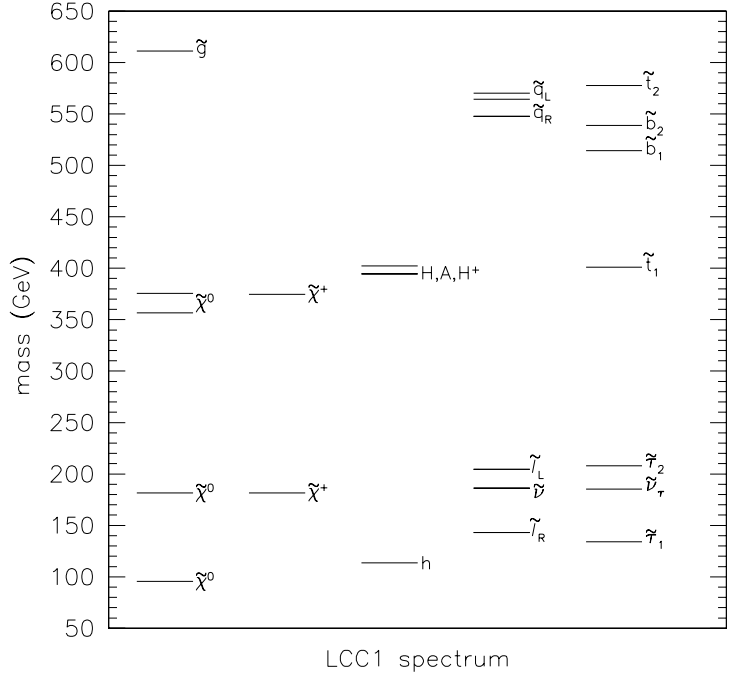,height=4.0in}
\caption{Particle spectrum for point LCC1.  The lightest neutralino is
predominantly bino, the second lightest neutralino and light chargino are
predominantly wino, while the heaviest two neutralinos and heavy chargino are
predominantly Higgsino.}
\label{fig:LCC1spectrum}
\end{center}
\end{figure}

We begin with the point
LCC1.  The supersymmetry spectrum at this point is shown in 
Fig.~\ref{fig:LCC1spectrum}. The model contains light sleptons, with masses
just above the mass of the lightest neutralino. 
The most important annihilation reactions 
for determining the relic density are 
those with $t$-channel slepton exchange.  To predict the relic density, 
we will need to make accurate measurements of the masses of the sleptons,
 including the $\s \tau$. 

Many other effects must be controlled to obtain the cross sections relevant
to WIMP detection.  The direct detection cross section is dominated by 
$H^0$ boson exchange, and the annihilation cross section at zero energy, which 
enters the rates for gamma ray and positron observations, is dominated by 
annihilation to $b\bar b$.  

At the same time that we are gathering data
to evaluate the dominant contributions to the WIMP
 cross sections, we must also 
gather data to prove that possible competing mechanisms from 
Fig.~\ref{fig:NNann} are truly subdominant.  In particular, we should 
use the data to show that neutralino annihilation does not take place 
close to an $A^0$ resonance, and that the wino and Higgsino content of
the lightest neutralino is
small enough that annihilation to vector bosons is not an important
process.  We will see that all of these factors can be controlled
from the prospective collider data for the point LCC1.

\subsection{Spectroscopy measurements}

The basic data for determining the properties of the neutralino WIMP come
from collider measurements of the SUSY spectroscopy.  The 
measurements we assume are summarized in  Tables~\ref{tab:LCC1masses}
and~\ref{tab:LCC1css}.   We will present similar tables for each 
of the four reference models. 
In general, we fix the central values for each observable to be 
equal to the prediction from the underlying model and fix the error
(1 standard deviation)
in accordance with the results of a detailed simulation study of the 
ability of the LHC and ILC detectors to measured the relevant 
SUSY masses, mass differences, or cross sections.
In Sections 4.1, 5.1, 6.1, and 7.1, we will briefly 
summarize the important experimental issues. However, for full details,
the reader should  go to the original source.  For the data in 
Tables~\ref{tab:LCC1masses}
and~\ref{tab:LCC1css}, the error estimates come from the studies 
reported in~\cite{LHCILC}.

\begin{table}
\centering
\begin{tabular}{lccccc}
   mass/mass splitting  & LCC1 Value & &  LHC  & ILC 500 &  ILC 1000\\ 
         \hline
  $m(\s\chi^0_1)$    &    95.5      & $\pm$ &    4.8   &   0.05      \\ 
  $m(\s\chi^0_2) - m(\s\chi^0_1)$ & 86.1 & $\pm $ &   1.2  &   0.07  \\
  $m(\s\chi^0_3) - m(\s\chi^0_1)$ &  261.2   & $\pm $ &    @$^a$  &   4.0 \\ 
  $m(\s\chi^0_4) - m(\s\chi^0_1)$ &  280.1   &  $\pm $ &   2.2$^a$ &   2.2 \\ 
  $m(\s\chi^+_1)     $ &  181.7            &  $\pm $ &  -  &   0.55 \\ 
  $m(\s\chi^+_2)     $ &  374.7             &  $\pm$ &   -  &    -   & 3.0 \\
       \hline
  $m(\s e_R)     $     &    143.1          &  $\pm$ &   -  &   0.05 \\ 
  $m(\s e_R) - m(\s\chi^0_1)$  &   47.6        & $\pm $ &   1.0  &   0.2  \\
  $m(\s \mu_R) - m(\s\chi^0_1)$   &  47.5        & $\pm $ &  1.0  &  0.2 \\ 
  $m(\s \tau_1) - m(\s\chi^0_1)$   & 38.6       &  $\pm $ &   5.0 &  0.3 \\ 
  $BR(\s\chi^0_2\to \s e e)/BR(\s\chi^0_2\to \s\tau \tau)$  &  0.077 & $\pm$ &
              0.008 &  &   \\ 
  $m(\s e_L) - m(\s\chi^0_1)$    &    109.1     & $\pm $ &   1.2  &   0.2  \\
  $m(\s \mu_L) - m(\s\chi^0_1)$  &    109.1     & $\pm $ &   1.2  &   1.0 \\ 
  $m(\s \tau_2) - m(\s\chi^0_1)$  &   112.3      &  $\pm $ &   - &   1.1 \\ 
  $m(\s \nu_e)$                 &    186.2           & $\pm $ & - & 1.2 \\
        \hline
  $m(h)     $    &   113.68              &  $\pm $ & 0.25  &   0.05 \\ 
  $m(A)     $   &     394.4        &  $\pm$ &   *  &   ($> 240$)     & 1.5 \\
       \hline
  $m(\s u_R)$, $m(\s d_R)$ &   548.        & $\pm $ &   19.0  &   16.0  \\
  $m(\s s_R)$, $m(\s c_R)$ &   548.        & $\pm $ &   19.0  &   16.0  \\
  $m(\s u_L)$, $m(\s d_L)$ &   564., 570.   & $\pm $ &   17.4  &   9.8  \\
  $m(\s s_L)$, $m(\s c_L)$ &   570., 564.     & $\pm $ &   17.4  &   9.8  \\
  $m(\s b_1)     $ &   514.               &  $\pm $ &    7.5  &   5.7 \\ 
  $m(\s b_2)     $ &   539.               &  $\pm$ &     7.9 &    6.2   \\
  $m(\s t_1)     $ &  401.      &  $\pm $ &  ($ > 270$)   & - & 2.0 \\ 
        \hline
  $m(\s g)$      &   611.       &  $\pm $ &   8.0     &   6.5 \\ 
\end{tabular}
\caption{Superparticle masses and their estimated errors or lower limits for 
       the parameter point LCC1. Lower limits are indicated in 
 parentheses. The ILC columns contain the 
   measurements added or improved by the ILC at that energy.  The symbol
   `-' denotes that the measurement is not yet available.  The symbol `*'
   denotes the formula:  $m_A > 200$ GeV, or $\tan \beta <  7.0 (m_A/200.0)$.
  The notation `@$^a$' indicates that the mass measurement marked with a 
superscript $^a$ could equally well be ascribed to this particle.
   All values are quoted in GeV.}
\label{tab:LCC1masses}
\end{table}

\begin{table}
\centering
\begin{tabular}{lcccrr}
   cross section
    &  & LCC1 Value (fb) & &  ILC 500 &  ILC 1000\\
         \hline
  $\sigma(\ee\to \s\chi^+_1\s\chi^-_1)$
      & LR & 431.5 (0.758)       & $\pm $  &    1.1\%$^*$ \\
     & RL & 13.1 (0.711)               & $\pm $    &    3.5\%$^*$ \\
  $\sigma(\ee\to \s\chi^0_1\s\chi^0_2)$
      & LR & 172.2                 & $\pm $   &    2.1\%$^*$ \\
     & RL & 20.6                 & $\pm $    &    7.5\%$^*$ \\
  $\sigma(\ee\to \s\chi^0_2\s\chi^0_2)$
      & LR & 189.9                 & $\pm $  &    2.0\%$^*$ \\
     & RL &  5.3                & $\pm $    &    10.2\%$^*$ \\
  $\sigma(\ee\to \s\tau^+_1\s\tau^-_1)$
      & LR & 45.6                 & $\pm $   &    7\% \\
     & RL & 142.1                 & $\pm $    &    4\% \\
  $\sigma(\ee\to \s e^+_R\s e^-_R)$
      & LR &  57.3 (0.696)          & $\pm $  &    6\% \\
     & RL &  879.9 (0.960)         & $\pm $   &    1.5\% \\
  $\sigma(\ee\to \s t_1\s {\bar t}_1)$
      & LR & 9.8                 & $\pm $   &  &  15\% \\
     & RL & 11.1                 & $\pm $   &   &  14\% \\
         \hline
\end{tabular}
\caption{SUSY cross sections and estimated errors for 
      the parameter point LCC1. The ILC columns indicate the center of
     mass energy of the measurement. All cross sections assume polarized
     beams.
 LR denotes left-handed $e^-$ polarization and 
     right-handed $e^+$ polarization; RL denotes the reverse. 
Cross section  values are quoted in fb; the error are expressed
   as a percentage of the value.   For chargino and selectron production
cross
      sections, the forward fraction is listed in parentheses.
 The errors labeled by $^*$ are
taken from the study \cite{LHCILC}; the others are estimated using
      \leqn{erroronsig}.}
\label{tab:LCC1css}
\end{table}

The point LCC1 is special in that it allows a large number of SUSY spectrum
parameters to be determined at the LHC from kinematic constraints.  From
Fig.~\ref{fig:LCC1spectrum}, we can see that the 
lighter sleptons have masses between  the masses of the $\s\chi^0_2$ and 
the $\s\chi^0_1$.  In this case, the decays of the $\s\chi^0_2$ will be 
dominated by 2-body decays to $\s \ell \ell$, followed by the 2-body 
decays $\s \ell \to \s\chi^0_1 \ell$.  The kinematic endpoints of the 
lepton spectra in these decays are simple functions of the SUSY particle 
masses.  The system is overconstrained, so it is possible to solve for the 
masses of the slepton and the two lightest neutralinos~\cite{PaigeHinch}. 
 The values of 
the heavier superparticle masses can be built up from this 
information.  Through this technique, the mass of the WIMP
can be determined to about 5\% accuracy, and the mass differences of the 
lighter neutralinos and sleptons can be measured to a 
few GeV.  

The 
squarks that are partners of left-handed quarks have large decay branching
fractions to $\s\chi^0_2$, so these mass differences can also be determined
quite well.  The partners of right-handed quarks can be studied using the
direct decay to $\s\chi^0_1$.   A similar analysis allows one to 
identify the right- and left-handed selectron and smuon 
states~\cite{LHCILC}.   For 
stau, one mass eigenstate is seen, and this is a mixture of $\s\tau_L$
and $\s\tau_R$.  The branching ratio of $\s \chi^0_2$ to $\s\tau$ is 
enhanced with respect to that to $\s e$, $\s \mu$ by the amplitude
for the wino component of $\s\chi^0_2$ to decay to the $\s\tau_L$ component
of $\s\tau_1$.  Thus, the measurement of this branching ratio gives 
information on the stau mixing angle~\cite{NPT}.

In the study \cite{LHCILC}, it was not possible to identify the 
top squark at the LHC.  However, this particle would have been visible 
through its decay $\s t_1 \to W^+ b \s \chi^0_1$ if the decay to 
$t \s\chi^0_1$ were kinematically forbidden.  We therefore place a limit
$m(\s t_1) > m_t + m(\s \chi^0_1)$.

The light Higgs boson in the model has a mass of  114 GeV and properties
very similar to those of
 the Standard Model Higgs boson.  It can be observed at the LHC in its
decay to $\gamma\gamma$.  This particle plays a relatively small
role in the properties of the WIMP.  However, the 
heavy Higgs boson $A^0$ can have a more important effect, since it 
potentially gives a resonant enhancement of the annihilation cross section.
 It is very significant, then, that
the LHC can constrain the properties of the $A^0$ boson.  If the $A^0$
is sufficiently light or if $\tan\beta$ is sufficiently large, the $A^0$
can be discovered at the LHC in its decays to $\tau^+\tau^-$ or (in a 
slightly smaller region) to $\mu^+\mu^-$.  At LCC1, the $A^0$ boson cannot be
discovered at the LHC, but the failure to observe this particle 
constrains the region in which the $A^0$ boson plays a role in the 
WIMP properties. 

At the ILC, we expect very precise mass measurements for all particles that
can be pair-produced at the machine's center of mass energy.  Measurements
of particle masses can come from kinematic fitting of pair-production reactions
and from dedicated threshold energy scans~\cite{LHCILC}.   
In Table~\ref{tab:LCC1masses}, mass measurements from kinematic fitting 
are given as mass differences from the lightest neutralino, and mass
measurements from threshold determinations are indicated as absolute mass
measurements.  Because the masses of the light charginos and neutralinos
enter the expressions for endpoints of kinematic distributions at the LHC,
the determination of squark and gluino masses are improved by the ILC 
data even though these particles are not observed directly at the ILC.  This
effect is reflected in Table~\ref{tab:LCC1masses}. 

The extension of the ILC to center of mass energies of 1000 GeV allows us
to observe several particles that were not accessible at the first stage
of 500 GeV.  These include the top squark and the second chargino.
A  particularly important addition is that of the $H^0$ and $A^0$ bosons, 
since the 
process $\ee\to H^0 A^0$ becomes kinematically
accessible above 900 GeV. We estimate
the accuracy 
of the $H^0$ and $A^0$ mass determinations from the results of the 
study \cite{BattagliaParis}.   This sharpened knowledge of the masses
of $A^0$ and $H^0$ will have an important effect on the
determination of several of the WIMP properties.

\subsection{Relic density}

\begin{figure}
\begin{center}
\epsfig{file=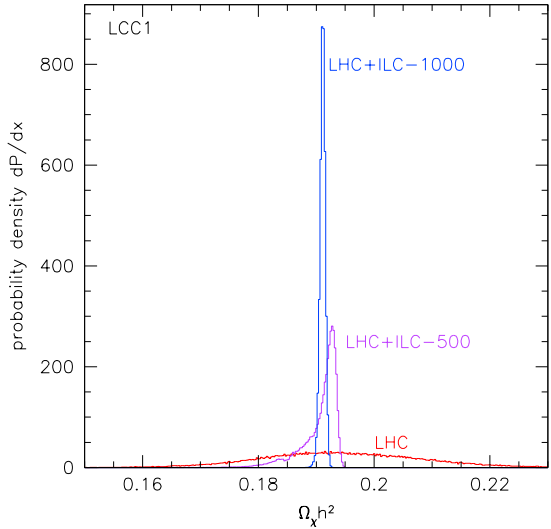,height=3.0in}
\caption{Relic density measurement for point LCC1.  Histograms 
in this and all
following figures give the probability distribution  $dP/dx$
of the quantity on the
$x$-axis, given the three different sets of accelerator constraints.
Where the $x$ axis is plotted logarithmically, the probability plotted is
actually $dP/d\log_{10} x$.  All histograms integrate to unity.  Results
for the LHC make use of the assumption that the underlying physics model
is supersymmetry.  This might not be clear from the LHC data
alone.}
\label{fig:LCC1relic}
\end{center}
\end{figure}

Now we can put the constraints shown in Tables~\ref{tab:LCC1masses} and
\ref{tab:LCC1css} into the machinery described in Section 3.2 to provide 
predictions of WIMP dark matter properties.  We begin with the WIMP relic
density, as computed from the spectrum using DarkSUSY 4.1.  To
see what the various data sets predict for this quantity, we generate
 the three likelihood distributions corresponding to the  three
sets of constraints in the Tables, including the cross section for the
two ILC points.   In  
Fig.~\ref{fig:LCC1relic}, we show the projections of these three likelihood
distributions onto the coordinate representing the relic density 
$\Omega_\chi h^2$.  The LHC data already gives a quantitative prediction 
for this density, with an accuracy of about 7\%.  The data from the 500 
GeV ILC sharpens this distribution to 2\% accuracy, though the
distribution has a long
tail toward lower values.  
  The extension 
of the ILC to 1000 GeV further sharpens the prediction to about 0.25\%
accuracy, comparable to the accuracy expected from the Planck CMB 
measurement.
 Comparison  of  two such accurate values of 
the dark matter relic density would provide a striking test of the WIMP
model, as we have noted already in Section 2.5.

\begin{figure}
\begin{center}
\epsfig{file=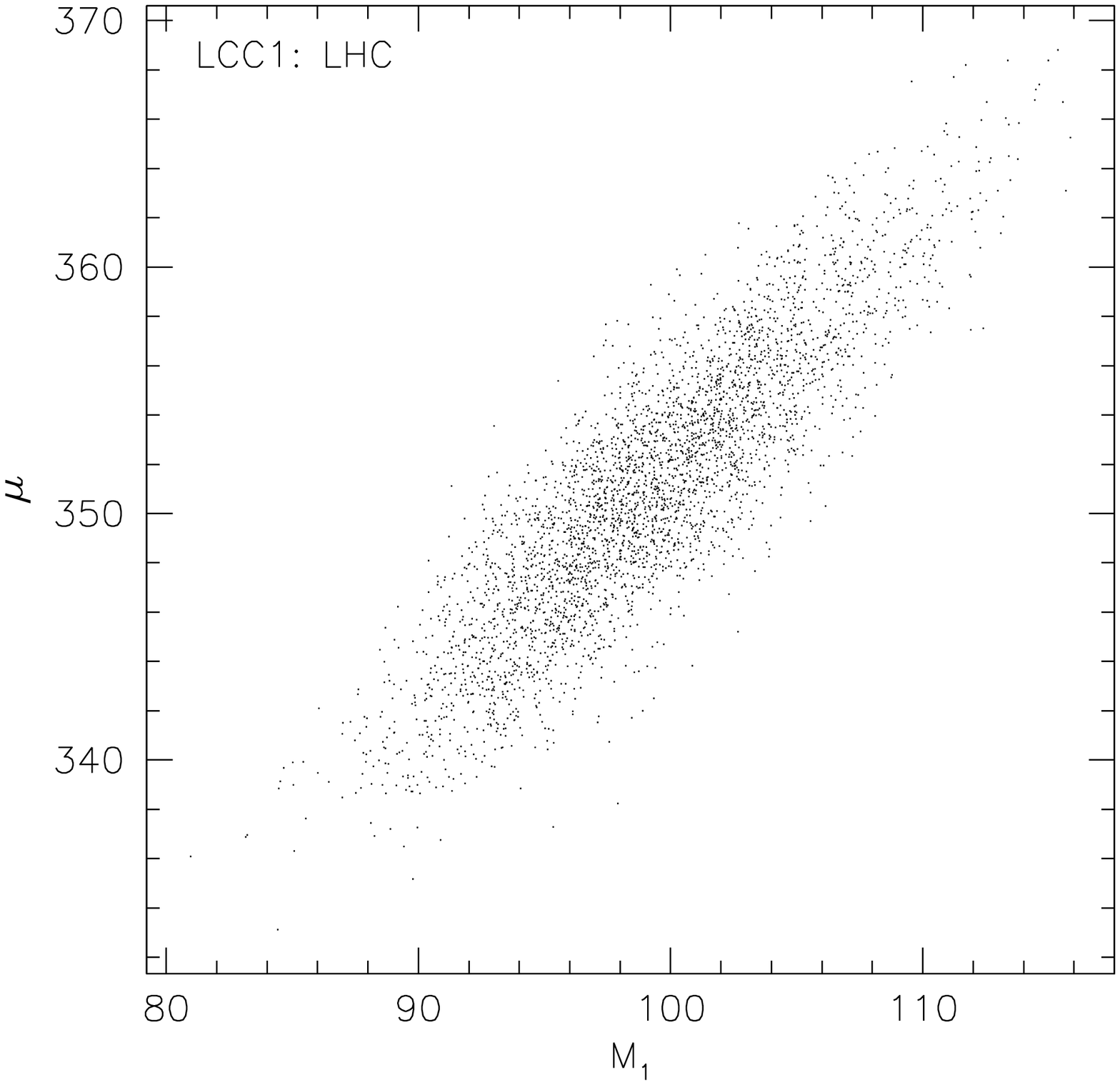,height=2.5in}
\epsfig{file=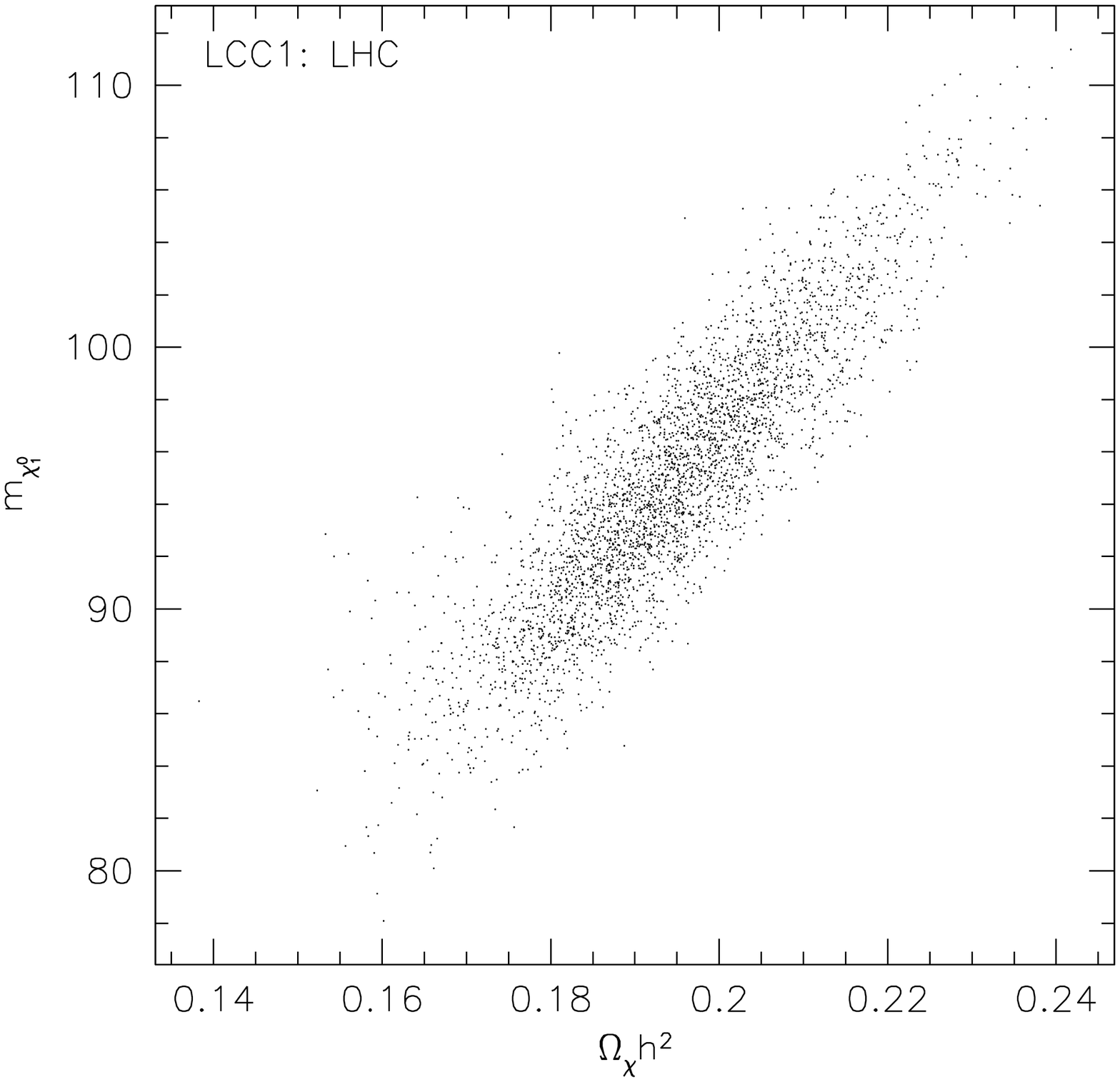,height=2.5in}\\
\epsfig{file=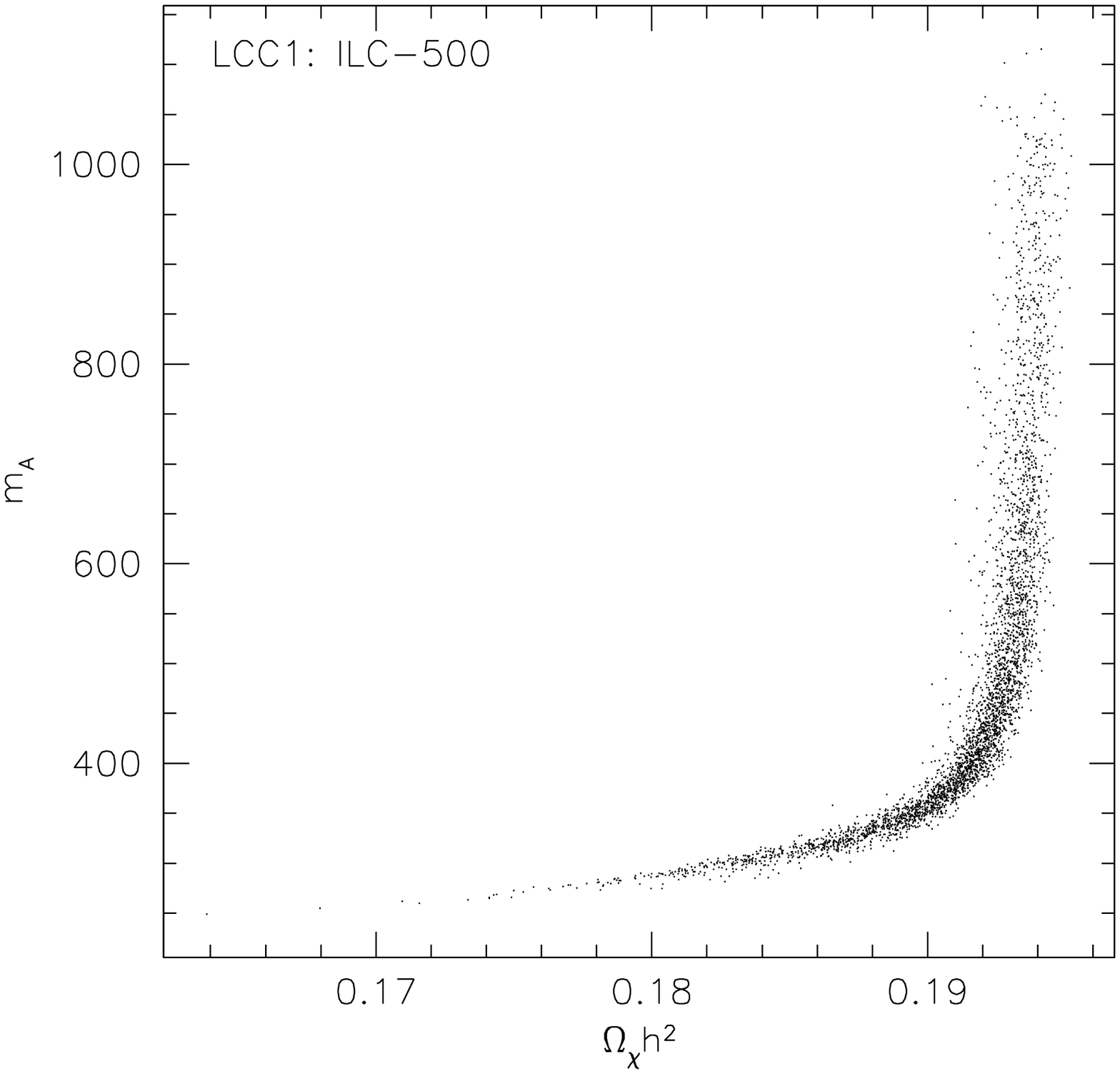,height=2.5in}
\caption{Scatter plots for point LCC1.  (a) Inferred mass parameters $m_1$ and
$\mu$ are shown to be well correlated in LHC data, with an uncertain overall
scale.  This reflects the fact that mass differences are measured more
precisely than absolute masses.  (b) The overall scale has a strong correlation
with relic density.
(c) For ILC-500, the mass of the $A^0$ boson
is still unknown.  For low values, there is a mild resonant enhancement to the
neutralino annihilation cross section, and a corresponding reduction in relic
density.}
\label{fig:LCC1relicscatter}
\end{center}
\end{figure}

One might have suspected that the large number of constraints from the 
LHC would already produce a very precise value of the relic density.
However, it is clear from the spread of the LHC predictions in 
Fig.~\ref{fig:LCC1relic} that some information is missing.  The scatter 
plots  in Fig.~\ref{fig:LCC1relicscatter}(a,b) show one of the effects 
that contributes.  Though the mass differences of the supersymmetry
particles are well fixed, the overall scale of masses is somewhat
uncertain.  This translates into the dominant uncertainty in the 
prediction of the relic density.

The tail in the  $\Omega_\chi h^2$ distribution from  
the 500 GeV ILC is also a surprise.  Its explanation is given in 
Fig.~\ref{fig:LCC1relicscatter}(c), which shows the correlation between 
$\Omega_\chi h^2$  and the mass of the $A^0$ boson in this likelihood 
function.  The $s$-channel process involving the $A^0$ boson actually 
makes a small contribution to the annihilation cross section at LCC1, and
this must be fixed to determine the cross section to a level below 1\%.
  When we learn the mass of the $A^0$ at the 1000 GeV
ILC, this uncertainty in the prediction of $\Omega_\chi h^2$ is 
removed.

\subsection{Relic density at SPS1a$'$}

Nojiri, Polesello, and Tovey (NPT)~\cite{NPT} have also 
estimated the capability of the LHC data to predict the neutralino 
relic density.  Their analysis was done at a point quite similar, but 
not identical to, LCC1.  By adjusting the parameters of our simulation, we
can make a direct comparison of our results to those of NPT. In this
section, we will describe our analysis of the point considered in \cite{NPT}
and show that our results are in good agreement.  Since NPT used a different
method to arrive at their estimate, this comparison provides a nontrivial
check of the two methods.

The value of $\Omega_\chi h^2$ at LCC1 is about a factor of 2 too large to 
be consistent with the CMB result \cite{cosmoreview}.  It has 
been suggested that one could
remedy this problem by moving the mass of the $\s\tau_1$ from 134 GeV at
LCC1 to 108 GeV, pushing the model into the stau coannihilation region.
The new point is called SPS1a$'$~\cite{SPA}. The shift of the $\s\tau_1$ mass
adds contributions to the annihilation cross section for supersymmetry
and therefore decreases the relic density.  The neutralino physics at this 
point is intermediate between that at LCC1 and LCC3.  Because coannihilation
is important, the point has increased sensitivity to the value of the 
$\s\tau_1$ mass, though this sensitivity is not as extreme as we will see in 
Section 6 in our discussion of LCC3.  

\begin{figure}
\begin{center}
\epsfig{file=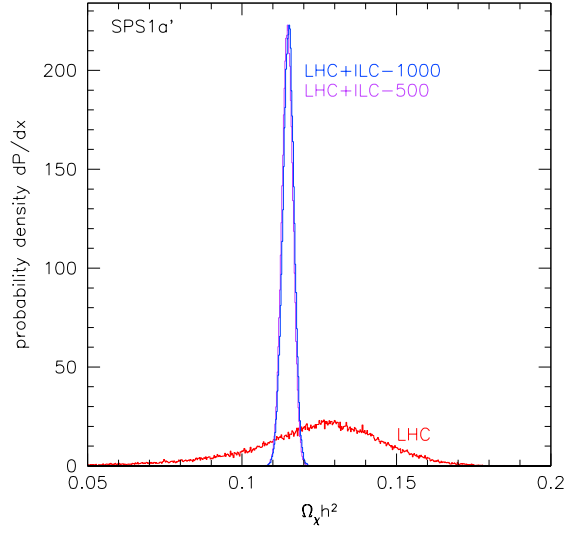,height=3.0in}
\caption{Relic density for point SPS1a$'$.  See Fig.~\ref{fig:LCC1relic} for
description of histograms.  In this case, the ILC-1000 provides no improvement
in the relic density measurement over ILC-500.}
\label{fig:LCC5relic}
\end{center}
\end{figure}

According to~\cite{SPA}, the table of measurement errors
at SPS1a$'$ should be the same as that of LCC1 (Tables 2 and 3), even though
the expected values are somewhat different.  We have therefore carried out
MCMC scans for SPS1a$'$ using the same constraints as for SPS1a (LCC1), only 
modifying the reference values of supersymmetry masses and cross sections.
  The results of 
that scan for the relic density are shown in Fig.~\ref{fig:LCC5relic}.   

The analysis of NPT ran as follows:  In the stau coannihilation region,
the neutralino relic density is mainly sensitive to 7 of the 24 MSSM
parameters, the four parameters of the neutralino mass matrix $m_1$, $m_2$, 
$\mu$, and $\tan\beta$ and the three parameters of the stau mass
matrix, which we can represent as the combinations $m(\s\tau_1)$, 
$m(\s\tau_2)$, $\theta_\tau$.  The LHC data is expected to give relatively
precise values for five quantities sensitive to these parameters, three
neutralino masses, the mass of the $\s\tau_1$, and the ratio of 
branching ratios $BR(\s\chi^0_2 \to e\s e)/BR(\s\chi^0_2\to \tau \s\tau_1)$.
At the true point, the gaugino-Higgsino and stau mixing angles are 
relatively small, so the actual dependence of the relic density 
on the mixing angles (and thus on 
$\tan\beta$) is relatively weak.  The dependence on $m(\s\tau_2)$ is also 
weak.  So it makes sense in this case to solve for the five parameters
$m_1$, $m_2$, $\mu$, $m(\s\tau_1)$, and $\theta_\tau$
from the measurements, propagate the measurement errors through to an
error on the prediction of $\Omega_\chi h^2$, and then add an estimate of
the additional uncertainty from variation of $\tan\beta$ and $m(\s\tau_2)$.
In solving for the five parameters, NPT chose the preferred solution and did 
not take account of the possibility of multiple solutions.

The final error on the relic density turned out to be dominated by the 
uncertainty in the mass of of the $\s\tau_1$.  This is tied to the error 
in measuring the endpoint of the dilepton mass spectrum in the decay 
$\s\chi^0_2 \to \tau^+\tau^- \s\chi^0_1$.  For the case of tau leptons, 
the endpoint is not a sharp feature and it has not been understood in detail
how well this point can be measured.  For the purpose of their analysis,
Polesello and Tovey, who are two of the experts on such measurements in 
the ATLAS collaboration, estimated that the error would be below 5 GeV
and expressed the hope that it could be brought down to 1 GeV.  This is
considerably more aggressive than the uncertainty presented in 
\cite{LHCILC}, which we have assumed in 
our study.

\begin{figure}
\begin{center}
\epsfig{file=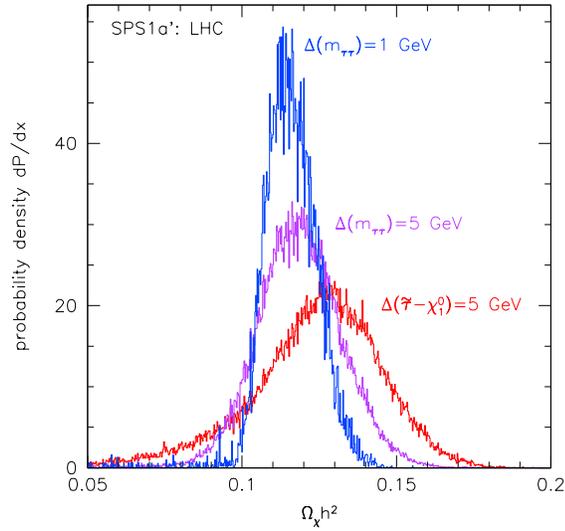,height=3.0in}\\
\caption{Relic density for point SPS1a$'$ at LHC.  A dominant uncertainty at
this point is the mass of the stau.  Results of taking the stau-neutralino mass
difference to have an error of 5 GeV are illustrated in red.  Alternatively,
results of taking the position of the tau-tau edge to have an error of 5 GeV
(purple) and 1 GeV (blue) are also illustrated.}
\label{fig:LCC5LHCrelic}
\end{center}
\end{figure}

\begin{figure}
\begin{center}
\epsfig{file=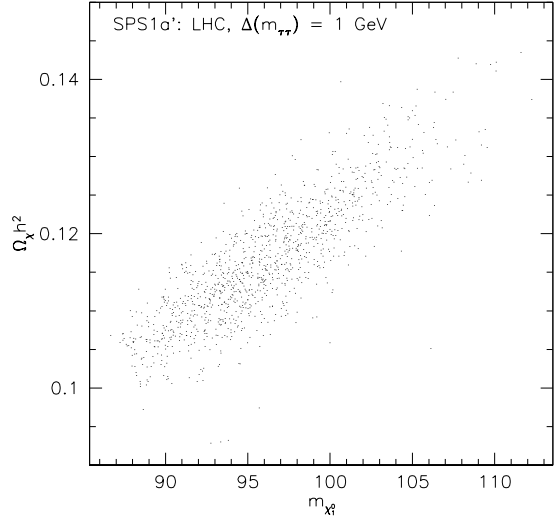,height=3.0in}
\caption{Scatter plot for point SPS1a$'$.  Relic density is plotted against
neutralino mass, for points where the position of the tau-tau edge is measured
to 1 GeV.}
\label{fig:LCC5Omm1scatter}
\end{center}
\end{figure}

In Fig.~\ref{fig:LCC5LHCrelic}, we show the histograms of $\Omega_\chi h^2$
from the LHC constraints
for our scan with parameters based on \cite{LHCILC} 
and imposing the stronger constraints that the
$\tau\tau$ endpoint can be measured to 5 GeV or to 1 GeV.  From the 
figure, it is clear that NPT are correct that this is the dominant source
of error.  The standard deviations for the three histograms correspond to
\beq
          \sigma(\Omega h^2)/\Omega h^2 =  18.6\%, \ 14.4\%, \ 11.7\%
\eeq{oursigmas}
with the last two cases to be compared to the results of NPT,
\beq
          \sigma(\Omega h^2)/\Omega h^2 =  19.\% \ , 10.5\%
\eeq{theirsigmas}
Their results correspond to a scan over 5 parameters, while our 
results scan over the full set of 24 MSSM parameters.  We thus verify that
the analysis of NPT does capture the strongest dependences of $\Omega h^2$
on the MSSM parameters, and also that our method for estimating the 
uncertainty in $\Omega_\chi h^2$ does agree with a more direct method in this
case where $\Omega_\chi h^2$ is strongly constrained.  In 
Fig.~\ref{fig:LCC5Omm1scatter}, we show 
the scatter plot of $\Omega_\chi h^2$ vs. $m(\s\chi^0_1)$ from 
our scan for the case of a 1 GeV measurement of the $\tau\tau$ endpoint.
This is, again, a scan over 24 parameters.  It is interesting to compare this
to Fig.~9 of NPT, which gives the result of a 5-parameter scan.

\begin{figure}
\begin{center}
\epsfig{file=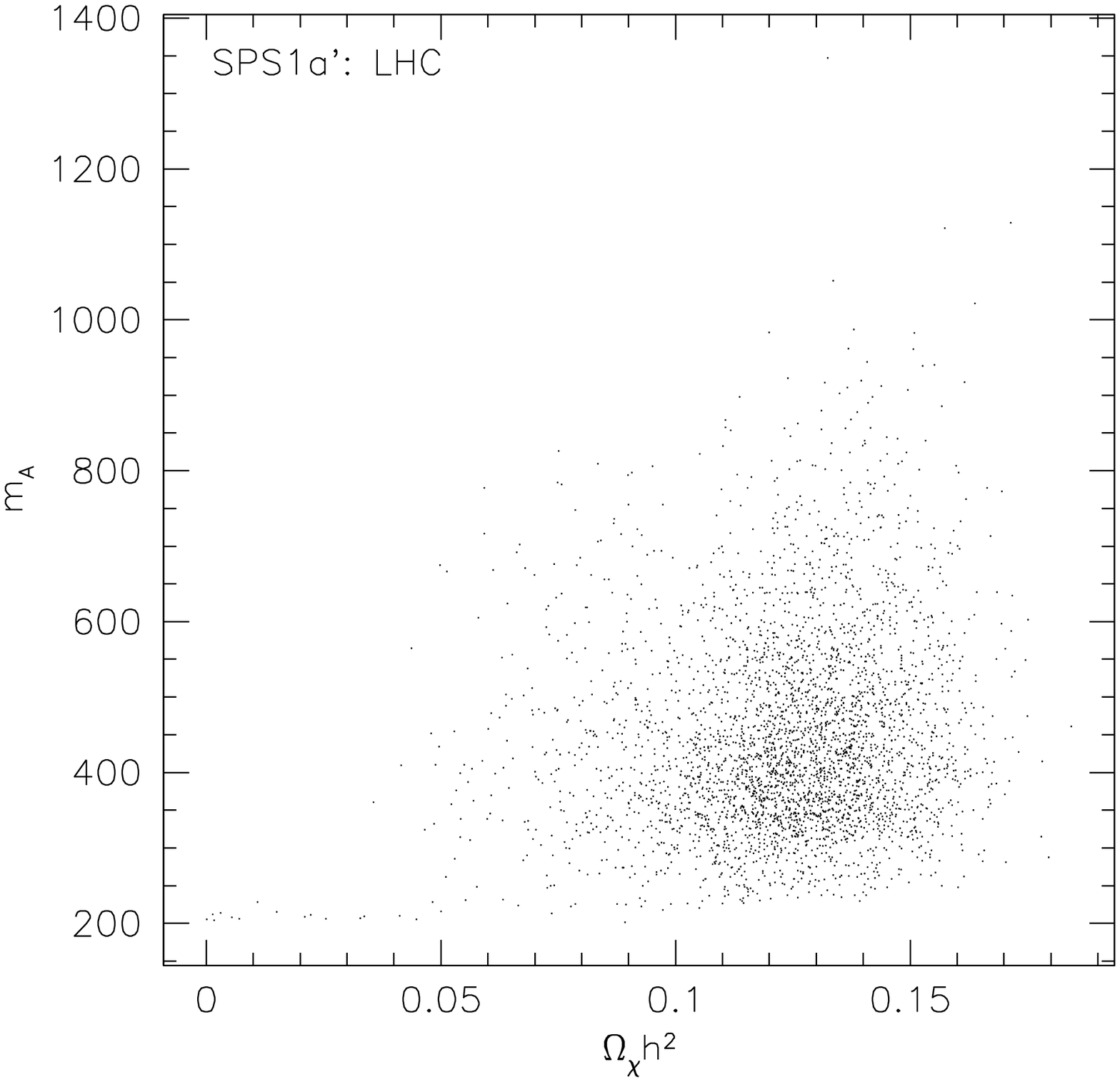,height=3.0in}
\caption{Scatter plot for point SPS1a$'$.  The $A^0$ mass is plotted against
relic density, illustrating the small allowed region where $A^0$ resonance
significantly decreases the relic density.}
\label{fig:LCC5scatters}
\end{center}
\end{figure}

Given the very strong constraints on the spectrum at SPS1a$'$, the 
ambiguities in the solutions play a rather small role.  in particular,
the constraint on the ratio of branching ratios of the $\s\chi^0_2$ almost
completely eliminates the solution in which the lighter $\s\tau$ is 
dominantly $\s\tau_L$.  Our data contains only a small influence  of the 
region in which the $A$  boson provides a resonant enhancement of
the neutralino annihilation.
Fig.~\ref{fig:LCC5scatters} shows a scatter plot $\Omega_\chi h^2$ vs
$m(A)$ for the SPS1a$'$ LHC data sample.
  We see a small branch leading to low $m(A)$
and low $\Omega_\chi h^2$, representing the effect of an 
$A$ boson resonance near the annihilation threshold.

\subsection{Annihilation cross section}

\begin{figure}
\begin{center}
\epsfig{file=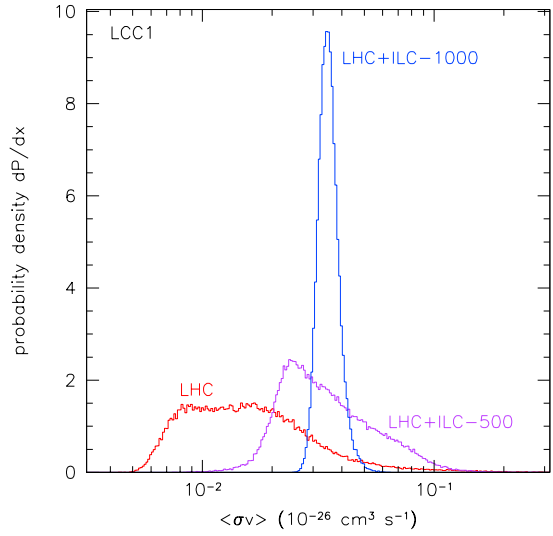,height=3.0in}
\caption{Annihilation cross section at threshold for point LCC1.  See
Fig.~\ref{fig:LCC1relic} for description of histograms.}
\label{fig:LCC1anncs}
\end{center}
\end{figure}

In Fig.~\ref{fig:LCC1anncs}, we show the likelihood distribution of the
neutralino pair annihilation cross section times velocity
$\sigma_{\chi\chi} v$, evaluated
at threshold.  As we have discussed in Section 2.7, this cross section 
is needed to interpret astrophysical signals of dark matter
annihilation.  We will discuss the determination of this quantity
from a more general point of view in Section 8.

At LCC1, the most important neutralino annihilation reactions are P-wave
processes whose cross section vanishes at threshold.   The important 
contributions to the threshold cross sections then come from annihilation 
to massive fermion pairs, $\tau^+\tau^-$ and $b\bar b$, with the latter 
final state dominating.  The most important parameters for $\s\chi^0_1
\s \chi^0_1 \to b\bar b$ are the $\s b $ masses; at LCC1, these
masses would be determined at the LHC.  The ILC at 500 GeV  determines the 
neutralino mixing angles, and this makes a small improvement in the 
prediction.  The marked improvement from the ILC at 1000 GeV comes from 
the determination of the $A^0$ mass.  Note that, because of the 
suppression of the P-wave channels, this annihilation cross section is
about 12 fb, as opposed to the 0.9 pb predicted by \leqn{findsigmav}.

\begin{figure}
\begin{center}
\epsfig{file=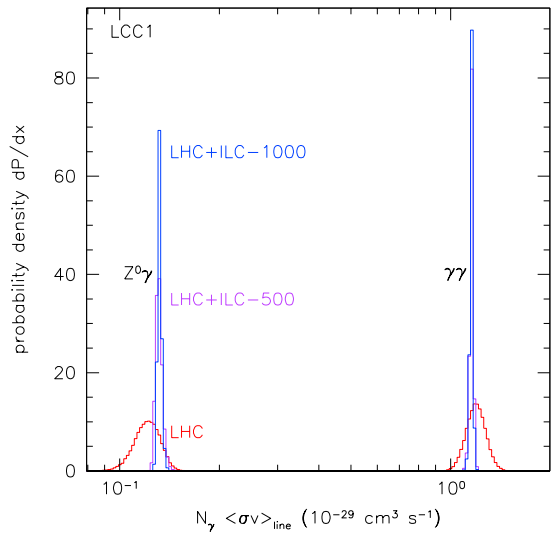,height=3.0in}
\caption{Gamma ray line annihilation cross section at threshold for point LCC1.
See Fig.~\ref{fig:LCC1relic} for description of histograms.}
\label{fig:LCC1twogamma}
\end{center}
\end{figure}

In  Fig.~\ref{fig:LCC1twogamma}, we show the likelihood distribution for the 
exclusive annihilation cross sections to $\gamma\gamma$ and $\gamma Z^0$.
These cross sections are dominated by squark loops and are determined 
quite well already from LHC data.  The cross sections are small, but, at 
LCC1, these processes might be visible as sharp lines in the gamma ray
spectrum.

\begin{figure}
\begin{center}
\epsfig{file=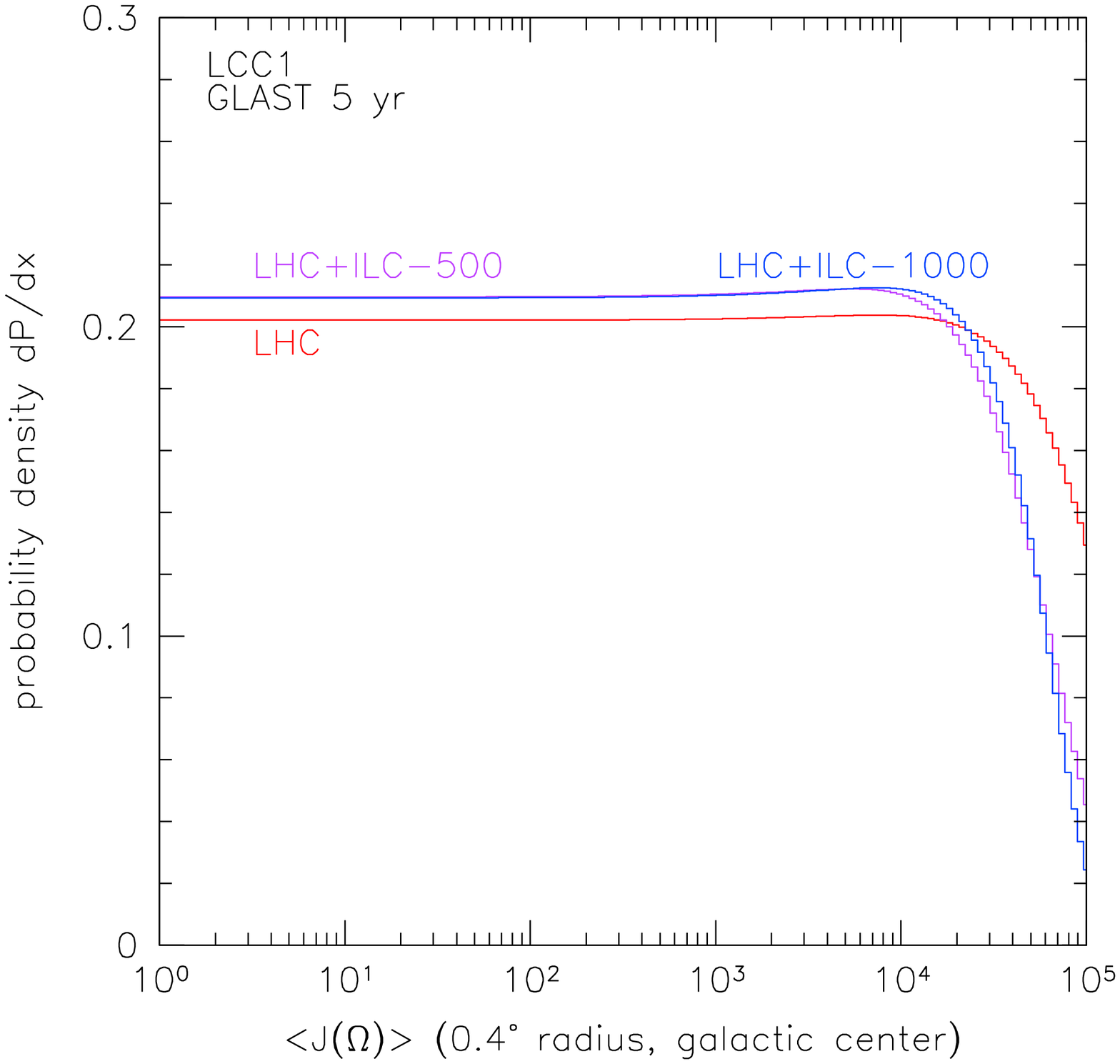,height=3.0in}\\
\epsfig{file=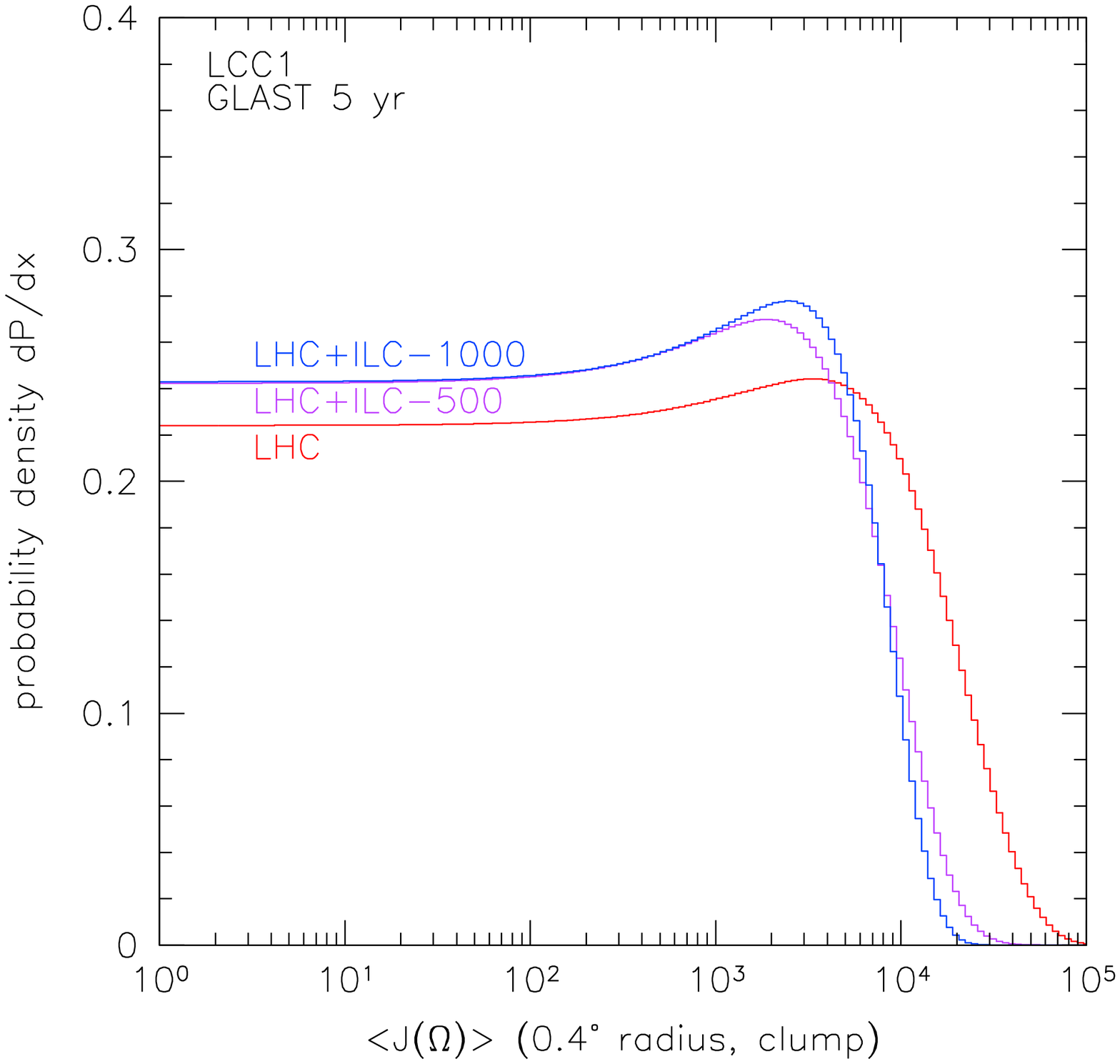,height=3.0in}\\
\caption{Halo density profiles for point LCC1: (a) galactic center,
(b) dark matter clump in the galactic halo.  Angle-averaged $J$ values as
measured by combining a 5-year all-sky dataset from GLAST with accelerator
measurements are shown.  See Fig.~\ref{fig:LCC1relic} for description of
histograms.}
\label{fig:LCC1halo}
\end{center}
\end{figure}

In Fig.~\ref{fig:LCC1halo}, we present two representative calculations 
that indicate how this information might be used.  We will postulate
two situations in which dark matter might be detected by gamma rays from 
annihilation, and we will work through the numbers to see how accurately
the absolute dark matter density can be obtained from these observations.
Because of the intrinsically small annihilation cross sections at LCC1,
the final results that we will show in this section will be rather marginal.
In later sections,
we will present the same calculations for the other three reference points.
At points LCC2 and LCC4, the annihilation cross section at 
threshold is 50 times larger, leading to much more optimistic expectations
for dark matter detection.

 First,
we consider the dark matter distribution at the galactic center, assuming
the NFW distribution shown in Fig.~\ref{fig:profiles}. We set the boost
factor $B = 1$.  Observation in
a circle of angular area $1.5 \times 10^{-4}$
 sr (angular radius 0.4$^\circ$) 
gives $\VEV{J(\Omega)} = 7700$. We use  this
value to predict the rate of gamma ray emission from this source.  
To do this, we use the spectrum of gammas expected for neutralino 
annihilation
at LCC1, scaled to the predicted total annihilation cross section.  (We will
describe this spectrum in Section 8.1.)  We impose a lower energy cutoff 
of $E_\gamma > 3.1$ GeV.  Background
from energetic processes at the galactic center is a major issue in this
analysis.  We have added to the counting rate the gamma ray background
in the same energy region according to the parametrization of 
Bergstrom, Ullio, and Buckley~\cite{UllioBergstrom}.  A five-year
observation by the GLAST gamma-ray observatory~\cite{GLAST}, in which the
galactic center is visible 20\% of the time, yields  an 
expectation of 5 signal photons over a background of 
360 background gamma ray photons.  This is a $S/\sqrt{B}$ less than 1, 
so we will only be able to quote upper bounds on $\VEV{J}$ for  this 
model.

We can now work backwards to determine the source strength $\VEV{J(\Omega)}$
from the observation.  For the purpose of this exercise, we assign the 
background rate a 5\% systematic uncertainty.  In pratice, it will be a 
challenge to obtain such a good understanding of the background.  Gamma 
rays come from the galactic center from many sources, and, in particular,
from  energetic
processes associated with the black hole at the center of galaxy.  However,
gamma rays from $\pi^0$'s produced in soft hadronic interactions have a 
soft power-law energy spectrum different from that predicted for neutralino 
annihilation.  In addition, much of the the observed gamma radiation
 seems to be 
associated with
molecular clouds that ring the galactic center rather than having a 
peak at the center itself.  This observation has already been 
used to 
improve the limits on  gamma rays from the galactic center from the EGRET
observatory~\cite{HooperDingus}.  However, recent work of 
the HESS Collaboration, while spatially resolving many gamma ray sources
from the galactic center, also shows that this center is the source
of a power law spectrum of very high-energy gamma rays, up to 10 TeV
in energy~\cite{HESS}.   For the purposes of the analysis in this paper, we 
assume that it will be possible to measure these effects and, to the 
extent allowed by statistics, separate them  from the
WIMP annihilation signal.  This might well be possible; we will know
better in a few years~\cite{Kamae}.

For the case of LCC1, even this aggressive estimate makes the uncertainty 
in the
background the dominant source of error. 
The  likelihood distribution for  $\VEV{J(\Omega)}$ that we obtain
from this analysis is shown in Fig.~\ref{fig:LCC1halo}(a).  
In this and later figures for likelihood distributions of $\VEV{J}$, 
we assume a prior distribution that is flat in 
$\log \VEV{J}$.    For LCC1, 
we do not obtain a prediction of $\VEV{J(\Omega)}$, but we do obtain a strong
upper limit that
excludes otherwise viable models predicting  large values of $J(\Omega)$.

In models in which the galaxy is build up from hierarchical 
dark matter clustering, one should expect to find localized clumps of
dark matter.  We choose a typical object from the semi-analytic 
simulation data of Taylor and Babul~\cite{TaylorBabul}, a 
dark matter clump with a mass of   $10^6 M_\odot$ and a scale radius 
of 500 pc, at a distance of 6 kpc, with its internal structure 
described by an NFW profile.  Observing in a 
disk of $1.5\times 10^{-4}$ sr, 
the object has $\VEV{J(\Omega)} = 2500$.   We impose a lower energy cutoff of
$E_\gamma > 1$ GeV   and add  
extra-galactic background  at the level measured by 
EGRET~\cite{EGRETbackground}.   Assuming 
a five-year observation by the GLAST observatory, in which this object would 
be visible 20\% of the time, we expect 4 signal photons, plus 60 photons from
the extra-galactic background.  Working backwards
from the observations, assuming that the extra-galactic background is 
well characterized from observations in other regions of the sky, we find 
the likelihood distribution of $\VEV{J(\Omega)}$
shown in Fig.~\ref{fig:LCC1halo}(b).  Given the ratio of signal/background,
we can only impose an upper limit on $\VEV{J}$, but this still has
the power to exclude models with very strongly peaked dark matter clumps.

Certainly, the discovery of a localized clump of dark matter in 
gamma rays would be remarkable in itself.  But it would be more remarkable
if we could obtain from particle physics a calibration of the absolute
scale of its dark matter density.  Fig.~\ref{fig:LCC1halo}(b) shows that 
this is possible in principle.  At   the other benchmark points, which 
give larger gamma ray fluxes from neutralino annihilation, we will see this 
idea realized more clearly.

\subsection{Direct detection cross section}

\begin{figure}
\begin{center}
\epsfig{file=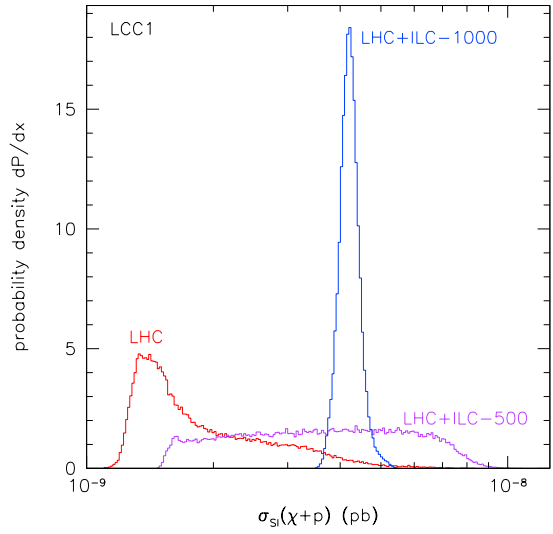,height=3.0in}
\caption{Spin-independent neutralino-proton direct
 detection cross section for
point LCC1.    See
Fig.~\ref{fig:LCC1relic} for description of histograms.}
\label{fig:LCC1direct}
\end{center}
\end{figure}

\begin{figure}
\begin{center}
\epsfig{file=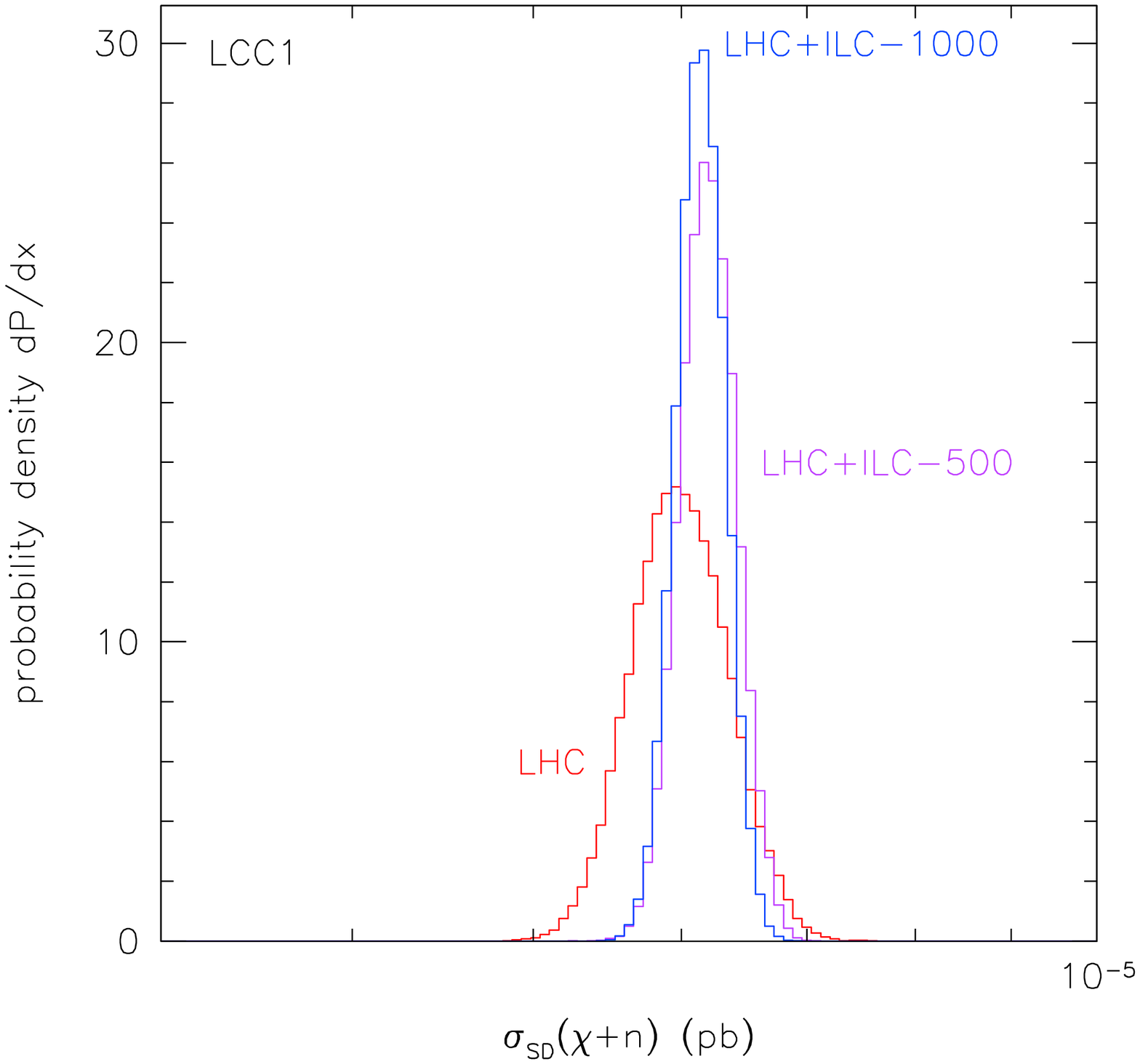,height=3.0in}
\caption{Spin-dependent
neutralino-neutron direct detection  cross section for point LCC1.  See
Fig.~\ref{fig:LCC1relic} for description of histograms.}
\label{fig:LCC1directsd}
\end{center}
\end{figure}

In Fig.~\ref{fig:LCC1direct}, we show the likelihood distributions of the
spin-averaged
neutralino-proton cross section $\sigma_{\chi p}$, evaluated
at threshold.  We remind the reader that, in our analysis in this 
section and in the corresponding analyses of the other models, we will
ignore the uncertainty in the evaluation of the  $\sigma_{\chi p}$
resulting from the uncertainty in the low-energy QCD parameter
$f_{Ts}$ discussed in Section 2.6.  The cross 
section  $\sigma_{\chi p}$ is sufficiently close to the 
spin-independent isoscalar cross section that it can be used to 
interpret signals of direct detection of dark matter in underground
detectors.  The corresponding distributions for the spin-dependent
part of the cross section $\sigma_{\chi n}$ are shown in 
Fig.~\ref{fig:LCC1directsd}.

\begin{figure}
\begin{center}
\epsfig{file=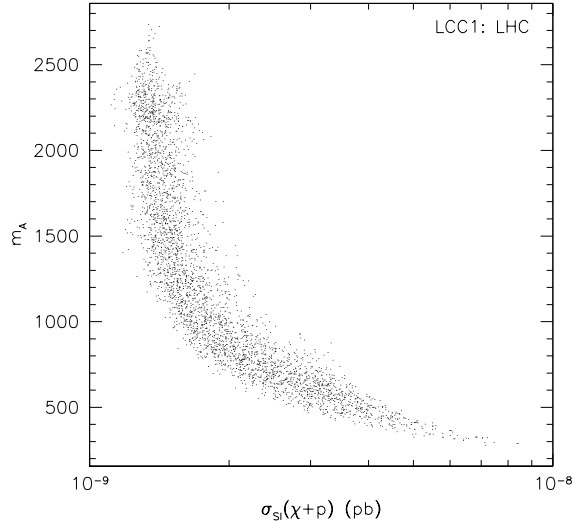,height=3.0in}
\caption{Scatter plot of the spin-independent direct detection 
cross section vs. $m(A)$ for point LCC1.  The strong dependence of the direct
detection cross section on the $A^0$ mass is clearly seen.} 
\label{fig:LCC1directscatter}
\end{center}
\end{figure}

From Fig.~\ref{fig:LCC1direct}, it is clear that neither the LHC nor the ILC at
500 GeV can make a particularly accurate prediction of this cross section. At
LCC1, the direct detection cross section is dominated by the $t$-channel
exchange of the heavy Higgs boson $H^0$. In the MSSM, the mass of the $H^0$ is
very close in mass to the $A^0$ unless both are light.  Even with the ILC data
at 500 GeV, this mass is essentially unconstrained. The sharp edge in the
distributions at $10^{-9}$ pb reflects the contribution of the light Higgs
$h^0$.  Fig.~\ref{fig:LCC1directscatter}, which displays the correlation in the
LHC likelihood function between the neutralino-proton cross section and the
heavy Higgs boson mass, shows clearly that this parameter is the essential
missing piece of information It also illustrates the fact that above about 1.5
TeV, the cross section is insensitive to the heavy Higgs mass as the light
Higgs contribution is of a comparable size. The slight improvement from the LHC
to the 500 GeV ILC reflects the determination of neutralino mixing angles,
which enter the neutralino-Higgs vertices.  Indeed, when the $H^0$ and $A^0$
particles are observed and measured at the 1000 GeV ILC, we obtain a prediction
of the direct detection cross section to about 20\% accuracy.

As in the previous section, it is illuminating to work through an example of
this application of this cross section determination to experimental data. At
the point LCC1, dark matter events would not be seen in the CDMS II experiment,
but the signal should be discovered in the next-generation detector
SuperCDMS or `SuperCDMS 25kg'~\cite{superCDMS}.  
This is a 26.67 kg detector on the model of CDMS,
with most of the mass in Germanium,
to be located in the low-background environment of the Sudbury mine.  Estimates
of the detection capabilities of SuperCDMS were computed assuming 2.5 years of
operation (ending in 2011) with a total exposure of close to 16,000 kg d.
Sensitivity calculations have assumed that the local halo density is 0.3
GeV cm$^{-3}$, the halo circular velocity is 220 km s$^{-1}$, the halo
escape velocity is 650 km s$^{-1}$, and that the velocity distribution is
Maxwellian.  The capabilities of heavy-noble-liquid detectors of the same mass
and exposure time should be similar.  Over the time scale required to realize
the ILC at 1000 GeV, we might expect to see a direct detection experiment
scaled up to 1 Ton size.  This would allow measurement of the annual flux
variation and the velocity distribution of directly detected WIMPs.  For our
analysis, however, we will concentrate on the implication of the total rate
measurement combined with collider data.

For the parameters that we have just described, the SuperCDMS group estimates
that they will see 16 signal events, assuming a uniform dark matter
distribution in the disk and the cross section predicted for LCC1, and
negligible background~\cite{Schnee}.  Dividing the observed rate by the
predicted cross section, we would obtain a direct measurement of the flux of
dark matter impinging on the detector.  We should note that
 this derived flux is an
effective quantity.   The detection efficiency for WIMPs depends on the
velocity distribution, and the simulation \cite{Schnee} used the 
efficiency calculated for the reference halo model described in the 
previous paragraph.  In presenting our results, we will refer to the 
derived quantity as $\Phi_{local}/\Phi_0$, the local flux of 
WIMPs divided by the flux in the reference model, assuming the same average
detection efficiency.

\begin{figure}
\begin{center}
\epsfig{file=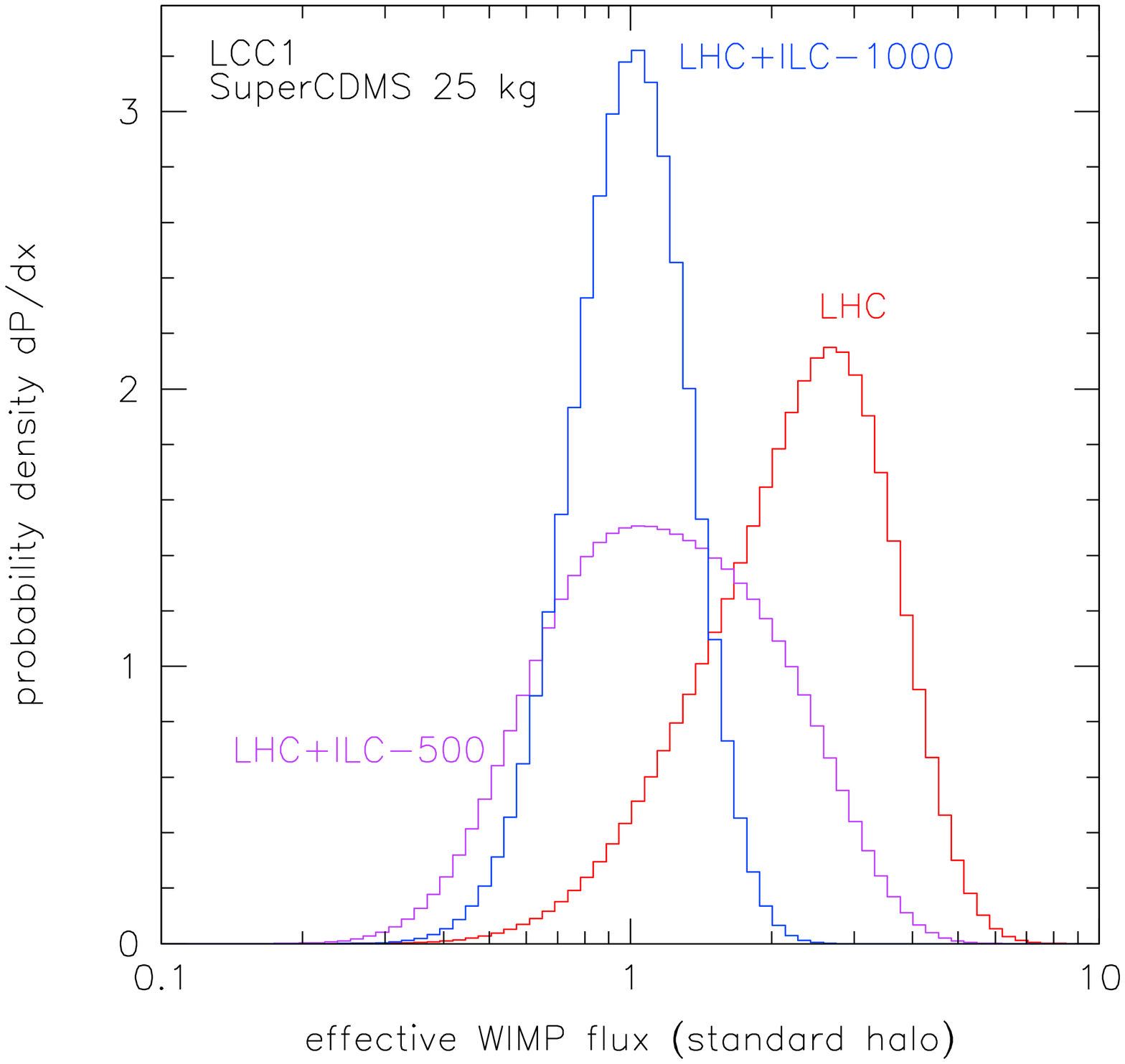,height=3.0in}
\caption{Effective local WIMP flux $\Phi_{local}/\Phi_0$ 
at the Earth for point LCC1.  The results assume the SuperCDMS measurement
described in the text.
 See Fig.~\ref{fig:LCC1relic} for description of histograms.}
\label{fig:LCC1localhalo}
\end{center}
\end{figure}

In Fig.~\ref{fig:LCC1localhalo}, we
show the likelihood distribution for the effective local flux,
 obtained by combining the
distribution of values of the cross section with the statistical uncertainty of
the direct detection measurement.  The distribution shown ignores
the uncertainty in the direct detection cross section from our poor
knowledge of low-energy QCD parameters. 
But, assuming that these parameters can be determined, we see from the 
figure that the combination of the data from the ILC at 1000 GeV and 
count rate from WIMP direct detection would measures the local flux of 
dark matter at the Earth to 28\%
accuracy.


\subsection{Constraints from relic density and direct detection}

If LCC1 is the correct theory of Nature, it is possible that, by the end of the
decade, the LHC will have observed missing energy events and a convincing
signal of dark matter from annihilation to gamma rays will also have been
observed.  Values of the WIMP mass will have been obtained from the LHC and
from the endpoint of the gamma ray spectrum, and these values will have been
seen to agree.  Underground direct detection experiments in the 25 kg range
such as SuperCDMS may also give the WIMP mass and flux at the Earth.  Further,
the Planck measurements of the CMB will have provided a very accurate
measurement of the cosmic density of dark matter.  Under these circumstances,
it would be very tempting to use the Planck and SuperCDMS measurements to
constrain the parameters of supersymmetry model.

This analysis would depend on very strong assumptions whose status would still
be open.  It would not be clear that the model leading to missing energy at the
LHC was in fact supersymmetry, or that the neutralino was the lightest
supersymmetric particle.  It would also be unclear whether the neutralino made
up 100\% of the dark matter, and whether unknown effects had diluted its
abundance.  The flux of dark matter particles at the Earth is also uncertain,
and depends on the halo model.  Nevertheless, by making these assumptions, we
could draw very strong conclusions that could later be checked by detailed
particle physics measurements.

\begin{figure}
\begin{center}
\epsfig{file=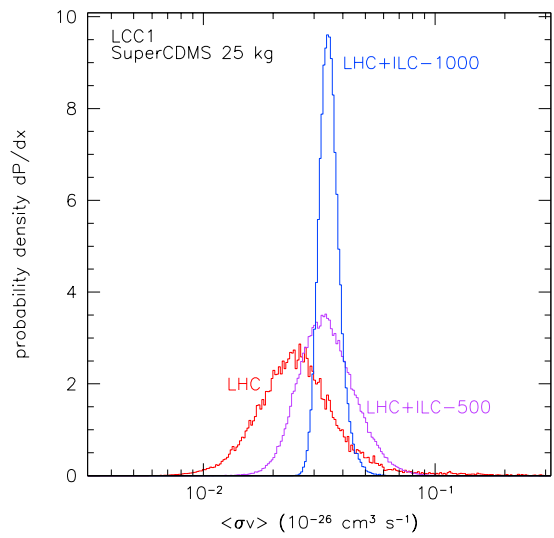,height=3.0in}
\caption{Annihilation cross section at threshold for point LCC1.  A direct
detection constraint from 2 years of a 25 kg SuperCDMS is applied.  Compared
with Fig.~\ref{fig:LCC1anncs}, it is clear that the direct detection constraint
significantly improves the measurement of the annihilation cross section in
advance of ILC-1000.  See Fig.~\ref{fig:LCC1relic} for description of
histograms.}
\label{fig:LCC1ddsigv}
\end{center}
\end{figure}

At this point, LHC data alone provide a prediction of the relic density 
to 7\% accuracy 
under the assumptions of the standard cosmology.  This is already quite
precise.  However, the annihilation and direct detection cross sections
would not be well determined, and information from a direct dectection
experiment could be used to improve our knowledge of the particle physics
model.    For example, applying a constraint from the direct detection
rate to the annihilation cross section improves the determination of 
this cross section both for the LHC and the ILC-500 data sets. As
shown in Fig.~\ref{fig:LCC1ddsigv}, the central values of the 
distributions shift much closer to the correct value, though, 
curiously,  the variances of the 
distributions are  not much improved. 

Further into the future, the ILC-500 will predict the relic density at the
1.5\% level.  It is thus unlikely that CMB or other cosmological measurements
of the relic density will greatly improve the situation, though if Planck can
measure relic density to 0.5\% as advertised, this would be a factor of 3
improvement.  The direct detection cross section measurement is now less
skewed, with 45\% errors.  SuperCDMS would give 16 events, for a
25\% error on the WIMP flux times cross section.  In a particular halo model,
one might hope to use this measurement, but the advantage would not be likely
to be significant. 

\begin{figure}
\begin{center}
\epsfig{file=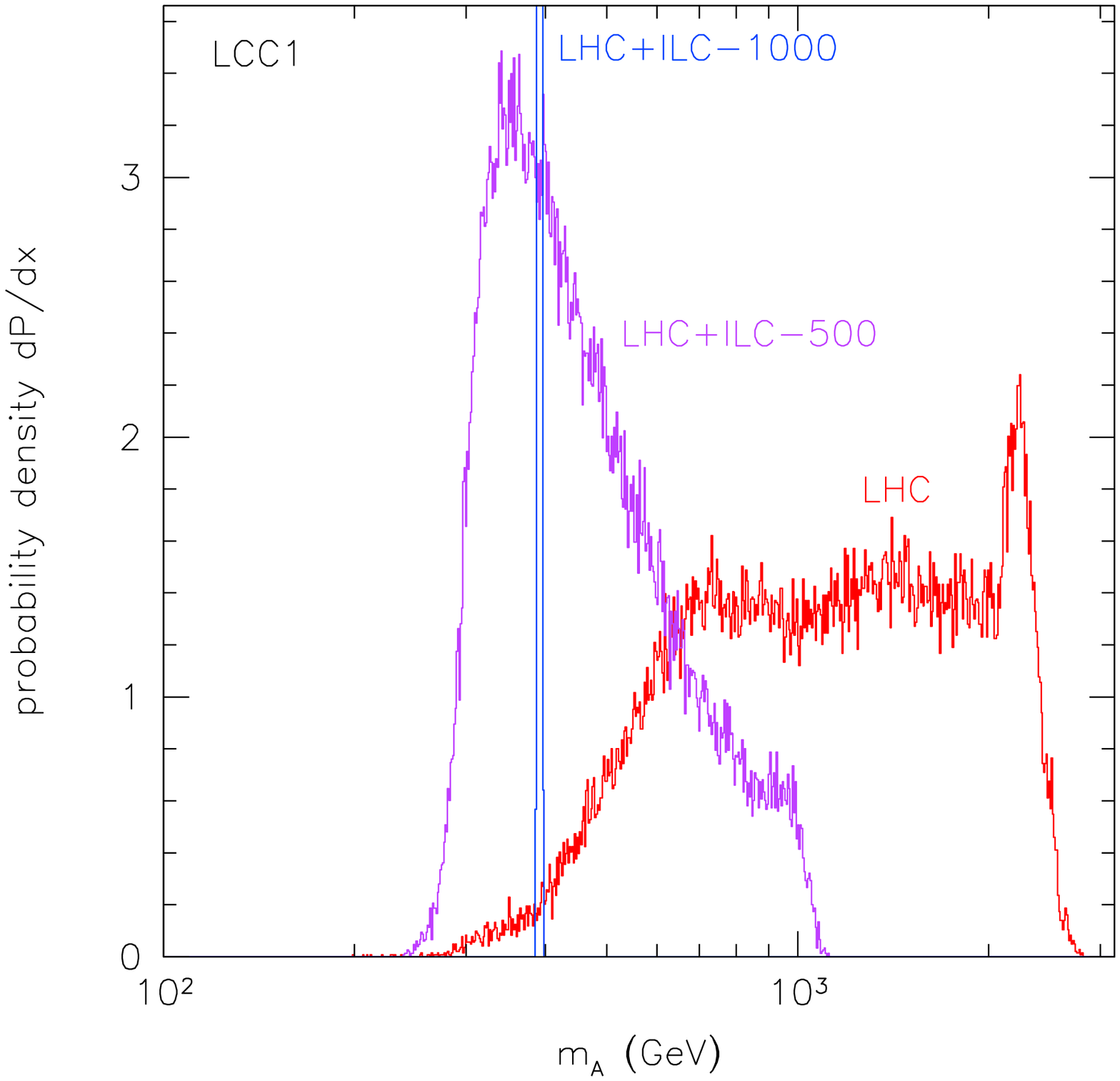,height=3.0in}
\epsfig{file=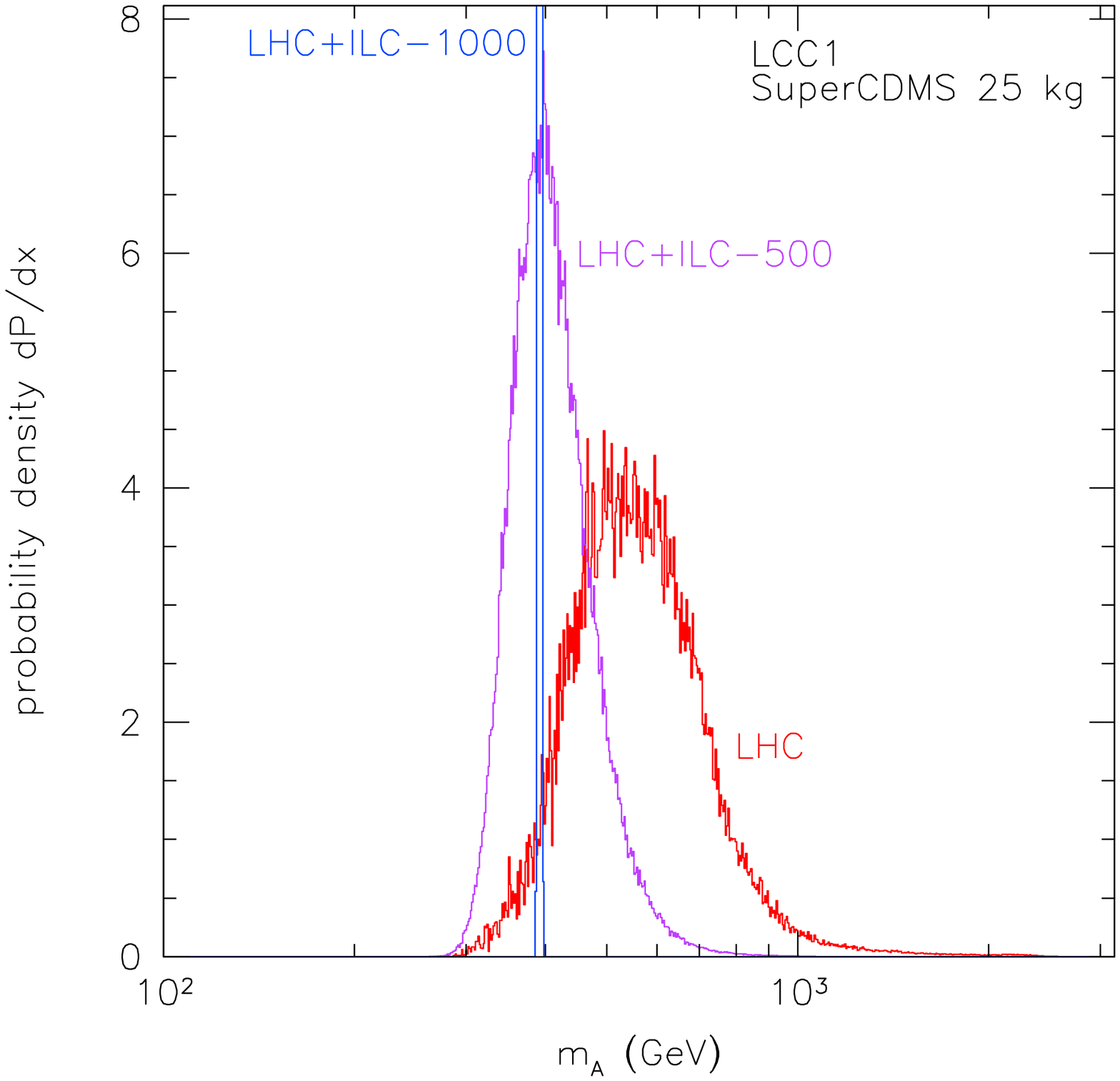,height=3.0in}
\caption{LCC1 heavy Higgs mass $m_A$, before and after a direct detection
constraint is applied.  The constraint allows a crude measurement of the $A^0$
mass in advance of the ILC-1000, which directly produces the $A^0$.  See
Fig.~\ref{fig:LCC1relic} for description of histograms.}
\label{fig:LCC1mA}
\end{center}
\end{figure}

Fundamentally, the large uncertainty in the direct detection cross section is
due to the unknown heavy Higgs mass.  Applying the direct detection constraint
greatly improves the shape of the distribution of this quantity at LHC and
ILC-500, where the heavy Higgs are unobserved, as illustrated in
Fig.~\ref{fig:LCC1mA}.

The ILC-1000 can predict the relic density at the 0.25\% level and the direct
detection cross section at the 5\% level.  This is the unique case under study
where the astrophysical measurements have no benefit beyond the crucial
consistency checks.

\clearpage

\section{Benchmark point LCC2}

In the next few sections, we will carry out the analysis that we have 
just described at LCC1 for the other three reference points.  The parameter
set 
LCC1 was specially chosen by the authors of \cite{LHCILC} as a point at which
most of particles of 
the supersymmetry spectrum could 
be observed at the LHC and then measured precisely at the ILC.  In more 
generic scenarios of supersymmetry, the information available to both 
colliders will be more limited.

At LCC2, for example, the squarks and sleptons are made extremely heavy, 
so heavy that it is unlikely that they could be observed at the LHC.  However,
the model still allows the discovery of supersymmetry at the LHC.
The  model contains a gluino at 850 GeV, giving a cross section of
about 10 pb for supersymmetry production at the LHC, and
charginos and neutralinos in the range 100--300 GeV.   The dominant mode
of neutralino annihilation is to $W^+W^-$ and $Z^0Z^0$. 

Although we cannot observe the heavy supersymmetric particles,
we might hope that the lighter ones, which we can observe, contain most of 
the information needed to predict the astrophysical cross sections needed
to analyze dark matter detection experiments.   The main contributions to 
neutralino annihilation come from diagrams in which neutralinos and 
charginos are exchanged.  The main contribution to the direct detection
cross section comes, in this case, from exchange of the light 
Higgs boson $h^0$.
Both reactions depend strongly on the gaugino-Higgsino mixing angles, and
so the measurement of these angles becomes the major issue for the 
interpretation of collider measurements in terms of the underlying 
spectrum parameters.

\begin{figure}
\begin{center}
\epsfig{file=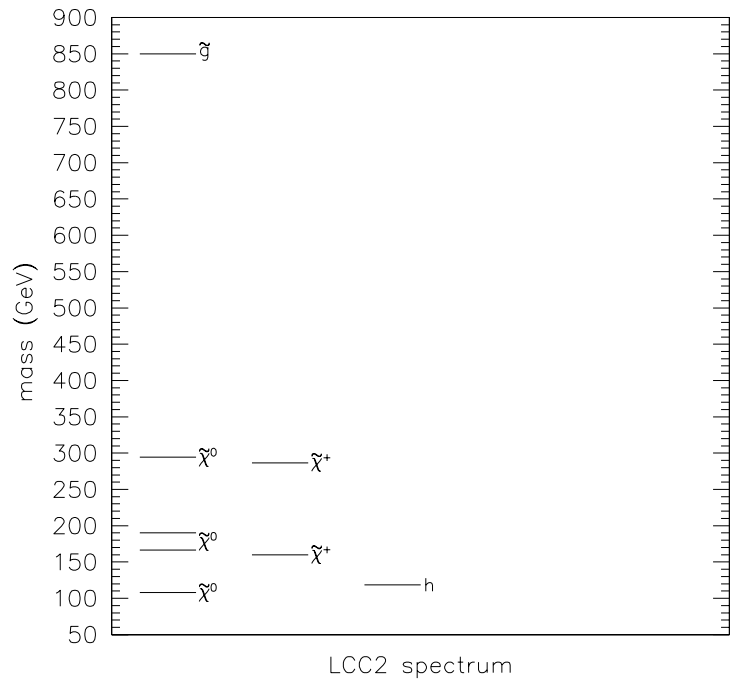,height=4.0in}
\caption{Particle spectrum for point LCC2.  All neutralinos and charginos are
mixed.  The most bino-like neutralino is the lightest one, and the most
wino-like neutralino is the heaviest one.  All scalars are above about 2 TeV.}
\label{fig:LCC2spectrum}
\end{center}
\end{figure}

\subsection{Spectroscopy measurements}

As we have already noted, the LHC will give a large sample of supersymmetry
events.  Most of these events will involve gluino pair production, followed
by gluino decays to $q \bar q \chi$, where $\chi$ is a neutralino or
chargino.  The subsequent evolution can be seen from the spectrum of the 
model, shown in Fig.~\ref{fig:LCC2spectrum}.  The mass difference of the 
first and third neutralinos is less than $\mz$.  Thus, both the second 
and the third neutralino will decay to the first neutralino through a 
virtual $Z^0$ that can be observed as a lepton pair,
\beq
        \s\chi^0_i \to \ell^+\ell^- \s \chi^0_1 \ ,
\eeq{neutralinotopairs}
for $i = 2,3$.    These processes will allow measurements of the two 
mass differences to an accuracy comparable to that with which the 
$\s\chi_2^0$--$\s\chi^0_1$ mass difference can be measured at LCC1.
In addition, the gluino mass should be determined to about 10\% accuracy
from the distribution of missing energy and visible transverse mass, and the 
mass of the lightest neutralino should be determined to about 10\% by 
more detailed kinematic fitting of the these events.
Unfortunately, it will be more difficult that at LCC1 to find the 
charginos or the fourth neutralino, so it is likely that these spectrum 
parameters are the only ones that can be determined from the LHC data.

Measurements at the ILC will map out much more of the chargino and 
neutralino spectrum~\cite{Gray}.   At a center of mass energy of 
500 GeV, the ILC will study the lighter states of this spectrum through
$\ee\to \s\chi_1^+\s\chi^-_1$ and $\ee \to \s\chi^0_2\s \chi^0_3$.  It will
also make a very accurate measurement of the mass of $\s\chi^+_1$
 by determination of the threshold for producing this
particle.   The $\ee$ production cross sections are sensitive to
the gaugino-Higgsino  mixing 
angles and can also be used to constrain the supersymmetry parameters.
A second stage of operation at center of mass energies of 1000 GeV can 
observe the reactions
 $\ee \to \s\chi_3^0 \s\chi^0_4$ and  $\ee \to \s\chi_2^+ \s\chi^-_2$ 
and measure the masses 
of the $\s\chi^0_4$ and $\s\chi^+_2$.

The complete list of spectrum constraints that we expect for this point
for the LHC and for each stage of the ILC is given in 
Tables~\ref{tab:LCC2masses}
and \ref{tab:LCC2css}.

\begin{table}
\centering
\begin{tabular}{lccccc}
   mass/mass splitting  & LCC2 value & &  LHC  & ILC 500 &  ILC 1000\\ 
         \hline
  $m(\s\chi^0_1)$    &  107.9      & $\pm$ &   10   &   1.0     \\ 
  $m(\s\chi^0_2) - m(\s\chi^0_1)$ &   58.5      & $\pm $ &   1.0  &   0.3  \\
  $m(\s\chi^0_3) - m(\s\chi^0_1)$ & 82.3          & $\pm $ &  1.0  &  0.2 \\ 
  $m(\s\chi^0_4) - m(\s\chi^0_1)$ &  186.3      &  $\pm $ &   - & - & 3.0 \\ 
  $m(\s\chi^+_1)     $ &   159.7             &  $\pm $ &  -  &   0.55 \\ 
  $m(\s\chi^+_1) - m(\s\chi^0_1)$ &   51.8       &  $\pm $ &   - &   0.25 \\ 
  $m(\s\chi^+_2)     $ &     286.7              &  $\pm$ &   -  &  - & 1.0 \\
       \hline
  $m(\s e_R)     $     &   3277.      &  $\pm$ &  ($> 350$) &  & ($>480$) \\ 
  $m(\s \mu_R)$   &     3277.     & $\pm $ &  ($> 350$)   &  &($>480$)  \\ 
  $m(\s \tau_1)$   &  3252. &  $\pm $ &   ($ > m(\chi^0_2)$)  & & ($>480$)\\ 
  $m(\s e_L)$    &    3280.   & $\pm $ & ($> 350$)   &  &($>480$)  \\
  $m(\s \mu_L) $  &    3280.  & $\pm $ &  ($> 350$)    &   &  ($>480$)  \\ 
  $m(\s \tau_2)$  & 3268.     &  $\pm $ &  &  & ($>480$) \\ 
        \hline
  $m(h)     $    &     118.68           &  $\pm $ & 0.25  &   0.05 \\ 
  $m(A)     $   &   3242.      &  $\pm$ &   *  &  ($>240$)      & ($>480$) \\
       \hline
  $m(\s u_R)$, $m(\s d_R)$ & 3312.   & $\pm $ &  ($> 2000$)      \\
  $m(\s s_R)$, $m(\s c_R)$ & 3312.    & $\pm $ &  ($> 1500$)       \\
  $m(\s u_L)$, $m(\s d_L)$ &  3301.      & $\pm $ &  ($> 2000$)       \\
  $m(\s s_L)$, $m(\s c_L)$ &  3301.    & $\pm $ &  ($> 1500$)     \\
  $m(\s b_1)     $ &   2710.         &  $\pm $ &  ($> 1500$)  \\ 
  $m(\s b_2)     $ &    3241.       &  $\pm$ &  ($> 1500$)    \\
  $m(\s t_1)     $ &  1976.             &  $\pm $ & ($> 1500$)  \\ 
        \hline
  $m(\s g)$      &  850.        &  $\pm $ &   85.     &   \\ 
\end{tabular}
\caption{Superparticle masses and their estimated errors or limits for 
       the parameter point LCC2.  
     The notation is as in Table~\ref{tab:LCC1masses}.}
\label{tab:LCC2masses}
\end{table}

\begin{table}
\centering
\begin{tabular}{lcccrr}
   cross section
    &  & LCC2 value & &  ILC 500 &  ILC 1000\\
         \hline
       \hline
  $\sigma(\ee\to \s\chi^+_1\s\chi^-_1)$
      & LR & 1364. (0.479)         & $\pm $  &    1\%$^*$ \\
     & RL & 145.6 (0.438)         & $\pm $    &    4\%$^*$ \\
  $\sigma(\ee\to \s\chi^0_2\s\chi^0_3)$
      & LR & 127.6              & $\pm $   &    4\%$^*$ \\
     & RL &  105.8              & $\pm $   &    5\%$^*$ \\
\end{tabular}
\caption{SUSY cross sections and estimated errors for
    the parameter point LCC2.  The notation is as in Table~\ref{tab:LCC1css}.
  The errors labeled by $^*$ are
taken from the study \cite{Gray}. }
  \label{tab:LCC2css}
\end{table}

\subsection{Relic density}

\begin{figure}
\begin{center}
\epsfig{file=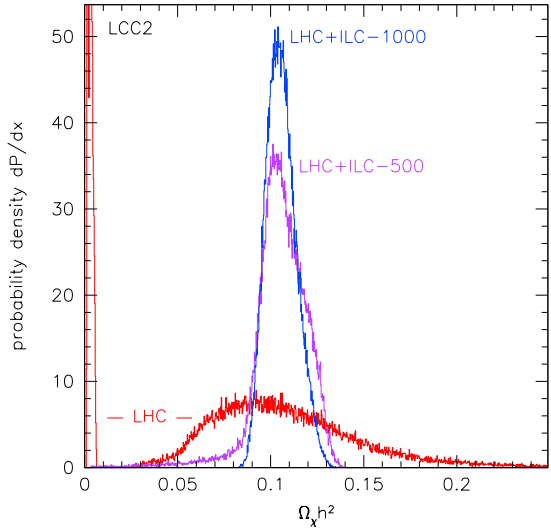,height=3.0in}
\caption{Relic density for point LCC2.  There are two overlapping very high
peaks at $\Omega_\chi h^2<0.01$, with maxima at $dP/dx = 122$ and $165$,
 due to the wino and Higgsino solutions to the
LHC constraints.  See Fig.~\ref{fig:LCC1relic} for description of histograms.}
\label{fig:LCC2relic}
\end{center}
\end{figure}

We can now use these constraints on the spectrum as the basis for an 
exploration of the allowed supersymmetry parameter space.  The results of 
the three Monte Carlo scans, projected onto the axis of the predicted 
WIMP relic density, are shown in Fig.~\ref{fig:LCC2relic}.  The distribution
from the LHC constraints is quite broad, with a standard deviation of about
40\% and also a significant secondary peak near
$\Omega_\chi h^2 = 0$.  The prediction of $\Omega_\chi h^2$ from the 
ILC data a 500 GeV has an accuracy of about 14\%, and this improves to 
about 8\% using the data from the ILC at 1000 GeV.

The main difficulty in determining the relic density from the LHC data is that
the two precision measurements of mass differences do not provide enough
information to fix the gaugino-Higgsino mixing angles.  Without an accurate
determination of the mixing angles, the relatively accurate determination
of the WIMP mass has little predictive power.   

\begin{figure}
\begin{center}
\epsfig{file=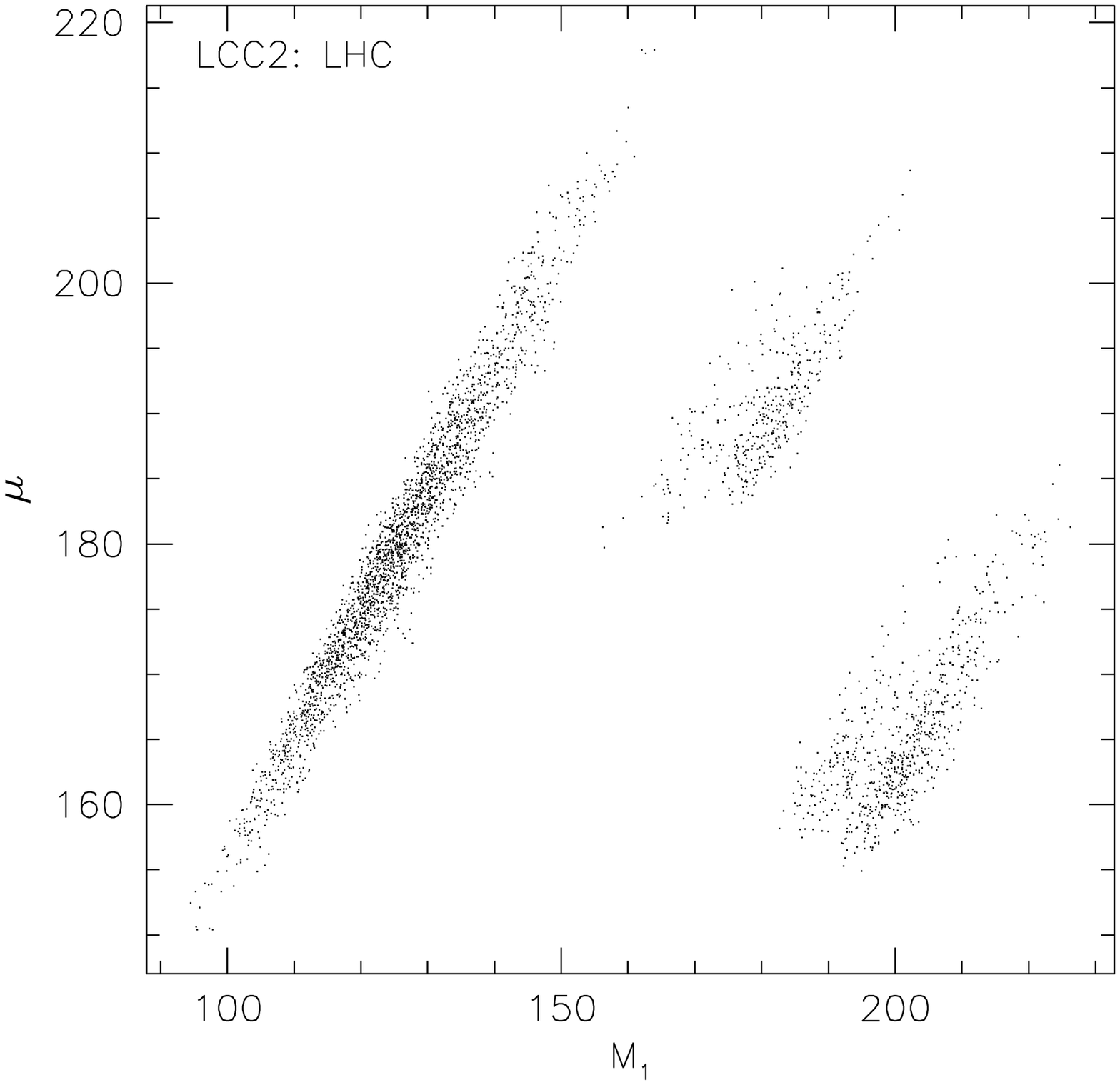,height=3.0in}\\
\epsfig{file=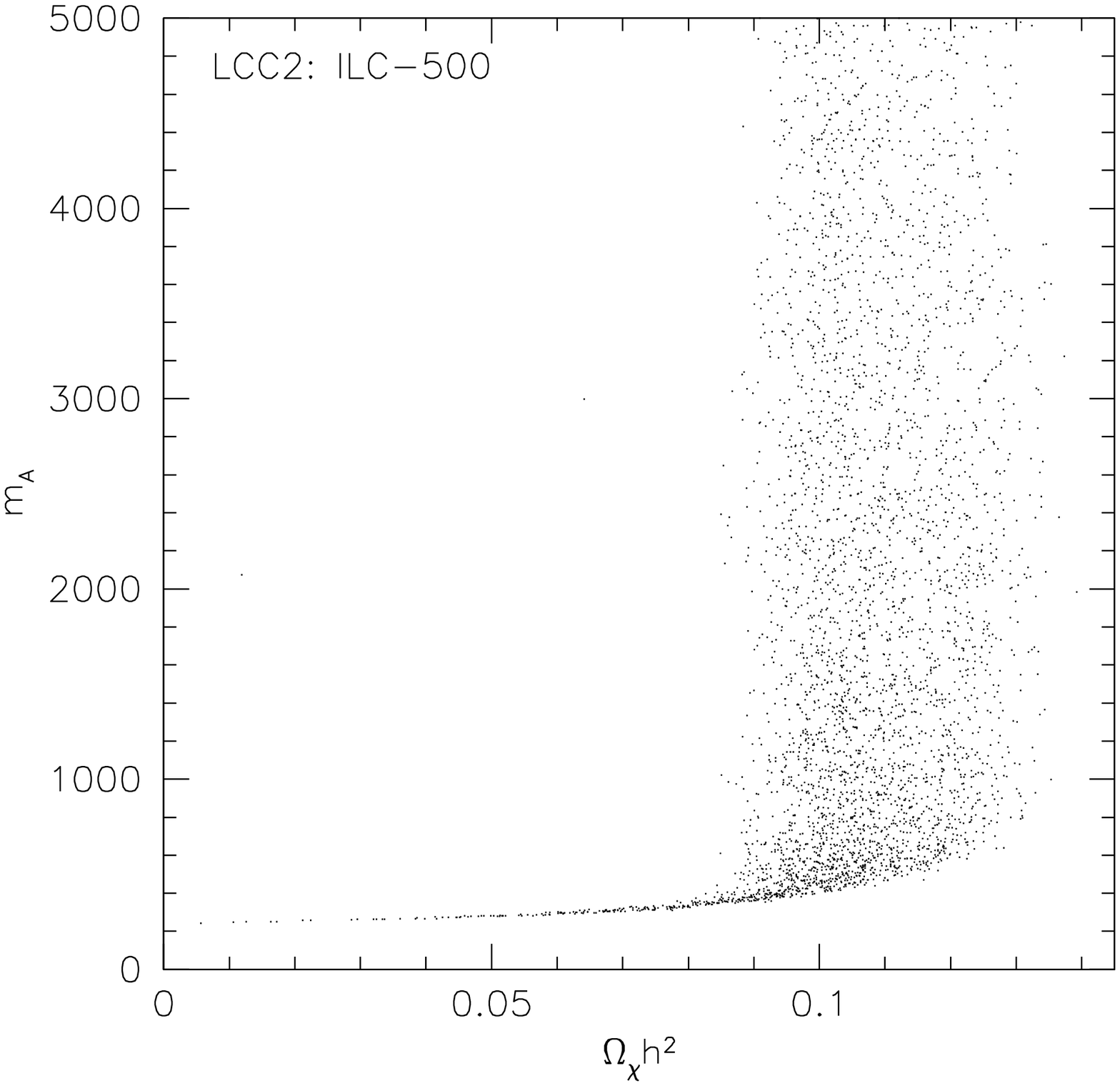,height=3.0in}\\
\caption{Scatter plots for point LCC2. (a) $m_1$ vs.\ $\mu$ distribution for
LHC, illustrating multiple solutions.  Left to right, these are bino (correct),
wino, and Higgsino.  (b) $m_A$ vs.\ $\Omega_\chi h^2$, showing the influence
of resonant annihilation.}
\label{fig:LCC2relicscatter}
\end{center}
\end{figure}

However, it is also true that the LHC data can be interpreted in multiple ways
in terms of the underlying parameters.  This is illustrated by making
scatter plots of the Monte Carlo data, as we show in
Fig.~\ref{fig:LCC2relicscatter}.  In Fig.~\ref{fig:LCC2relicscatter}(a), we
show the scatter plot of the data with LHC constraints in the plane of $m_1$
vs. $\mu$.  The data clearly shows three solutions, corresponding to a bino-,
wino-, and Higgsino-like lightest neutralino.  The phase space for the latter
two solutions is restricted by our constraint that $m(\s\chi^+_1)$ should be
greater than 125 GeV.  These incorrect solutions are responsible for the peak
in the relic density likelihood function at LHC for very small values.  

In
Fig.~\ref{fig:LCC2relicscatter}(b), we show the correlation between
$\Omega_\chi h^2$ and $m(A)$ at the 500 GeV ILC.  Although most of the
distribution forms a smooth wide cloud, reflecting the uncertainty in the
mixing angles, one sees also a dense corridor at the low end of the allowed
range for $m(A)$, leading to vanishingly small relic density.  This reflects
the influence of the $A^0$ pole.  At LCC2, the true mass of the $A^0$ boson is
about 3200 GeV; however, we must rely on data to exclude the $A^0$ from being
in a mass region in which its pole influences the annihilation cross section.

\begin{figure}
\begin{center}
\epsfig{file=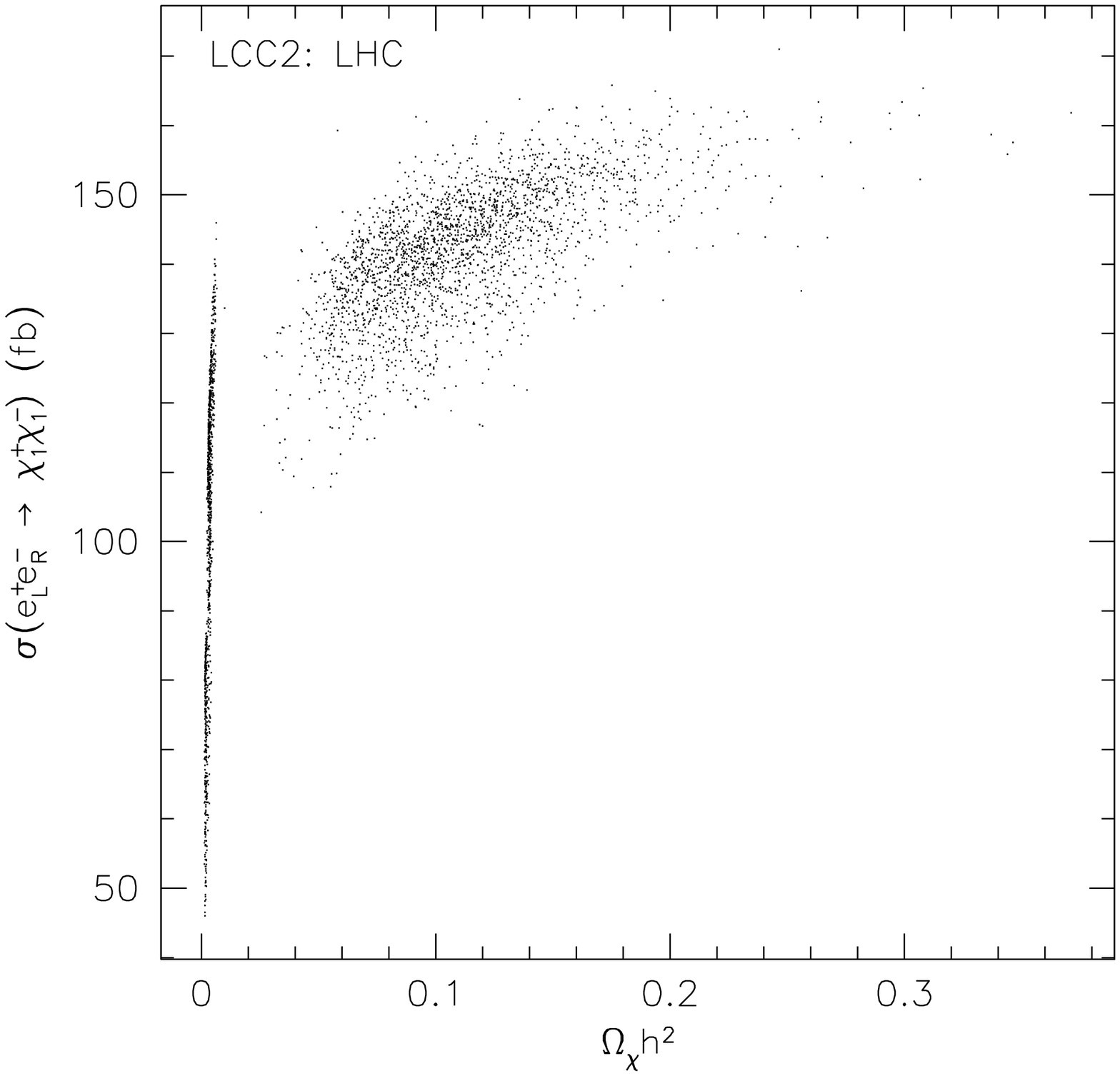,height=3.0in}\\
\epsfig{file=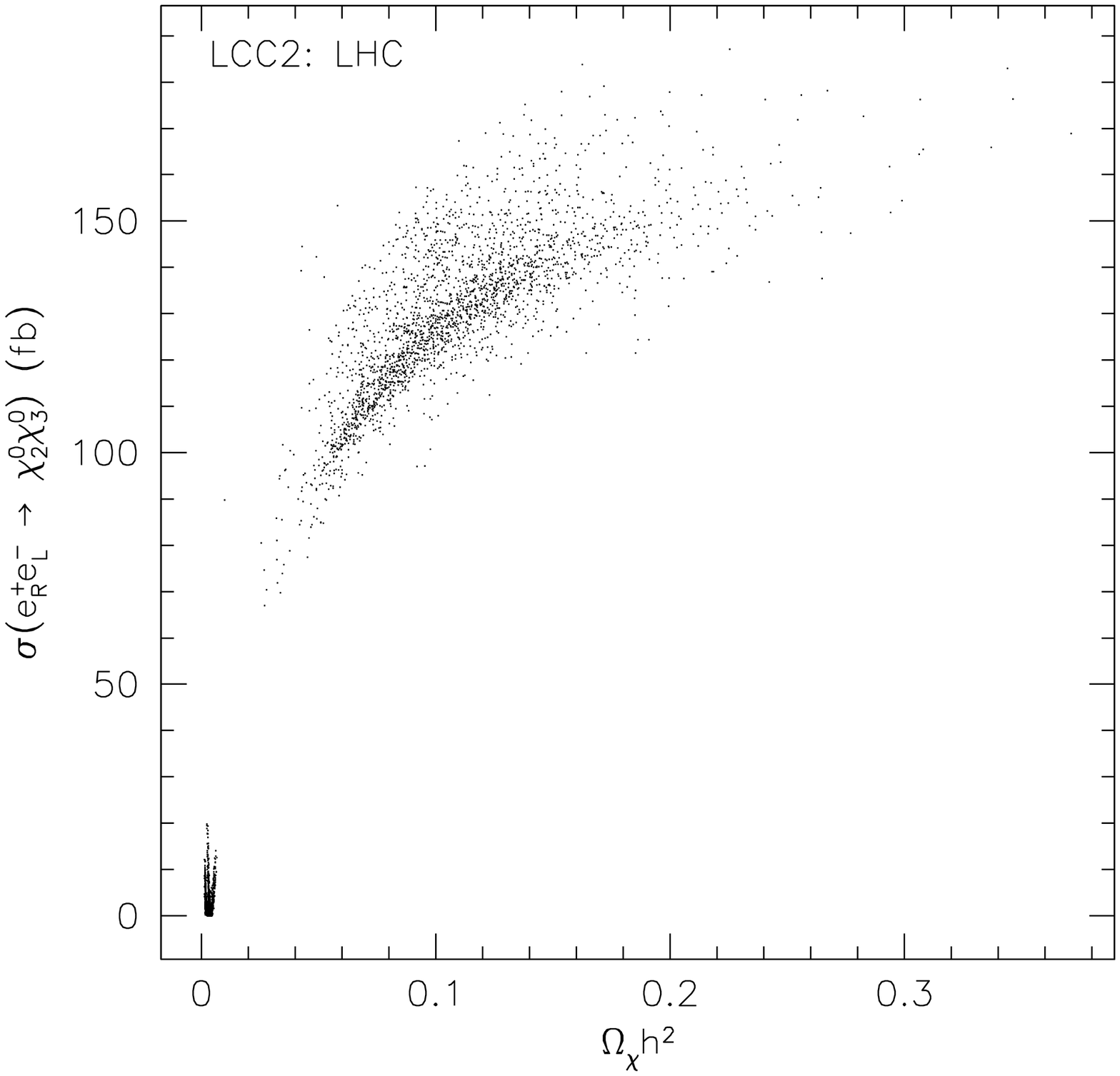,height=3.0in}\\
\caption{Scatter plots of the LHC data 
for LCC2 illustrating the effects of $e^+e^-$ cross
sections in defining the relic density. The wino and Higgsino islands in 
Fig.~\ref{fig:LCC2relicscatter}(a) are mapped to the lines on the left side
of the figures at very low
values of $\Omega_\chi h^2$. Thus they are removed by the $\ee$ cross section
constraints.}
\label{fig:LCC2cseffect}
\end{center}
\end{figure}

In Fig.~\ref{fig:LCC2cseffect}, we illustrate the power of measuring 
the polarized neutralino
and chargino production cross sections at the ILC.  Thje cross sections
depend strongly on the gaugino-Higgsino mixing angles; thus, measurement
of the cross sections allows us to fix the values of these
angles.  From the figures, it is clear that there is a significant
correlation between these cross sections and the relic density.  Furthermore,
measuring the cross sections removes the  ambiguities in the identities of the
particles.

The general form of the three curves in Fig.~\ref{fig:LCC2relic} is, we
believe, closer than that of  Fig.~\ref{fig:LCC1relic} to the generic
situation.   The LHC, despite its ability to make precision measurements of
the supersymmetry spectrum, has little capability to predict $\Omega_\chi h^2$.
The predictions from the ILC are about at the level of the current 
determination from WMAP and will not be at the level of the determination 
from Planck.  Nevertheless, the ILC prediction, based only on microscopic
data, is quite sufficiently accurate that its agreement or disagreement with 
the relic density found by Planck would be a striking test of the assertion
that the neutralino is the sole source of WIMP dark matter.

\subsection{Annihilation cross section}

\begin{figure}
\begin{center}
\epsfig{file=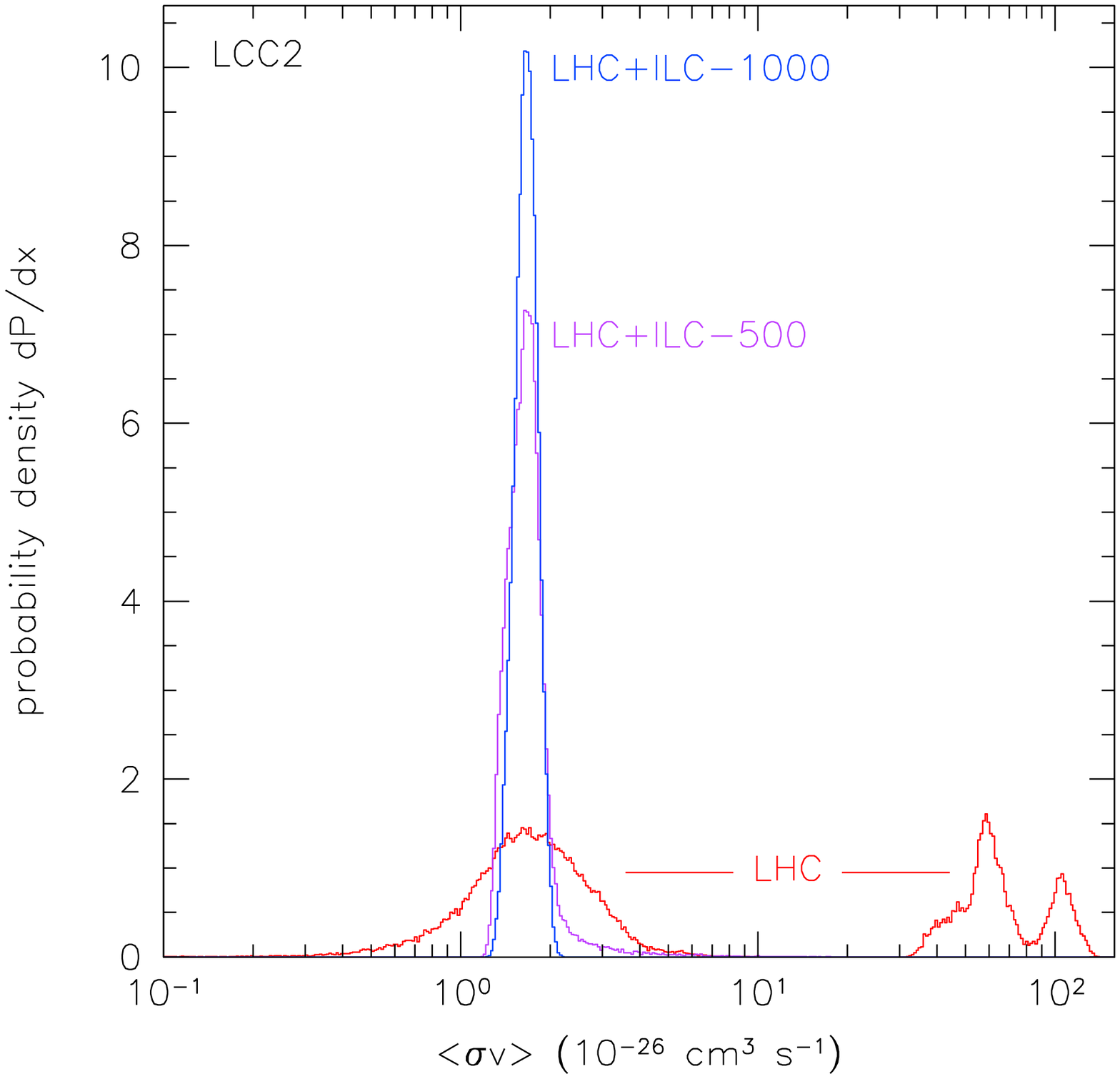,height=3.0in}
\caption{Annihilation cross section at threshold for point LCC2.  The wino and
Higgsino solutions giving very high cross section are clearly visible.  See
Fig.~\ref{fig:LCC1relic} for description of histograms.}
\label{fig:LCC2anncs}
\end{center}
\end{figure}

In Fig.~\ref{fig:LCC2anncs}, we show the prediction of our likelihood analysis
for the neutralino pair annihilation cross section at threshold.  Unlike the
situation for LCC1, this figure is very similar in form to the set of
predictions for the relic density, with a relatively broad distribution from
the LHC constraints and progressively narrower distributions from the two
stages of the ILC.  This is not so surprising.  At LCC2, the dominant
annihilation processes that determine the relic density are the decays to
$W^+W^-$ and $Z^0Z^0$.  At low energy, these processes go mainly in the S-wave,
yielding cross sections that are similar at threshold and at the freeze-out
temperature $T/m_\chi\sim 1/25$.  Thus, since the relic density is proportional
to the inverse of the annihilation cross section at freeze-out, the likelihood
distribution for the relic density should be just the reflection of that for
the annihilation cross section.  And so it is.

\begin{figure}
\begin{center}
\epsfig{file=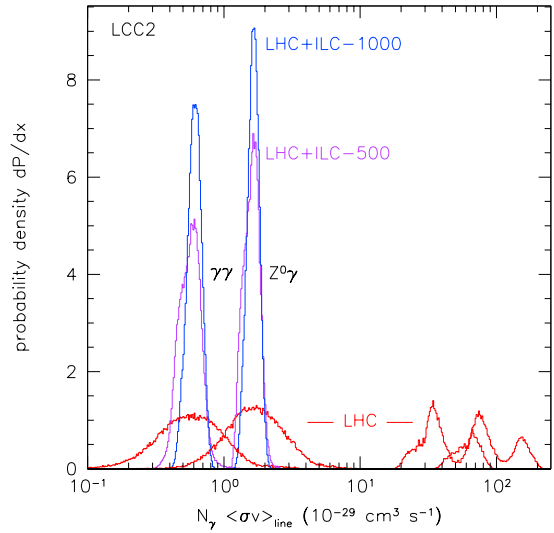,height=3.0in}
\caption{Gamma ray line annihilation cross section at threshold for point LCC2.
The wino and Higgsino solutions are clear.  See Fig.~\ref{fig:LCC1relic} for
description of histograms.}
\label{fig:LCC2twogamma}
\end{center}
\end{figure}

In Fig.~\ref{fig:LCC2twogamma}, we show the likelihood distributions for the
exclusive annihilation cross sections to $\gamma\gamma$ and $Z^0\gamma$.
Unlike the case of LCC1, where the loop diagrams that produce these 
reactions are dominated by squark exchange, here the dominant effects come
from chargino and neutralino exchange.  The cross sections thus depend on the
gaugino-Higgsino mixing angles in a way similar to the relic density.

\begin{figure}
\begin{center}
\epsfig{file=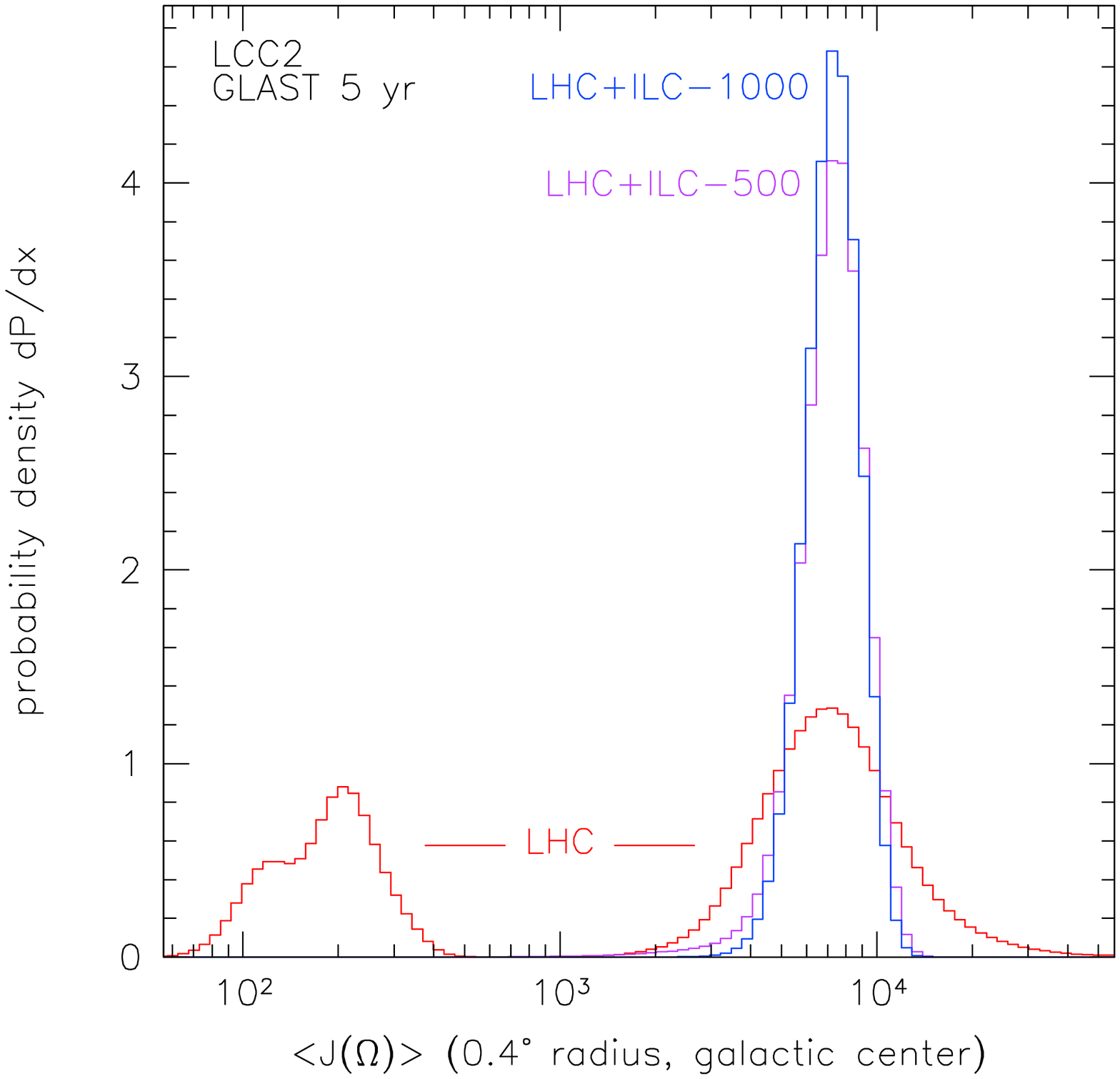,height=3.0in}\\
\epsfig{file=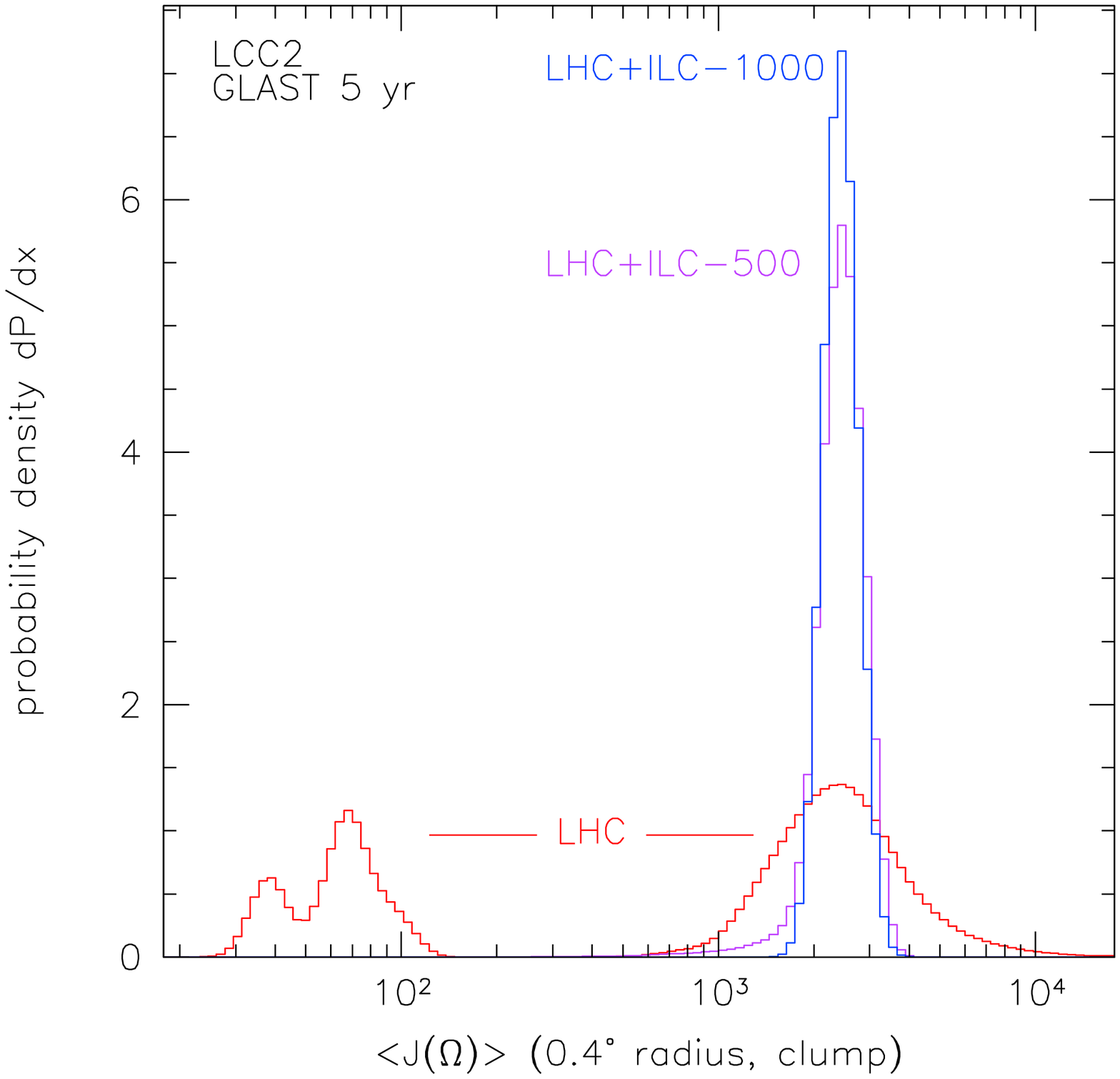,height=3.0in}\\
\caption{Halo density profiles for point LCC2: (a) galactic center,
(b) dark matter clump in the galactic halo.  Angle-averaged $J$ values as
measured by combining a 5-year all-sky dataset from GLAST with accelerator
measurements are shown.  The influence of the incorrect solutions is clearly
seen as subsidiary peaks at low $J$ values.  See Fig.~\ref{fig:LCC1relic} for
description of histograms.}
\label{fig:LCC2halo}
\end{center}
\end{figure}

The microscopic determination of the annihilation cross section allows us
to interpret observations of gamma rays from dark matter annihilation and 
to directly measure the density distribution $\VEV{J(\Omega)}$ for a 
source of dark matter.  In Section 4.4, we described some specific exercises
based on the capabilities of GLAST.  At LCC2, the annihilation cross section 
is about 50 times larger, leading to 172 signal photons (over 360
background) in the GLAST 
observation of the galactic center and 168 signal photons (over 60 background)
in the GLAST observation of the reference subhalo dark matter clump.
 Folding the photon statistics with the 
likelihood distributions from Fig.~\ref{fig:LCC2anncs}, and including a 
5\% uncertainty in the background from the galactic center, we find for
LCC2 the predictions shown in Fig.~\ref{fig:LCC2halo} for the reconstructed
values of $\VEV{J(\Omega)}$.  For the large annihilation cross section 
characteristic of LCC2, we obtain measurements of $\VEV{J(\Omega)}$ at the
10\% level.  Such measurements would be very powerful constraints on
models of dark matter clustering and galaxy formation.

\subsection{Direct detection cross section}

\begin{figure}
\begin{center}
\epsfig{file=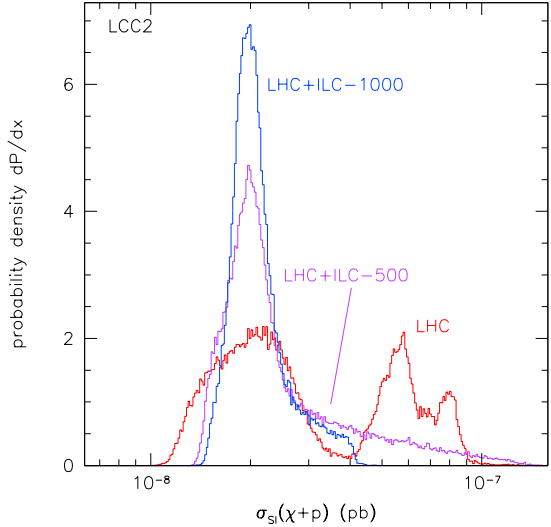,height=3.0in}
\caption{Spin-independent neutralino-proton direct
 detection cross section for
point LCC2.   The wino and
Higgsino solutions have large cross sections, easily seen.  See
Fig.~\ref{fig:LCC1relic} for description of histograms.}
\label{fig:LCC2direct}
\end{center}
\end{figure}

\begin{figure}
\begin{center}
\epsfig{file=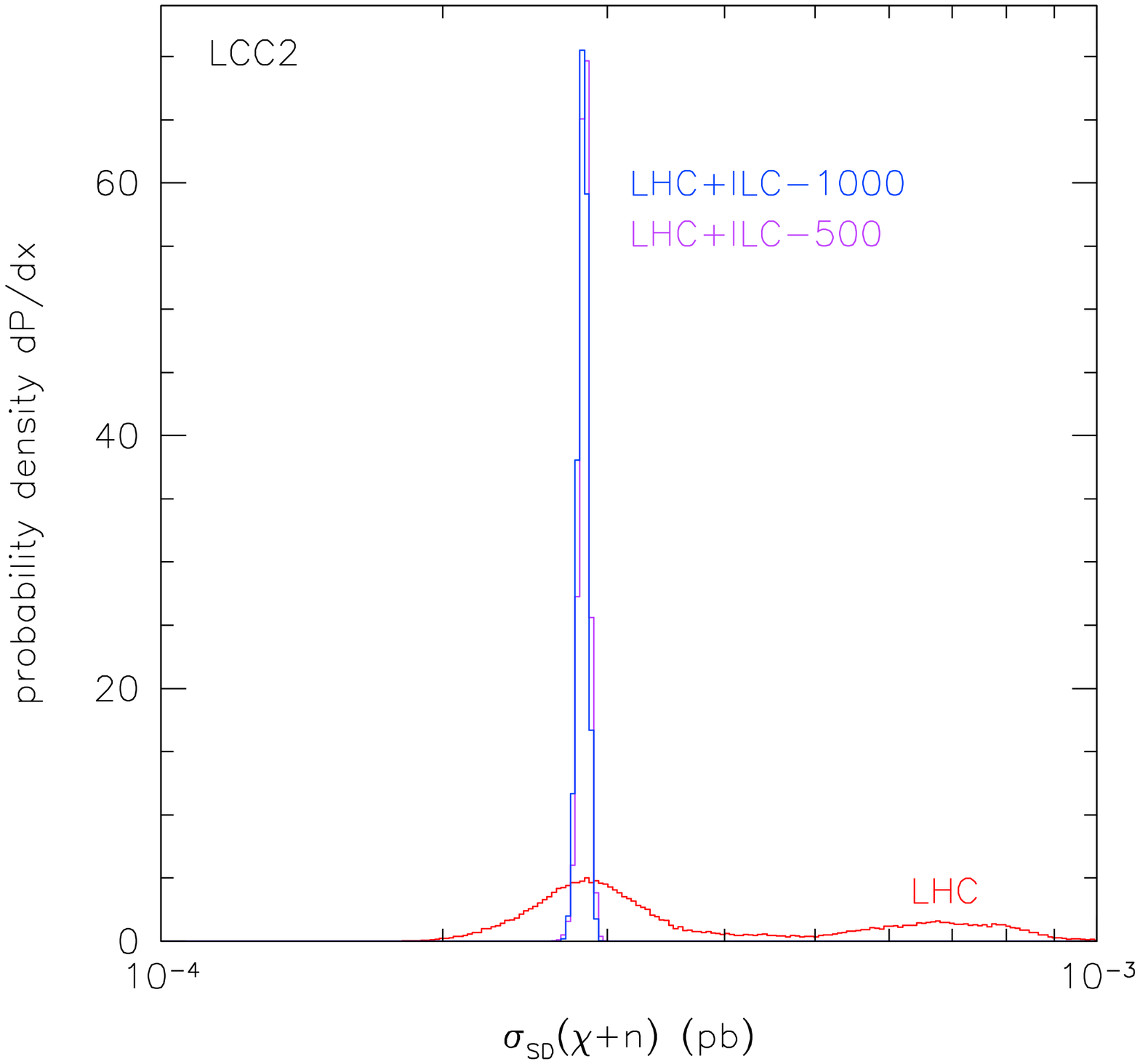,height=3.0in}
\caption{Spin-dependent
neutralino-neutron direct detection  cross section for point LCC2.  See
Fig.~\ref{fig:LCC1relic} for description of histograms.}
\label{fig:LCC2directsd}
\end{center}
\end{figure}

In a similar way, we can repeat the analysis of Section 4.3 for the 
direct detection cross section.  The likelihood distribution of
the cross section values given by our Monte Carlo scans is 
shown in Fig.~\ref{fig:LCC2direct}.  At LCC2, the spin-independent
neutralino-proton cross 
section is dominated by $t$-channel exchange of the light Higgs boson
$h^0$.  The mass of this particle is expected to be measured well
at the LHC in the decay $h^0\to \gamma\gamma$.  So the LHC constraints 
already give a reasonably precise estimate of the direct detection 
cross section, although, as we see from the figure, the ambiguity in the
solution leads to additional peaks at high values of the cross section.
The ILC measurements sharpen the determination of the mixing angles
and also remove the alternative solutions. 
The corresponding predictions for the spin-dependent part of the 
neutralino-neutron cross section are shown in Fig~\ref{fig:LCC2directsd}.

\begin{figure}
\begin{center}
\epsfig{file=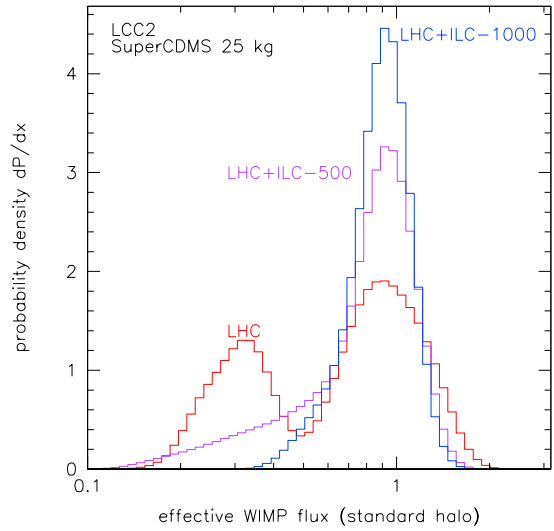,height=3.0in}
\caption{Effective local 
WIMP flux at the Earth for point LCC2.  The results assume 
the SuperCDMS measurement
described in the text.
  The second peak at low effective flux is due to the high cross
section wino and Higgsino solutions.  See Fig.~\ref{fig:LCC1relic} for
description of histograms.}
\label{fig:LCC2localhalo}
\end{center}
\end{figure}

As in Section 4.4, we can combine our microscopic knowledge of the detection
cross section with the expected event yield from the SuperCDMS detector
to estimate our ability to directly measure the local flux of dark matter
at the earth.  As before, our analysis omits the uncertainty from 
low-energy QCD parameters.
 For this case, we expect a signal of 67   events in SuperCDMS.  
Folding the statistical uncertainty with the uncertainty in the cross section,
we find the determination of the effective local 
halo flux shown in Fig.~\ref{fig:LCC2localhalo}.


\subsection{Constraints from relic density and direct detection}

LCC2 is a special case in that the identity of the lightest neutralino at LHC
is unknown, with discrete possibilities: the bino (correct), wino and Higgsino
solutions discussed previously.  Both incorrect solutions are confined to
regions quite far from the central values of both relic density and direct
detection cross section.  Thus, we find that {\em either} the direct detection
or relic density constraint completely eliminates the wino and Higgsino
solutions for the lightest neutralino.  The elimination of these islands has
very significant effects on the measurements at LHC, including the annihilation
cross section and branching ratios.  In fact, the entire structure of the
neutralino mass matrix is greatly improved.  With {\em either} the direct
detection {\em or} relic density constraint removing the incorrect islands, the
bino and wino fractions of {\em every} neutralino are measured to better that
15\% at the LHC.

\begin{figure}
\begin{center}
\epsfig{file=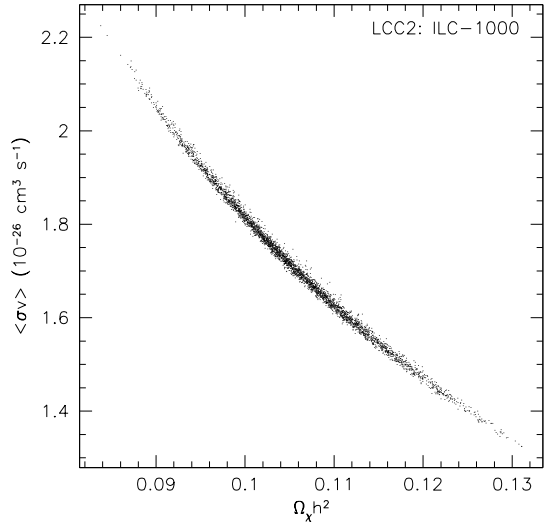,height=3.0in}
\caption{Scatter plot of annihilation cross section against relic density for
point LCC2.  Given ILC-1000 data, the two quantities are very well 
correlated.}
\label{fig:LCC2oh2sigv}
\end{center}
\end{figure}

Another unique feature of LCC2 is that the annihilation cross section at $v=0$
is tightly coupled to the relic density.  This is because the dominant
annihilation channels are to gauge boson pairs, and coannihilations are
unimportant.  The gauge boson channels are not helicity suppressed, thus the
annihilation cross section for typical freeze-out velocities is close to 
the $v=0$ cross section.  At the LHC, any relic density constraint
first removes the incorrect solutions which have very large annihilation cross
sections, and further constrains the annihilation cross section within the
correct solution.  The 1\% constraint provides a 4\% measurement of the
annihilation cross section.  At either stage of the ILC, applying a relic
density constraint gives a constraint on annihilation cross section at the same
level of precision.  At ILC-1000, both of these are measured at the 8\% level,
so even here, relic density gives a quite powerful constraint.  This tight
correlation is illustrated in Fig.~\ref{fig:LCC2oh2sigv}.

A third unique feature of LCC2 is that the dominant process in direct detection
is the exchange of the {\em light} Higgs boson, because the heavy Higgses are
very heavy, at 3.2 TeV.  Because the lightest neutralino is mixed, the cross
section is quite large.  If the heavy Higgs were lighter, the cross section
would be even larger.  Using the direct detection constraint thus places a
lower limit of 1 TeV on the mass of the heavy Higgs boson, which is much 
stronger that any direct lower limit that can be obtained from LHC or ILC.

\section{Benchmark point LCC3}

In Sections 6 and 7 we turn to points illustrating special circumstances 
in which the neutralino relic density depends on accidental relationships
among particle masses in the theory.  At LCC1, we would obtain the correct
neutralino
relic density if sleptons exchanged in the $t$-channel were light enough;
at LCC2, we would obtain the correct neutralino relic density if the 
gaugino-Higgsino mixing angles were large enough.  These are generic 
constraints valid in large regions of the MSSM parameter space.  At LCC3
and LCC4, we require more specific tuning of particle masses against 
one another.

The essential physics of the neutralino relic density at LCC3 is coannihilation
of the neutralino with the stau lepton.  That is, the dominant annihilation
reactions that determine the relic density are $\s\tau \chi^0_1\to \tau 
\gamma$
and $\s\tau\s\tau \to \tau\tau$, both of which can proceed in the S-wave.
The relative density of $\s\tau$ particles relative to neutralinos during 
the annihilation process is~\cite{Griest}
\beq
  \exp \left[ - {m(\s\tau ) - m(\chi)\over T} \right] \approx
 \exp \left[ - 25\left( {m(\s\tau ) - m(\chi)\over m(\chi)}\right) \right] \ .
\eeq{coanntuning}
Thus, this mechanism works only a a narrow region of parameter space, one 
that becomes increasingly constrained as the mass of the neutralino 
increases.

\begin{figure}
\begin{center}
\epsfig{file=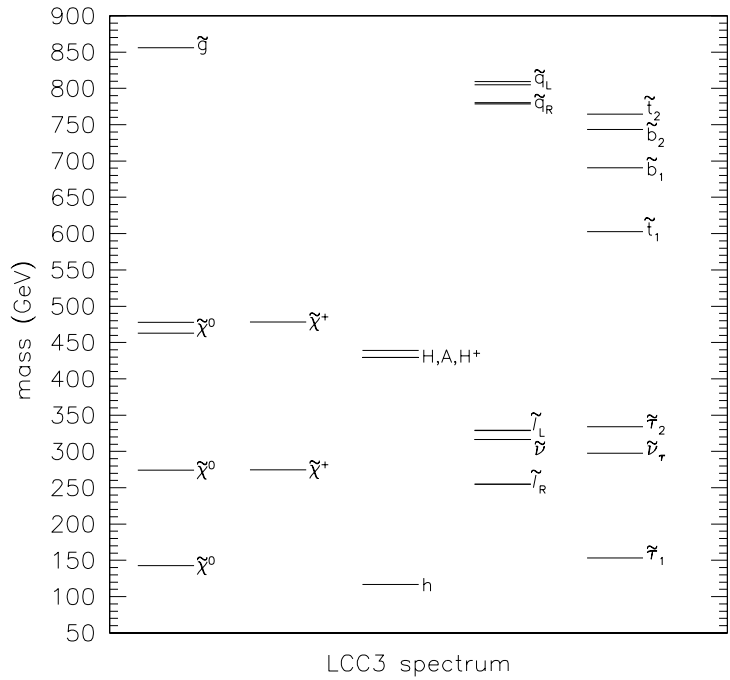,height=4.0in}
\caption{Particle spectrum for point LCC3.  The stau-neutralino mass 
splitting
is 10.8 GeV.  The lightest neutralino is predominantly bino, the second
neutralino and light chargino are predominantly wino, and the heavy 
neutralinos
and chargino are predominantly Higgsino.}
\label{fig:LCC3spectrum}
\end{center}
\end{figure}

\subsection{Spectroscopy measurements}

The supersymmetry spectrum at the point LCC3 is shown in 
Fig.~\ref{fig:LCC3spectrum}.  The point is 
one at which the LHC will be able to obtain a 
reasonably complete overview of the supersymmetry spectrum.  The squarks
and the gluino have masses of about 800 GeV, leading to cross sections
of tens of 
pb for supersymmetry production.  Effects enhanced by the 
large value of $\tan\beta$ move the lighter top and bottom squarks down
to about 600 GeV.  The $H^0$ and $A^0$ bosons can be observed in their
decay to $\tau^+\tau^-$. 

Unfortunately, we expect that the very precise spectrum measurements
that would be available at LCC1 and LCC2  will not be possible here.
The dominant decay mode of the $\s\chi^0_2$ is $\s\chi^0_2 \to 
\s\tau_1 \tau$, with decays also by $\s\chi^0_2 \to h^0 \s\chi^0_1$ (3\%)
and  $\s\chi^0_2 \to Z^0 \s\chi^0_1$ (1\%).
The  right-handed $\s e$ and $\s\mu$  are  lighter than 
the $\s \chi^0_2$, so that the decay chains  $\s \chi^0_2 \to 
\ell \s \ell \to \ell^+ \ell^- \s\chi^0_1$ are kinematically allowed.
However, for the case of $\s e_R$ and $\s\mu_R$, the branching 
fractions of $\s\chi^0_2$  to these channels are only $4 \times 10^{-4}$,
giving too small an event sample to be useful.
The decay chain involving $\s\tau_1$ is difficult to  study because
the mass gap
between the $\s\tau_1$ and the $\s\chi^0_1$ is only 10 GeV.  This
small mass gap is chosen precisely so that the coannihilation 
channels will dominate in the computation of the relic density.
 In the study~\cite{Bhaskar}, it is shown
that the the soft taus from this decay can be observed at the LHC
and used to estimated the mass gap.  What is directly observed is
the endpoint of the $\tau^+\tau^-$ mass spectrum, and this could
be measured to an accuracy of $\pm 5$ GeV~\cite{ourmod}.  In our analysis,
we have assumed that the 
mass of the $\s\chi^0_1$ can be obtained to 10\% accuracy and the 
mass of the $\s\chi^0_2$ to 15\% accuracy by general  kinematic fitting of
supersymmetry events, and that the masses of the  $\s e_R$ and $\s\mu_R$
can be found to 20\% accuracy using slepton pair production.

At the ILC, the studies \cite{Dutta,Bambade} have shown that it is
possible to observe the stau decay to tau and to measure the mass splitting.
The study \cite{Dutta} obtains an error on the mass splitting of about
1 GeV from kinematic fitting of the $\ee\to \s\tau^+\s\tau^-$ events.
The study \cite{Bambade} obtains an error of about 0.5 GeV on the stau
mass from a scan of the $\s\tau^+\s\tau^-$ threshold.  This must be 
combined with a precision measurement of the neutralino mass, which can 
be obtained from the analysis of the reaction   
$\ee\to \s\chi^0_1\s\chi^0_2$.  The value of the cross section
for $\ee\to \s\tau^+\s\tau^-$ events measures the stau mixing angle.

However, this is the extent of the information that can be found at this
point at the 500 GeV ILC.  To fix the gaugino-Higgsino mixing angles and
$\tan\beta$, more information is needed.  These additional constraints can 
be provided by the 1000 GeV ILC, from two sources.  First, the remaining
states of the chargino and neutralino system can be discovered and studied
in the reactions $e^+e^- \to \s\chi^0_3 \s \chi^0_4$ and 
$e^+e^- \to \s\chi^+_1 \s \chi^-_2$.
Second, because the mass of the $A^0$ Higgs boson at LCC3 is 428 GeV,
this particle can be observed in the process $\ee\to H^0 A^0$.  The
mass of the $A^0$ can be important for the detection cross sections, as
we have seen, but also the total width of the $A^0$ is interesting as a
way to measure $\tan\beta$.  We have estimated the error on the $\Gamma_A$
following \cite{BattagliaParis} and used this as a constraint.

The complete list of spectrum constraints that we expect for this point
for the LHC and for each stage of the ILC is given in 
Tables~\ref{tab:LCC3masses}
and \ref{tab:LCC3css}.

\begin{table}
\centering
\begin{tabular}{lccrrr}
   mass/mass splitting  & LCC3 value & &  LHC  & ILC 500 &  ILC 1000\\
         \hline
  $m(\s\chi^0_1)$    &   142.6       & $\pm$ &    14.   &   0.1      \\
  $m(\s\chi^0_2)$ &     274.2        & $\pm $ &   41.  \\
  $m(\s\chi^0_2) - m(\s\chi^0_1)$ &   131.5       & $\pm $ &   -  &
  0.5 \\
  $m(\s\chi^0_3) - m(\s\chi^0_1)$ &    320.2      & $\pm $ &  -   & -  &
      2.0 \\
  $m(\s\chi^0_4) - m(\s\chi^0_1)$ &   335.4        &  $\pm $ &   - &   -
                 & 2.0 \\
  $m(\s\chi^+_1)     $ & 274.5    &  $\pm $ &  -  &   -  &  0.7 \\
  $m(\s\chi^+_2)     $ &  478.2      &  $\pm$ &   -  &    -   & 2.0 \\
       \hline
  $m(\s e_R)     $     &    254.9       &  $\pm$ &  50.$^a$  &   - &
  1.0 \\
  $m(\s \mu_R)   $   & 254.7   & $\pm $ &   50.$^b$  & \\
  $m(\s e_R)-m(\s\chi^0_1)$     &   112.3    &  $\pm$ &  -  &   - &  0.2
  \\
  $m(\s \mu_R)-m(\s\chi^0_1)$     &   112.1   &  $\pm$ &  -  &   - &
  0.2 \\
  $m(\s \tau_1)$   &  153.4      &  $\pm $ &  -  &  0.5 \\
  $m(\s \tau_1)- m(\s\chi^0_1)$   &  10.8        &  $\pm $ & - &  1.0 &
  \\
  $m(\s e_L) $    &  328.9            & $\pm $ &   @$^a$  &  \\
  $m(\s \mu_L) $  &  329.1           & $\pm $ &    @$^b$   & \\
  $m(\s e_L)-m(\s\chi^1_0) $    &  186.3 & $\pm $ &  -  &  - & 1.0 \\
  $m(\s \mu_L)-m(\s\chi^1_0) $  &  186.5  & $\pm $ &  -   & - & 1.0 \\
  $m(\s \tau_2)-m(\s\chi^1_0) $  &  191.3 &  $\pm $ &   - &   - & 3.0 \\
        \hline
  $m(h)     $    & 116.58            &  $\pm $ & 0.25  &   0.05 \\
  $m(A)     $   &  429.5            &  $\pm$ & 1.5 * &        & 0.8  \\
  $\Gamma(A)     $   &  9.1             &  $\pm$ &   &        &  1.0 \\
       \hline
  $m(\s u_R)$, $m(\s d_R)$ &  780., 778.        & $\pm $ &   78.  &
  \\
  $m(\s s_R)$, $m(\s c_R)$ &  778., 780.       & $\pm $ &   78.  &
  \\
  $m(\s u_L)$, $m(\s d_L)$ & 805., 809.          & $\pm $ &   121.  &
  \\
  $m(\s s_L)$, $m(\s c_L)$ &  809., 805.         & $\pm $ &   121.  &
  \\
  $m(\s b_1)     $ &  690.                   &  $\pm $ &     35.  &   \\
  $m(\s b_2)     $ & 743.                    &  $\pm$ &     74. &     \\
  $m(\s t_1)     $ &  603.                   &  $\pm $ &  ($ > 315$)
  &  \\
        \hline
  $m(\s g)$      &  856.        &  $\pm $ & 171. \\
\end{tabular}
\caption{Superparticle masses and their estimated errors or limits for
       the parameter point LCC3.
     The notation is as in Table~\ref{tab:LCC1masses}.
The LHC measurements of slepton masses apply to the lighter of $\s e_R$,
    $\s e_L$ and the lighter of $\s\mu_R$, $\s\mu_L$.  The symbol
 `*' indicates that, because the $A^0$ can be seen at LHC,
     $\tan\beta > 7 (m_A/200)$.}
\label{tab:LCC3masses}
\end{table}

\begin{table}
\centering
\begin{tabular}{lcccrr}
   cross section
    &  & LCC3 value & &  ILC 500 &  ILC 1000\\
         \hline
   minimal set\\
       \hline
  $\sigma(\ee\to \s\chi^0_1\s\chi^0_2)$
      & LR & 34.4                 & $\pm $   &    8\%$^*$ \\
     & RL & 2.1                 & $\pm $   &    - \\
  $\sigma(\ee\to \s\tau^+_1\s\tau^-_1)$
      & LR & 45.6                 & $\pm $   &    100\%$^*$ \\
     & RL & 103.4                 & $\pm $    &    4\%$^*$ \\
   $\sigma(\ee\to \s\chi^+_1\s\chi^-_1)$
      & LR & 212.3 (0.808)        & $\pm $  &  &    3\% \\
     & RL &  6.3 (0.774)               & $\pm $    &   &   - \\
  $\sigma(\ee\to \s\chi^0_2\s\chi^0_2)$
      & LR & 88.7              & $\pm $  & &   5\% \\
     & RL & 2.5                 & $\pm $    &  & - \\
  $\sigma(\ee\to \s e^+_R\s e^-_R)$
      & LR & 19.5 (0.735)            & $\pm $  &    10\% \\
     & RL & 350.5 (0.971)             & $\pm $   &    3\% \\
         \hline
\end{tabular}
\caption{SUSY cross sections and estimated errors for
    the parameter point LCC3.  The notation is as in
    Table~\ref{tab:LCC1css}.
      The symbol
   (-) denotes that the cross section is less than 10 fb or is otherwise
        not measurable.  The errors labeled by $^*$ are
taken from the study \cite{Dutta,ourchange}; the others are estimated using
\leqn{erroronsig}.}
  \label{tab:LCC3css}
\end{table}

\subsection{Relic density}

\begin{figure}
\begin{center}
\epsfig{file=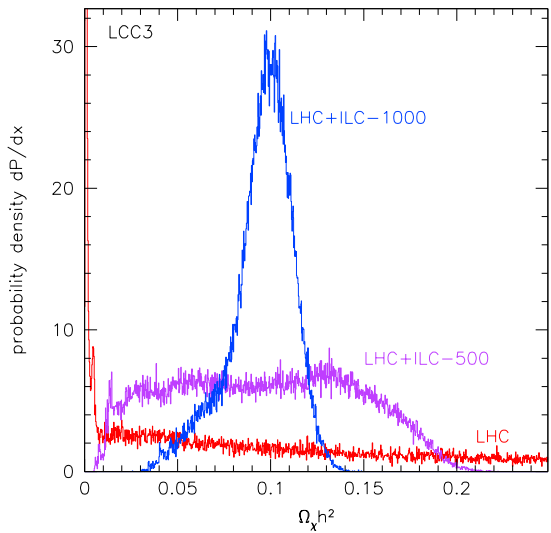,height=3.0in}
\caption{Relic density measurement for point LCC3.  The wino peak at very small
relic density is clear.  See Fig.~\ref{fig:LCC1relic} for description of
histograms.}
\label{fig:LCC3relic}
\end{center}
\end{figure}

We can use these constraints on the spectrum as the basis for an 
exploration of the allowed supersymmetry parameter space.  The results from 
the three Monte Carlo scans, projected onto the axis of the predicted 
WIMP relic density, are shown in Fig.~\ref{fig:LCC3relic}.  Because the 
LHC data is not sensitive to the stau-neutralino mass difference, that
set of constraints leads to essentially no information about the relic 
density, yielding a distribution that stretches well to the right of the
region plotted.  The information from the 500 GeV ILC can only check that
we are in the stau coannihilation region, predicting the relic density
only within a factor of two.  However, when the information from the 
1000 GeV ILC is added, with strong constraints on the mixing angles and
on $\tan\beta$, the relic density is predicted to about 18\%, an error
three times as large as the current WMAP determination.

\begin{figure}
\begin{center}
\epsfig{file=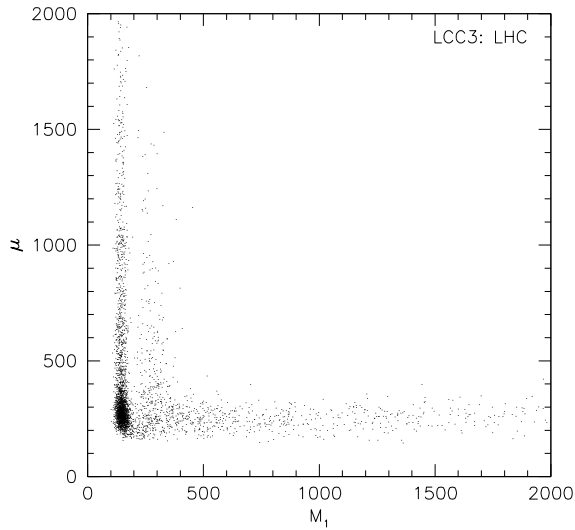,height=3.0in}
\caption{Scatter plot of $\mu$ vs.\ $m_1$ for LHC data at point LCC3.  The
``F'' structure indicates the fact that the lightest neutralino is either bino
or wino, while the second could be anything.}
\label{fig:LCC3relicscatter}
\end{center}
\end{figure}

Though this is not the main issue for the LHC, the scan data does show quite
clearly the presence of continuous ambiguities in the interpretation of the
spectrum data in terms of underlying parameters.  In
Fig.~\ref{fig:LCC3relicscatter}, we show the correlation of the parameters
$m_1$ and $\mu$.  The plot shows three regions, 
two horizontal and one vertical,
corresponding to models in which the lightest neutralino is mainly bino, wino,
and Higgsino, respectively.

\begin{figure}
\begin{center}
\epsfig{file=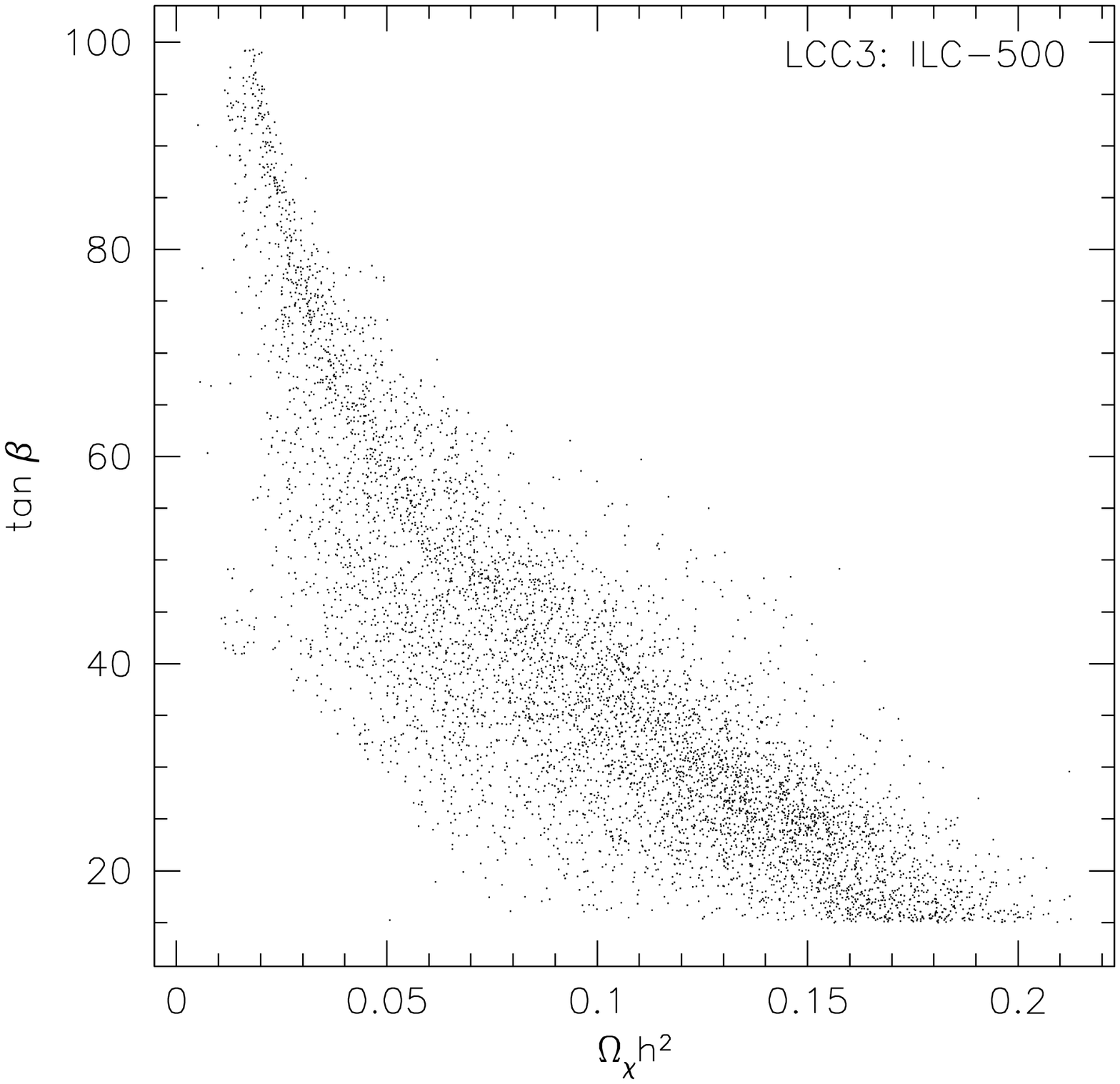,height=3.0in}\\
\epsfig{file=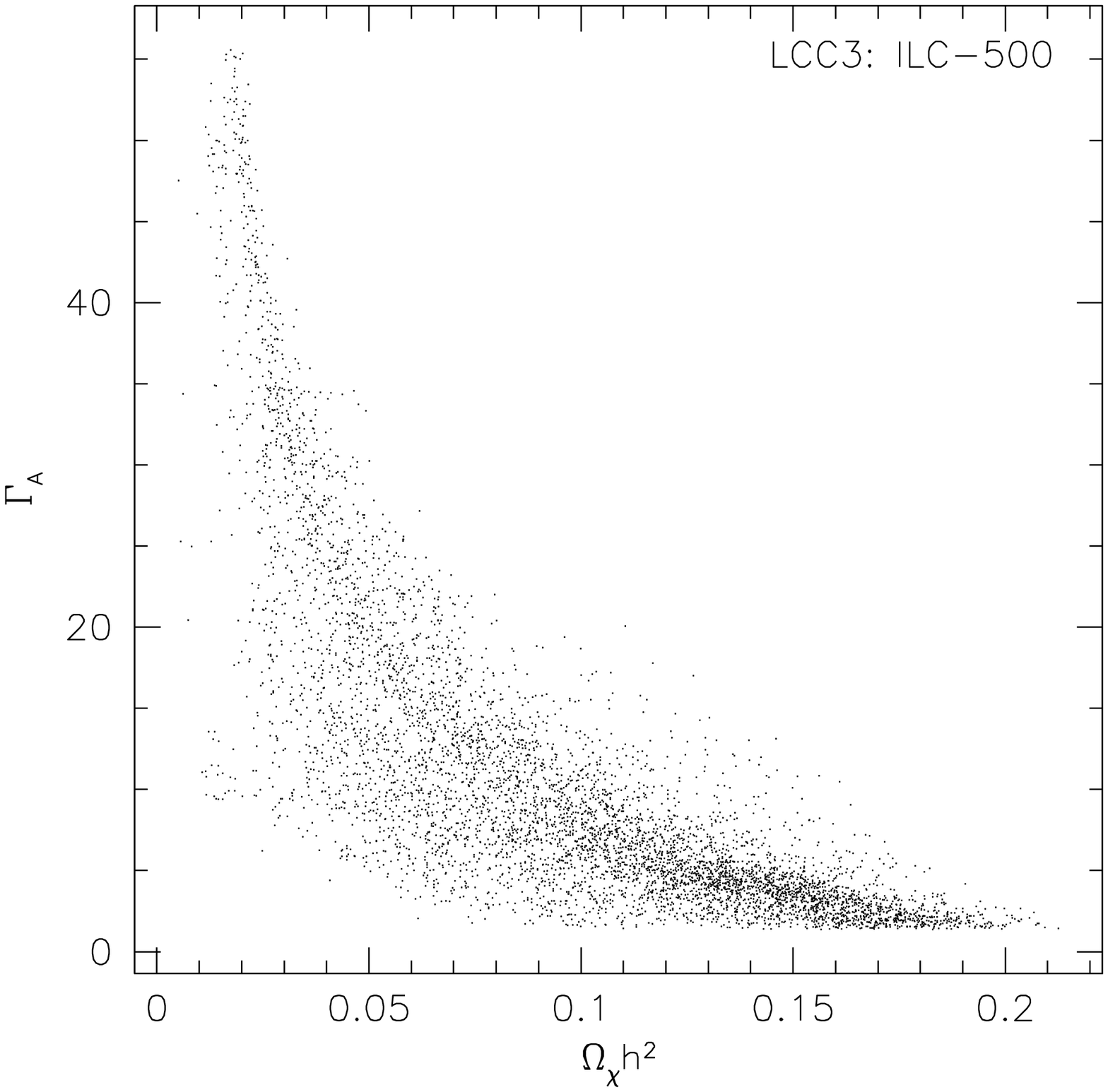,height=3.0in}
\caption{Scatter plots of both $\tan\beta$ and $\Gamma_A$ vs.\ relic density
for point LCC3.  There is a significant correlation, which can be resolved at
ILC-1000.  In fact, the correlation between $\tan\beta$ and $\Gamma_A$ is quite
close.}
\label{fig:LCC3gammaeffect}
\end{center}
\end{figure}

\begin{figure}
\begin{center}
\epsfig{file=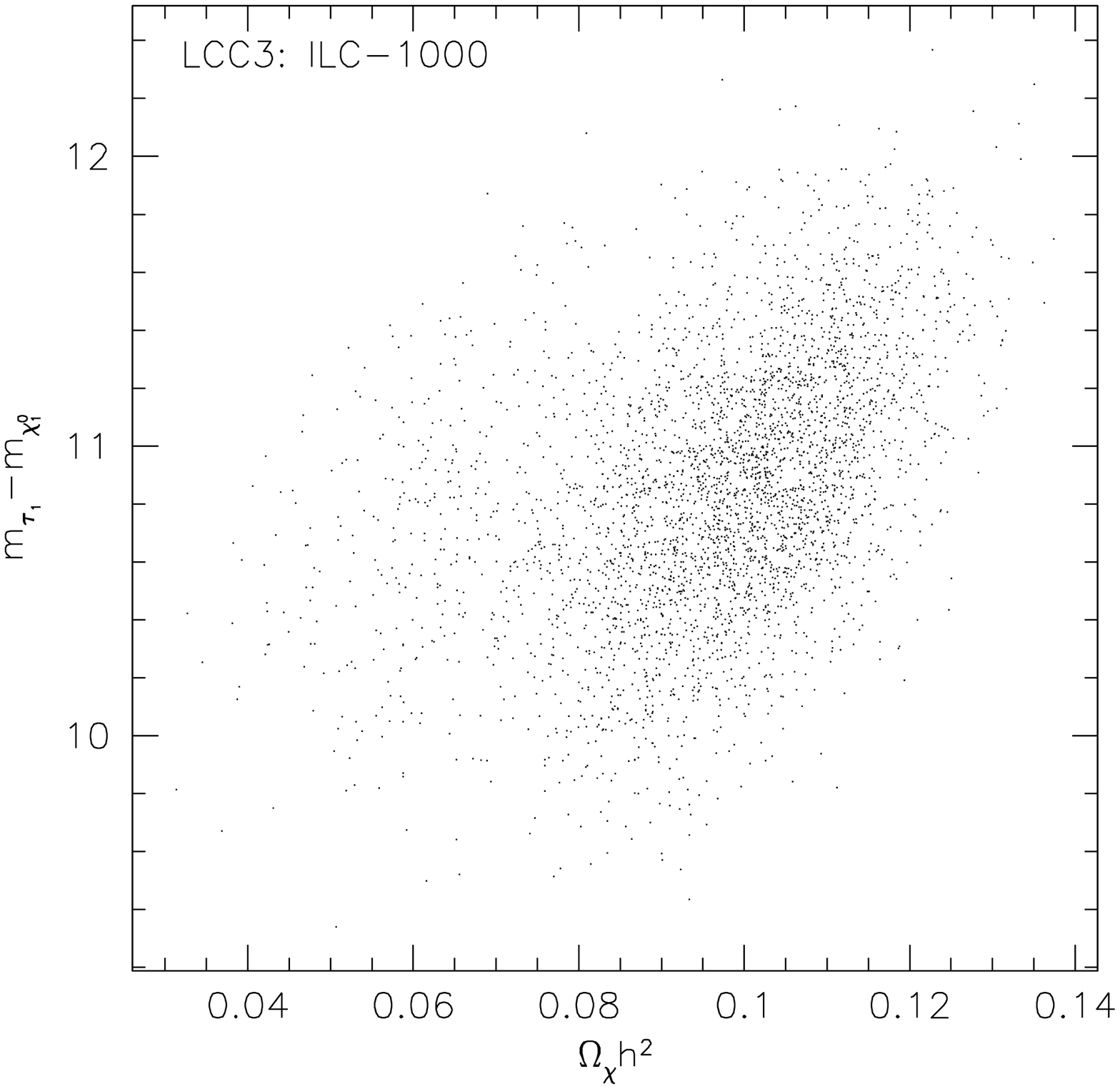,height=3.0in}
\caption{Scatter plot of $m_{\tilde{\tau}}-m_{\chi^0_1}$ for point LCC3,
ILC-1000 sample.  A correlation exists, but it is not strong.}
\label{fig:LCC3deltameffect}
\end{center}
\end{figure}

The dependence of the coannihilation cross sections on parameters that are not
fixed at the 500 GeV ILC is illustrated by the scatter plot shown in
Fig.~\ref{fig:LCC3gammaeffect}.  This plot shows the correlation between
$\Omega_\chi h^2$, $\tan\beta$, and $\Gamma_A$.  The relic density is mainly
determined by annihilations of the stau.
This plot makes clear how the measurement of $\Gamma_A$, which fixes
$\tan\beta$ to about $\pm 2$, has such an important effect.  At the ILC-1000,
there remains a weak correlation between the stau-neutralino mass splitting and
relic density, illustrated in Fig.~\ref{fig:LCC3deltameffect}.

\subsection{Annihilation cross section}

\begin{figure}
\begin{center}
\epsfig{file=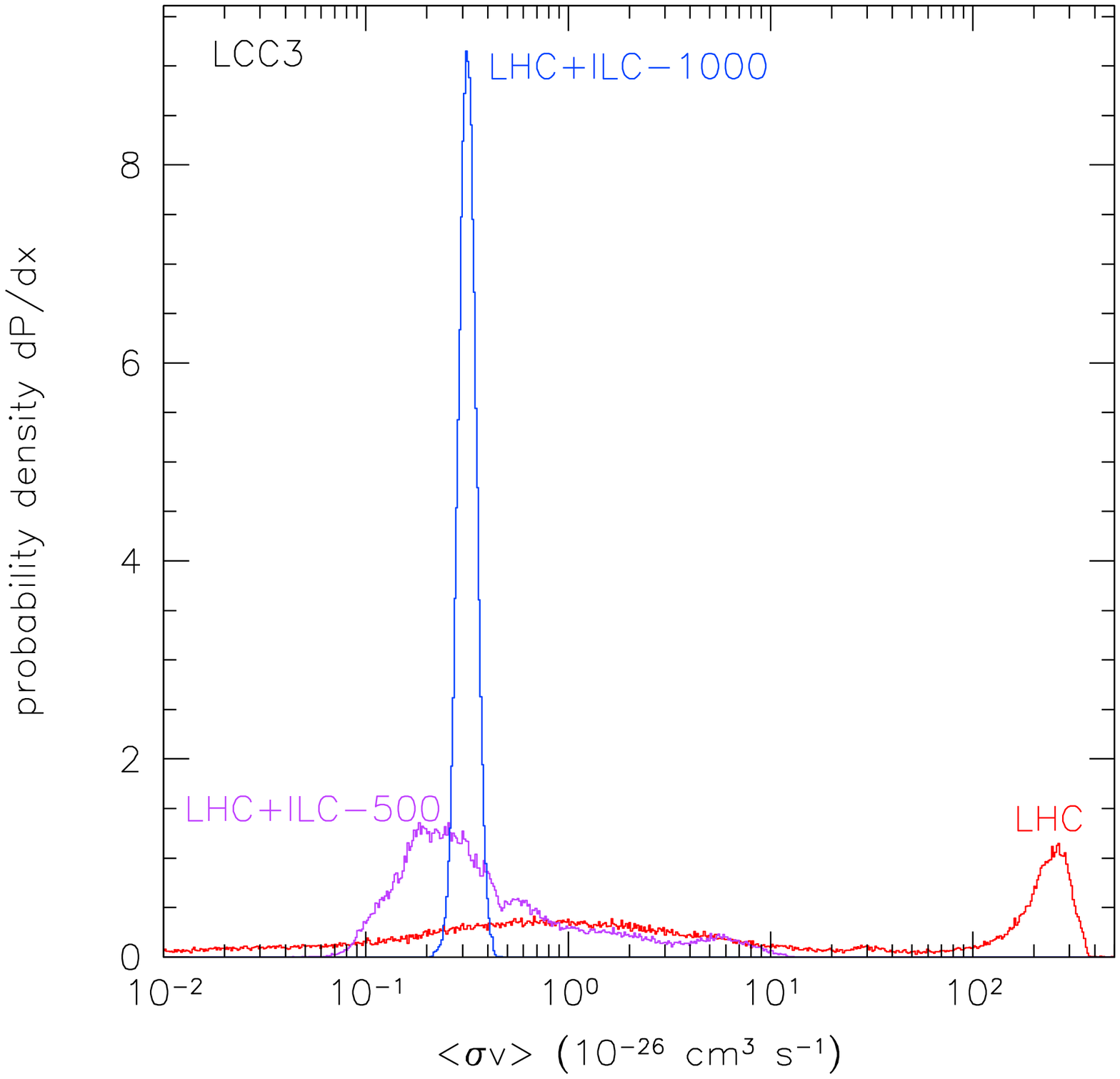,height=3.0in}
\caption{Annihilation cross section at threshold for point LCC3.  The wino
solution giving very high cross section is clearly visible.  See
Fig.~\ref{fig:LCC1relic} for description of histograms.}
\label{fig:LCC3anncs}
\end{center}
\end{figure}

\begin{figure}
\begin{center}
  \epsfig{file=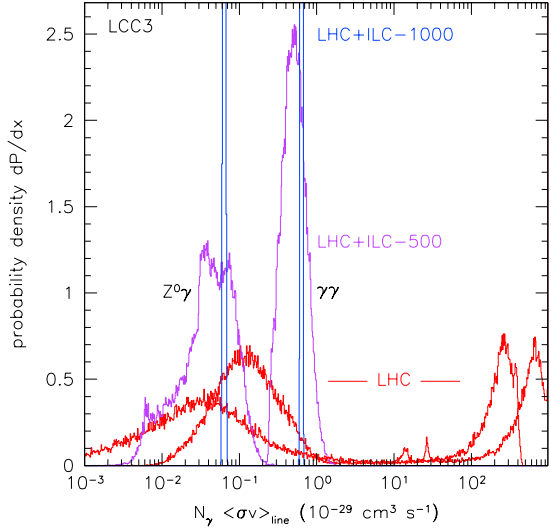,height=3.0in}
\caption{Gamma ray line annihilation cross section at threshold for point LCC3.
The wino solution for LHC data at large cross section is clearly seen.  See
Fig.~\ref{fig:LCC1relic} for description of histograms.}
\label{fig:LCC3twogamma}
\end{center}
\end{figure}

In Fig.~\ref{fig:LCC3anncs}, we show the prediction of our likelihood
analysis for the neutralino pair annihilation cross section at 
threshold.   The dominant annihilation processes contributing to the 
relic density at LCC3 are actually coannihilation reactions, and these 
are not longer available, because all primordial staus have decayed long
ago.  So we have a situation similar to that of LCC1, in which a subdominant
annihilation reaction for the relic density becomes the most important
one for the threshold cross section.  The relevant reaction is the same
one that was important at LCC1, $\chi\chi \to b\bar b$, by $t$-channel
$\s b$ exchange, but also getting a contribution from the $A^0$ 
$s$-channel resonance.  The relative influence of these two contributions
is reflected in the sharpening of the distribution after the $A^0$ 
is determined at the 1000 GeV ILC.  In the LHC distribution, we again see
a  subsidiary
peak at high values of the cross section that reflects the 
possibility of solutions in which the lightest neutralino is wino-
or Higgsino-like.  In Fig.~\ref{fig:LCC3twogamma}, we
 show the similar evolution for the exclusive annihilation cross sections
to $\gamma\gamma$ and $\gamma Z$.

\begin{figure}
\begin{center}
\epsfig{file=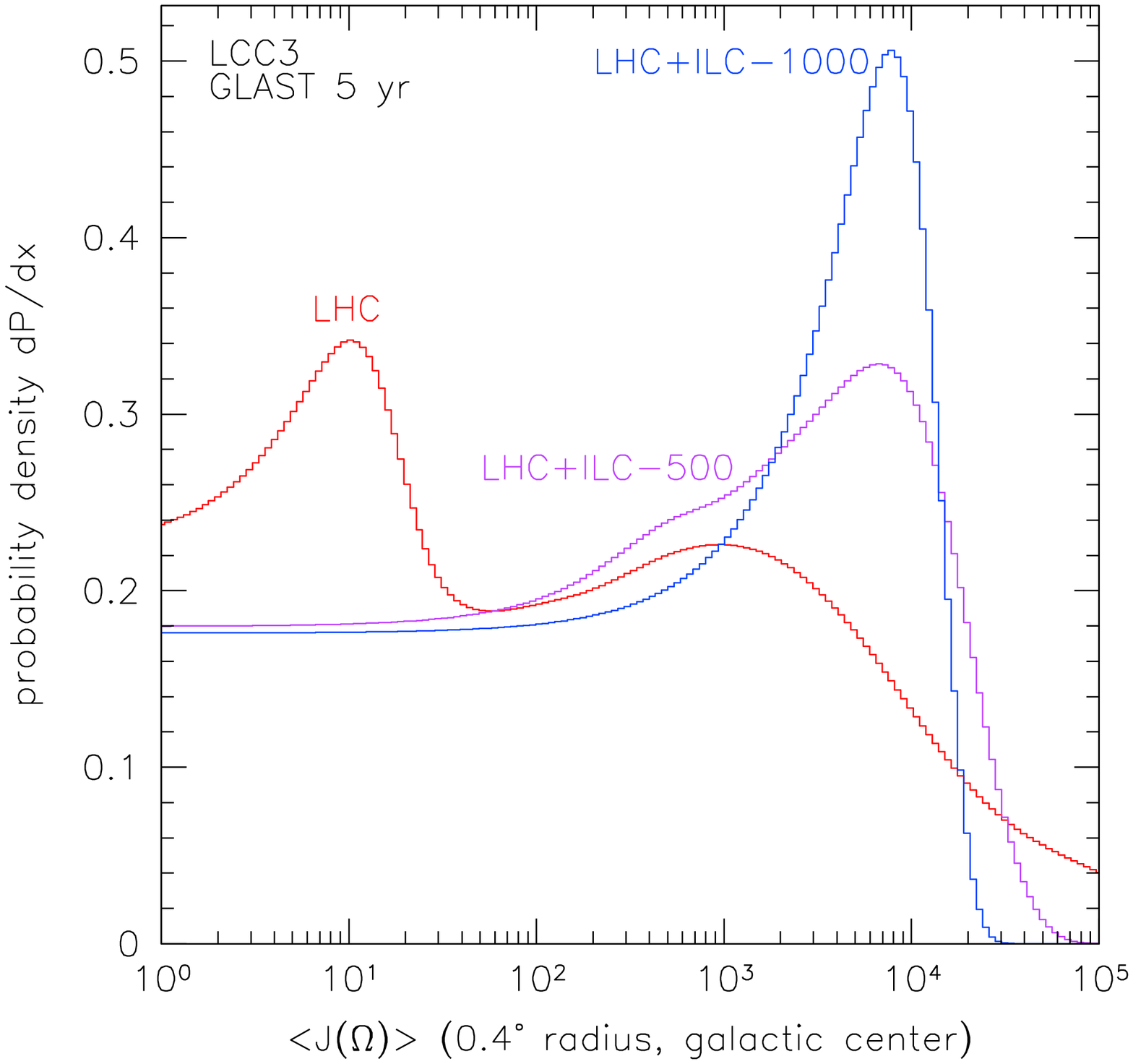,height=3.0in}\\
\epsfig{file=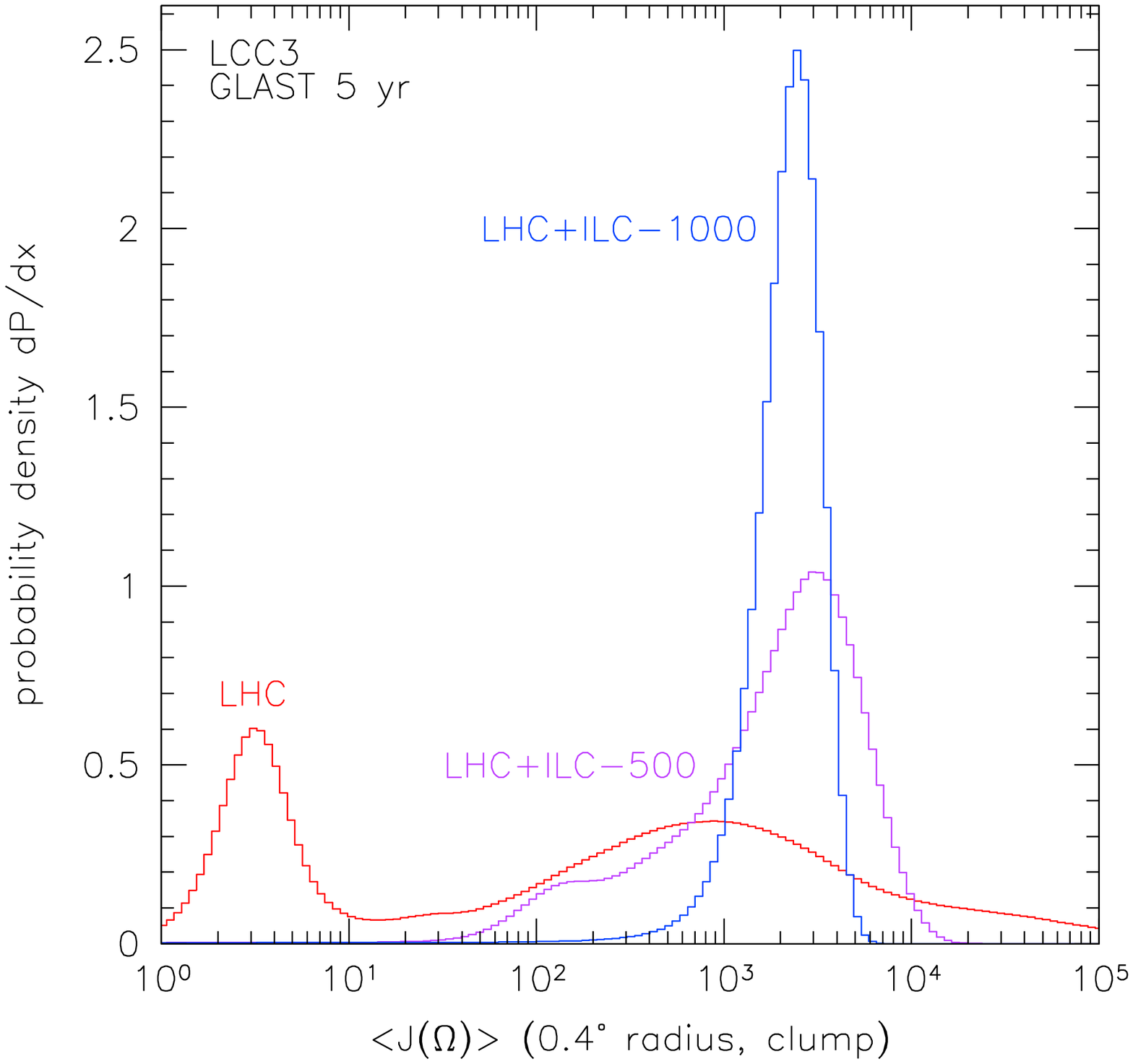,height=3.0in}
\caption{Halo density profiles for point LCC3: (a) galactic center,
(b) dark matter clump in the galactic halo.  Angle-averaged $J$ values as
measured by combining a 5-year all-sky dataset from GLAST with accelerator
measurements are shown.  For LHC, the wino peak at low $J$ is clearly visible.
See Fig.~\ref{fig:LCC1relic} for description of histograms.}
\label{fig:LCC3halo}
\end{center}
\end{figure}

In Fig~\ref{fig:LCC3halo}, we show the result of carrying out the exercise
described in Section 4.4 in which we combine collider data with the 
annihilation gamma
ray signal that should be found by GLAST.  For the galactic center,
we expect 38 signal photons, over 360 background; for 
the canonical halo object, we expect 31 signal photons, over 60 background.
Using  these counting rates and the determination of the annihilation 
cross section that we have described, we obtain the predictions for 
$\VEV{J(\Omega)}$ shown in Fig.~\ref{fig:LCC3halo}.  The quality of the
determinations is somewhat better than in the case of LCC1, For the 
halo object, we already obtain a 20\% measurement with this relatively
small signal.

\subsection{Direct detection cross section}

\begin{figure}
\begin{center}
\epsfig{file=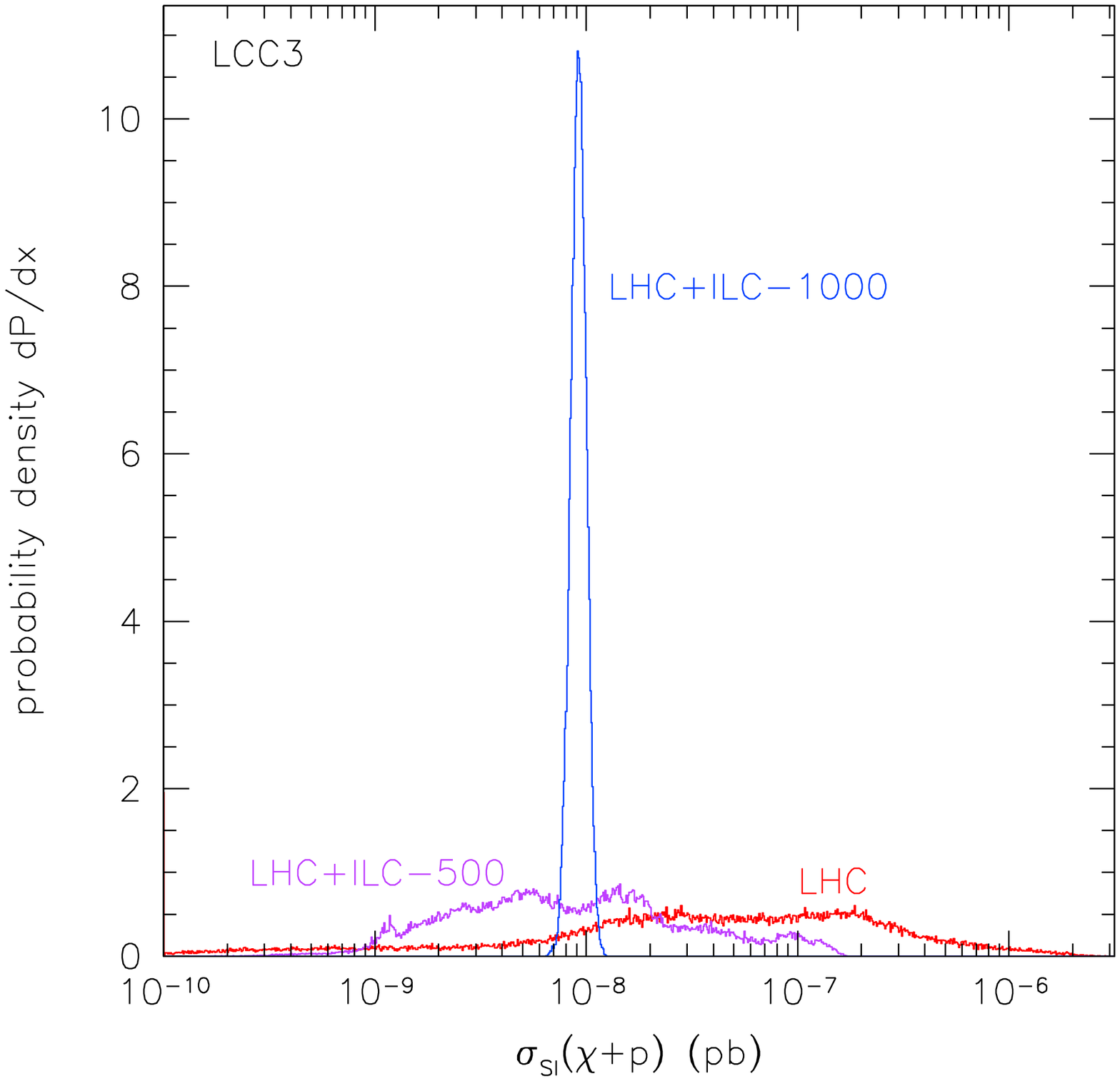,height=3.0in}
\caption{Spin-independent neutralino-proton direct
 detection cross section for
point LCC3.   See
Fig.~\ref{fig:LCC1relic} for description of histograms.}
\label{fig:LCC3direct}
\end{center}
\end{figure}

\begin{figure}
\begin{center}
\epsfig{file=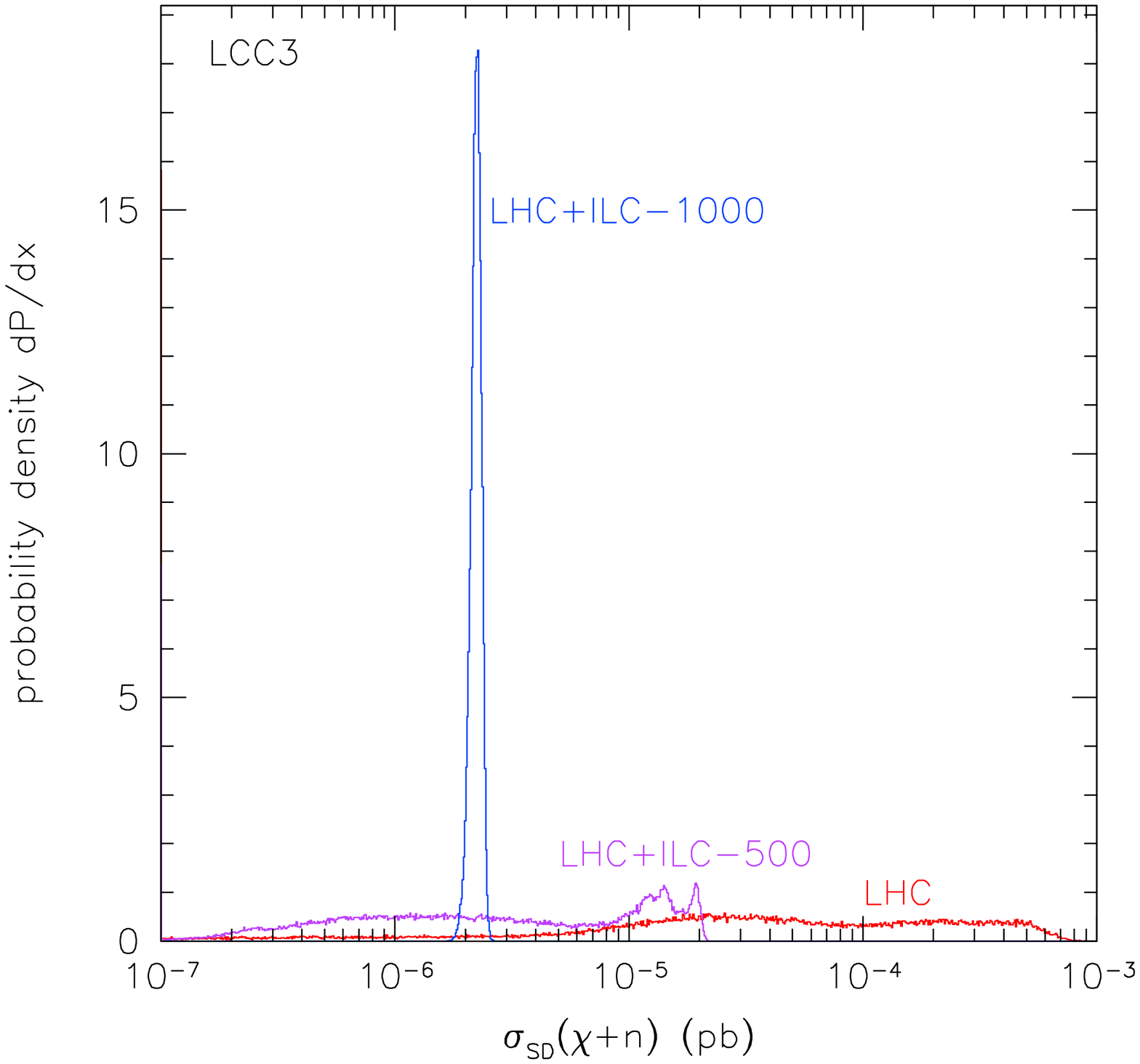,height=3.0in}
\caption{Spin-dependent
neutralino-neutron direct detection  cross section for point LCC3.  See
Fig.~\ref{fig:LCC1relic} for description of histograms.}
\label{fig:LCC3directsd}
\end{center}
\end{figure}

At LCC3, we return to the situation seen at LCC1 in which the 
direct detection cross section is dominated by the $t$-channel 
exchange of the heavy Higgs boson $H^0$.   The direct detection 
cross section is poorly determined by the LHC data, even though the
mass of the $H^0$ is known.  However, the progressive clarification of
the supersymmetry mixing angles and the value of $\tan\beta$ at the 
500 GeV and 1000 GeV ILC yields a fairly precise determination.
The evolution is shown in  Fig.~\ref{fig:LCC3direct}.
The corresponding predictions for the spin-dependent part of the 
neutralino-neutron cross section are shown in Fig~\ref{fig:LCC3directsd}.

\begin{figure}
\begin{center}
\epsfig{file=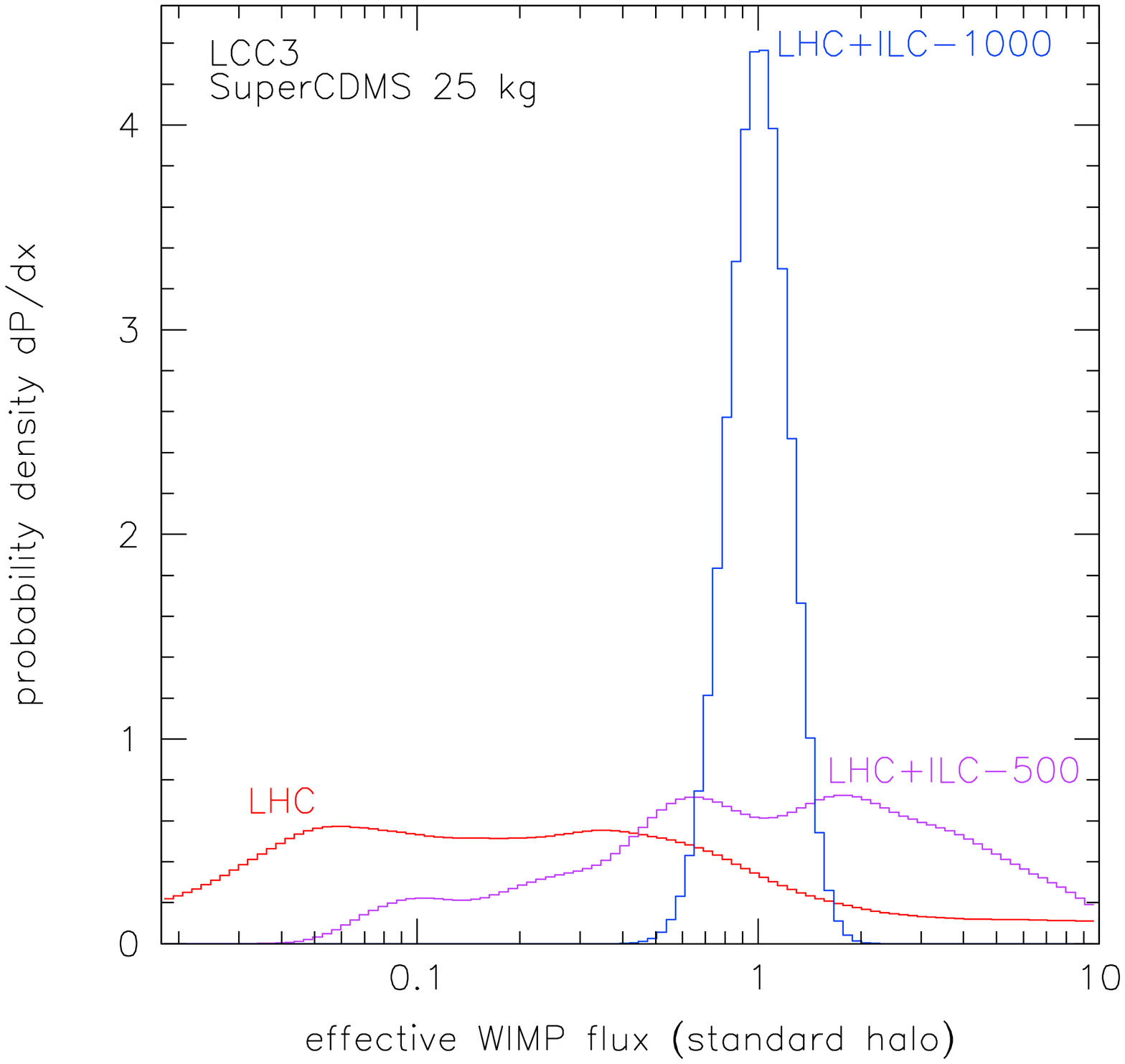,height=3.0in}
\caption{Effective local 
WIMP flux at the Earth for point LCC3.  The results assume 
the SuperCDMS measurement
described in the text.
  See Fig.~\ref{fig:LCC1relic} for description of histograms.}
\label{fig:LCC3localhalo}
\end{center}
\end{figure}

As before, we can combine our determination of the detection cross section with
the expected event yield from the SuperCDMS detector to estimate our ability to
directly measure the local flux of dark matter at the earth.  Again, 
our analysis omits the uncertainty from 
low-energy QCD parameters.  For this case, we
expect a signal of 27 events in SuperCDMS. This gives the determination of the
effective local flux of neutralinos at the Earth that is shown in
Fig.~\ref{fig:LCC3localhalo}.


\subsection{Constraints from relic density and direct detection}

\label{section:lcc3Astro}
Point LCC3 illustrates a scenario that in some sense is quite likely.  The LHC
will measure some part of the low energy spectrum of supersymmetric particles,
but the mechanism for establishing relic density will be completely unknown.
In this case, coannihilations with the light stau reduce the relic density by a
large amount.  LHC data give no hint of this as the stau is unobserved.
Furthermore, only two neutralinos are observed at LHC, and no chargino is
observed.  Thus, the composition of the neutralinos is almost completely
unknown.

Applying a relic density constraint to LHC data is a technical challenge as the
range allowed is so large.  Applying even the current WMAP constraint of 6\%
reduces the effective number of samples by a factor of 40.  Nevertheless, we
can make a general statement about the composition of the lightest neutralino
in that if it has too large a wino or Higgsino fraction, the relic density will
be too low.  Thus, the relic density constraint selects the 
pure bino solution.

The direct detection cross section is governed by the heavy Higgs boson, seen
at the LHC, and the gaugino-Higgsino mixing of the neutralino.  From the
cross section measured by SuperCDMS  and the LHC measurement of the 
$H$ boson mass, it could
immediately be inferred that the WIMP was a fairly pure gaugino or Higgsino
($Z_g$ or $Z_h>0.94$).

\begin{figure}[p]
\begin{center}
\epsfig{file=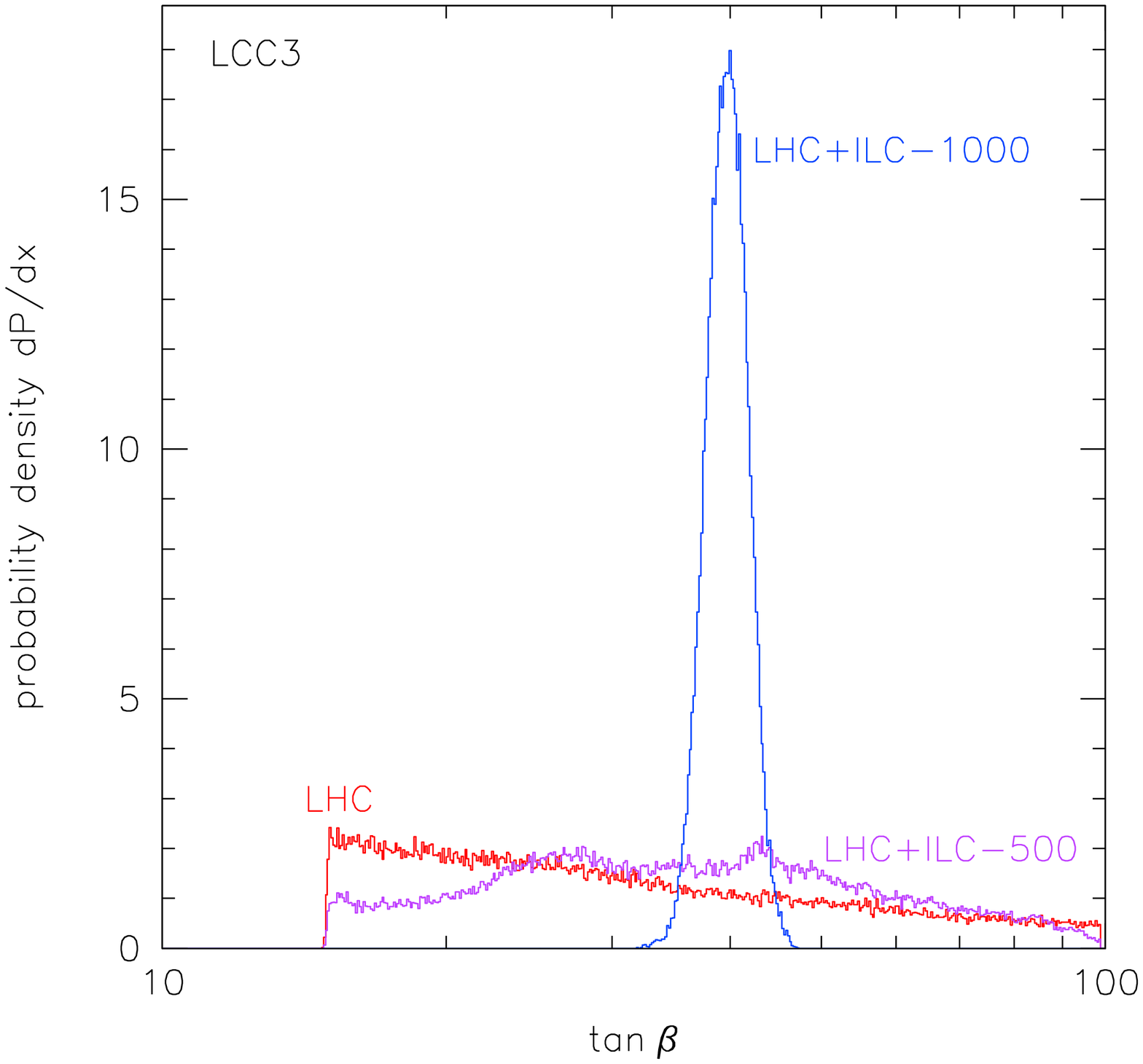,height=2.5in}
\epsfig{file=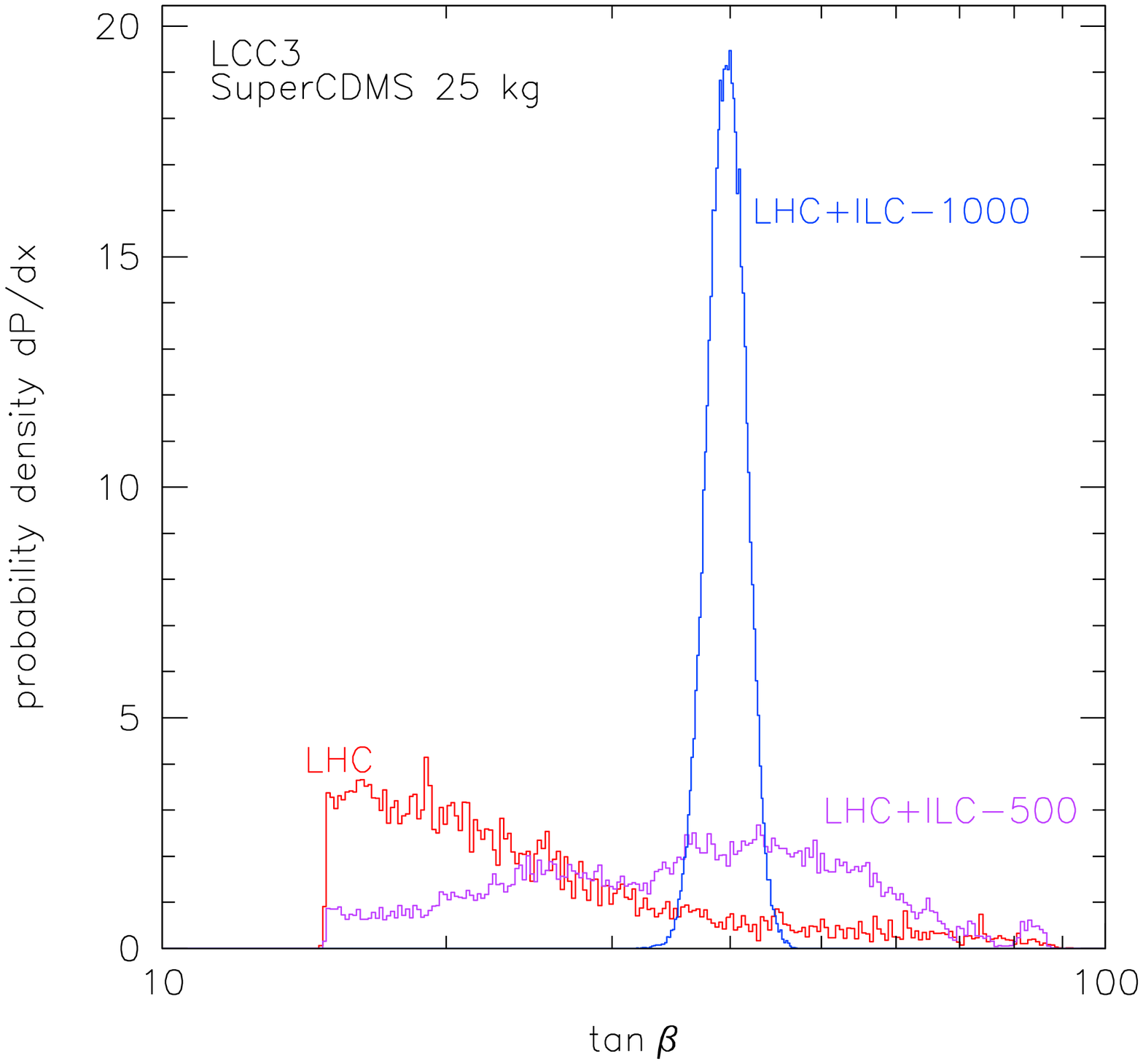,height=2.5in}
\epsfig{file=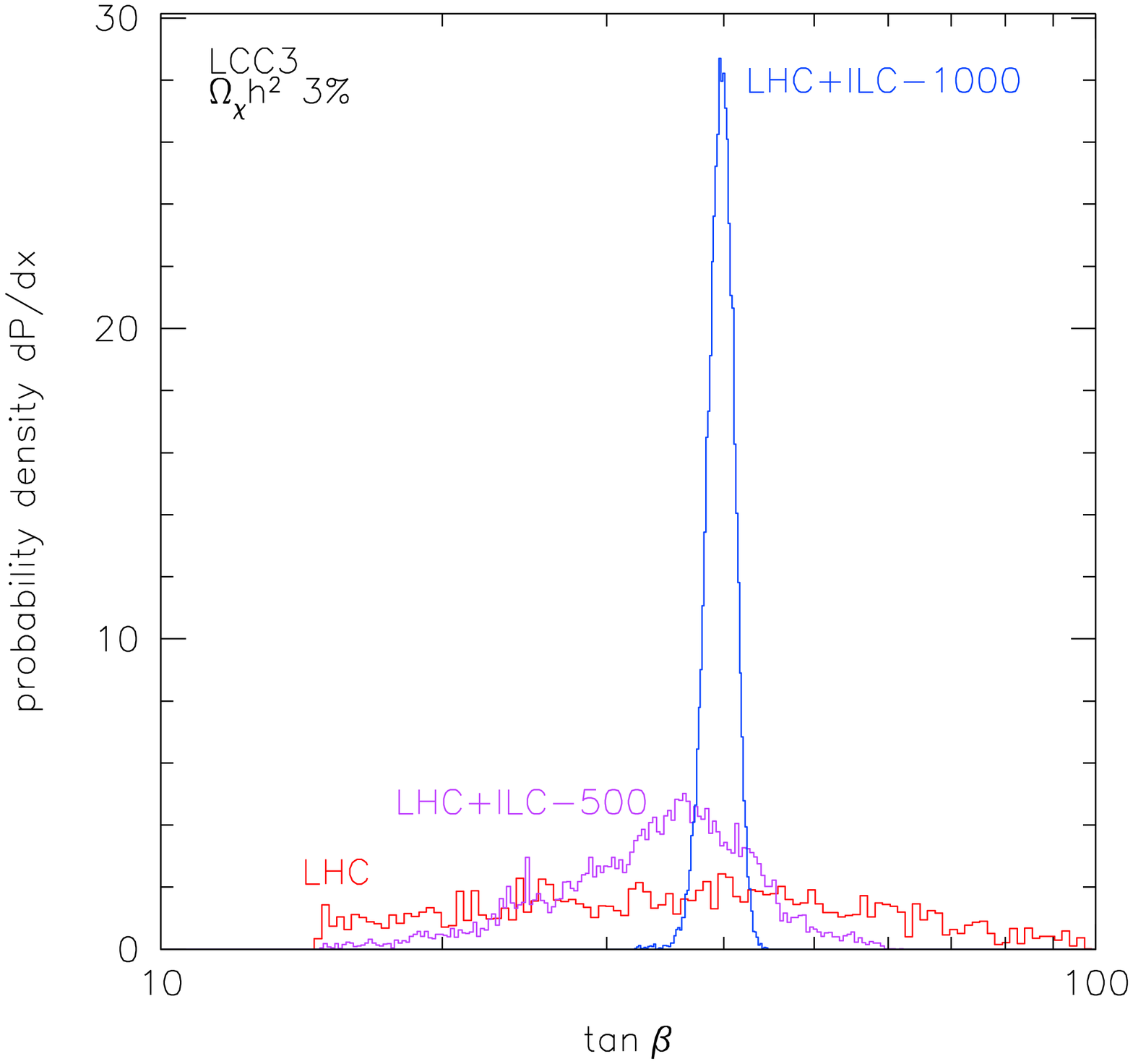,height=2.5in}
\epsfig{file=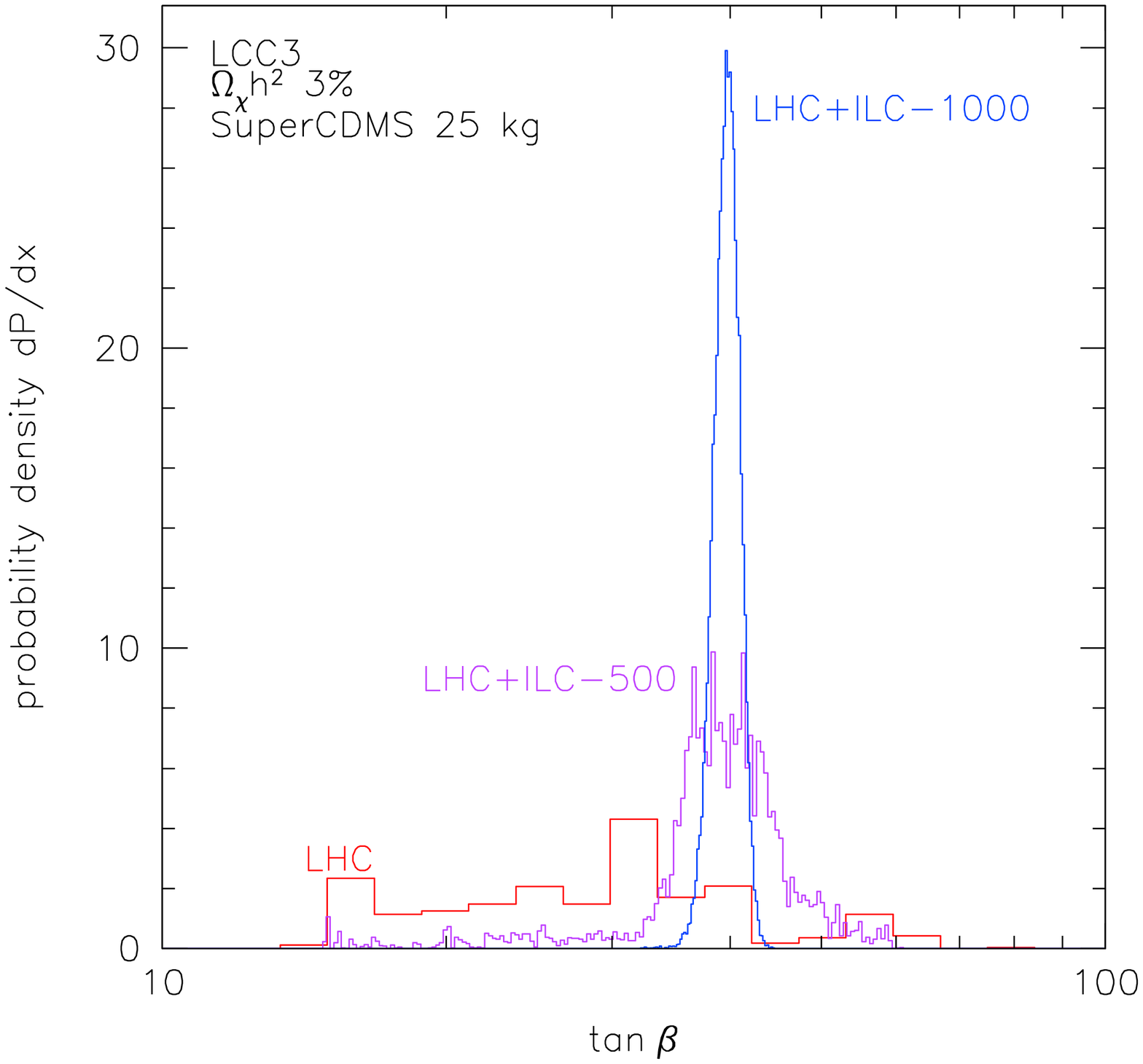,height=2.5in}
\caption{Measurements of $\tan\,\beta$ for point LCC3.  Various astrophysical
constraint sets are included.  The relic density constraint improves even the
ILC-1000 measurement, while the ILC-500 measurement is greatly helped by
including both direct detection and relic density.}
\label{fig:LCC3tanbeta}
\end{center}
\end{figure}

The ILC-500 measurements of relic density and direct detection cross section
are still quite poor.  We can illustrate an interesting possibility here, that
the combination of relic density and direct detection constraints can be more
powerful than either alone.  In particular, we study the distribution of
$\tan\,\beta$.  As shown in Fig.~\ref{fig:LCC3tanbeta}, the direct detection
constraint alone has little effect.  Applying the relic density constraint
tightens the ILC-500 measurement considerably (and in fact reduces the ILC-1000
errors from 5\% to 3\%).  Applying the direct detection constraint does have
significant power if the relic density constraint is also applied, especially
in the ILC-500 sample.

The ILC-1000 measures the direct detection cross section to 9\%, more
accurately that it can be determined by SuperCDMS.
But the  relic density can only be measured to 18\%, so
there is useful information in a cosmological relic density constraint.
However, we have not been able to identify a clear beneficiary for this
information.  Somewhat surprisingly, the stau-neutralino mass
splitting has only a weak  correlation with relic density, given the ILC-1000
measurements (see Fig.~\ref{fig:LCC3deltameffect}).

\section{Benchmark point LCC4}

The general properties of the supersymmetry spectrum at the benchmark
point LCC4 are very similar to those at LCC3. 
The two points differ completely, however, in the physics that establishes
 the neutralino relic density.  At LCC3, the stau and neutralino were 
sufficiently close in mass that coannihilation dominated the annihilation
of supersymmetric particles in the early universe.  At LCC4, the parameters
of the model are adjusted in a different way, such that the $A^0$
boson creates a resonance in neutralino annihilation near threshold.

The $A^0$ resonance has the potential to increase the neutralino annihilation
cross section by three orders of magnitude. This implies that the influence
of the $A^0$ typically creates a `funnel' in the parameter space.  When 
$m(A^0) = 2 m(\chi)$ precisely, the annihilation cross section is very large
and the neutralino relic density is essentially zero.  As the $A^0$ moves
away from the neutralino pair threshold, the annihilation cross section 
reverts to its typical small value.  In two intermediate regions, one on
each side of the threshold, the effect of the resonance is just right to
produce an S-wave cross section of about 1 pb.  The 
location of this region depends on the $A^0$ mass but also on the $A^0$
width.  The width of the $A^0$ is sensitive to many parameters of the 
theory, and especially to the value of $\tan\beta$, which is difficult
to determine independently.   Thus, as we will see, none of the properties
of the neutralino are determined particularly well at this point until 
the $A^0$ is measured in $\ee$ annihilation. However,  once that is 
done, the quantities needed for the study of dark matter snap into place.

\begin{figure}
\begin{center}
\epsfig{file=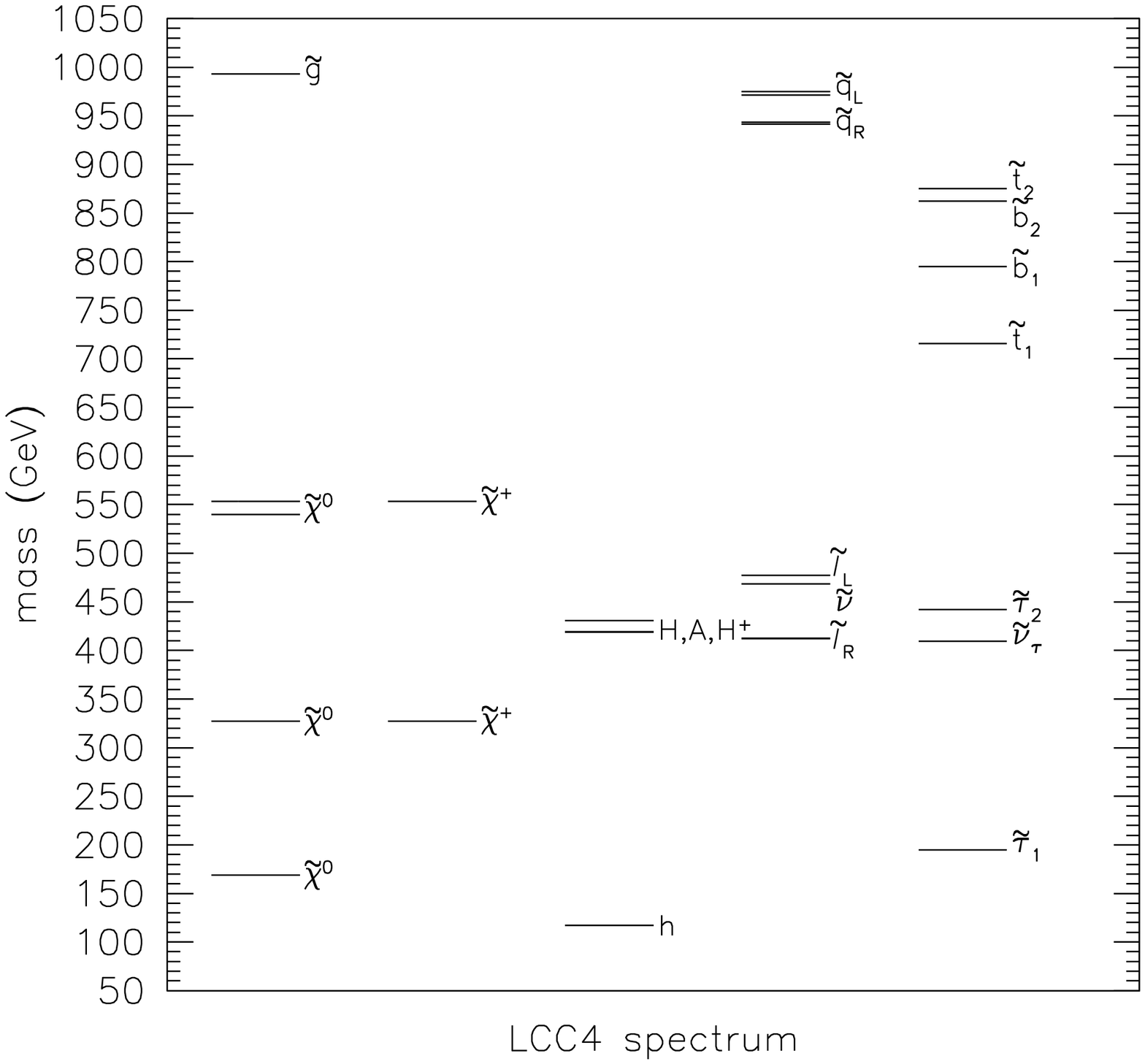,height=4.0in}
\caption{Particle spectrum for point LCC4.  The lightest neutralino is
predominantly bino, the second neutralino and light chargino are predominantly
wino, and the heavy neutralinos and chargino are predominantly
Higgsino.}
\label{fig:LCC4spectrum}
\end{center}
\end{figure}

\subsection{Spectroscopy measurements}

The supersymmetry spectrum at the point LCC4 is shown in 
Fig.~\ref{fig:LCC4spectrum}. 
The properties of the supersymmetry spectrum at LCC4 visible to the 
LHC and to the 500 GeV ILC are very similar to those discussed for 
LCC3.  The squarks and the 
gluino all lie below 1 TeV, so these states should be seen and 
characterized at the LHC. The two lightest states in the neutralino
spectrum should also be visible.  The parameter set of LCC4
 has a large value of $\tan\beta$, making the $A^0$ and $H^0$ visible
at the LHC in their $\tau^+\tau^-$ decay modes.  The large value of 
$\tan\beta$ also leads to 
large downward shifts of the masses of the lighter tau 
slepton and the bottom squark.  In this case, the lighter stau is the only 
slepton with mass below 350 GeV, and thus it will be difficult to see
any slepton at the LHC.   The point LCC4 shares with LCC3 the property 
that only the few lightest  supersymmetric
states will be visible at the 500 GeV stage of the ILC.

At the 1000 GeV ILC, new features of the model can be studied in detail.
The process $\ee\to H^0 A^0$ becomes accessible, leading to striking
events with four $b$ jets. The analysis of these events has been 
studied in \cite{BattagliaParis}. 
 By combining the jets in pairs, it is possible
to reconstruct the mass peak of the heavy Higgs bosons.  The widths of the
$H^0$ and $A^0$ can be measured from this mass distribution.  The two 
peaks substantially overlap, since the widths are about 14 GeV while the 
mass splitting is about 1 GeV.  Since the splitting and the width are
controlled by the common MSSM parameters $\tan\beta$ and $m(A^0)$, 
it is possible to fit for those parameters and improve the accuracy of the
$A^0$ width determination.  

This is the only place in the paper where we
make strong use of 
 the fact that we are restricting ourselves to the MSSM rather
than working in a still more general supersymmetric model.  However,
in a model with a more general Higgs structure, the 
masses of the $H^0$ and $A^0$ would typically be distinct, and the two sets 
of masses and widths could be obtained separately by kinematic fitting 
of 4-jet events.

The higher energy running of the ILC also makes it possible to observe
the gauginos $\s\chi_1^+$ and  $\s\chi^0_3$ and to measure their 
masses.  This gives enough information to determine the parameter $\mu$
and thus to fix the gaugino-Higgsino mixing angles.

The complete list of spectrum constraints that we expect for this point
for the LHC and for each stage of the ILC is given in 
Tables~\ref{tab:LCC4masses}
and \ref{tab:LCC4css}.

\begin{table}
\centering
\begin{tabular}{lccccc}
   mass/mass splitting  & LCC4 value & &  LHC  & ILC 500 &  ILC 1000\\ 
         \hline
  $m(\s\chi^0_1)$    &  169.1       & $\pm$ &    17.0  &  - & 1.4      \\ 
  $m(\s\chi^0_2)$    &  327.1      & $\pm$ &    49.    \\ 
  $m(\s\chi^0_2) - m(\s\chi^0_1)$ & 158.0     & $\pm $ &   -  & - &  1.8  \\
  $m(\s\chi^0_3) - m(\s\chi^0_1)$ & 370.6     & $\pm $ &   - &  -   &  2.0 \\ 
  $m(\s\chi^+_1)$  &  327.5    &  $\pm $ &  -  &  - & 0.6 \\ 
  $m(\s\chi^+_1)-m(\s\chi^0_1)$  &  158.4   &  $\pm $ &  -  &  - & 2.0  \\ 
  $m(\s\chi^+_2) - m(\s\chi^+_1) $ &  225.8  &  $\pm$ &   -  &    -   & 2.0 \\
       \hline
  $m(\s e_R) - m(\s\chi^0_1)$  &  243.2       & $\pm $ &   - & -   &   0.5  \\
  $m(\s \mu_R) - m(\s\chi^0_1)$   & 243.0     & $\pm $ &    - &-   &   0.5 \\ 
  $m(\s \tau_1)$   &  194.8       &  $\pm $ &   -  &  0.9 \\ 
  $m(\s \tau_1) - m(\s\chi^0_1)$   & 25.7       &  $\pm $ &   -  &  1.0 \\ 
        \hline
  $m(h)     $    &      117.31           &  $\pm $ & 0.25  &   0.05 \\ 
  $m(A)     $   &       419.3          &  $\pm$ &   1.5 *  &  - & 0.8      \\
  $\Gamma(A)     $   &   14.8           &  $\pm$ &   -   &  - & 1.2     \\
       \hline
  $m(\s u_R)$, $m(\s d_R)$ &  944.,941.        & $\pm $ &   94.  &   \\
  $m(\s s_R)$, $m(\s c_R)$ &  941., 944.          & $\pm $ &   97.  &     \\
  $m(\s u_L)$, $m(\s d_L)$ &  971., 975.          & $\pm $ &   141.  &    \\
  $m(\s s_L)$, $m(\s c_L)$ &  975., 971.         & $\pm $ &   146. &    \\
  $m(\s b_1)     $ &     795.                &  $\pm $ &   40.  &   \\ 
  $m(\s b_2)     $ &     862.                &  $\pm$ &   86.   &     \\
  $m(\s t_1)     $ &    716.                 &  $\pm $ &  ($> 345$)   &  \\ 
        \hline
  $m(\s g)$      &  993.        &  $\pm $ &   199.     &    \\ 
\end{tabular}
\caption{Superparticle masses and their estimated errors or limits for 
       the parameter point LCC4.  
     The notation is as in Table~\ref{tab:LCC3masses}.}
\label{tab:LCC4masses}
\end{table}

\begin{table}
\centering
\begin{tabular}{lcccrr}
   cross section
    &  & LCC4 value & &  ILC 500 &  ILC 1000\\
         \hline
   minimal set\\
       \hline
  $\sigma(\ee\to \s\tau^+_1\s\tau^-_1)$
      & LR &  21.1                & $\pm $   &    100\% \\
     & RL &  54.2                & $\pm $    &    7\% \\
  $\sigma(\ee\to \s\chi^+_1\s\chi^-_1)$
      & LR & 137.3 (0.786)                 & $\pm $  &    & 4\% \\
     & RL &  4.0   (0.760)             & $\pm $    &   &  - \\
  $\sigma(\ee\to \s\chi^0_1\s\chi^0_2)$
      & LR &  43.9             & $\pm $   &   &  7\%\\
     & RL &  1.8              & $\pm $   &     &  - \\
  $\sigma(\ee\to \s e^+_R\s e^-_R)$
      & LR & 5.1  (0.683)             & $\pm $  &    & -  \\
     & RL & 84.7 (0.899)              & $\pm $   &    & 5\% \\
         \hline
\end{tabular}
\caption{SUSY cross sections and estimated errors for
    the parameter point LCC4.  The notation is as in Table~\ref{tab:LCC1css}.
         The symbol
   (-) denotes that the cross section is less than 10 fb or is otherwise
        not measurable.  The errors   are estimated using
                  \leqn{erroronsig}.}
  \label{tab:LCC4css}
\end{table}

\subsection{Relic density}

\begin{figure}
\begin{center}
\epsfig{file=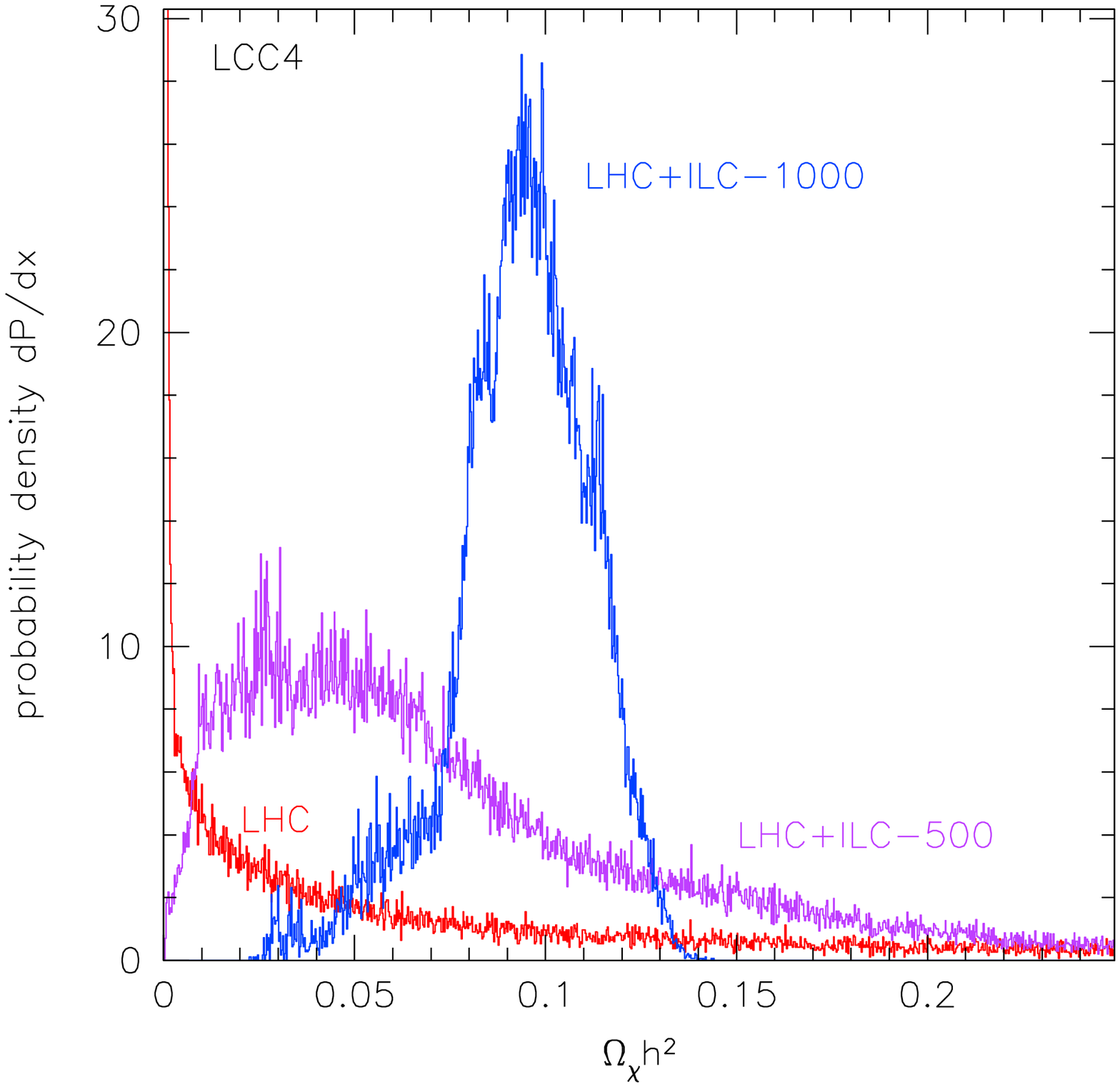,height=3.0in}
\caption{Relic density measurement for point LCC4.  The wino peak at very small
relic density is clear.  See Fig.~\ref{fig:LCC1relic} for description of
histograms.} 
\label{fig:LCC4relic}
\end{center}
\end{figure}

\begin{figure}
\begin{center}
\epsfig{file=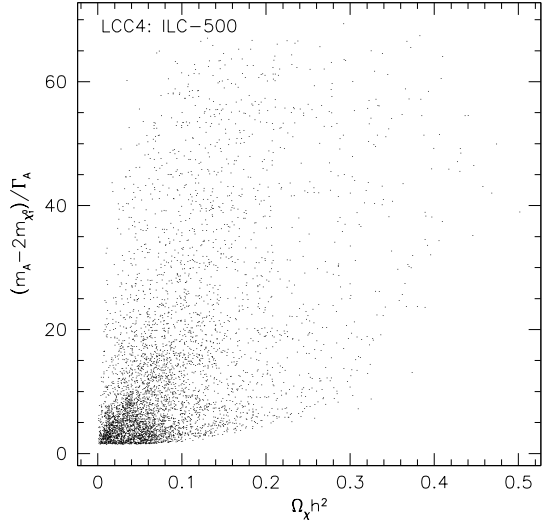,height=3.0in}
\caption{Scatter plot of the distance from the $A^0$ resonance vs.\ relic
density for point LCC4.  The quantity plotted is actually the distance in
widths, with true value 5.5.}
\label{fig:LCC4dmaeffect}
\end{center}
\end{figure}

The evaluation of the relic density at LCC4 from collider data reflects
the fact that the result depends sensitively on both the mass and width
of the $A^0$ boson.  At the LHC and the 500 GeV ILC, the qualitative
features of the supersymmetry spectrum are known, and the mass of the 
lightest neutralino and the $A^0$ are both known to some precision.  
However, this still leaves almost complete uncertainty as to the 
predicted value of the relic density.  It is only when the width of the
$A^0$ is measured that the actual picture begins to come into focus.
This evolution is shown clearly in Fig.~\ref{fig:LCC4relic}.  The dependence of
relic density on the $A^0$ width is easily seen in the ILC-500 data,
illustrated in Fig.~\ref{fig:LCC4dmaeffect}.

The predicted value of the relic density also depends on the gaugino-Higgsino
mixing parameters.   The measurement of the mass of the $\s\chi_3^0$, in 
particular, allows one to determine the $\mu$ parameter.  This significantly
improves the determination shown in Fig.~\ref{fig:LCC4relic} over scans
done without this information.   With the full set of collider data,
we find that the relic density can be predicted to 19\% accuracy.

\subsection{Annihilation cross section}

\begin{figure}
\begin{center}
\epsfig{file=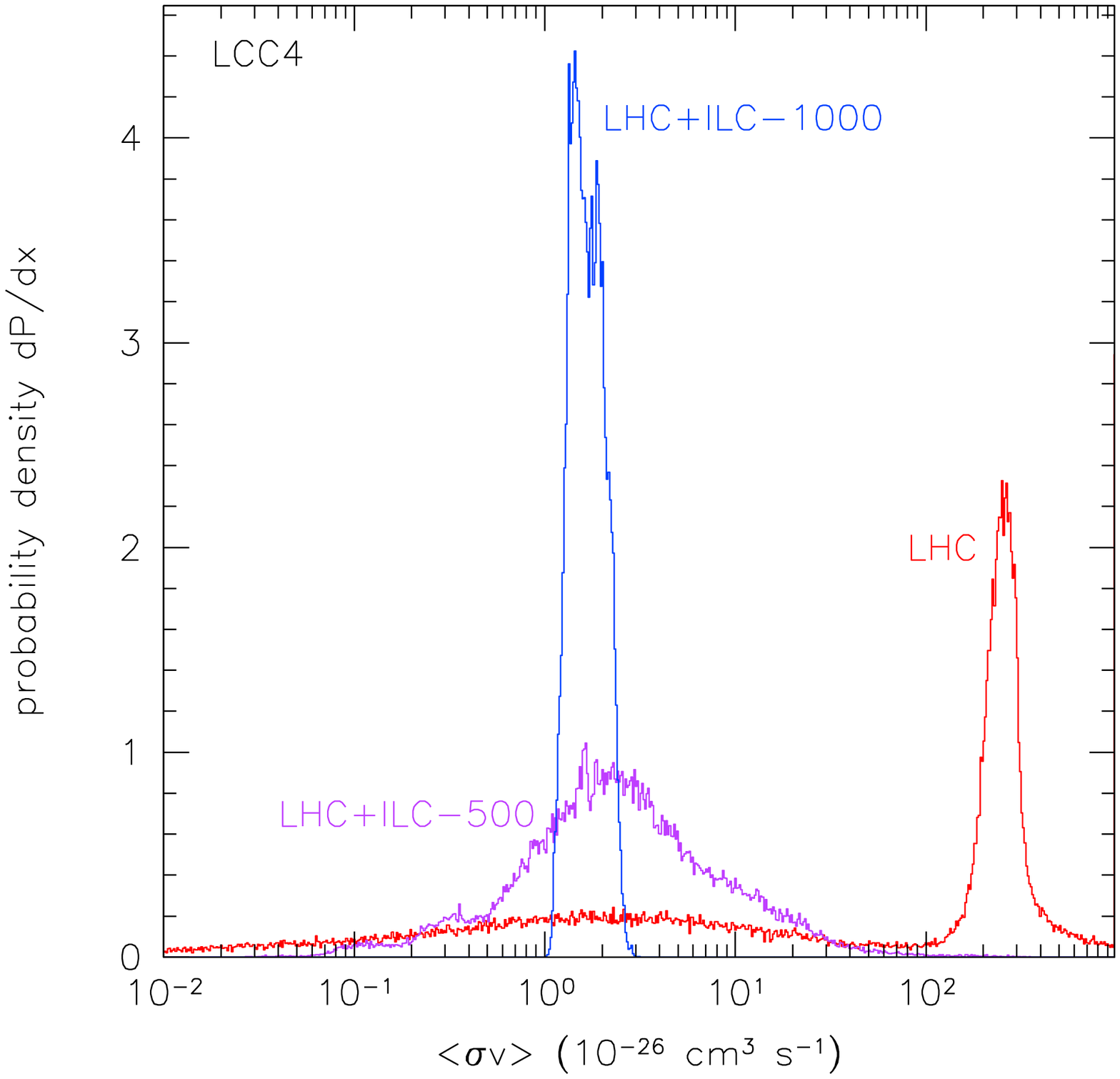,height=3.0in}
\caption{Annihilation cross section at threshold for point LCC4.  The wino
solution giving very high cross section is clearly visible.  See
Fig.~\ref{fig:LCC1relic} for description of histograms.}
\label{fig:LCC4anncs}
\end{center}
\end{figure}

\begin{figure}
\begin{center}
\epsfig{file=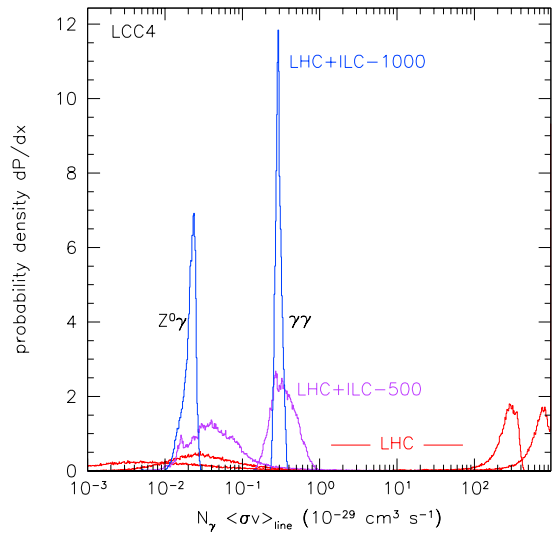,height=3.0in}
\caption{Gamma ray line annihilation cross section at threshold for point LCC4.
Again, the wino solution at large cross section is clear.  See
Fig.~\ref{fig:LCC1relic} for description of histograms.}
\label{fig:LCC4twogamma}
\end{center}
\end{figure}

In Fig.~\ref{fig:LCC4anncs}, we show the prediction of the neutralino
annihilation
cross section at threshold from collider data. Because the resonant
annihilation through the $A^0$ is a simple S-wave process, this cross
section is highly correlated with the predicted relic density discussed
in the previous section.  The form of the prediction is, again, 
complete ignorance
until the $A^0$ width is measured, and a sharp value thereafter.
The prediction of the exclusive annihilation cross sections to $\gamma\gamma$
and $\gamma Z$ follow the same pattern; this is shown in 
Fig.~\ref{fig:LCC4twogamma}.

\begin{figure}
\begin{center}
\epsfig{file=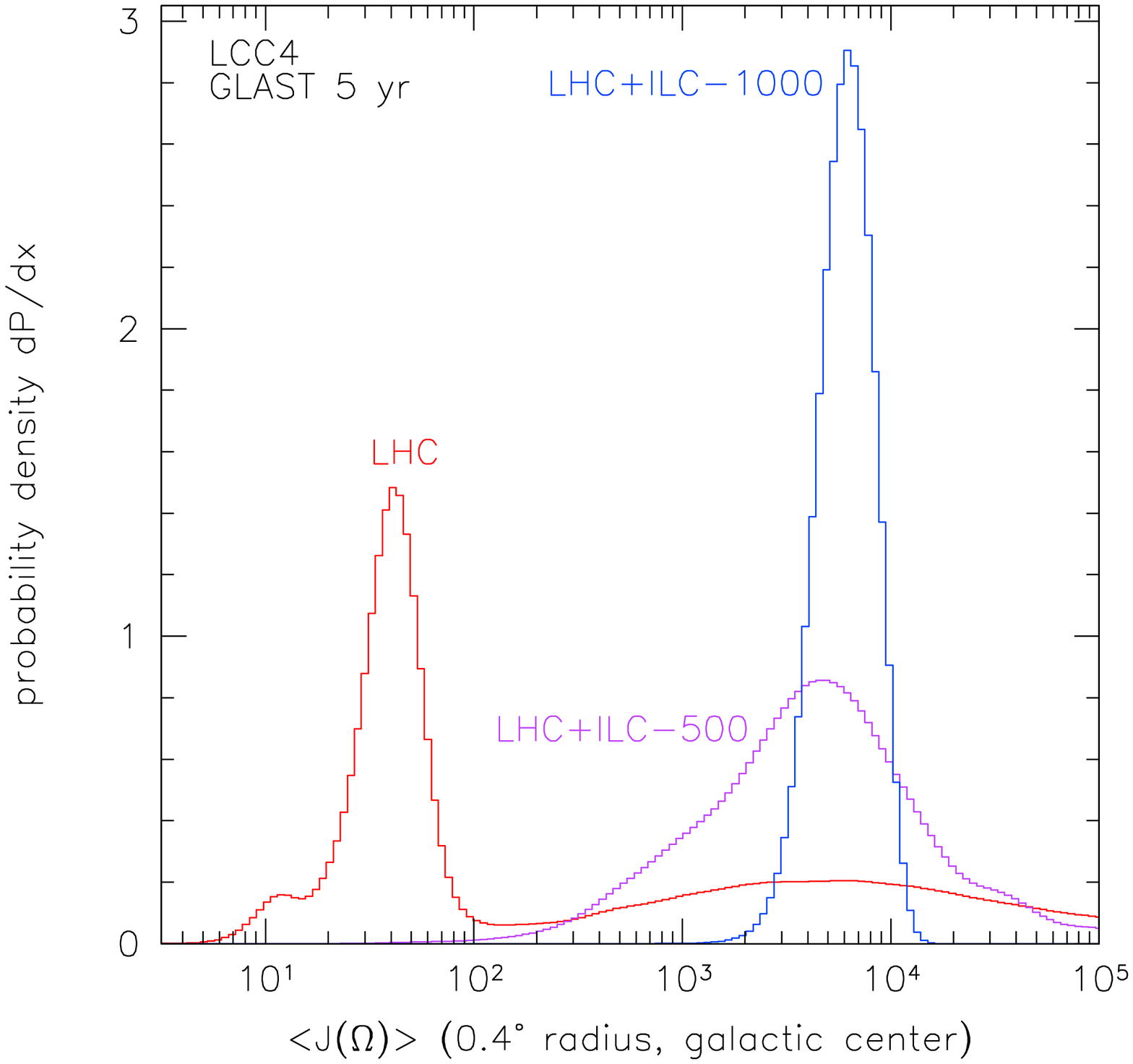,height=3.0in}\\
\epsfig{file=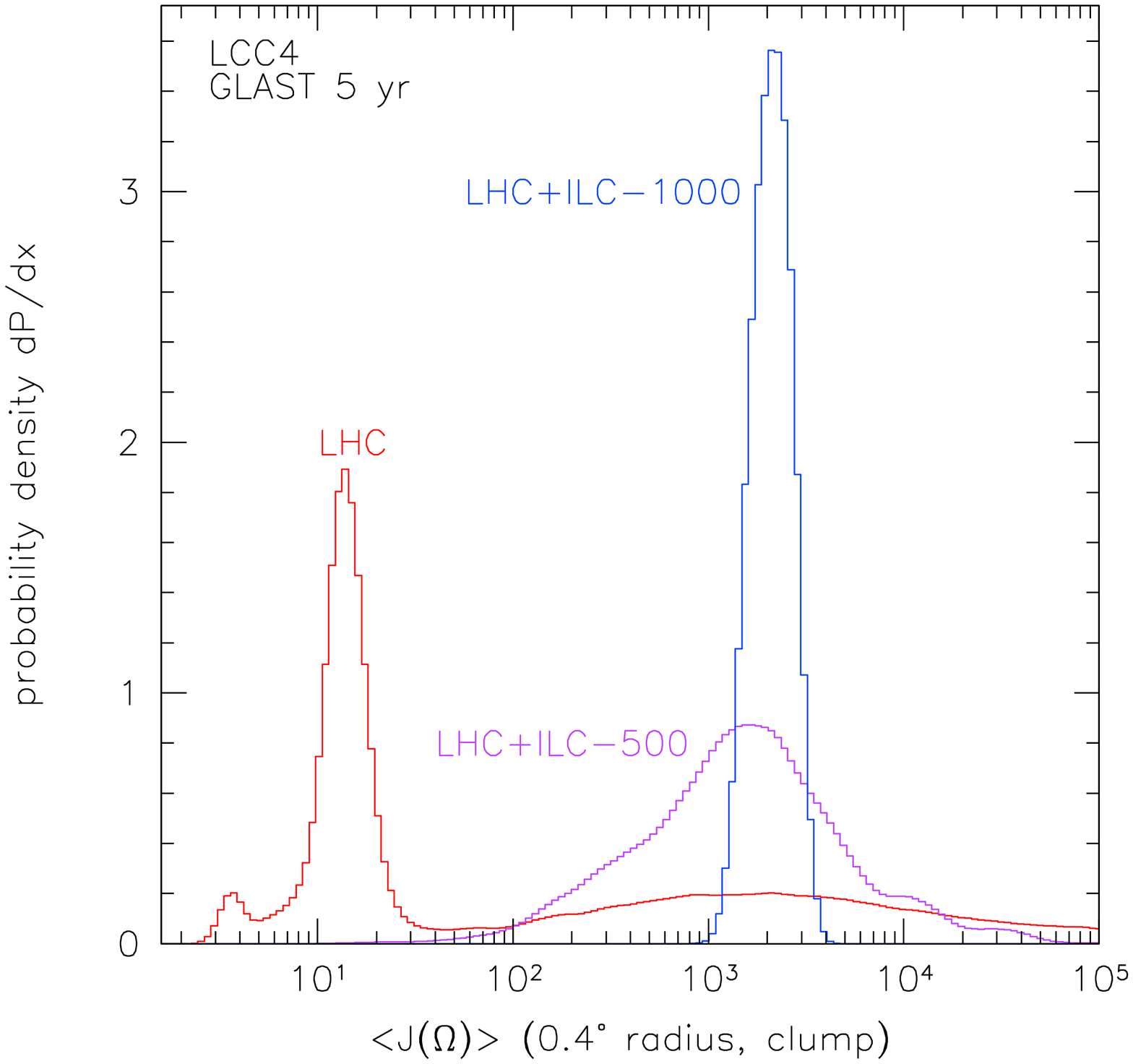,height=3.0in}\\
\caption{Halo density profiles for point LCC4: (a) galactic center,
(b) dark matter clump in the galactic halo.  Angle-averaged $J$ values as
measured by combining a 5-year all-sky dataset from GLAST with accelerator
measurements are shown.  See Fig.~\ref{fig:LCC1relic} for description of
histograms.} 
\label{fig:LCC4halo}
\end{center}
\end{figure}

Because the annihilation cross section at threshold is large at LCC4,
the collider data gives us a significant ability to interpret the counting
rates from experiments that measure gamma rays from neutralino annihilation.
In 
Fig.~\ref{fig:LCC4halo}, we show the determinations of $\VEV{J(\Omega)}$ for 
the galactic center and the canonical halo object described in Section 4.4.
For the galactic center,
we expect 128 signal photons, over 360 background; for 
the canonical halo object, we expect 100 signal photons, over 60 background.
These signals are similar to those obtained at LCC2, and--using the
very well determined annihilation cross section provided by the data
from the 1000 GeV ILC---we obtain similarly powerful results on the 
clustering and halo profile of dark matter.

\subsection{Direct detection cross section}

\begin{figure}
\begin{center}
\epsfig{file=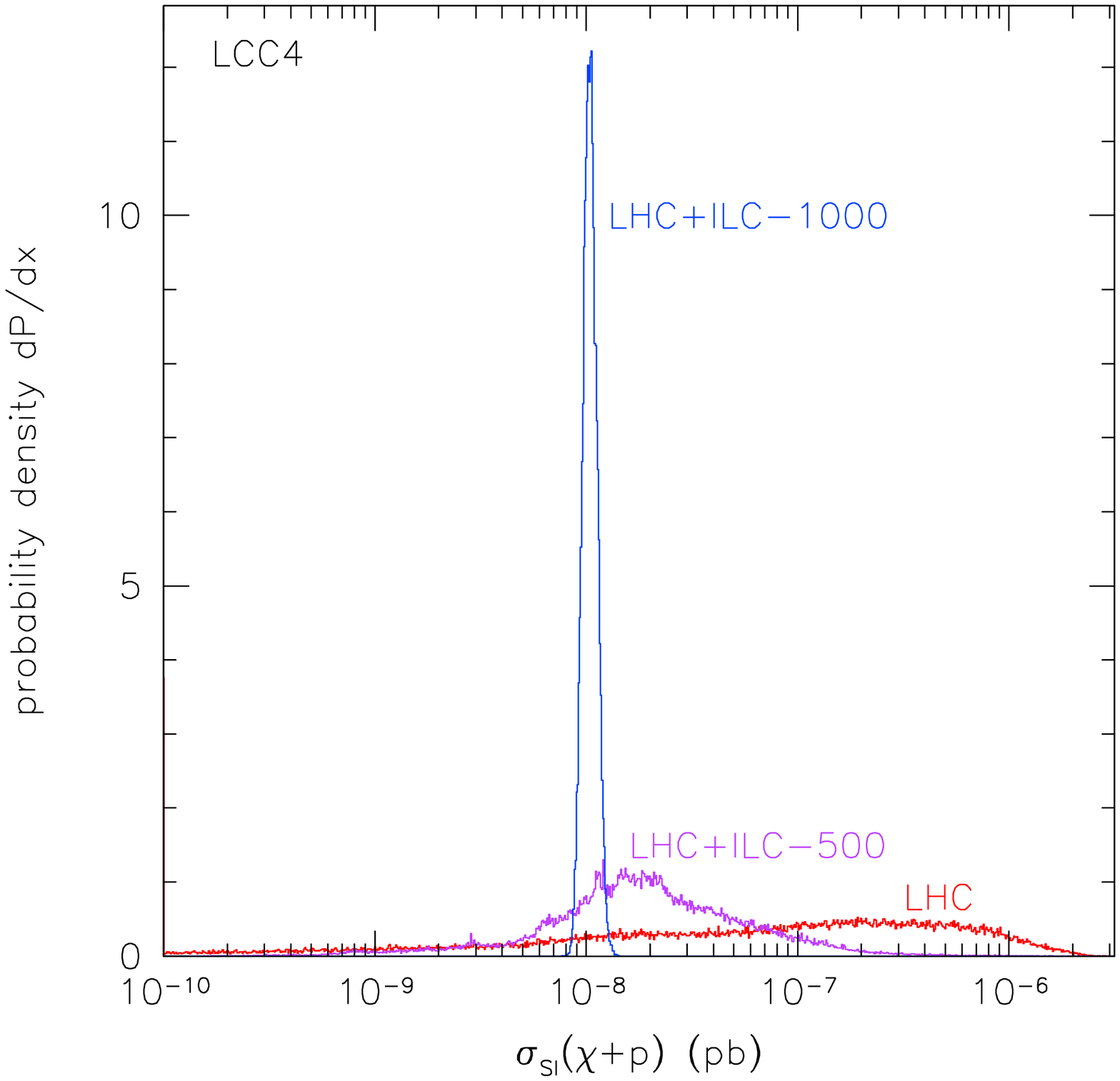,height=3.0in}
\caption{Spin-independent neutralino-proton direct
 detection cross section for
point LCC4.   See
Fig.~\ref{fig:LCC1relic} for description of histograms.} 
\label{fig:LCC4direct}
\end{center}
\end{figure}

\begin{figure}
\begin{center}
\epsfig{file=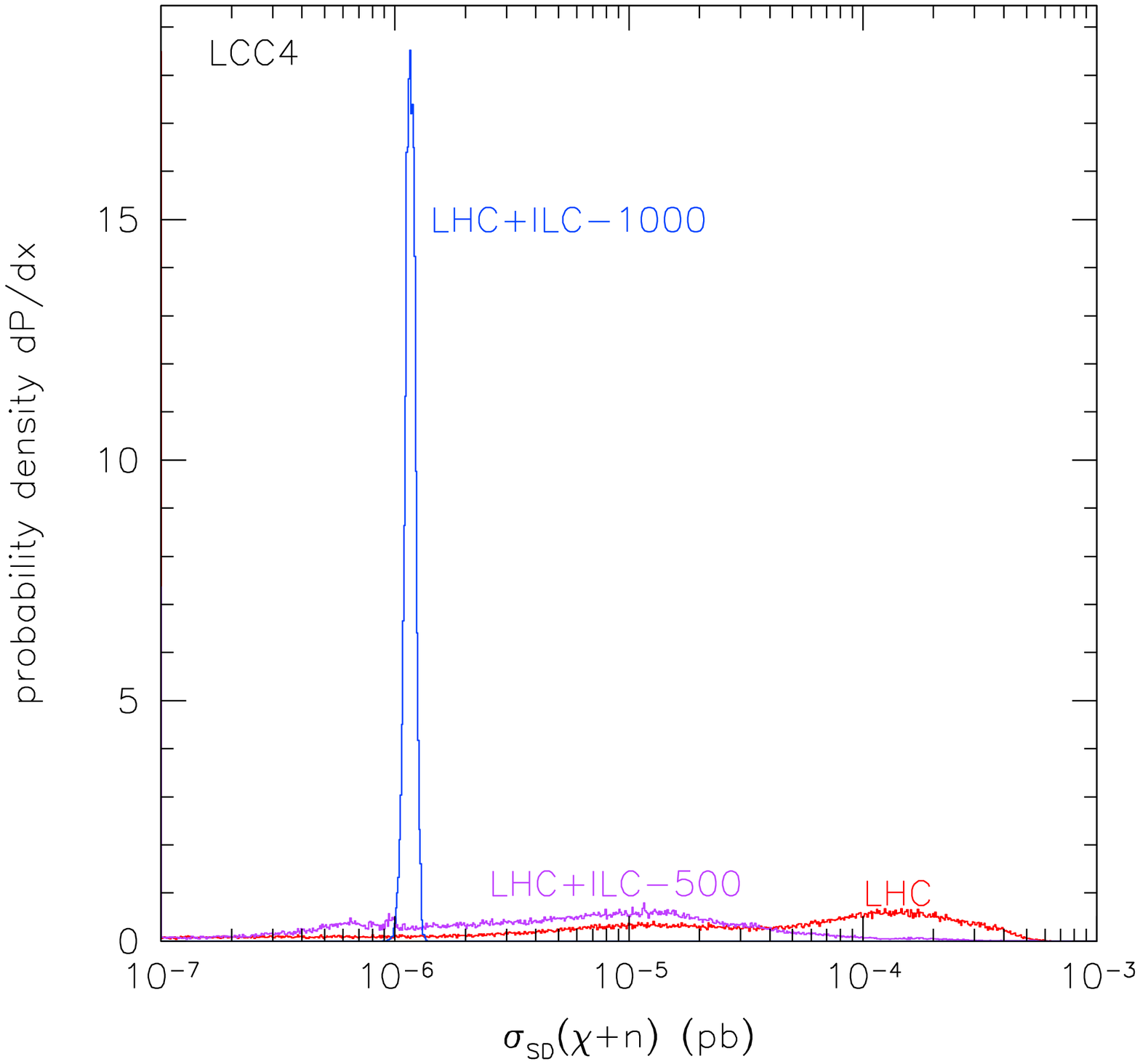,height=3.0in}
\caption{Spin-dependent
neutralino-neutron direct detection  cross section for point LCC4.  See
Fig.~\ref{fig:LCC1relic} for description of histograms.}
\label{fig:LCC4directsd}
\end{center}
\end{figure}

Again at LCC4, the direct detection cross section is dominated by 
the exchange of the heavy Higgs boson $H^0$.  The determination of 
this cross section from the data at the three colliders is shown 
in Fig.~\ref{fig:LCC4direct}.  The evolution is, if anything, more 
striking than that displayed in the previous two sections.  The 
determination of this cross section relies crucially on the information
from the 1000 GeV ILC.  The determinations of the gaugino-Higgsino
mixing angles and $\tan\beta$ are crucial here as elsewhere in fixing
the neutralino-Higgs couplings.
The corresponding predictions for the spin-dependent part of the 
neutralino-neutron cross section are shown in Fig~\ref{fig:LCC4directsd};
these show a similar behavior. 

\begin{figure}
\begin{center}
\epsfig{file=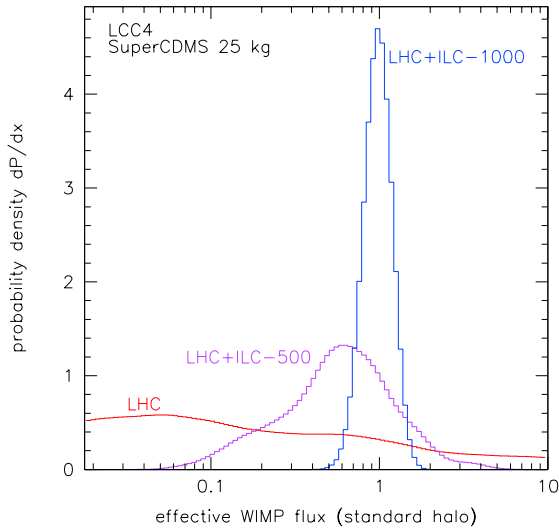,height=3.0in}
\caption{Effective local WIMP flux at the Earth for point LCC4. The results 
assume the SuperCDMS measurement
described in the text. 
  See Fig.~\ref{fig:LCC1relic} for description of histograms.} 
\label{fig:LCC4localhalo}
\end{center}
\end{figure}

As before, we can combine our determination of the detection cross section with
the expected event yield from the SuperCDMS detector to estimate our ability to
directly measure the local flux of dark matter at the earth.  Again, 
our analysis omits the uncertainty from 
low-energy QCD parameters.  The direct
detection cross section is the same as at LCC3, and again we expect a signal of
27 events in SuperCDMS. This gives the determination of the effective
local flux of
neutralinos at the Earth that is shown in Fig.~\ref{fig:LCC4localhalo}.

\subsection{Constraints from relic density and direct detection}

Point LCC4 is quite similar to point LCC3, so much of the discussion of
section~\ref{section:lcc3Astro} relevant to the LHC applies.  In particular,
any relic density constraint applies fixes the neutralino to be almost pure
bino.  However, like LCC3, there is little hint in LHC data of the mechanism
for establishing the correct relic density, in this case resonant annihilation
through the CP-odd Higgs $A$.

The ILC-1000 will measure the direct detection cross section to 7.5\% accuracy,
essentially perfect accuracy 
given astrophysical uncertainties.  It measures the relic
density at the 19\% level.  But, as we found 
at LCC3, there is no clear fundamental parameter
estimate that would benefit greatly from a cosmological constraint.  There is a
mild correlation between relic density and the parameter
$(m_A-2m_{\chi^0_1})/\Gamma_A$ (distance in widths from the $A$ resonance), but
it is not spectacular.  One thing that is greatly helped is the annihilation
cross section at $v=0$.  Since the annihilation is resonant, the cross section
at freeze-out velocities is not much different, though the correlation is not
as tight as with LCC2.  Applying a 1\% relic density constraint changes the
accuracy of the 
annihilation cross section estimate from 20\% to 5\%, a quite significant
improvement.

\begin{figure}
\begin{center}
\epsfig{file=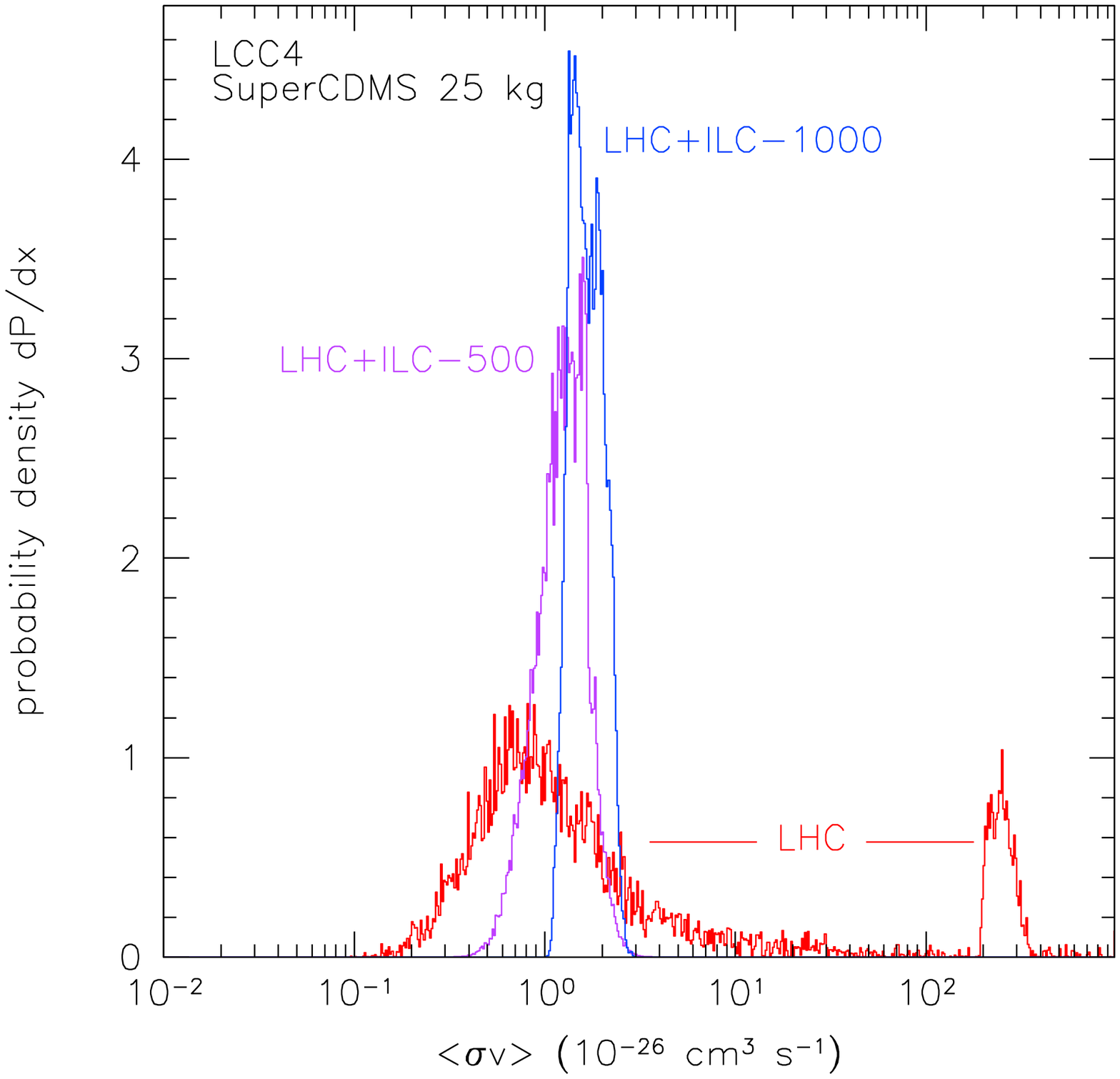,height=3.0in}
\epsfig{file=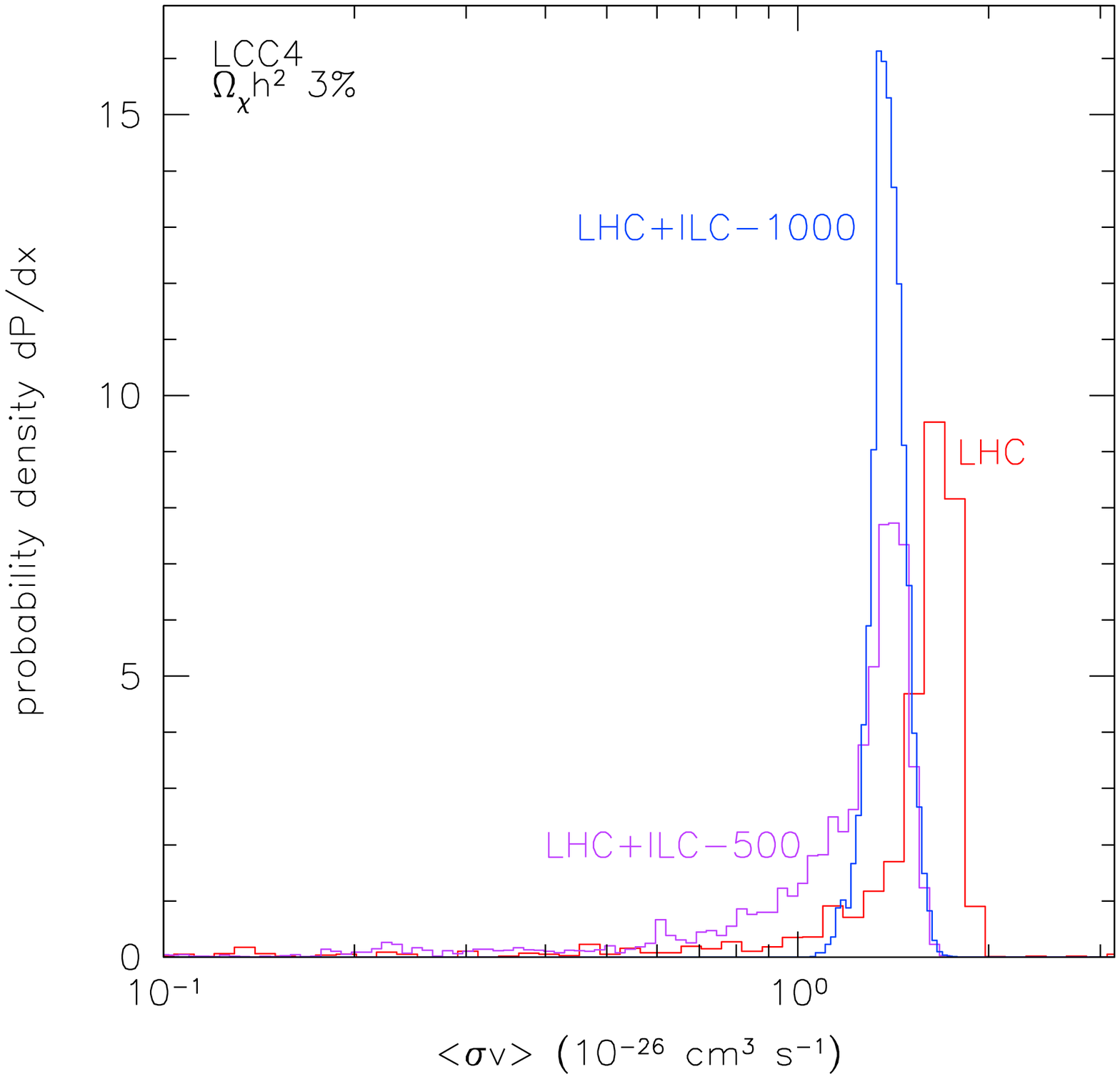,height=3.0in}
\caption{Annihilation cross section at threshold for point LCC4, including
various astrophysical constraints.  Direct detection reduces the weight of the
wino solution, and relic density completely eliminates this possibility.}
\label{fig:LCC4sigvastro}
\end{center}
\end{figure}

Astrophysical constraints can provide a significant information on the
annihilation cross section.  Direct detection can reduce the significance of
the wino peak, while a relic density constraint completely removes it.  We
illustrate this point in Fig.~\ref{fig:LCC4sigvastro}.

\section{Neutralino annihilation products}

In this section, we consider in more detail the computation of
indirect signals of dark matter annihilation from observations of 
gamma rays and  positrons that are produced in  neutralino pair
annihilation.  We leave discussion of neutrinos, antiprotons, and 
antideuterons to future
work.  

In the previous few sections, we presented estimates
of the neutralino annihilation cross section that might be obtained 
from collider data on the SUSY spectrum.  However, as we have remarked 
in Section 2.7, there is a simpler
way to obtain what would seem to be an acceptably accurate prediction.
We start from a  value of the neutralino mass obtained from LHC data.  
If we assume that the neutralino makes up the bulk of the dark matter,
we can use the cross section
\leqn{findsigmav} derived from the relic density as an estimate of the 
astrophysical annihilation cross section.  We assume that the neutralinos
annihilate to hadronic jets, either in direct decays to quarks or through
decays to $W$ and $Z$ bosons.  This gives a roughly universal spectrum of 
energies for the annihilation products, scaling with the neutralino mass.
The arguments are robust and simple to implement.  But are they correct?

\subsection{Gamma ray spectra}

\begin{figure}
\begin{center}
\epsfig{file=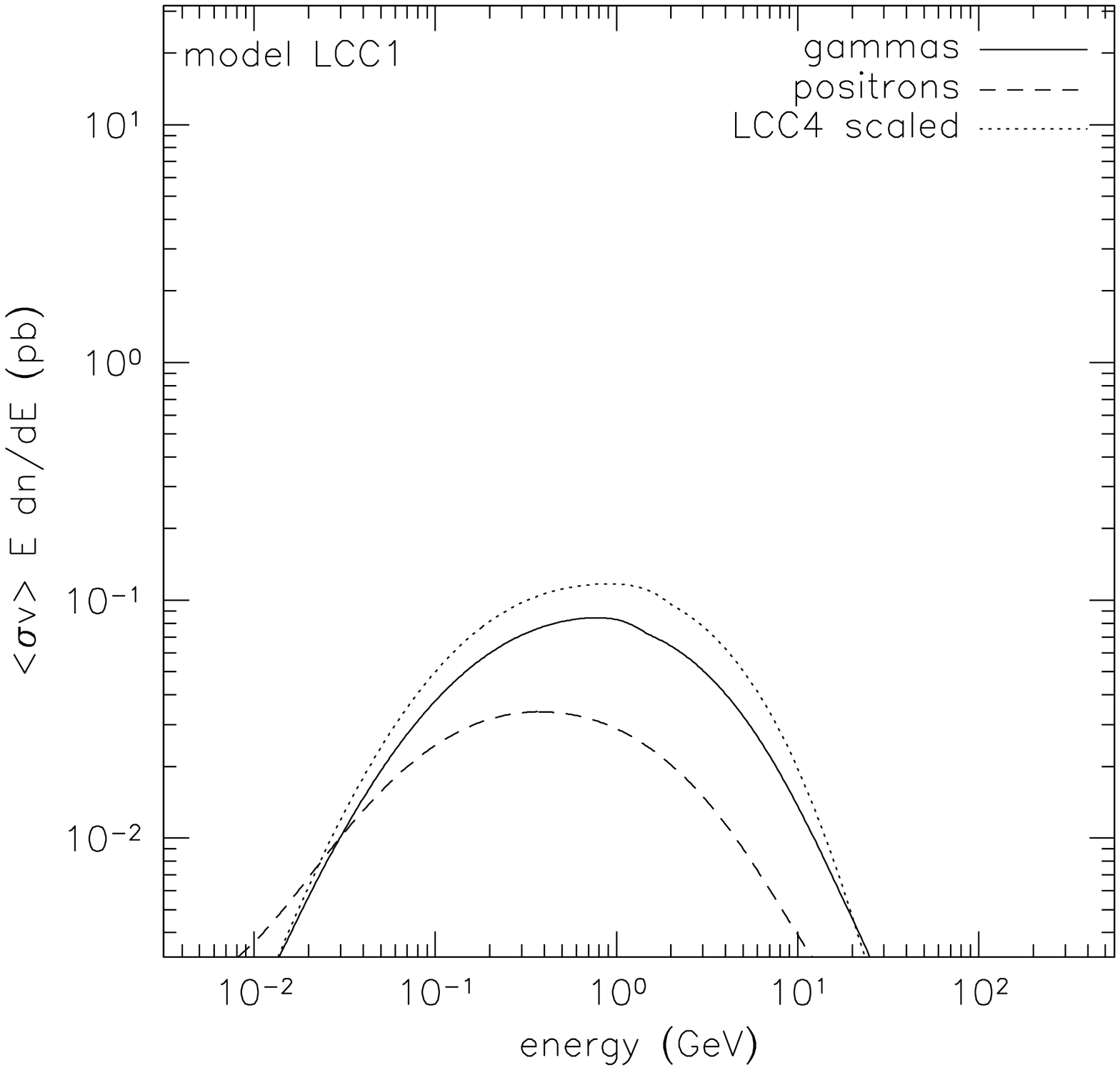,height=2.5in}\qquad
\epsfig{file=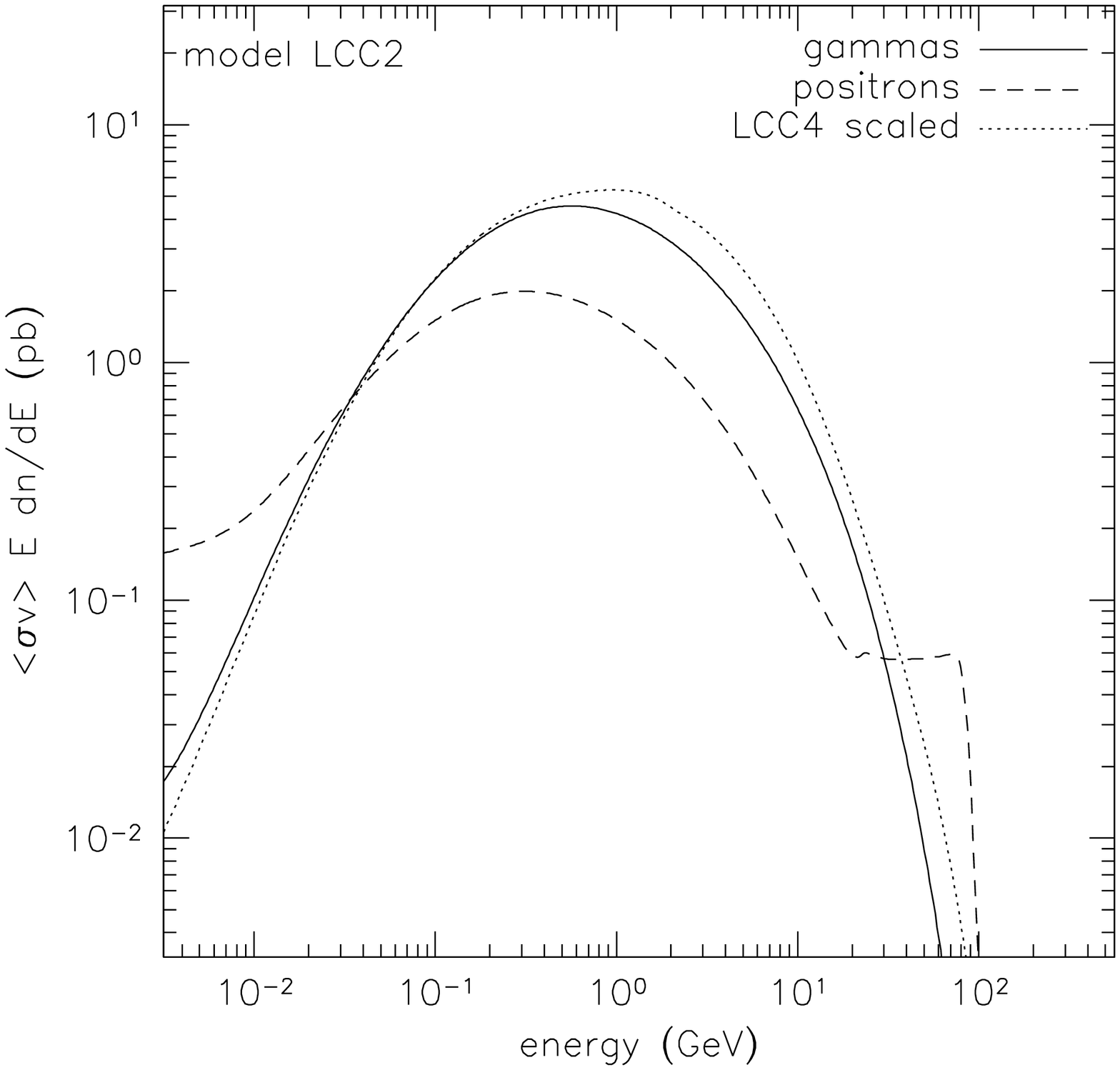,height=2.5in}\\
\epsfig{file=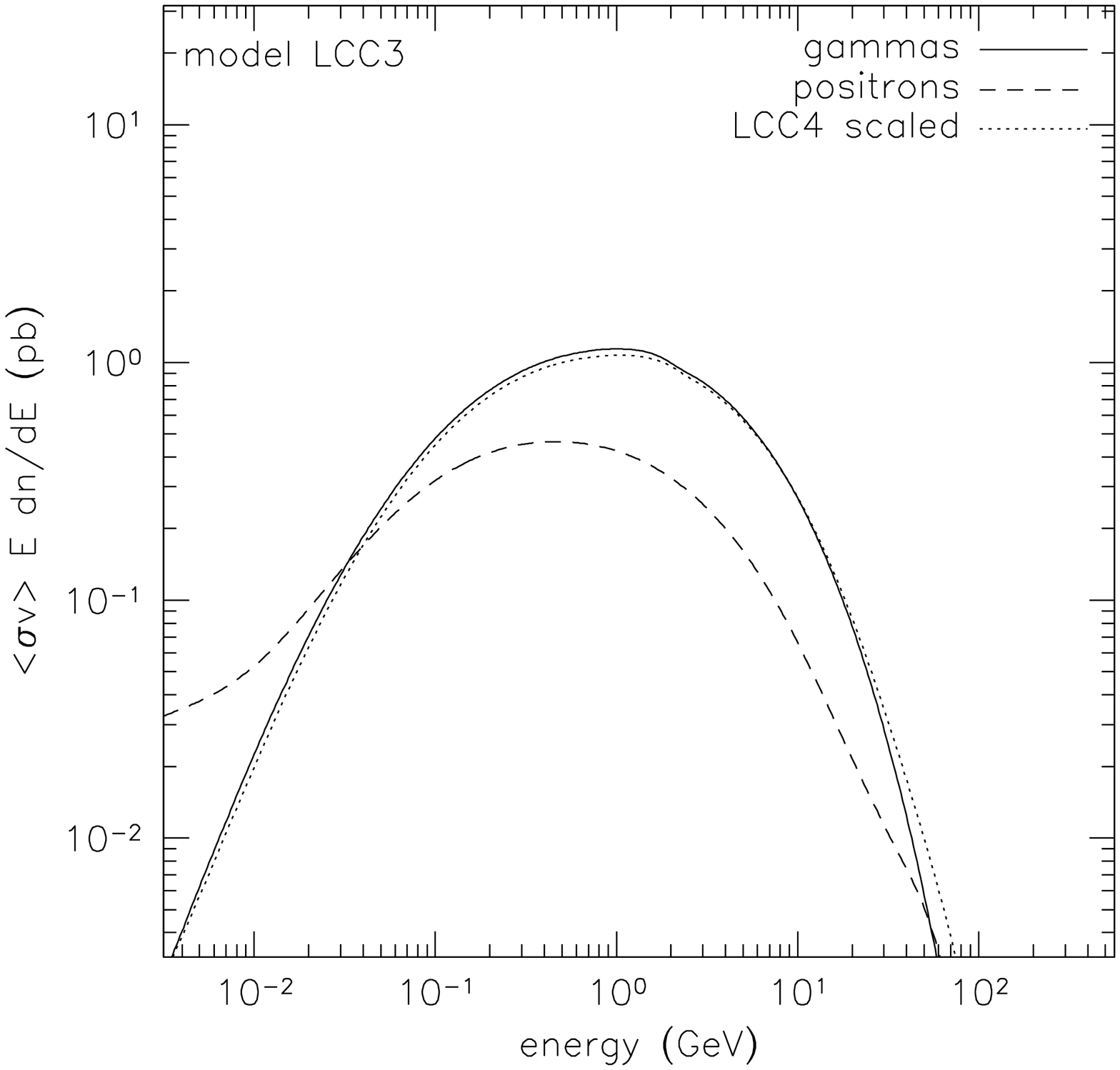,height=2.5in}\qquad
\epsfig{file=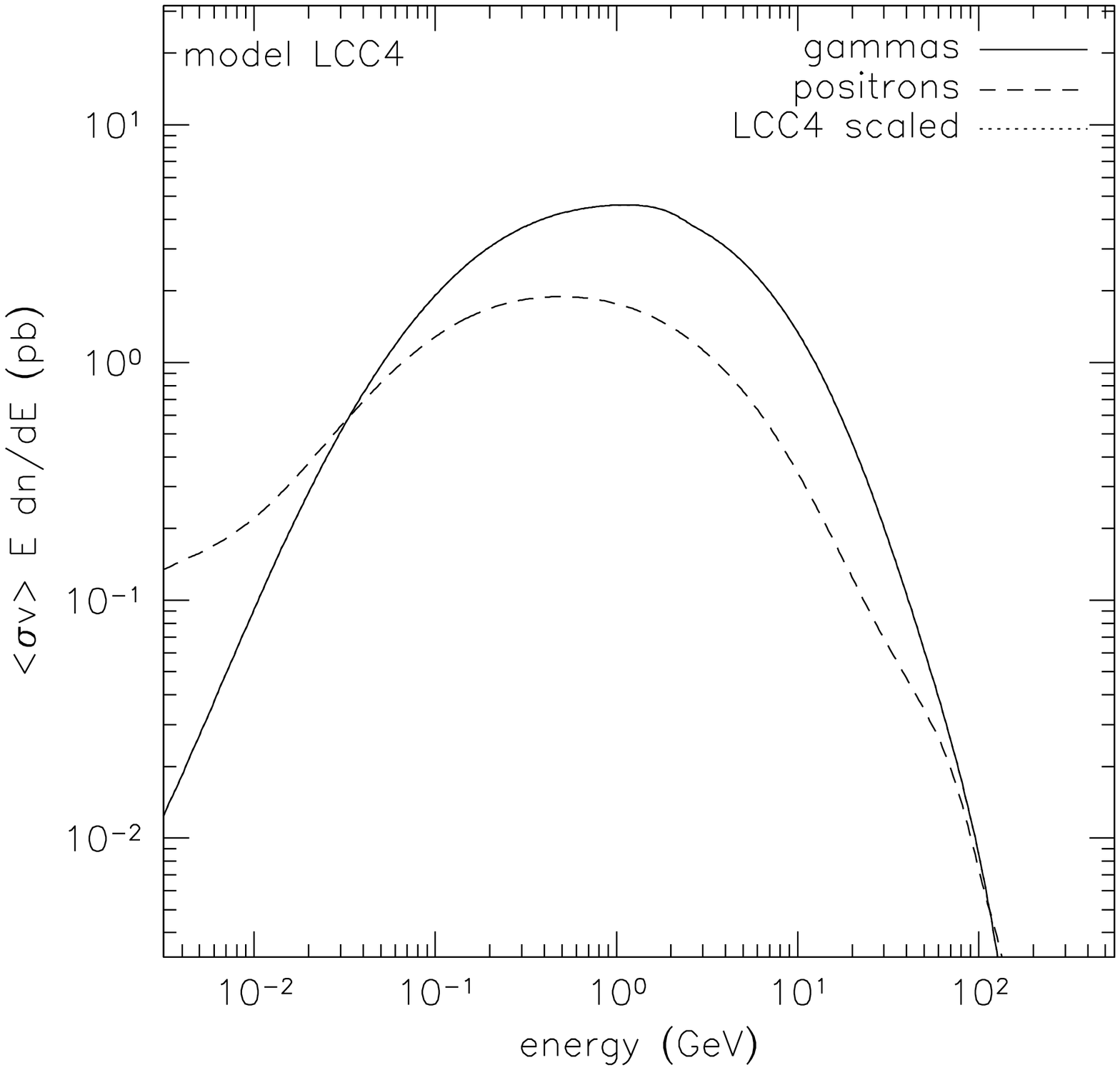,height=2.5in}
\caption{Spectra of gamma rays and positrons from neutralino annihilation at
threshold for points LCC1-4.  Solid curves are gammas, dashed curves are
positrons, and dotted curves show the gamma spectrum of LCC4, scaled to each
endpoint and total cross section.  The scaled LCC4 spectrum is a good match in
every case.  The positron spectrum for LCC2 exhibits a shelf due to direct
decays of gauge bosons, e.g. $\chi\chi\rightarrow W^+W^-\rightarrow e^+\nu
\bar{u}d$.}
\label{fig:gammaspecs}
\end{center}
\end{figure}

The argument we have just given works best for the gamma ray spectra from dark
matter annihilation.  We have computed the spectrum of gamma rays from
neutralino annihilation for each of our four models using results from the
Monte Carlo program PYTHIA~\cite{PYTHIA} (as tabulated in
DarkSUSY~\cite{DarkSUSY}); this code
is expected to give an excellent description
of hadronic final states in the energy region of 100 GeV.  In
Fig.~\ref{fig:gammaspecs}, we show the gamma ray spectra computed from the
model parameters at the four LCC points.  The four spectra are compared to
reference spectra with the shape of the LCC4 spectrum, scaled horizontally to
the correct endpoint at $E_\gamma = m(\chi^0_1)$, and normalized so that the
peak cross section is proportional to the total annihilation cross section.  We
have taken the reference shape from LCC4 because, in this case, the dominant
annihilation reaction goes to the 2-jet final state $b\bar b$.  We see that the
approximation is an excellent one for all four models.

It is not surprising that the shapes of the spectra are similar for 
the cases of LCC1, LCC3, and LCC4.  In all four cases, the annihilation
at threshold is dominated by the process  $\chi\chi \to b\bar b$, 
with subsequent evolution of the jets into $\pi^0$'s and the decay of these
to gammas. However, it is quite surprising that the spectrum for LCC2, 
which is dominated by the processes
$\chi\chi \to W^+W^-, Z^0 Z^0$, yields the same spectrum.
Apparently, the gamma ray spectra from hadronic jets, whatever their
origin, do have the 
universal form assumed in our argument above.

In addition, the discrepancy in the normalizations is readily understood.
The relic density of WIMP dark matter is established at a small but 
nonzero temperature, $T/m \sim 1/25$.  The cross section relevant to 
gamma ray detection of WIMP is that almost precisely at threshold.  In 
cases in which the dominant modes of annihilation proceed in the S-wave,
\leqn{findsigmav} is a reasonable approximation to the correct cross section.
We find for the neutralino
annihilation cross section at threshold $\sigma v = 0.55$ pb for LCC2
and $\sigma v = 0.48$ pb for LCC4. 
However, at LCC1, the dominant modes of annihilation for the purpose
of computing the relic density are P-wave 
annihilations to $\ell^+\ell^-$.   These cross sections are very small
at threshold; the  dominant process just at threshold is the subdominant
reaction $\chi\chi\to b\bar b$.  At the point LCC3, the relic density is
set by coannihilation processes such as $\s\tau^- \s \tau^- \to \tau^-
\tau^-$.   In present astrophysical conditions, all of the $\s\tau$'s have
decayed away.  At  both LCC1 and LCC3, 
$b \bar b$ appears because this is the fermion-antifermion final state 
with the least amount of helicity suppression.
We find, for the neutralino
annihilation cross section at threshold for these points,
 $\sigma v = 0.012$ pb for LCC1
and $\sigma v = 0.11$ pb for LCC3.

If more incisive probes of WIMP annihilation are available, it might be 
important to know what are the fractions of the total annihilation rate
that go to the various possible final states.  In Table~\ref{tab:annBRs}, 
we present the branching fractions to the most important final states
for the four reference points, and the estimates of these branching 
fractions that we obtain from the three sets of collider constraints.

\begin{table}
\centering
\begin{tabular}{llccrrr}
  Process & & & & LHC & ILC-500 & ILC-1000\\ \hline
 LCC1:\\
 $ \chi\chi \to$ & $ b\bar b$ &   0.629   & $\pm$ & 46\% & 46\%  & 15\%
 \\
   & $ \tau^+\tau^-$ &     0.282  & $\pm$ &  79\%  & 71\% & 43\%  \\
   & $ W^+W^- $ &   0.046          & $\pm$ & 65\% & 40\% & 13\%     \\
   &  $\gamma\gamma$ &    0.016       & $\pm$ &  54\%  & 42\%  & 10\% \\
   &   $gg $  &    0.013      & $\pm$ &  52\%    & 41\%  & 11\%  \\
 LCC2:\\
 $ \chi\chi \to $ & $  W^+W^- $ &   0.868& $\pm$ & 11\% & 11\%  &
 1.4\% \\
   &  $  Z^0 Z^0 $ & 0.114   & $\pm$ & 66\%  &  12\%  & 3.7\%  \\
  LCC3:\\
  $ \chi\chi \to $ & $  b\bar b $ & 0.974 & $\pm$ & 70\% & 36\% &
  1.3\% \\
   &  $ \tau^+\tau^- $ & 0.014       & $\pm$ & 113\% & 93\%  & 92\%  \\
  LCC4:\\
 $ \chi\chi\to  $ & $ b\bar b $ & 0.889 & $\pm$ & 70\% & 16\% & 16\% \\
  &  $ \tau^+\tau^- $ & 0.103 & $\pm$ & 90\% & 50\% & 52\% \\
\end{tabular}
\caption{Branching ratios in neutralino pair annihilation. The last three
columns give the fractional error ($\sigma$/mean) from the MCMC scans.}
\label{tab:annBRs}
\end{table}

\subsection{Positron spectra}

We have just shown that
 the gamma ray spectrum resulting from neutralino annihilation is
remarkably independent of the model.  It is controlled almost entirely 
by the total neutralino pair annihilation cross section, and, through this,
is often determined if the physics of neutralino annihilation is understood
even qualitatively.   For other annihilation products, however, the 
story can be quite different.  In this section, we will discuss the visibility
of the positron signal of dark matter annihilation in our four models.
Similar considerations apply to neutrino signals.  However, since in all 
four of our models the neutralino mass is below 200 GeV, neutrinos from 
neutralino annihilation will be produced at too low an energy to be visible
above the thresholds of cosmic-ray neutrino detectors such as ICE-CUBE.

Unlike gamma rays, which fly directly from the source to a detector on 
earth, positrons execute a random walk in the galactic magnetic field,
losing energy continually along the way.  Thus, positrons from WIMP 
annihilation that are observed at the earth must originate inside the galaxy,
within a few kpc. The annihilation rate depends on $\rho^2(x)$ averaged
over this volume, a quantity closely related
 to the local halo density discussed in 
relation to direct detection, but possibly enhanced by local clumpiness
parametrized by the boost factor~\leqn{boostfactor}.  The energy loss in 
propagation favors the highest-energy positrons in the spectrum.

\begin{figure}
\begin{center}
\epsfig{file=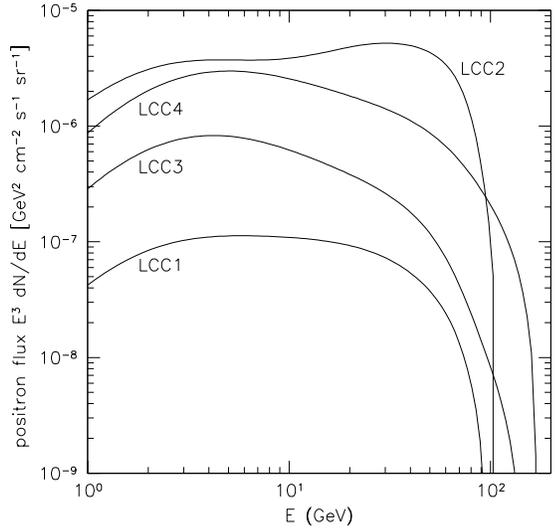,height=3.0in}
\caption{Spectrum of positrons for all models after galactic propagation
effects are accounted for.  The ``interstellar'' spectrum is illustrated.
Solar modulation is neglected.  Above a few GeV solar modulation effects are
negligible.}
\label{fig:allpositrons}
\end{center}
\end{figure}

The propagation of positrons through the galaxy has been modeled 
quantitatively in~\cite{moskalenko,baltzedsjo}.  We have found it an 
interesting exercise to fold the positron spectra for our four models 
with the smearing predicted by the model of~\cite{baltzedsjo}.  The 
results are shown in Fig.~\ref{fig:allpositrons}.   

\begin{figure}
\begin{center}
\epsfig{file=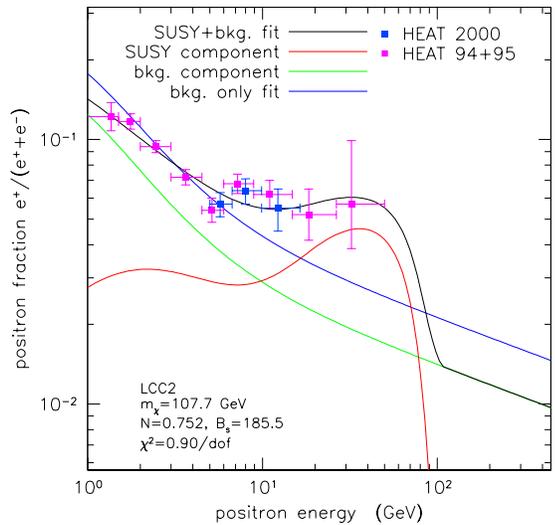,height=3.0in}
\caption{Positrons observed on earth, as a fraction of electrons, for LCC2.
The HEAT data are plotted as well, indicating the possibility of an excess.
The positron signal from a smooth halo has been boosted by a factor of 185 in
order to fit the data.}
\label{fig:BEpositrons}
\end{center}
\end{figure}

The signal from LCC2
is larger than the others,
 for several reasons.  First of all, this point has a full 
strength total annihilation cross section.  In addition, as one can see from
Fig.~\ref{fig:gammaspecs}, the positron spectrum at this point has a 
component with a flat distribution extending to the kinematic endpoint.
This results from the fact that the dominant annihilation reactions
 at LCC2 are to $W^+W^-$ and $Z^0Z^0$, that is, to vector bosons that have 
direct two-body decays to positrons.  The sharp feature at the extreme right
of Fig.~\ref{fig:gammaspecs}(b) is transformed into the peak at the 
high-energy edge of the LCC2 curve in Fig.~\ref{fig:allpositrons}.

We can compare this spectrum to the recent measurement of the cosmic-ray
positron spectrum by the HEAT experiment.  The comparison is shown in
Fig.~\ref{fig:BEpositrons}.  The HEAT data shows an anomaly at its upper edge,
with a cross section that is roughly flat in a region in which the background
is expected to be decreasing.  As the figure shows, we can fit this with the
positron spectrum from LCC2.  To obtain the correct normalization, we must
enhance the production over that for a smooth distribution of dark matter with
density $\rho_0 = 0.3$ GeV cm$^{-3}$ by
assuming a large boost factor $B = 190$.

Is this a correct way to interpret the data?  We ourselves are 
skeptical.  Nevertheless, if we knew from collider measurements that the 
WIMP mass was 100 GeV and that we were in a region of the parameter space
that favored annihilation through $\chi\chi \to W^+W^-$, this fit to the data
might be considered compelling evidence for neutralino annihilation and for
a significantly clumpy distribution of WIMPs in the galaxy.

\section{Recap: Collider determination of WIMP properties}

Our analysis in this paper has presented a series of worked examples that
illustrate the interactions we might expect over the coming years
between astrophysical dark matter detection experiments  and high-energy
physics experiments that will measure the spectrum of new particles in 
the hundred-GeV mass range.  In this section, we will assemble the 
results that we have obtained  and   give our interpretation of  their
implications.

\subsection{Summary of results: cross sections}

In Table~\ref{tab:HEPresults}, we summarize the results of our Monte Carlo
scans for the most important WIMP parameters that are determined by 
collider measurements of supersymmetry spectroscopy:  the relic density
$\Omega_\chi h^2$, the annihilation cross section at threshold $\sigma v$ 
(in pb),
and the spin-independent scattering cross section on protons $\sigma(\chi p)$
(in units of $10^{-8}$ pb).
The results are given for each of the four benchmark models and for each 
of the three sets of collider measurements discussed in Sections 4-7.
The results are quoted as a percentage error, defined as the standard
deviation divided
by the mean, computed from the statistical sample generated by our 
Monte Carlo
process.  A large variance indicates a poor determination of the
model parameters and might also indicate the presence of multiple
solutions.  For the correct interpretation, one should look at the 
detailed shapes of the probability distributions shown in the figures in 
Sections 4-7. For reference, the mean values of the Monte Carlo samples are 
listed in the last three columns of the table when these differ by 10\% or 
more from the nominal value.  

We emphasize that all of these cross section determinations are
`model-in\-de\-pen\-dent' 
in the following sense:  We assume that it has been shown
from collider data that the model of new physics at the hundred GeV scale is
supersymmetry.  We then fit the collider data on new particles to the Minimal
Supersymmetric Standard Model, studying the model in complete generality with
24 free parameters.  We have shown in many examples given in Section 4-7 that
this parametrization is sufficiently general to allow all of the possible
physical mechanisms that are expected in this general class of models to be
considered in the fitting procedure.  Thus we believe that we have justified
the claim made in Section 3.2 that cross section estimates
obtained from collider data through an analysis such as ours can be used 
without undue qualification to analyze astrophysical
measurements.

\begin{table}
\centering
 \begin{tabular}{l|r|rrr||rrr}
  &    &   LHC &   ILC-500 & ILC-1000  & LHC & ILC-500 & ILC-1000 \\
   &     $\Omega h^2$ &   & & &   (mean) &  \\ \hline
LCC1 &   0.192   &  7.2\%   &    1.8\%   &   0.24\%  &       \\
LCC2 &   0.109   &  82.\%   &    14.\%   &   7.6\% &  0.074 \\
LCC3 &   0.101   &  167\%   &    50.\%   &   18.\%  & 0.24 & \\
LCC4 &   0.114   &  405\%   &    85.\%    &   19.\% & 0.26 & 0.083 &
 0.094 \\
\hline
 \\
   &     $\sigma v$ &   & & &   (mean) \\ \hline
LCC1 &   0.0121  &  165.\%   &    54.\%   &   11.\%  &   0.0069    \\
LCC2 &   0.547  &  143.\%   &    32.\%   &   8.7\% &  8.47 &  \\
LCC3 &   0.109   &  154.\%   &    178.\%   & 10.\%  & 24.2 &  0.311 & \\
LCC4 &   0.475   &  557.\%   &     228.\%    &   20.\% & 82.5 & 1.83 &
 0.57 \\
\hline
 \\
   &     $\sigma(\chi p)$ &   & & &   (mean) \\ \hline
LCC1 &   0.418  &  44.\%   &    45.\%   &   5.7\%  &   0.20     \\
LCC2 &   1.866  &  62.\%   &    63.\%   &   22.\% &  3.57 & 2.82 & 2.19
 \\
LCC3 &   0.925   &  184.\%   &    146.\%   &   8.6\%  & 13.2 &  1.86 &
 \\
LCC4 &   1.046   &  150.\%   &    190.\%    & 7.5\% & 23.2 & 3.59 &   \\
\hline
\end{tabular}

\caption{Fractional errors in the determination of the most important
microscopic WIMP parameters derived from the MCMC scans: $\Omega h^2$,
the predicted relic density, $\sigma v$, the annihilation cross section
at threshold (in pb), and $\sigma(\chi p)$, the spin-independent
neutralino-proton cross section (in units of $10^{-8}$ pb).  The second
column lists the values predicted by the benchmark models.  Columns 3--5
give the fractional error ($\sigma$/mean) from the MCMC scans.  Columns
6-8 give the mean value found from the MCMC data when this deviated by
more than 10\% from the nominal value in column 2.  As discussed in Appendix A,
the quoted errors are accurate to 10\% or better, e.g. a 20\% error is $20\%\pm
2\%$.}
\label{tab:HEPresults}
\end{table}

The table shows that, for all of the quantities listed, carrying out the full
program of collider measurements that we have presented results in 
a clear
determination of the microscopic cross sections that can be used to interpret
astrophysical dark matter observations.  The precise quality of the
determination depends on the specific scenario.  In scenarios in which the
dominant annihilation mechanism is through simple annihilation to leptons, we
find that the accuracy of the prediction of the relic density is at the level
of a fraction of a percent, comparable to the best measurements of the dark
matter density from the cosmic microwave background expected in that era.  In
scenarios in which special relations among the superparticle masses are
necessary for rapid enough annihilation, the quality of the microscopic
prediction decreases and we find an accuracy of about 20\%.  In all cases, 
the
agreement of the microscopic and astrophysical determinations within the 
errors
would be nontrivial and would provide striking evidence that the particle
identified at colliders is indeed the dominant component of astrophysical 
dark matter.

We believe that the accuracies we have listed are the best ones that can be
obtained at the current level of our understanding of the experimental 
capabilities of the next-generation colliders.  That is, we could not find
additional measurements beyond the ones we have listed that would 
significantly 
improve the accuracies we have quoted for the  neutralino cross sections.
We note, though, that, in the linear collider studies, relatively little 
effort has gone into analyses aimed at very accurate cross section 
measurements.   We have seen that the ILC cross sections can be very 
important in fixing the gaugino-Higgsino mixing angles and $\tan\beta$, 
parameters that provide a
major uncertainty in the neutralino properties.  If cross sections could be 
measured at the ILC with errors considerably smaller than the estimate
\leqn{erroronsig}, the output accuracies would improve.  To do this,
however, it is necessary to confront the Standard Model backgrounds to the 
observations of supersymmetric particles and to model and subtract these
backgrounds with very high precision.

We have also shown that the collider data predicts the WIMP 
annihilation cross
section and the cross section relevant to direct detection 
at the 20\% level or
better in all of the cases that we have considered.  This is quite sufficient
to use the microscopically-determined cross sections to provide strong
constraints on the distribution of the observed WIMPs in the galaxy.
For the case of the direct detection cross section, this conclusion 
does require that the uncertainty from  low-energy  QCD matrix elements,
discussed in Section 2.6, can be brought under control.

\subsection{Summary of result: astrophysics}

In Table~\ref{tab:Astroresults}  we summarize the constraints
for the three specific 
model problems that we have considered in this paper, 
the determination
by GLAST 
of the dark matter density integral  $J \sim \int dz \rho^2$
near the galactic center,
the determination by GLAST 
of the value of $J$ for a representative clump of dark matter
in the halo of the galaxy, and the determination by SuperCDMS 
of the local flux of dark matter at a location on Earth.  The 
specific parameters assumed for the detectors and sources were detailed
in Section 4.4 and 4.5.  We emphasize that we have chosen these 
specific experiments as representative examples meant to illustrate the 
implications of the collider measurements for the 
many dark matter observation experiments that will be carried out over
the next decade.

\begin{table}
 \centering
\begin{tabular}{l|r|rrr||rrr}
  &    &   LHC &   ILC-500 & ILC-1000  & LHC & ILC-500 & ILC-1000 \\ 
   &     $\VEV{J}$(gc) &   & & &    limits & (95\% CL) \\ \hline
LCC1 &   7760  &    &     &   &   $<$ 50,000 &  $<$ 34,000 & 
                              $<$ 34,000 \\ 
LCC2 &   7760   &   51.\%  $^*$  &    25.\%   &   20.\% &  \\ 
LCC3 &   7760   &    &       &   & $<$ 16900  & $<$ 16900  & $<$ 11500
                                                              \\ 
LCC4 &   7760   &  160.\% $^*$  &    122.\%   &   32.\%  &  &  & \\ \hline
 \\ 
   &     $\VEV{J}$(clump)  &   & & &   limits & (95\% CL)\\ \hline
LCC1 &   2500  &   &     &    &  $<$ 14500    &$<$  6700 &$<$ 5700 \\ 
LCC2 &   2500  &  51.\% $^*$   &   21.\%   &   13.\% &   &  \\ 
LCC3 &   2500 &   152.\%  $^*$  &   131.\%   &  81.\%  &  & & \\ 
LCC4 &   2500   &  180.\%  $^*$  &   124.\%    &   23.\% &  & & \\
\hline
 \\ 
   &     $\Phi_{local}/\Phi_0$ &   & & & limits &  (95\% CL) \\ \hline
LCC1 &    1.  &     &    54.\%   &   29.\% &$>$ 0.95        \\ 
LCC2 &    1.  & 28.\% $^*$  &    46.\%   &   24.\% &   \\ 
LCC3 &    1.   &   &    126.\%   &   20.\% &  $>$ 0.016  \\ 
LCC4 &    1.   &     &    115.\%    &  20.\% &  $>$  0.010  \\ 
\hline
\end{tabular}
\caption{Errors or limits for  the illustrative astrophysical measurements
 that we have presented in Section 4-7.  The quantities considered are:
$\VEV{J}_{gc}$, the average of the density integral $J\sim \int dz \rho^2$
near the galactic center, $\VEV{J}_{clump}$, the average of $J$ in a
small circle around the center of a typical clump of dark matter in the 
galactic halo, and $\Phi_{local}$, the effective
flux of dark matter impinging on
a direct detection experiment on earth, normalized to 
a standard halo distribution.  The output values 
quoted include the expected experimental errors. More details of the 
assumptions involved in these analyses are 
given in Sections 4.4 and 4.5. The second column
lists the assumed astrophysical values.  Columns 3--5   give the 
standard deviation of $\ln \VEV{J}$ 
from the MCMC scans.  Columns 6-8 give 
95\% upper or lower confidence limits based on the MCMC data, as
 appropriate.   Upper limits on $J$ assume $\VEV{J} > 1$ as a prior.
For each situation, 
either an error or a limit is quoted.  In the cases labeled by 
$^*$, the MCMC data includes multiple solutions, and we restrict 
our calculation to points in the neighborhood of the correct solution.
See the text for further discussion of all of these points.
As discussed in Appendix A, the quoted errors for the microscopic quantities
(annihilation cross section, direct detection cross section) are accurate to
10\% or better.}
\label{tab:Astroresults}
\end{table}

The results given in the table for determinations of the density integral
$\VEV{J}$ reflect several factors in addition to the accuracies given 
in Table~\ref{tab:HEPresults} with which the microscopic WIMP cross 
section will be known.  They include also the statistics of the observation,
and the error from the expected uncertainty in the background.
These two factors depend in turn on the underlying physics scenario
and on the value of the relevant WIMP cross section.
If the cross section is small, the statistical error will be relatively
large, and the background will be more important relative to the signal.

These contributions are reflected in our results on $\VEV{J}$ from the
galactic center in the following way.  The points LCC1 and LCC3 have 
suppressed annihilation cross sections at threshold.  In these cases,
 the background always
dominates, and we can only set upper limits on the density integral.
However, the lower bounds on the cross sections
that we obtain from the collider data
allow these upper limits to significantly exclude dark matter distributions
that are highly peaked at the galactic center.  For LCC2 and LCC4,
the annihilation cross section is at the full strength expected from 
\leqn{findsigmav}, and a robust signal is expected above background.
For the LHC data set, ambiguities arising from multiple solutions still
make it  difficult to pin down the value of $\VEV{J}$. We have quoted
the error on $\VEV{J}$ obtained when we resolve this ambiguity in 
favor of the correct solution by removing the subsidiary peaks in the
likelihood distributions for $\sigma v$ at high cross section values.
The variances are still quite large.
These   ambiguities are resolved by the ILC data, and for those
cases we give the fractional error on $\VEV{J}$ computed from the 
full scan data.  Thus, at LCC2 and at LCC4, we find that
$\VEV{J}$ is given to 30\% accuracy. 
  These accuracies 
should be compared to the current astrophysical estimates of 
$\VEV{J}$ at the galactic center, which 
range over many orders of magnitude.  A measurement of the $\VEV{J}$
correct to the level we have shown, or even an upper limit  
on 
$\VEV{J}$ free of astrophysical assumptions, 
would represent a major improvement in our knowledge.

For our model dark matter clump in the galactic halo, the  situation is
somewhat better.  The background still dominates for LCC1, but not
in the other cases.  For the LHC data sets, the problem of 
ambiguities in the solution is still present, and again we 
restrict our estimates to MCMC points in the peak corresponding to 
the correct solution.    But, using the ILC data,
we will find quantitative  measurements of $\VEV{J}$.  At LCC2, we can 
achieve a 20\% measurement already with the 500 GeV data.  At LCC4 also, 
 the data from the 1000 GeV stage
determines $\VEV{J}$ an accuracy of about 20\%.  

For the local flux of neutralino dark matter, we find a similar
picture.  We first remind the reader that our estimates ignore
uncertainty from the evaluation of low-energy QCD matrix elements.
With the assumption that these matrix elements can be determined,
we draw the following conclusions:
In all cases except for 
LCC2, the LHC data set provides  only a weak determination of the 
relevant WIMP cross section.  We can at best quote an upper limit
on the cross section and thus a lower limit on the local flux. 
For the 
LHC data at LCC2, we have quoted a measurement error after removing
the peak arising from the incorrect solutions.  
The
data from the 500 GeV ILC constrains the direct detection cross sections
to some extent, and for these entries we have quoted the fractional
error, which is however large in all cases.  
The measurements at the 1000 GeV ILC finally fix the mass of the heavy
Higgs bosons and other spectroscopic
parameters involved in  the most important contributions
 to the direct detection 
cross section.   From this data, the local flux can be determined 
to 20-30\%  accuracy for all four benchmark models. 
  Our current knowledge of the galactic
halo constrains this flux only to  a factor of 2, and even then
only if we assume
that the halo has a smooth distribution both in position and in momentum 
space.  Here too, precise microscopic information can have a large 
impact.

In all three examples, what would be determined would be the density or
flux of that component of dark matter corresponding to the WIMP observed
in collider experiments.  Other possible components of dark matter
such as axions or very heavy weakly-interacting particles give 
negligible signals in direct detection experiments and experiments on 
dark matter annihilation.   Thus, first of all, the direct and 
indirect detection experiments would demonstrate concretely that
the WIMP seen in particle physics is present in the structure of the
galaxy.  At the next stage, the consistency of the overall 
picture that results from these experiments could give additional insight,
beyond what is gained from analysis of the overall relic density, on the
broad question of whether the observed WIMP is the dominant component of
dark matter.

\subsection{LHC and astrophysical measurements}

Up to this point, we have been discussing the comparison to astrophysical
data of the full set of results that we will obtain from the 
next-generation colliders.  It will take some time, of course, for the 
collider data to become available.  The LHC experiments will begin in 
just another year.   The 500 GeV stage of ILC  may begin within ten years
from now.  The 1000 GeV stage of the ILC would be an upgrade to the 
basic facility and would produce data, at the earliest, at the end of the
next decade.  Astrophysical dark matter experiments are  also spaced out
in time through the next decade.   It is interesting, then, to look at
the implications of colliders in terms of this timeline and see what
results can be expected at each stage.  In this and the next two sections,
we will discuss some aspects of this evolution, emphasizing the the 
new information that we will obtain as each collider presents its 
results.   

There are many possibilities for what will happen in the
future.  In the discussion of the next few sections, we will 
assume that underlying physics 
model is a supersymmetric model of dark matter similar to those we 
have analyzed in this paper.  Then we will be able to discuss the
evolution in a very concrete way.

We begin with the situation as it might appear in 2012.  In the 
scenarios discussed in this paper, we will by that time have  the 
discovery of supersymmetry in the 
LHC experiments and the first positive results 
from direct detection experiments and searches for WIMP annihilation.

At this stage, it should already be possible to compare three observed
masses relevant to dark matter: (1) the mass of the escaping neutral 
particle produced at the LHC in missing energy events, (2) the mass of the
directly detected dark matter particle obtained from the recoil energy 
spectrum, and (3) the mass of the annihilating dark matter particle, obtained
from the endpoint of the gamma ray spectrum.  In the best case, all three 
masses should be determined to better than 20\% accuracy.  Their agreement
will provide a nontrivial test that the particle being
produced at the LHC is indeed a dominant component of cosmic dark matter.

It will be difficult to learn more about the dark matter
particle without additional assumptions.
As we have discussed in Section 2.3, it will be very difficult from the LHC
data alone to narrow the possible explanations of missing-energy events and
new particles to a single model.  Nevertheless, we might proceed by 
assuming a particular model (for example, supersymmetry or even a restricted
model of supersymmetry) and examining its consequences for astrophysics.

In some scenarios (for example, the benchmark points LCC1 and SPS1a$'$
discussed in Section 4), we would be able to combine the assumption that the
new physics is supersymmetry with the detailed spectroscopy measurements
available from the LHC to give a quite accurate prediction of the WIMP relic
density. In this situation, it is tempting to assume that the observed WIMP
is the sole component of dark matter, fix the relic density to the value
measured from the cosmic microwave background, and make higher-precision
predictions for the supersymmetry mass spectrum.  In addition, because the
heavy Higgs bosons typically give the dominant contribution to the
spin-independent direct detection cross section, one can assume that the 
local
flux of dark matter is near its nominal value and use the direct 
detection rate
to fix the heavy Higgs boson mass.  In both cases, the later stages of the
collider physics program on supersymmetry spectroscopy will test these
predictions and thus confirm or refute the astrophysical assumptions.

In most scenarios, however, it is not possible to derive a definite 
prediction
for the WIMP relic density from the LHC data even if supersymmetry is 
assumed to be the underlying theory.  The precise point in supersymmetry
parameter space might not be determined uniquely from the data, or the
data might not select sufficiently precisely the special mechanism of 
neutralino annihilation.  We have seen examples of the first difficulty 
at LCC2 and of the second at LCC3 and LCC4.   In these cases, the LHC
would give us only a first tantalizing glimpse of the particle physics 
origin of dark matter, leaving many questions to be resolved by the ILC
experiments.

We have also noted a circumstance in which qualitative information from the 
LHC can be bootstrapped into quantitative information for astrophysics.
The particle physics cross section needed for the interpretation of 
gamma ray data must be close to the value \leqn{findsigmav} if the observed 
WIMP is the dominant component of dark matter and  if the 
relic density of WIMPs is set primarily by WIMP annihilation 
in the S-wave
without coannihilation.  The LHC data could point to a qualitative
scenario (one similar to LCC2, for example) in which S-wave annihilation
would be expected.  We would then have a quantitative estimate of the 
annihilation cross section that could be used to analyze astrophysical
gamma ray spectra.

\subsection{ILC at 500 GeV}

In all of the scenarios we have discussed,
 the estimates of the neutralino properties from the 
LHC would be dramatically improved when the neutralino and the other 
light particles in its sector are observed in $\ee$ annihilation.  

First 
of all, the measurements of $\ee$ annihilation cross sections and 
angular distributions will give the spins and Standard Model quantum
numbers of the lightest states in the new particle sector, allowing
definite identification of the model that is giving rise to the 
stable WIMP.  This identification is a prerequisite for any 
`model-independent' estimation of the WIMP cross sections.

Second, measurements in $\ee$ annihilation can  
improve the accuracy on the WIMP mass from 5--10\% at the LHC
to a fraction of a percent.  The value of $\Omega_\chi h^2$ is 
directly correlated with the WIMP mass in all of our models, so this 
is crucial information for obtaining an accurate prediction of the 
relic density.

Third, measurements in $\ee$ annihilation can 
identify all light partners of the WIMP with electromagnetic 
or weak charge, including some ({\it e.g.}, in supersymmetry, the 
charginos and the staus) that are difficult to study at hadron 
colliders.  The masses of these particles would also be measured to 
a fraction of a percent.

Fourth, measurements of production cross sections in $\ee$ annihilation
give direct sensitivity to the mixing angles that define the mass 
eigenstates of the new particles.  In supersymmetry, the cross sections
for chargino and neutralino pair production fix the gaugino-Higgsino
mixing angles, and the cross section for stau pair production fix
the stau mixing angles.  At a linear collider, the cross sections with
polarized beams can be measured, and the individual cross sections from 
left- and right-handed polarized electrons provide complementary
information. We have seen one special case (at LCC1) where the 
measurement of a ratio of branching ratios at the LHC can substitute
some of this information.  But the technique of extracting these 
angles from polarized cross section measurements is general and 
much more effective.

Finally, the information from spectra and cross sections obtained in 
$\ee$ annihilation allows one to resolve ambiguities that arise in 
imposing the constraints from the LHC data.  Here we refer both to 
the question of multiple solutions to the various eigenstate mixing 
problems and to the question of whether special situations such as
a Higgs boson resonance or coannihilation are present.
  In some cases, as we have discussed, these
predictions can be sharpened by adding constraints from the relic
density or from direct detection rates.

Looking back at the tables, we see that, in almost all cases, we do not 
obtain precise
determinations of the basic WIMP cross sections until we have data from 
$\ee$ experiments.  

In this paper, we have intentionally chosen models in which the lightest
states of the new particle spectrum can be explored at the ILC at a 
center of mass energy of 500 GeV.   This is our expectation, based on the
idea that models of WIMPs arise naturally from models of electroweak 
symmetry breaking.  We will learn very soon from the LHC whether this 
assumption is justified.  If the spectrum of new particles associated with
the WIMP is out of reach of the 500 GeV ILC, we will still need $\ee$
data to understand the dark matter problem.  We will just have to wait 
longer to obtain it.

\subsection{ILC at 1000 GeV}

Although in all of our models the ILC at 500 GeV gives
 a great improvement in the information available from colliders, we
always found an advantage in doing additional $\ee$ experiments at 
higher energy.  In all of the models except LCC2, this higher-energy
running was particularly important for one specific reason:  The WIMP 
cross sections are sensitive to the masses and couplings of the 
heavy Higgs bosons, and we needed accurate values of the Higgs boson
masses and the mixing angle $\tan\beta$ to determine these cross section
accurately.  The spin-independent direct detection cross section, 
in particular, is
typically dominated by $t$-channel Higgs boson exchange.  If the heavy
Higgs
bosons are not seen directly, there is no strong prediction for this
cross section;   once these Higgs bosons are seen, there is  suddenly 
a precise determination from the collider data.

In supersymmetric models, the heavy Higgs bosons are typically heavier
than the lightest superpartners.  In this range of parameters, they 
are pair-produced in the  process $\ee \to H^0 A^0$.  In our study,
it was important for $\ee$ experiments to reach the threshold for this 
process in order to complete the set of experiments needed to predict the
WIMP properties.

\subsection{Conclusions}

Finally, we return to the larger picture.  In this paper, we have shown
that the experimental results from the 
hadron and $\ee$ colliders of the next generation can be used to 
determine the basic particle physics cross sections of WIMP dark matter
particles.  Using this information, we will be able to test whether 
particles seen at high-energy accelerators make up cosmic dark matter.
If this is so, we can apply the cross sections determined from collider
data to astrophysical dark matter experiments and use them to study 
in a very general way the distribution of dark matter in the galaxy.

We expect that the interplay between measurements from high-energy 
colliders and measurements of dark matter in astrophysics will become
a major theme of both subjects in their evolution over the next ten to 
fifteen years.  It will lead us to learn much more about the structure 
of the galaxy and of the universe, and also about the underlying 
structure of the elementary particles and their laws.  To carry out the
full program will take persistence and it will require 
collaboration across the field of physics.  But if we can see the goal,
we can reach it.

\appendix

\section{Markov Chain Monte Carlo}

In this appendix we detail the Markov Chain Monte Carlo technique used to
explore the $D=24$ dimensional parameter space described in this paper.  We
then describe tests of the Markov chains' convergence.

\subsection{Adaptive Metropolis-Hastings algorithm}

The Metropolis-Hastings algorithm is simple.  Consider a point $\vec{p}_i$ in
the parameter space.  From this point, a new point $\vec{q}$ is proposed, with
probability density $P(\vec{q},\vec{p}_i)$.  Note that the simple algorithm
requires that that the density $P$ be symmetric in its arguments.  If the
likelihood of the proposed point is larger than that at the current point,
${\cal L}(\vec{q})\ge{\cal L}(\vec{p}_i)$, then set $\vec{p}_{i+1}=\vec{q}$.
If the likelihood at the proposed point is lower, set $\vec{p}_{i+1}=\vec{q}$
with probability ${\cal L}(\vec{q})/{\cal L}(\vec{p}_i)$, otherwise set
$\vec{p}_{i+1}=\vec{p}_i$.  The set of points $\{\vec{p}_i\}$ then converges to
the correct target distribution, independent of the proposal density $P$.  The
difficulty in implementing this algorithm lies entirely in choosing a proposal
$P$ which allows an efficient exploration of the target distribution.

We can use the covariance matrix to construct an efficient proposal.  From a
sample set of $N$ points $\vec{p}_i$, we construct the mean and covariance
matrix,
\begin{eqnarray}
\vec{\mu}&=&\frac{1}{N}\sum_{i=1}^N\vec{p}_i,\\
\mathbf{C}&=&\frac{1}{N}\sum_{i=1}^N\left(\vec{p}_i-\vec{\mu}\right)
\left(\vec{p}_i-\vec{\mu}\right)^T.
\end{eqnarray}
Note that for $\mathbf{C}$ to be a positive definite matrix, $N$ must be at
least $D+1$.  This matrix is essentially the variance; it can be thought of as
$\mbox{\boldmath$\sigma$}^2$.  If the sample set reasonably covers the region
of high likelihood, then the shape of the region, complete with degenerate
directions, etc., is encoded in $\mathbf{C}$.  Is is natural to use a Gaussian
proposal distribution based on the covariance matrix and scaled by an
efficiency factor $f$.  The proposal is $\vec{q}=\vec{p_i}+\vec{y}$, where
$\vec{y}$ is distributed as
\begin{equation}
P(\vec{y})=\frac{1}{\sqrt{(2\pi f^2)^D\det\mathbf{C}}}
\exp\left(-\frac{1}{2f^2}\,\vec{y}^T\mathbf{C}^{-1}\vec{y}\right).
\end{equation}
This can be implemented by choosing a vector $\vec{x}$ where each element has a
Gaussian distribution with zero mean and unit variance and taking
$\vec{y}=f\sqrt{\mathbf{C}}\,\vec{x}$.  By $\sqrt{\mathbf{C}}$ we mean any
matrix $\mathbf{L}$ such that $\mathbf{L}\mathbf{L}^T=\mathbf{C}$.  It is
convenient to take $\mathbf{L}$ to be lower triangular: this is the Cholesky
decomposition of $\mathbf{C}$.  It can be shown that for a Gaussian target
distribution, the most efficient step is $f=2.381/\sqrt{D}$ in the limit of
large $D$.  This prefactor is fairly optimal even for $D=1$~\cite{Dunkley}.
Alternatively, we can use a Cauchy-Lorentz proposal distribution.  This has the
advantage of occasionally allowing very long steps, though for a Gaussian
target distribution, it is less efficient.  The Cauchy-Lorentz distribution in
$D$ dimensions is given by
\begin{equation}
P(\vec{y})=\frac{(2/\!f)^D\,\Gamma\left(\frac{D+1}{2}\right)}
{\sqrt{\pi^{D+1}\det\mathbf{C}}}\left(1+\frac{4}{f^2}\,
\vec{y}^T\mathbf{C}^{-1}\vec{y}\right)^{-(D+1)/2}.
\end{equation}
This can be implemented analogously by taking $\vec{x}$ to have a
Cauchy-Lorentz distribution with a unit full-width at half maximum as follows,
\begin{equation}
x_i=\frac{t_i}{2\sqrt{i}}\,\left(1+4\sum_{j=1}^{i-1}x_j^2\right)^{1/2},\;\;
i\in\{1\dots D\},
\end{equation}
where $t_i$ has a $t$-distribution with $i$ degrees of freedom.  As before,
$\vec{y}=f\sqrt{\mathbf{C}}\,\vec{x}$.

The Metropolis-Hastings algorithm can be made adaptive, by updating the matrix
$\mathbf{C}$.  A key consequence of this is that the proposal distribution is
no longer symmetric, which must be accounted for.  We chose a method that uses
$N\ge D+1$ Markov chains in parallel.  The covariance matrix used in the
proposal is constructed from the current point of each chain.  At each step, a
chain is chosen at random, and a proposal is made for that chain.  If accepted,
the one chain is advanced, and now of course the covariance matrix changes,
$\mathbf{C}\rightarrow\mathbf{C}'$.  This means that the probability density to
return to the previous state is not the same as the probability density to
arrive at the current state from the previous state.  If this is not corrected,
detailed balance is violated.  As a shorthand, call the two probability
densities $P(\mathbf{C})$ and $P(\mathbf{C}')$.  Detailed balance is restored
with the following acceptance probability,
\begin{equation}
P({\rm accept})=\min\left[1,\frac{P(\mathbf{C}'){\cal L}(\vec{q})}
{P(\mathbf{C}){\cal L}(\vec{p}_i)}\right].
\end{equation}
The step $\vec{x}'$ required to return is different than the step $\vec{x}$,
but both are unit-Gaussian distributed.  For the Gaussian case, it is easy to
see that
\begin{eqnarray}
\frac{P(\mathbf{C}')}{P(\mathbf{C})}&=&
\sqrt{\frac{\det\mathbf{C}}{\det\mathbf{C}'}}\,
\exp\left[-\frac{1}{2}\left(x'^2-x^2\right)\right],\\
&=&\sqrt{\frac{\det\mathbf{C}}{\det\mathbf{C}'}}\,
\exp\left[-\frac{1}{2f^2}\vec{y}^T\left(\mathbf{C}'^{-1}-
\mathbf{C}^{-1}\right)\vec{y}\right].
\label{eq:MCMCcorrectionfactor}
\end{eqnarray}
We now need to know $\mathbf{C}'$ and its inverse and determinant.  Assume that
the point $\vec{p}_j$ is the one to be updated,
$\vec{p}_j\rightarrow\vec{p}_j+\vec{y}$, we find
\begin{eqnarray}
\vec{\mu}'&=&\vec{\mu}+\frac{1}{N}\vec{y},\\
\mathbf{C}'&=&\frac{1}{N}\sum_{i=1}^N
\left[\vec{p}_i-\vec{\mu}+\left(\delta_{ij}-\frac{1}{N}\right)\vec{y}\right]
\left[\vec{p}_i-\vec{\mu}+\left(\delta_{ij}-\frac{1}{N}\right)
\vec{y}\right]^T,\\
&=&\mathbf{C}+
\frac{1}{N}\,\left(\vec{p}_j-\vec{\mu}\right)\vec{y}^T+
\frac{1}{N}\,\vec{y}\left(\vec{p}_j-\vec{\mu}\right)^T+
\frac{N-1}{N^2}\,\vec{y}\,\vec{y}^T.
\end{eqnarray}
The covariance matrix has been adjusted by the addition of dyad products, and
only two of them (collecting terms in $\vec{y}^T$).  It is possible to invert
such a matrix analytically if the inverse of the base matrix is known.
\begin{eqnarray}
\mathbf{C}'&=&\mathbf{C}+\sum_i\vec{a}_i\vec{b}^T_i,\\
\lambda_{ij}&=&\vec{b}_i^T\mathbf{C}^{-1}\vec{a}_j,\\
\mathbf{C}'^{-1}&=&\mathbf{C}^{-1}-
\sum_{i,j}\left(\mathbf{1}+\mbox{\boldmath$\lambda$}\right)^{-1}_{ij}
\mathbf{C}^{-1}\vec{a}_i\vec{b}^T_j\mathbf{C}^{-1},\label{eq:dyadinvert}\\
\det\mathbf{C}'&=&\det\mathbf{C}\,\det\left(\mathbf{1}+
\mbox{\boldmath$\lambda$}\right)\label{eq:detCprime}.
\end{eqnarray}
There are only two dyads in our case, thus the matrix inversion in
\leqn{eq:dyadinvert} is trivial.  In this way, we do not need to perform a
$D\times D$ matrix inversion for each proposal.  If the proposal is accepted,
we do need to recompute the Cholesky decomposition as there is no such formula
to update it.  This does save time, as we typically want acceptance
probabilities around 25\%.  Notice here that both probability densities
$P(\mathbf{C})$ and $P(\mathbf{C}')$ have an implicit factor of $1/N$ insuring
that it is in fact chain $j$ that is being updated.  In evaluating
\leqn{eq:MCMCcorrectionfactor}, we need
$\mathbf{C}^{-1}=\mathbf{L}^{-1,T}\mathbf{L}^{-1}$.  Since $\mathbf{L}$ is
lower triangular, its inverse multiplied by a vector is trivially obtained by
back-substitution.  Then \leqn{eq:detCprime} is easily verified for the case of
adding a single dyad.  The result for an arbitrary number of dyads follows
inductively: if true for $n-1$ dyads, the case of $n$ dyads is shown to be the
Laplace expansion for the $n\times n$ determinant.

For a Cauchy-Lorentz proposal, the correction factor is quite similar.  Using
the results of the previous paragraph,
\begin{equation}
x'^2-x^2=\Delta x^2=
-\frac{1}{f^2}\left(\mathbf{1}+\mbox{\boldmath$\lambda$}\right)^{-1}_{ij}
\vec{y}^T\mathbf{C}^{-1}\vec{a}_i\vec{b}_j\mathbf{C}^{-1}\vec{y},
\end{equation}
we find
\begin{equation}
\frac{P(\mathbf{C}')}{P(\mathbf{C})}=
\frac{1}{\sqrt{\det\left(\mathbf{1}+\mbox{\boldmath$\lambda$}\right)}}\,
\left(1+\frac{4\Delta x^2}{1+4 x^2}\right)^{-(D+1)/2}.
\end{equation}

The efficiency of this adaptive proposal relies on the fact that the target
distribution is Gaussian.  If it is non-Gaussian, which is always the case for
the distributions studied in this paper, the efficiency may be different.  We
take a stepsize $f=2.381\,\epsilon/\sqrt{D}$, where $\epsilon=1$ is most
efficient in the Gaussian case.  For the various cases studied, we used
efficiencies $\epsilon\in[0.15,0.5]$ as required to have acceptance
probabilities that were not too small (above 5\%).

\subsection{Exploring the distributions}

For each of the 12 cases (four LCC benchmark points, three colliders), we have
run Markov Chains as follows.  Our fiducial runs have $N=25$ chains in
parallel, the minimal number that gives a positive definite covariance matrix.
The total number of samples taken is $4\times10^6$, so roughly $1.6\times
10^5$
per chain.
Starting at the beginning, some fraction of each chain, up to 10\%, is
used for ``burn-in'' to find the region of acceptable likelihood.  For the
first half of the burn-in period, we apply a ``cooling'' technique where the
``temperature'' is gradually lowered to the correct value.  This means that we
take the likelihood function to be ${\cal L}^\lambda$.  We update $\lambda=1/T$
geometrically by $\lambda_i=\lambda_0^{1-i/n}$, where $n$ is the total number
of burn-in points, and we take $\lambda_0=0.01$.  Note that the current
temperature is used for {\em both} likelihoods in the acceptance probability.

\subsection{Thinning the chains}

\begin{figure}
\begin{center}
\epsfig{file=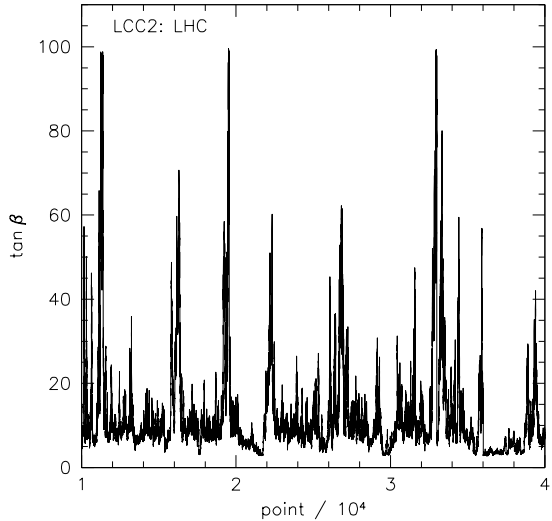,height=3.0in}
\caption{Section of the Markov chain for $\tan\beta$ at LCC2-LHC.  The repeated
brief excursions to large values are clearly seen.}
\label{fig:tanbetaseries}
\end{center}
\end{figure}

We do not compute relic density at every point due to the computational
expense.  The chains are correlated at short distances so this is not even
necessary to achieve good statistics.  We instead thin the chains by some
factor, usually taken to be $t=50$.  This means we construct a new chain from
every 50th point of the current chain.  We only compute relic density for the
points of the thinned chains.  Fig.~\ref{fig:tanbetaseries} shows a 
section of a thinned Markov chain.

\subsection{Convergence test}

\begin{figure}
\begin{center}
\epsfig{file=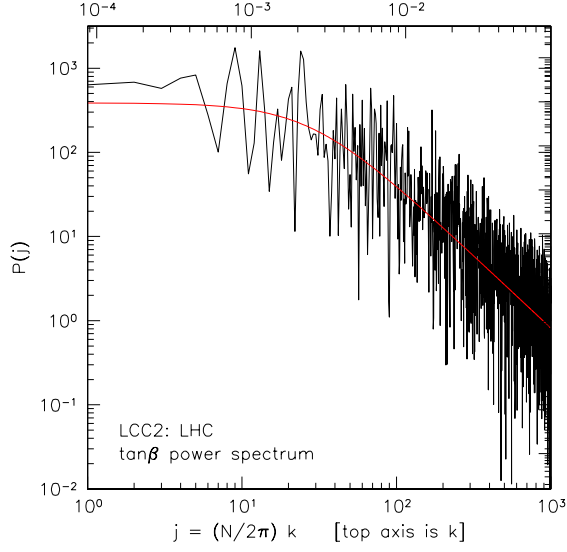,height=3.0in}
\caption{Fourier transform of the Markov chain for $\tan\beta$ at LCC2-LHC.  A
constant rolling into a power law fit~\cite{Dunkley} is shown.  This
illustrates the white noise (flat) spectrum for low $k$ indicating
decorrelation at large distances in the chain and the random walk ($k^{-2}$)
spectrum at short distances (large $k$).}
\label{fig:tanbetaFFT}
\end{center}
\end{figure}
\begin{figure}
\begin{center}
\epsfig{file=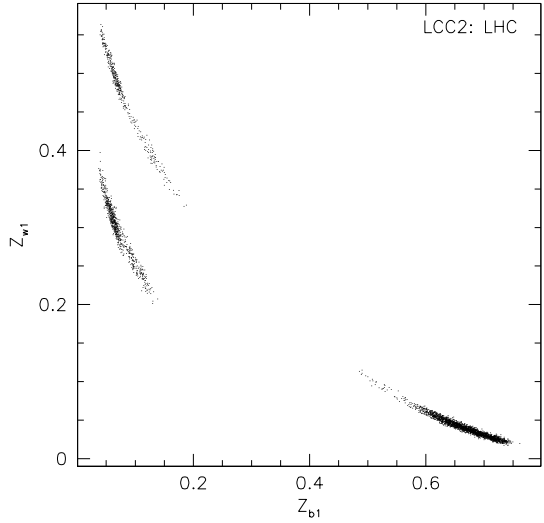,height=3.0in}
\caption{Illustration of MCMC bridging the gap between islands in parameter
space.  These are solutions for LCC2 with LHC data, as in
Fig.~\ref{fig:LCC2relicscatter}.  The true solution has the largest bino
fraction $Z_{b1}$.}
\label{fig:LCC2islands}
\end{center}
\end{figure}

\begin{figure}
\begin{center}
\epsfig{file=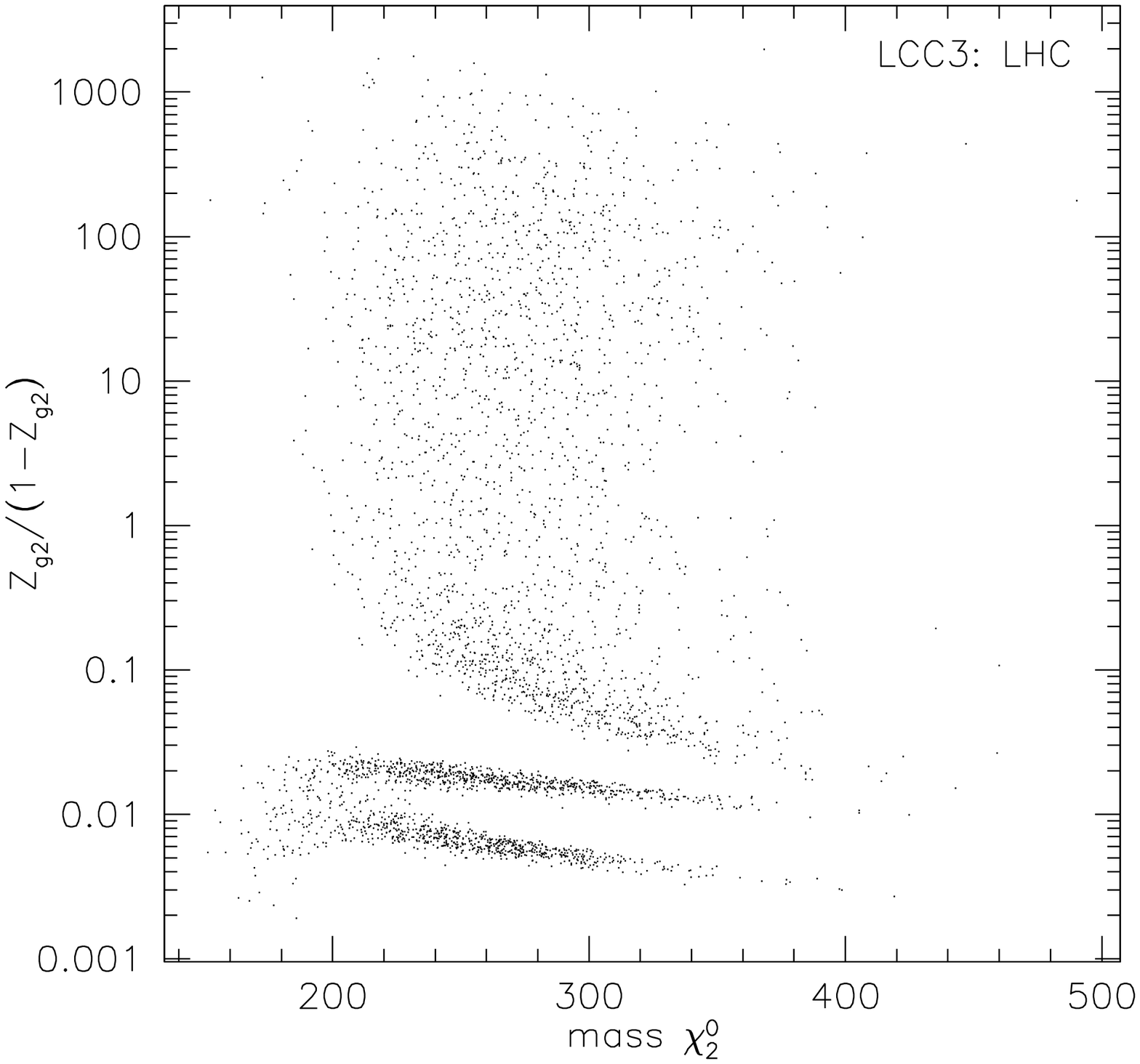,height=3.0in}
\caption{Illustration of MCMC exploring very different regions of parameter
space.  These regions are actually connected, as shown in
Fig.~\ref{fig:LCC3relicscatter}, but allow vastly different properties for
$\chi^0_2$, as shown here.}
\label{fig:LCC3bridge}
\end{center}
\end{figure}

We apply the technique of \cite{Dunkley} to determine whether the Markov
Chains have converged.  In a given run, the $N$ chains are concatenated (after
removing the burn-in periods) to construct one long chain of length $N_{\rm
tot}$.  For each of the 24 parameters, we normalize to unit variance and take
the Fourier transform.  We then fit the power spectrum $P(k)=|a_j|^2$, where
the $a_j$ are the Fourier components (with $j$ ranging between $-N_{\rm tot}/2$
and $N_{\rm tot}/2$) and $k=2\pi j/N_{\rm tot}$.  If the elements of the 
chain are 
uncorrelated at large separation, then 
  for small $k$ (long distance in the chain), we expect to see
a white noise spectrum, that is, a power spectrum that is approximately
constant.   At short distances in the
chain (large $k$) we expect the power spectrum to behave like $k^{-2}$, the
random walk spectrum.  Following \cite{Dunkley}, we require that the break
between these two behaviors occurs beyond $j=20$.  When this is true, the fit
to $P(0)$ can be trusted, and $r=P(0)/N_{\rm tot}$ indicates the fractional
error made in the variance of the parameter.  We require $r<0.01$, meaning at
most 10\% errors made in the {\em error} of a given parameter, based on the
Markov chain.

The power spectrum test can be applied either to the full chain or the thinned
chain.  The results are insensitive as long as the scales removed are in the
random walk regime.  For chains of length $N_{\rm tot}$, thinning by a factor
of $t$ has little effect unless the turnover is beyond $j=N_{\rm tot}/t$.  We
typically find $j$ values at turnover of $20-200$, so this is not an issue.
An illustration of the power spectrum test is given in 
Fig.~\ref{fig:tanbetaFFT}.

\section{Disconnected families of supersymmetry parameters}

In this appendix we describe the conditions where multiple discrete regions of
parameter space would be consistent with collider measurements.  Such
ambiguities occur for all of the benchmark points under study.  The simplest
alternate solutions involve only changing the signs of $\mu$ and $m_1$.  This
four-fold degeneracy is evident in every model given only LHC data. Since these
signs are relevant to the mixing parameters of the neutralinos and charginos,
it is essential to measure the mixed pair production cross sections at the
ILC,
{\it e.g.},
 $e^+e^-\to \s\chi^0_1 \s\chi^0_2$.  There are more subtle degeneracies as
well, as we discuss in the following sections.

\subsection{Benchmark point LCC1}
At the LHC, three neutralinos are visible: the lightest (bino) at 95.5 GeV, the
next lightest (wino) at 181.6 GeV, and the heaviest (heavier Higgsino) at 375.6
GeV.  No chargino is seen, implying a mass $> 125$ GeV.  Assuming that the
identities of the neutralinos are unknown at the LHC, it is possible that the
values of the parameters $|\mu|$ and $m_2$ are exchanged.  This implies that
the second and third neutralinos are Higgsinos, and the heaviest is mostly
wino.  This solution gives an acceptable mass spectrum for the LHC errors.  In
addition, there is the four-fold degeneracy due to the uncertain signs of $\mu$
and $m_1$.  There are thus 8 solutions for LHC data.  Notice that we can be
confident that the lightest neutralino is mostly bino-like; its mass is 95 GeV,
and a chargino at this mass would be visible.

The ILC-500 completely resolves the degeneracies of this model.  The identities
of $\s\chi^0_2$ and $\s\chi^+_1$ as wino-like are determined by the pair
production cross sections, and the sign ambiguities are completely resolved by
the $\s\chi^0_1\s\chi^0_2$ production cross section: changing the signs has a
large effect on the (small) bino fraction of $\s\chi^0_2$.

At this point, the 8 islands are very well separated in parameter space.  For
the LHC, we have shown the results of the correct island only.  Including the
other solutions would not make a large difference, as they only affect the
subdominant wino and Higgsino admixtures in the lightest neutralino.

\subsection{Benchmark point LCC2}

The three lightest neutralinos are visible at the LHC, while there is only the
usual limit of 125 GeV on the lightest chargino mass.  All of the neutralinos
are quite mixed in composition, thus the permutation ambiguities in
$m_1,\,m_2,\,\mu$ are somewhat ambiguous.  As shown in
Fig.~\ref{fig:LCC2islands}, there are essentially three solutions for positive
$\mu$ and $m_1$.  When plotted as islands in neutralino composition, the
solutions are clearly separated.  The incorrect solutions have the lightest
neutralino being mostly wino-like or Higgsino-like rather than mostly
bino-like.  In fact, the wino and Higgsino cases have two peaks, not well
separated, as seen in Fig.~\ref{fig:LCC2relicscatter}.  There is an additional
bino solution as well (not shown), with $M_2$ larger than 1 TeV, but its
properties are similar to the proper bino solution, except that the wino
fraction is very much smaller.

The ILC-500 removes the degeneracies.  The $\s\chi^0_2\s\chi^0_3$ cross section
identifies the neutralinos and also their mixing angles, thus the negative
$\mu$ solution at least is gone.  The ILC-1000 allows no incorrect solution.

At this point only, the disjoint islands at positive $\mu$ and $m_1$ are close
together in parameter space.  Our LHC results show all islands with the correct
signs.  During the cooling and burn-in period, Markov chains find an island at
random, and remain there for the remainder of the run.  We find that burn-in is
slow enough that the number of chains that finds a solution is proportional to
the likelihood of the solution.  We have verified this with a ``stepping
stone'' technique, where the space between the islands, normally having
vanishing likelihood, is assigned a small likelihood.  Markov chains can now
step away from an island, into this ``sea'', and thus travel back and forth
between islands.  Points in the ``sea'' can be removed at the end with
importance sampling.  We have verified the relative weights of the islands with
just such a simulation.

\subsection{Benchmark points LCC3 and LCC4}

Benchmark points LCC3 and LCC4 have similar structure for discrete solutions,
thus we discuss them together.  Both have the two lightest neutralinos visible
at the LHC.  Assuming that none have been missed between the two, the ``F''
structure in the $m_1$ vs.\ ($m_2$ or $\mu$) plot appears naturally.  This is
understood physically as follows.  Given the large mass splitting between the
two visible neutralinos, and the assumption that none have been missed, the
lightest neutralino is either bino or wino, and the heavier of the two visible
neutralinos can be anything.  As all neutralinos at these benchmark points are
heavier than 125 GeV, the non-observation of charginos gives no additional
information.  The ``F'' structure is repeated for each of the four sign
combinations for $m_1$ and $\mu$.  In Fig.~\ref{fig:LCC3bridge} we illustrate
an interesting effect of the ``F'' structure, in that there are three possible
solutions to the composition of the second lightest neutralino.

For each of these points, ILC-500 data would collapse the ``F'' structure
somewhat, but the sign ambiguities remain.  ILC-1000 is required to remove the
last ambiguities, with the neutralino and chargino pair production cross
sections.

Unfortunately, the sign ambiguity in $\mu$ remains for point LCC4, even taking
ILC-1000 data into account.  However, since
 this point has large $\tan\beta$, its
correction to the muon $g-2$ is large, and has a sign equal to the sign of
$\mu$.  The BNL measurement \cite{muongminus2} rules out the negative $\mu$
solution, even with a very conservative constraint allowing the union of the
3$\sigma$ regions from the $e^+e^-$ and $\tau$ decay evaluations of the
hadronic vacuum polarization \cite{DEHZ1,DEHZ2}.

For these two points, we have only illustrated the correct signs.  The ``F''
patterns are too far apart in parameter space for easy exploration by single
Markov chains.

\Acknowledgements
Our perspective on dark matter has been shaped by discussions with many 
people over the past few years.  We are particularly grateful to  Genevieve
Belanger,  Andreas 
Birkedal, Jonathan Feng, Paolo Gondolo, Konstantin Matchev, 
and Mark Trodden for sharing their insights. We thank Mihoko Nojiri
and Giacomo Polesello for an helpful correspondence.
EAB thanks Richard Schnee, Blas Cabrera and Dan Akerib for many useful
discussions on the capabilities of direct detection experiments, and Larry Wai
and Tune Kamae for useful discussions relating to gamma ray detection.
 MEP thanks Jim Alexander,  Richard Gray, Bhaskar Dutta, and Teruki Kamon 
for discussions of their ILC
simulation results.  He is also grateful to  Ee Hou Yong and
Wu-Yen Chuang for educating him about neutralino dark matter.
 The work of MB 
was supported by the US Department of Energy under Contract No. 
DE--AC02-05CH11231 and used resources of the National Energy Research
Scientific Computing Center, supported by Contract No. DE-AC03-76SF0098.
The work of EAB, MEP, and TW was supported by the US Department of Energy 
under Contract No.
DE--AC02--76SF00515.


\begin{thebibliography}{99}
\bibitem{DAMAOK} 
  Specific particle physics scenarios that partially resolve 
    the contradiction between
     DAMA and the other experiments listed are reviewed in \cite{Gaitskell}.
\bibitem{complete}
A complete expression for the WIMP-nucleon cross section in 
 SUSY models is given in \cite{JKG}.
\bibitem{moresCDMS}
  The parameters of the SuperCDMS experiment will be discussed
            in Section 4.5.
\bibitem{logdivergence}
  The divergence is logarithmic and thus not very sensitive to the 
      value of a cutoff.  In Fig.~\ref{fig:profiles}, we put the cutoff at the
       Schwartzschild radius of the black hole at the center of the 
          galaxy.
\bibitem{UEDdarkmatter}
Recently, the relic density calculation
in the Universal Extra Dimension model of WIMP dark 
matter~\cite{UEDDM,moreUEDDM} has been analyzed over a comparably large
parameter space.  See \cite{exactUED,exactUEDtwo}.
\bibitem{ourchange}
  To be precise, \cite{Dutta} studied a set of closely related points
        with $m(\s\tau_1) - m(\chi^0_1)$ = 4.7, 9.5, 19.0 GeV, 
         defined with a presciption slightly different from ours for the 
        supersymmetry spectrum calculation.  Our LCC3,
  which has  $m(\s\tau_1) - m(\chi^0_1)$ = 10.5 GeV, is sufficiently close
   to the point 2 defined in \cite{Dutta} to make use of the simulations
      done in that study.
\bibitem{morebenchi}
Additional benchmark points for supersymmetry studies 
are defined in the papers~\cite{BattBench}.  Certain of these points are
now under study with detailed simulations by the CMS Collaboration
(A. de Roeck, personal communication).
\bibitem{ourmod}
  The MCMC runs for LCC3 displayed in Section 6 did not include this
      constraint on the $\tau\tau$ edge position, as 
        Table~\ref{tab:LCC3masses} indicates.   After the first 
        paper in \cite{Bhaskar} appeared, we redid the LHC runs 
        adding this constraint, but found negligible difference
        in the results.
\end{thebibliography}
\end{document}